\documentclass[prd,aps,twocolumn,preprintnumbers, showpacs, nofootinbib,superscriptaddress,notitlepage]{revtex4-2}
\usepackage{mathrsfs}
\usepackage{amsfonts}
\usepackage{amsmath}
\usepackage{slashed}
\usepackage{array}
\usepackage{verbatim}
\usepackage{epsfig}
\usepackage{graphicx}
\usepackage{color}
\usepackage{hyperref}
\usepackage[dvipsnames]{xcolor}
\usepackage{subfigure}
\usepackage{float}

\newcommand{\beq}{\begin{eqnarray}}
\newcommand{\eeq}{\end{eqnarray}}

\def\be{\begin{equation}}
\def\ee{\end{equation}}
\def\bea{\begin{eqnarray}}
\def\eea{\end{eqnarray}}

\begin{document}

\title{Self-Renormalization of Quasi-Light-Front Correlators on the Lattice}

\collaboration{\bf{Lattice Parton Collaboration ($\rm {\bf LPC}$)}}

\author{Yi-Kai Huo}
\email{yh3285@columbia.edu}
\affiliation{Zhiyuan College, Shanghai Jiao Tong University, Shanghai 200240, China}
\affiliation{Physics Department, Columbia University, New York, NY 10027}

\author{Yushan Su}
\email{ysu12345@umd.edu}
\affiliation{Department of Physics, University of Maryland, College Park, Maryland 20742, USA} 

\author{Long-Cheng Gui}
\affiliation{Department of Physics, Hunan Normal University, Changsha, 410081, China}

\author{Xiangdong Ji}
\affiliation{Center for Nuclear Femtography, SURA, 1201 New York Ave. NW, Washington, DC 20005, USA}
\affiliation{Maryland Center for Fundamental Physics, Department of Physics, University of Maryland, College Park, Maryland 20742, USA}

\author{Yuan-Yuan Li}
\affiliation{Nanjing Normal University, Nanjing, Jiangsu, 210023, China}

\author{Yizhuang Liu}
\affiliation{Institut f\"{u}r Theoretische Physik, Universit\"{a}t Regensburg, D-93040 Regensburg, Germany}

\author{Andreas Sch\"afer}
\affiliation{Institut f\"{u}r Theoretische Physik, Universit\"{a}t Regensburg, D-93040 Regensburg, Germany}

\author{Maximilian Schlemmer}
\affiliation{Institut f\"{u}r Theoretische Physik, Universit\"{a}t Regensburg, D-93040 Regensburg, Germany}

\author{Peng Sun}
\affiliation{Nanjing Normal University, Nanjing, Jiangsu, 210023, China}

\author{Wei Wang}
\affiliation{INPAC,  Key Laboratory for Particle Astrophysics and Cosmology (MOE), Shanghai Key Laboratory for Particle Physics and Cosmology, School of Physics and Astronomy, Shanghai Jiao Tong University, Shanghai 200240, China}

\author{Yi-Bo Yang}
\affiliation{CAS Key Laboratory of Theoretical Physics, Institute of Theoretical Physics, Chinese Academy of Sciences, Beijing 100190, China}
\affiliation{School of Fundamental Physics and Mathematical Sciences, Hangzhou Institute for Advanced Study, UCAS, Hangzhou 310024, China}
\affiliation{International Centre for Theoretical Physics Asia-Pacific, Beijing/Hangzhou, China}

\author{Jian-Hui Zhang}
\affiliation{Center of Advanced Quantum Studies, Department of Physics, Beijing Normal University, Beijing 100875, China}

\author{Kuan Zhang}
\affiliation{University of Chinese Academy of Sciences, School of Physical Sciences, Beijing 100049, China}
\affiliation{CAS Key Laboratory of Theoretical Physics, Institute of Theoretical Physics, Chinese Academy of Sciences, Beijing 100190, China}

\date{\today}

\begin{abstract}
In applying large-momentum effective theory, renormalization of 
the Euclidean correlators in lattice regularization is a challenge due to 
linear divergences in the self-energy of Wilson lines.
Based on lattice QCD matrix elements of the quasi-PDF operator at lattice spacing $a$= 0.03\,fm $\sim$ 0.12\,fm with clover and overlap valence quarks on staggered and domain-wall sea, we design a strategy to disentangle the divergent renormalization factors from finite physics matrix elements, which can be matched to a continuum scheme at short distance such as dimensional regularization and minimal subtraction. Our results
indicate that the renormalization factors are universal in the hadron state matrix elements. Moreover, the physical matrix elements appear independent of the valence fermion formulations.
These conclusions remain valid even with HYP smearing which reduces the statistical errors albeit reducing control of the renormalization procedure. 
Moreover, we find a large non-perturbative effect in the popular RI/MOM and 
ratio renormalization scheme, suggesting favor of the hybrid renormalization procedure proposed recently. 

\end{abstract}

\maketitle

\section{Introduction}\label{sec:intro}
Parton distribution functions (PDFs) provide an effective description of quarks and gluons inside a light-travelling nucleon~\cite{Ellis:1991qj,Thomas:2001kw}.
They play an essential role in calculating hardronic cross sections involving the nucleons~\cite{Gao:2017yyd}, and their uncertainties from phenomenological extractions have been one of the major
sources of errors in the theoretical predictions at Large Hadron Collider (LHC) and other hadron facilities. Thus, a precise knowledge of PDFs is very important both for the accurate tests of the Standard Model (SM) and for the search of new physics beyond the SM. On the other hand, the densities of partons in the nucleon provide direct information on its intrinsic properties such as the origin of the nucleon spin and mass, as well as the role of sea quarks for various physical quantities~\cite{Accardi:2012qut}. 
  
However, directly calculating parton physics from first principles of quantum chromodynamics (QCD) has been a difficult task. A review on various efforts of doing so can be found in Ref.~\cite{Cichy:2018mum}. The proposal of the large-momentum effective field theory (LaMET) has been an important step toward meeting the challenge ~\cite{Ji:2013dva,Ji:2014hxa,Ji:2020ect}. So far, LaMET has been widely used in calculating quark isovector distribution functions~\cite{Lin:2014zya,Alexandrou:2015rja,Chen:2016utp,Alexandrou:2016jqi,Alexandrou:2018pbm,Chen:2018xof,Lin:2018pvv,Liu:2018uuj,Alexandrou:2018eet,Liu:2018hxv,Chen:2018fwa,Izubuchi:2019lyk,Shugert:2020tgq,Chai:2020nxw,Lin:2020ssv,Fan:2020nzz}, generalized parton
distributions~\cite{Chen:2019lcm,Alexandrou:2019dax}, distribution amplitudes (DAs)~\cite{Zhang:2017bzy,Chen:2017gck,Zhang:2020gaj}, and transverse-momentum-dependent distributions~\cite{Shanahan:2019zcq,Shanahan:2020zxr,Zhang:2020dbb}. 

LaMET suggests to calculate time-independent physical distributions in a finite momentum nucleon, which
are Euclidean observables. Such finite-momentum quantities can then be matched to light-front (LF)
parton properties using effective field theory techniques~\cite{Ji:2020byp}. For example, for the collinear quark distributions, it has been suggested to first compute a Euclidean space correlation (or quasi-light-front (quasi-LF) correlation) on the lattice, 
\bea\label{eq:quasi}
\tilde{h}(z, P_z) = \langle P| O_{\gamma_t}(z)|P \rangle,
\eea
where $|P\rangle$ is the nucleon state with a large momentum $P$ and the non-local operator is
\bea\label{eq:operator}
O_{\Gamma}(z)=\bar{\psi}(0) \Gamma U(0,z) \psi (z),
\eea
where $\psi$, $\bar{\psi}$ denote the bare quark field, $\Gamma$ is a Dirac structure,  and $U(0, z) = \exp(-ig\int_0^{z} dz' A_z(z'))$ is the Wilson link along the direction $z$, where $A^\mu$ is the gluon gauge potential. After renormalizing $\tilde{h}$ properly and matching it to some continuum scheme such as
dimensional regularization and (modified) minimal subtraction ($\overline{\rm MS}$), one obtains the so-called quark quasi-PDF via a Fourier transform, from which the quark PDF can be extracted through perturbative QCD matching. 

Renormalizing the quasi-LF correlation $\tilde{h}$ under lattice regularization is a non-trivial task. To see this, let us take, as a simple example, the matrix element of $O_\Gamma(z)$ in an off-shell quark state $|q\rangle$ with large Euclidean momentum $p^2$ (actually an off-shell truncated Green's function). It has the following form at 1-loop level~\cite{Alexandrou:2017huk},
\bea\label{eq:1-loop}
\langle O_{\Gamma}(z)\rangle =\Gamma\left(1+\gamma g^2 \textrm{log}(z^2/a^2)-m_{-1}\frac{z}{a}+……\right),\nonumber\\
\eea
where $g$ is the bare gauge coupling and $a$ is the lattice spacing. Unlike the coefficient of the logarithm $\gamma$, the coefficient of the linear divergence $m_{-1}$ can be sensitive to the details of the fermion and gauge actions on the lattice as those actions themselves define the lattice regularization. The linear-divergence term is proportional to the Wilson link length $z/a$ in lattice units, and it can be exponentially large at large $z$ or small $a$, when higher-order corrections are included. Thus, the linear divergence effect has to be removed before the lattice calculated quasi-PDF can be extrapolated to the continuum limit $a\rightarrow0$ and eventually matched to the continuum-scheme PDF.

Theoretical studies so far have shown that the quasi-PDF operator is multiplicatively renormalizable~\cite{Ji:2015jwa, Ji:2017oey,Ishikawa:2017faj,Green:2017xeu} in a continuum theory. On the lattice, due to non-commutativity of the limit $z\to 0$ and $a\to 0$, it has been suggested recently~\cite{Ji:2020brr} to use
a hybrid scheme to renormalize
the large and short distance correlations separately, which has the advantage of avoiding certain discretization effects and undesired infrared effects introduced in the renormalization stage. At small $z$ where
a short distance expansion is valid, one can use various ratio schemes, such as dividing by an off-shell quark matrix element of the quasi-PDF operator in the regularization-independent momentum subtraction (RI/MOM) scheme~\cite{Green:2017xeu,Chen:2017mzz,Alexandrou:2017huk} or by a hadron matrix element in the rest frame $\tilde{h}(z, P_z=0)$~\cite{Orginos:2017kos,Izubuchi:2018srq}. At large distance, one can
directly remove the Wilson line self-energy effect by subtracting the
power divergence~\cite{Chen:2016fxx,Ji:2017oey,Ishikawa:2017faj,Green:2017xeu}. Apart from using the RI/MOM factor or the rest-frame 
hadron matrix element, other methods have been suggested to extract the 
linear divergence, including Wilson loop~\cite{Chen:2016fxx,Zhang:2017bzy}, 
vacuum expectation values $\langle O_{\Gamma}(z)\rangle $~\cite{Braun:2018brg}, and gauge fixed Wilson link~\cite{Ji:2020ect}, and so on~\cite{Ji:2020brr}. 

However, numerically subtracting the linear divergence is an extremely delicate exercise. First, linear divergence could be sensitive to computational systematics in lattice calculations. In data, there may be slight differences between two matrix elements with the same linear divergence. These differences may lead to a failure of the suggested methods, especially for small lattice spacings. Second, some of the renormalization factors have other problems. For example, the leading contribution to the vacuum expectation value of $\langle O_{\Gamma}(z)\rangle $ at short distance vanishes and therefore, it is numerically very challenging to obtain the relevant linear divergence. Finally, as we shall see, linearly-divergent chiral symmetry breaking effects for Wilson fermions may render the linear divergence non-universal~\cite{Zhang:2020rsx}. 

To understand better the effects of the linear divergence in LaMET applications, we study systematically the linear divergences of matrix elements for sets of lattice data, generated with different lattice actions. We propose a self-renormalization method to eliminate all divergences and discretization errors when data for several different lattice spacings are available. The idea of this method is to extract the renormalization factor and the residual contribution directly from the matrix element we want to renormalize, without using an additional matrix element for renormalization. We then match the empirically renormalized 
matrix elements to those in the continuum. Our method is largely model-independent, and each term of our fitting functions is motivated by physics considerations. Tests show that our method works well for all data sets considered, which include matrix elements calculated with the lattice spacings from $0.03$ fm to $0.12$ fm using the valence clover and overlap actions on the MILC~\cite{Bazavov:2012xda} $N_f=2+1+1$  and RBC~\cite{Blum:2014tka} $N_f=2+1$ configurations. We also find that the method works for matrix elements after applying smearing which is needed to improve the statistical precision.

The rest of the paper is organized as follows. We present the theory of perturbative renormalization in Sec.~\ref{sec:theory}. The definition of the matrix elements we calculate and the simulation setup can be found in Sec.~\ref{sec:setup}. In Sec.~\ref{sec:test} the linear divergence is analysed for the different ensembles. Sec.~\ref{sec:StrtoRe} presents our strategy to self-renormalize a matrix element and collects the results for the $O_{\gamma_t}(z)$ matrix elements in different states and calculated with different valence fermion actions. Sec.~\ref{sec:highorder} discusses other fitting options, which mainly differ in the treatment of higher-order terms. In Sec.~\ref{sec:test},~\ref{sec:StrtoRe}, and~\ref{sec:highorder}, we only analyze matrix elements without HYP smearing. In Sec.~\ref{sec:smearing}, we extend our method to HYP smeared cases. In Sec.~\ref{sec:otherresult}, we test the linear divergence in some vacuum state matrix elements, including Wilson loop, quasi-PDF operator, and gauge fixed Wilson link, for the HYP smeared cases. In Sec.~\ref{sec:summary}, we summarize the results and discuss issues to be addressed in further studies.

\section{Renormalization in perturbation theory}\label{sec:theory}

According to the standard renormalization in local quantum field theories~\cite{Collins:1984xc}, the renormalization of the matrix elements of an operator does not depend on the external states but is only related to 
the short-distance property of the operator itself. Therefore, one 
can study the renormalization property of the operator in perturbative Green's functions, 
or off-shell quark and gluon matrix elements. For LaMET applications to PDFs, 
we are interested in the non-local operator $O_{\Gamma}(z)=\bar{\psi}(0) \Gamma U(0,z) \psi (z)$ in Eq.~(\ref{eq:operator}). There
are two types of divergences: The linear divergence associated with the Wilson link (its self-energy) and the logarithmic divergences associated with the renormalization of the vertices involving 
the Wilson line and light quark. 
The renormalization is multiplicative, and only the linear divergence has a (linear) $z$ dependence. Therefore the renormalized operator is~\cite{Ji:2017oey,Ishikawa:2017faj,Green:2017xeu, Ji:2020brr}
\bea\label{eq:Reoperator}
O_{\Gamma}(z)_{R}=Z_{O}^{-1} e^{\delta \bar{m} z} O_\Gamma(z),
\eea
where $e^{\delta \bar{m} z}$ contains the linear divergence and 
$Z_{O}$ the logarithmic ones. $\delta \bar{m}$ is not uniquely defined apart
from the linear divergence, introducing a subtraction
scheme dependence which affects the $z$-dependence of the renormalized operator~\cite{Ji:2020brr}.

The linearly-divergent mass renormalization can be calculated in perturbation theory. At one-loop order, the result is independent of lattice action~\cite{Chen:2016fxx}, 
\bea\label{eq:delta_m}
\delta \bar{m} = \frac{2 \pi}{3 a}(\alpha_s+{\cal O}(\alpha_s^2)+...) \ .
\eea
The energy scale of $\alpha_{s}$ can be chosen as $1/a$ to match the lattice results,
\bea\label{eq:alpha_s}
\alpha_{s}(1/a, \Lambda_{\rm QCD}) = \frac{2 \pi}{b_{0} \ln[1/(a \Lambda_{\rm QCD})]} \ ,
\eea
where $b_{0}=11-\frac{2}{3}n_f$ (we will take $n_f$=3) is the QCD $\beta$ function at one-loop order with $n_f$ species of fermions. $\Lambda_{\rm QCD}$ is the non-perturbative QCD scale. 
Including higher-orders in
the $\beta$ function will lead to a more complicated expression. Different choices
of energy scale amount to including high-order corrections.
$\Lambda_{\rm QCD}$ and higher-order $\alpha_s$ terms in $\delta \bar{m}$ also depend on the lattice action. 

The perturbation theory does not converge due to infrared renormalons, see from example~\cite{Ji:1995tm,Beneke:1998ui,Bauer:2011ws,Bali:2013pla}. The existence of the renormalons signals a non-perturbative term in $\delta \bar m$ which is independent of $a$, 
\begin{equation}\label{eq:m0}
      \delta\bar m = m_{-1}(a)/a - m_0 \ , 
\end{equation}
where the minus sign is just a convention.
The uncertainty in summing the perturbation series to get $m_{-1}(a)$
is compensated by the same uncertainty in the non-perturbative $m_0$, leaving the total independent of the renormalon uncertainty. 

Additional uncertainty in $m_0$ comes from the subtraction scheme, or equivalently from the matching to the continuum scheme.  
To reduce the subtraction scheme dependence, we require the renormalized lattice correlation function to be consistent with the $\overline{\rm MS}$ result from continuum perturbation theory at short-distances $z$. To be more concrete, we determine $m_0$ by matching the lattice result to the $\overline{\rm MS}$ one 
within a window $a \ll z < z_S$, where $z_S < 1/\mu$ is the point beyond which perturbation theory ceases to work and $\mu$ is a perturbative scale. The condition $a \ll z$ ensures that the discretization effect on lattice results is small. For this special choice $m_{0c}$, the LaMET expansion does not have a linear term in $1/P^z$~\cite{Ji:2020brr}.

At one-loop order and in dimensional regularization, the renormalization factor of the logarithimic divergence is~\cite{Ji:2020ect,Constantinou:2017sej},
\bea\label{eq:ZO}
Z_{O} = 1+\frac{3 C_{F} \alpha_{s}}{2 \pi} \frac{1}{4-d},
\eea
where $C_{F}=4/3$ is the Casimir operator for the fundamental representation of $SU(3)$ and $d$ is the space-time dimension.
One can resum the logarithimic divergence through solving the renormalization group equation,
\bea\label{eq:ZO_RGE}
\frac{d Z_{O}(\epsilon, \mu)}{d \ln (\mu)}=\gamma Z_{O}(\epsilon, \mu),
\eea
where $\epsilon=(4-d)/2$ and $\gamma=-\frac{3 C_{F}}{2 \pi} \alpha_{s}(\mu, \Lambda_{\rm QCD})$ 
is the leading anomalous dimension, which is independent of the regularization method. 
The form of the leading-order solution of Eq.~(\ref{eq:ZO_RGE}) is independent of regularization scheme and, on the lattice, is
\bea\label{eq:ZO_RS}
Z_{O}(1/a, \mu)=\left(\frac{\ln[1 /(a \Lambda_{\rm QCD})]}{\ln[\mu / \Lambda_{\rm QCD}]}\right)^{\frac{3 C_{F}}{b_0}},
\eea
where the bare ultraviolet (UV) cut-off is $1/a$, and the renormalization scale is $\mu$. 

If one considers the contributions from sub-leading logarithms to the anomalous dimension $\gamma$ when solving the renormalization group equation Eq.~(\ref{eq:ZO_RGE}), we obtain~\cite{Ji:1991pr}
\bea\label{eq:ZO_RS_higher}
Z_{O}(1/a, \mu)=\left(\frac{\ln [1 /(a \Lambda_{\rm QCD})]}{\ln [\mu / \Lambda_{\rm QCD}]}\right)^{\frac{3 C_{F}}{b_0}}\left(1+\frac{d}{\ln[a \Lambda_{\rm QCD}]}\right),\nonumber\\
\eea
where $\Lambda_{\rm QCD}$ and the constant $d$ depend on the specific lattice action.

\section{Lattice matrix elements and simulation setup}\label{sec:setup}

The standard non-perturbative renormalization of lattice matrix elements is through 
calculating some auxiliary matrix elements of the operator (which can also be computed in perturbation theory) and using them as the 
renormalization factor to approach the continuum limit. 
In this paper, we focus on the study of the auxiliary matrix elements 
potentially useful for the renormalization of quasi-LF correlations. We
mainly consider the matrix elements of the quasi-PDF operators in the 
following cases: 1) in a large Euclidean momentum quark state, 2) in the
physical zero-momentum state of a hadron such as the pion or nucleon.
We will also study the matrix element in the vacuum~\cite{Braun:2018brg} (case 3), and show that it is not a good choice for renormalization because at small $z$ such a  matrix element is suppressed as can be shown by operator
product expansion (OPE). 
Moreover, since the $z$-dependent renormalization is mainly associated 
with the Wilson link, we shall also consider the matrix elements of Wilson loop as well as the Wilson link in a fixed gauge (case 4).  

We calculate the pion and nucleon matrix elements $\tilde{h}^{0}_{H}(z)=\langle H| O_{\gamma_t}(z)|H \rangle_{\vec{P}=0}$ in the rest frame evaluating the following three-point function,
\begin{align}\label{eq:hadron}
&R_{H}(t_2,t,z)\equiv\frac{\langle O_H(t_2)\sum_{\vec{x}}O_{\gamma_t}(z;(\vec{x},t))O_H^{\dagger}(0)\rangle}{\langle O_H(t_2)O_H^{\dagger}(0)\rangle}\nonumber\\
&=\langle H| O_{\gamma_t}(z)|H \rangle\nonumber\\
&+{\cal O}(e^{-\delta m t})+{\cal O}(e^{-\delta m (t_2-t)})+{\cal O}(e^{-\delta m t_2}),
\end{align}
where $O_{\gamma_t}(z,(\vec{x},t))=\bar{\psi}(\vec{x},t) \gamma_t U((\vec{x},t),(\vec{x}+\hat{n}_z z,t)) \psi (\vec{x}+\hat{n}_z z,t)$, $\hat{n}_z$ is the unitary vector along the $z$ direction, and $O_H$ is the interpolation field of a hadron such as a pion and a nucleon.  $t_2$ corresponds to the source-sink separation. 

To obtain $\tilde{h}^0_{H}(z)$ accurately, we need both $t$ and $t_2-t$ to be large enough to suppress excited-state contaminations, or fit the three point function with a proper parametrization on the excited-state contaminations. We can use the first strategy for the pion case since the signal-to-noise ratio will not decay for the pion in the rest frame; while the second strategy is essential for the other cases where the statistical uncertainty increases exponentially with $t$ and $t_2$. Due to (anti)periodic boundary conditions we can use at most $t_2=T/2$ which is larger than 2 fm on most modern lattice ensembles. Since the mass gap $\delta m\sim $ 1 GeV between the ground and first exited state of pion  $\langle \tilde{h}^0_{\pi}(z)\rangle$ can be extracted with sufficient  accuracy. At small perturbative $z$, $\langle \tilde{h}^0_{\pi}(z)\rangle^{-1}$ acts as a renormalization factor up to certain ${\cal O}(\Lambda_{\rm QCD}^2z^2)$ power corrections~\cite{Orginos:2017kos}.

A common non-perturbative renormalization method for lattice QCD matrix elements with local operators is the RI/MOM scheme~\cite{Martinelli:1994ty} and its modified versions~\cite{Aoki:2007xm}. One can calculate the bare matrix element with given operators in an off-shell quark state, for both the lattice and dimensional regularizations, and renormalize their difference in the $\overline{\textrm{MS}}$ scheme to get the renormalization factor for the bare quantities. For the quasi-PDF, such a renormalization constant (the RI/MOM renormalization factor) is defined through the $O_{\gamma_t}(z)$ matrix element in the off-shell quark state with given momentum $p^2$~\cite{Green:2017xeu,Chen:2017mzz,Alexandrou:2017huk,Stewart:2017tvs}:
\begin{align}\label{eq:Z_ri}
&Z^{RI}(z,\mu_R)=\frac{1}{4N_c}\textrm{Tr}[\gamma_t\langle q|O_{\gamma^t}(z)|q\rangle]|_{p^2=-\mu_R^2,p_z=p_t=0},
\end{align}
where $N_c=3$ is the number of colors, $\mu_R$ is the RI/MOM renormalization scale. We will take $\mu_R$ = 3 GeV since the RI/MOM factor has little dependence on the renormalizaiton scale in a certain range~\cite{Zhang:2020rsx}. We work with Landau gauge fixed configurations, and choose $p_z=p_t=0$ to eliminate the difference between the projections~\cite{Liu:2018uuj}. 

Factor $Z^{RI}$ can be calculated non-perturbatively for any lattice regularization, or equivalently for any quark and gluon action defined on the lattice. However, the current lattice QCD calculation is limited to the Landau gauge. The perturbative matching with off-shell states in Landau gauge can be complicated beyond one-loop level~\cite{Chen:2020ody}.

The $z$-dependent part of renormalization comes from the Wilson line, therefore 
one may extract the linear divergences from matrix elements of pure Wilson lines.
First one can consider a Wilson loop~\cite{Chen:2016fxx,Zhang:2017bzy,Musch:2010ka,Green:2017xeu,Chen:2017gck},
\begin{equation}
{\cal U}(r,t)=\langle U(\vec{r},t;\vec{r},0)U(\vec{r},0;\vec{0},0)U(\vec{0},0;\vec{0},t)U(\vec{0},t;\vec{r},t)\rangle , 
\end{equation}
Ref.~\cite{Chen:2016fxx} proposed this scheme to renormalize the pion DA, and the linear divergence coefficient was extracted precisely based on the calculation with several lattice spacings  in the following Kaon DA study~\cite{Zhang:2017bzy}. But most of subsequent lattice QCD calculations have switched to the RI/MOM scheme or used a hadronic matrix element.

 One can also consider the matrix element of a single Wilson Link $\langle U(0,z)\rangle$ in a fixed gauge such as the Landau gauge. As the simplest choice without any external state, $\langle U(0,z)\rangle$ can provide a reference to identify whether the linearly divergent behavior is sensitive to the existence of the external state, as suggested in the multiplicative renormalizability studies of the quasi-PDF operator~\cite{Ji:2015jwa, Ji:2017oey,Ishikawa:2017faj,Green:2017xeu}.
 
\begin{table}[htbp]
  \centering
  \begin{tabular}{c|ccrc}
  \toprule
 symbol & $6/g^2$ & $L$ & $T$  & $a(\mathrm{fm})$ \\
\hline
a12m310 &  3.60 & 24 & 64  & 0.1213(9) \\
\hline
a09m310 &  3.78 & 32 & 96 & 0.0882(8)  \\
\hline
a06m310 &  4.03 & 48 & 144 & 0.0574(5) \\
\hline
a04m310 &  4.20 & 64 & 192 & 0.0425(4)  \\
\hline
a03m310 &  4.37 & 96 & 288 & 0.0318(3) \\
\hline
  \end{tabular}
  \caption{Setup of the MILC ensembles, including the bare coupling constant $g$, lattice size $L^3\times T$, and lattice spacing $a$.}
  \label{tab:lattice}
\end{table}

\begin{table}[htbp]
  \centering
  \begin{tabular}{c|ccrccc}
  \toprule
 symbol & $6/g^2$ & $L$ & $T$ &  $a(\mathrm{fm})$\\
\hline
DW11 & 2.13 &24 & 64  & 0.1105(3) \\
\hline
DW08 & 2.25 &32 & 64  & 0.0828(3) \\
\hline
DW06 & 2.37 &32 & 64  & 0.0627(3) \\
\hline
  \end{tabular}
  \caption{Setup of the RBC ensembles, including the bare coupling constant $g$, lattice size $L^3\times T$, and lattice spacing $a$.}
  \label{tab:lattice2}
\end{table}

  
Our lattice calculations are performed using the Chroma software suite~\cite{Edwards:2004sx} and QUDA~\cite{Clark:2009wm,Babich:2011np,Clark:2016rdz} in the HIP programming model~\cite{Bi:2020wpt}. We use 2+1+1 flavors (degenerate up and down, strange, and charm degrees of freedom) of highly improved staggered quarks (HISQ)~\cite{Follana:2006rc} ensembles generated by the MILC Collaboration~\cite{Bazavov:2012xda} at five lattice spacings, and 2+1 
flavor domain wall (DW) quarks and Iwasaki gauge
ensembles from the RBC/UKQCD collaboration~\cite{Blum:2014tka} at
three lattice spacings. The pion mass for all ensembles are tuned to be roughly 310 MeV based on the pion mass of the light sea quark mass on the corresponding ensemble. The lattice spacings for the MILC ensembles are determined using Wilson flow based on the parameters determined by Ref.~\cite{Miller:2020evg}.

We use the matrix elements without hyper-cubic (HYP) smearing in order
to test the perturbative renormalization analysis (Sec.~\ref{sec:test},~\ref{sec:StrtoRe}, and~\ref{sec:highorder}). 
However, since in practical calculations the results without smearing can 
be rather noisy, it is standard to use some type of smearing in lattice simulations. It is unclear to us how strongly the smearing will interfere with renormalization. Therefore, we regard our method as a purely phenomenological approach for the time being to analyze data with one step HYP smearing~\cite{Hasenfratz:2001hp} in Sec.~\ref{sec:smearing} and~\ref{sec:otherresult}, hoping that the gain in statistical precision outweighs the additional systematic uncertainty 
from moderate smearing.

We use two kinds of fermion actions for the valence quarks: clover and overlap fermions. Clover fermions break chiral symmetry and the action is defined by
\begin{align}\label{eq:clover}
S^{clv}_{q}=&\sum_{x}\bar{\psi}(x)\bigg[\frac{1-\gamma_{\mu}}{2a}U_{\mu}(x,x+\hat{n}_{\mu})\psi(x+\hat{n}_{\mu})\nonumber\\
&+\frac{1+\gamma_{\mu}}{2a}U_{\mu}(x,x-\hat{n}_{\mu})\psi(x-\hat{n}_{\mu})\nonumber\\
&-\left(\frac{4}{a}+c_{sw}\sigma_{\mu\nu} F_{\mu\nu}(x)a+m_q^0\right)\psi(x)\bigg],
\end{align}
where $\hat{n}_{\mu}$ is the unit vector along the $\mu$ direction, the clover coefficient $c_{sw}$ is the tadpole improved tree level value, and $m^0_{q}a$ is the bare quark mass which is determined by requiring the pion mass to be roughly 310 MeV.  Parameters $c_{sw}$ and $m^0_{q}a$ should approach 1 and 0 respectively in the continuum, while the lattice spacing dependence is weaker than the ${\cal O}(a)$ discretization effect in the present range of $a$, and closer to that of the gauge coupling $g^2$ as predicted by lattice perturbative theory. 

The overlap action preserves chiral symmetry but the simulation is rather expensive. We use it to test
the dependence of renormalization on the fermion action. 
The overlap fermion action is defined by~\cite{Chiu:1998eu, Liu:2002qu} 
\begin{align}
S^{\textrm{ov}}_{q}&=\sum_{x,y}\bar{\psi}(x) D_{\textrm{ov}}(x,y)\psi(y),\\
 D_{\textrm{ov}}&=\rho\Big(1+\frac{D_\textrm{w}(-\rho)}{\sqrt{D^{\dagger}_\textrm{w}(-\rho)D_\textrm{w}(-\rho)}}\Big),\nonumber
\end{align}
where
\bea
&D_\textrm{w}(m^{\textrm{w}}_q;x,y)=\frac{1-\gamma_{\mu}}{2a}U(x,x+\hat{n}_{\mu})\delta_{x+\hat{n}_{\mu},y}\nonumber\\
&+\frac{1+\gamma_{\mu}}{2a}U(x,x-\hat{n}_{\mu})\delta_{x-\hat{n}_{\mu},y}-(\frac{4}{a}+m^{\textrm{w}}_q)\delta_{x,y},
\eeq
and $-\rho$ should be smaller than the bare quark mass for which vanishes the pion mass to make $D_{\textrm{ov}}$ to be the same as the standard Dirac operator in the continuum limit. We choose $-\rho=1.5$.

Although different active and sea fermion formulations
will in general introduce non-unitarity, the effect
shall vanish in the continuum limit. However, at finite $a$,
the difference will generate systematic uncertainties 
which can affect the final results. 

 \begin{figure}[tbp]
  \centering
  \includegraphics[width=8cm]{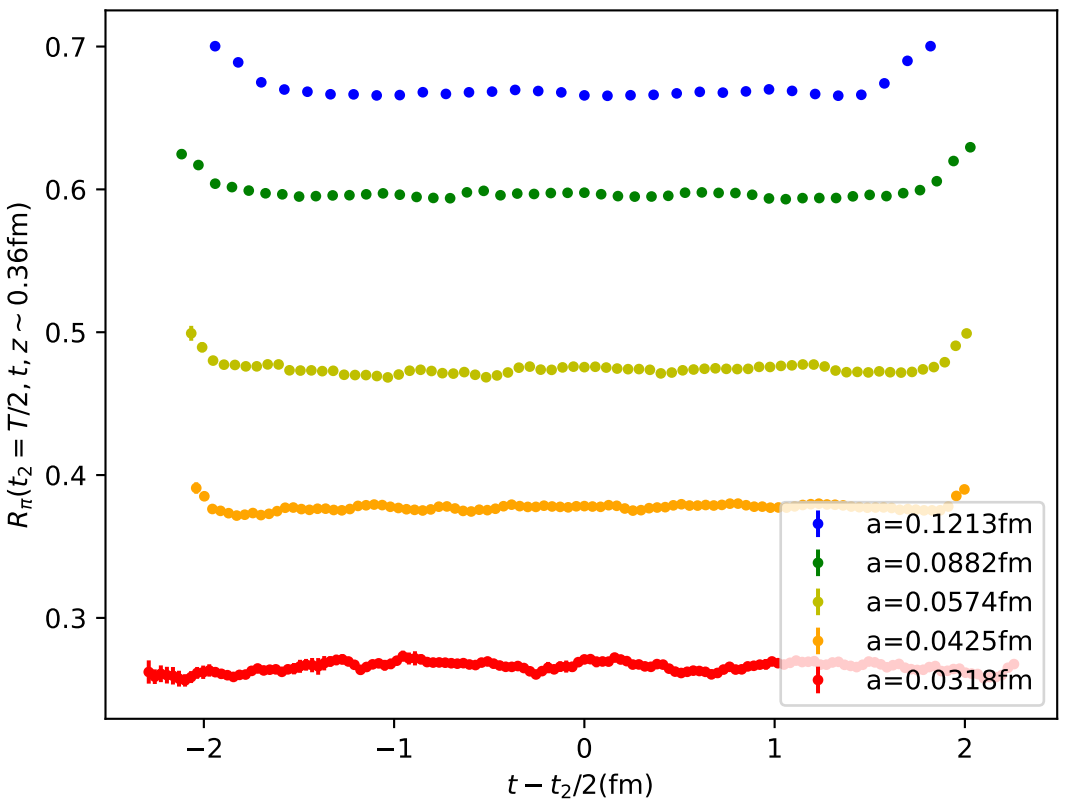}
   \includegraphics[width=8cm]{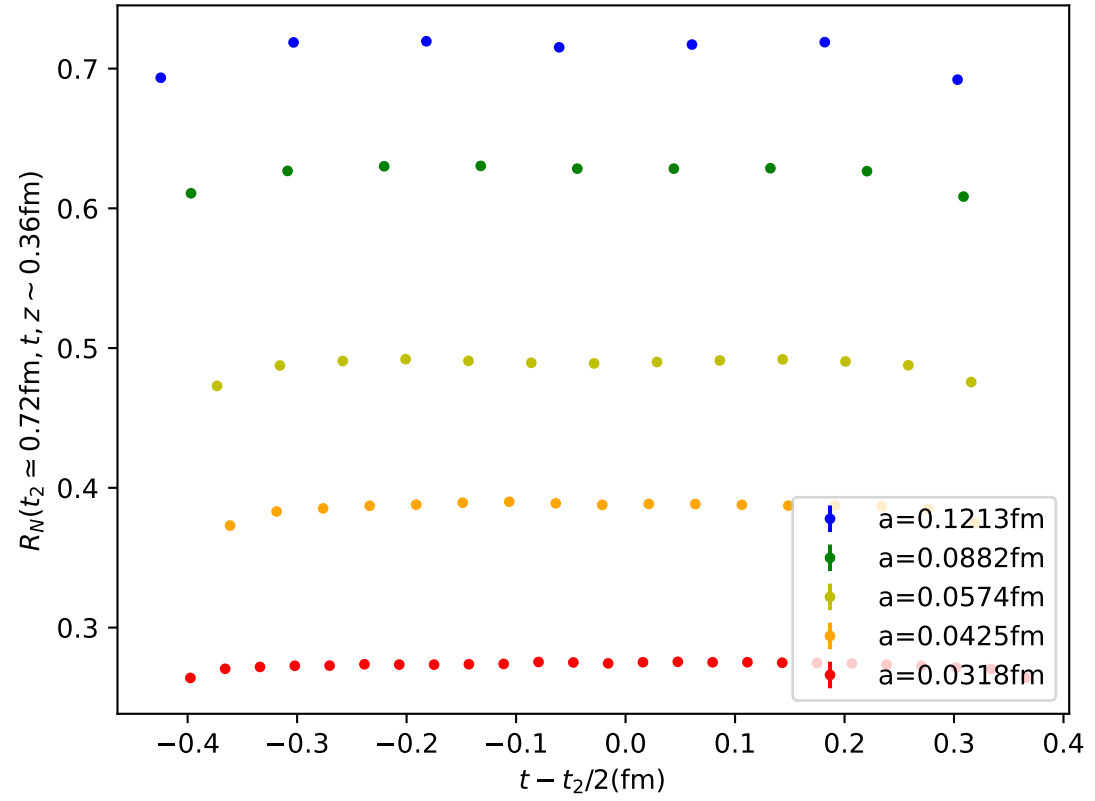}
  \caption{The bare ratio $R_{\pi}(t_2=T/2,t,z)$ for the pion (upper panel) and $R_{N}(t_2\simeq 0.72 \textrm{fm},t,z)$ for the nucleon (lower panel) with $z\sim$ 0.36 fm. For the pion case, we can see that the data in the region between $t-t_2/2\in[-1,1]$ fm are consistent with each other up to statistical fluctuations. For the nucleon case the ratio is also flat around $t\sim t_2/2$.}
  \label{fig:pion_plateau}
  \end{figure}

We start with nucleon and pion matrix elements for the MILC ensembles and clover action. For the pion matrix element, We average the $R_{\pi}(t_2,t,z)$ data with $t_2=T/2$ and $t\in[T/8,3T/8]$ to get a precise estimate of the ground state matrix element, and it should be also precise since $T$ is between 7.7 and 9.1 fm in the ensembles we used, as shown in the upper panel of Fig.~\ref{fig:pion_plateau}. Note that $R_{\pi}(T/2,T/4,z)$ equals to $\langle \pi| O_{\gamma_t}(z)|\pi \rangle/2$ in such a case. The additional factor 1/2 corrects for the fact that there are forward and backward propagating states for (anti)periodic boundary conditions. The nucleon case is known to be much noisier especially at large $t_2$. Thus, we just consider $t_2=2t\simeq 0.72$ fm in all cases, and plot $R_{N}(t_2,t,z)$ in the lower panel of Fig.~\ref{fig:pion_plateau}.

 \begin{figure}[tbp]
  \centering
  \includegraphics[width=8.5cm]{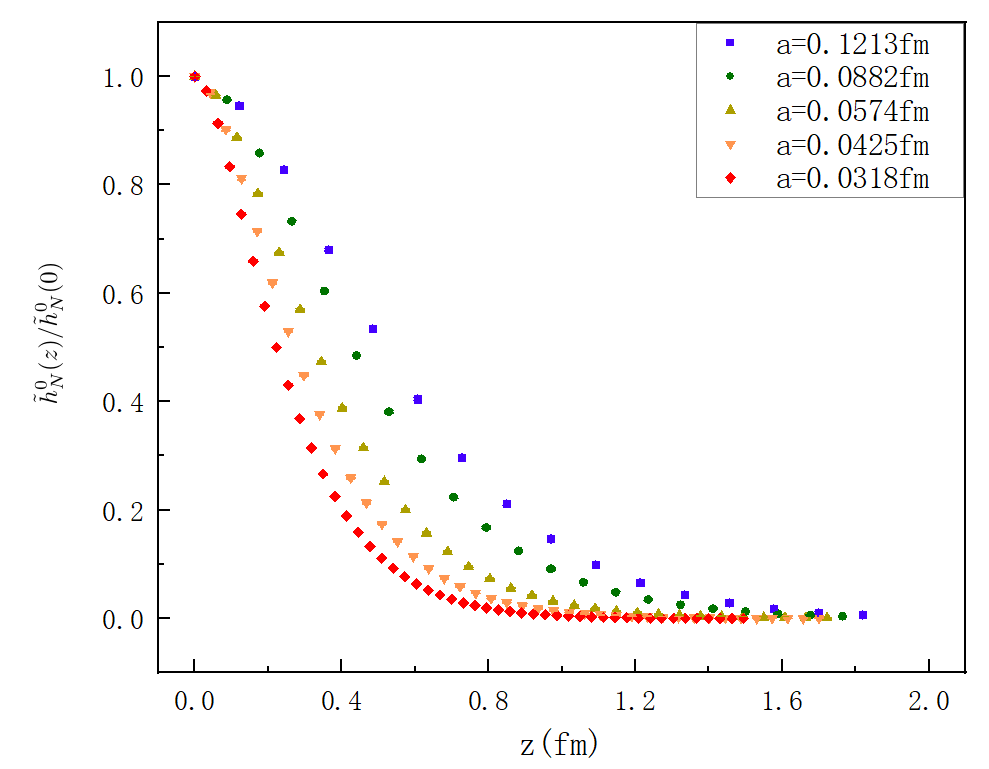}
  \caption{The normalized bare nucleon matrix element in the rest frame, $\tilde{h}^0_{N}(z)/\tilde{h}^0_{N}(0)$ as the function of the Wilson link length $z$. The matrix element has a fast exponential decay at large $z$ and the decay process accelerates with smaller
  lattice spacings.}
  \label{fig:ratio_had}
  \end{figure}

The normalized bare $\tilde{h}^0_{N}(z)/\tilde{h}^0_{N}(0)$ where $\tilde{h}^0_{N}(z)$ is approximated by $R_{N}(t_2,t,z)$ with $t_2=2t\simeq 0.72$ fm is plotted in Fig.~\ref{fig:ratio_had}, with a normalization factor $1/\tilde{h}^0_{N}(0)$ using jackknife resampling to make it to be exactly one at $z=0$ at all lattice spacings. As in the figure, the linear divergence makes $\tilde{h}^0_{N}(z)/\tilde{h}^0_{N}(0)$ decay exponentially with both $z$ and $a$, and thus one cannot obtain any meaningful continuum limit when $a\to 0$. A renormalization is necessary to recover a good continuum limit.

\section{Simple Test of Linear Divergence}\label{sec:test}

To study the renormalization properties of the quasi-PDF operator, we need to calculate matrix elements at several different lattice spacings. To ensure the data
to be useful for a refined high-precision analysis, we start by testing whether they approximately show the linear divergence  predicted by perturbation theory. In particular, one needs to show that the $z$-dependence of the linearly divergent term is linear. 

To achieve this, we first extract the factor $e^{-\delta \bar{m} z}$ from the bare matrix element to test the $1/a$ dependence in $\delta \bar{m}$ in Eq.~(\ref{eq:delta_m}). 
Based on Eqs.~(\ref{eq:Reoperator}),~(\ref{eq:delta_m}), and~(\ref{eq:alpha_s}), if we take the natural log on a bare matrix element $\mathcal{M}$, we have,
\bea\label{eq:logM_s}
\ln \mathcal{M}(z,a) = \frac{e(z)}{a \ln[a \Lambda_{\rm QCD}]} + g(z),
\eea
where the first term is linearly divergent and $g(z)$ is the residual. We have ignored the logarithmic divergence and discretization error here because these terms are much smaller and thus have little influence on this simple test of the linear divergence.

\begin{figure}[tbp]
\centering
\includegraphics[width=8cm]{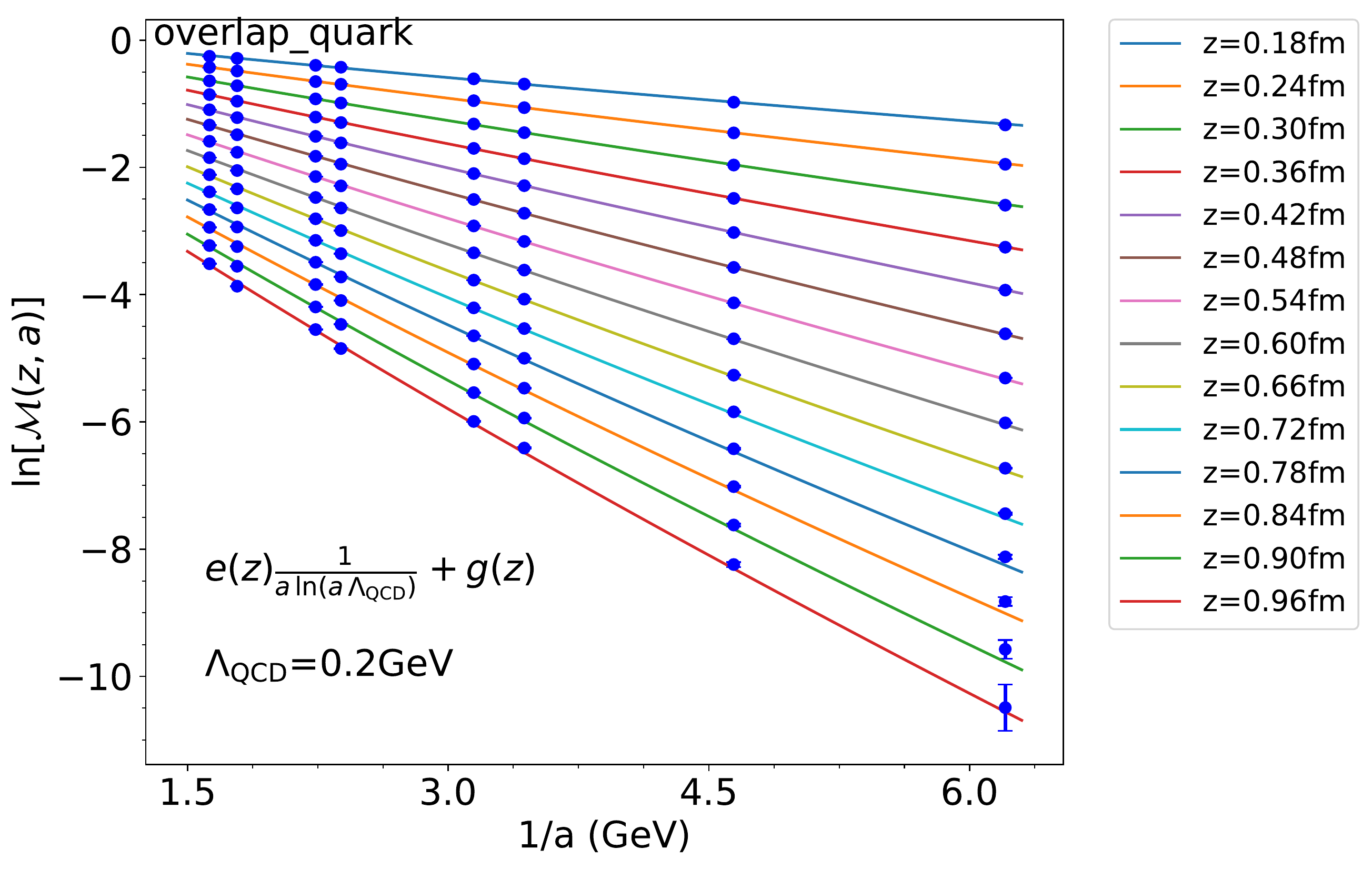}
\includegraphics[width=8cm]{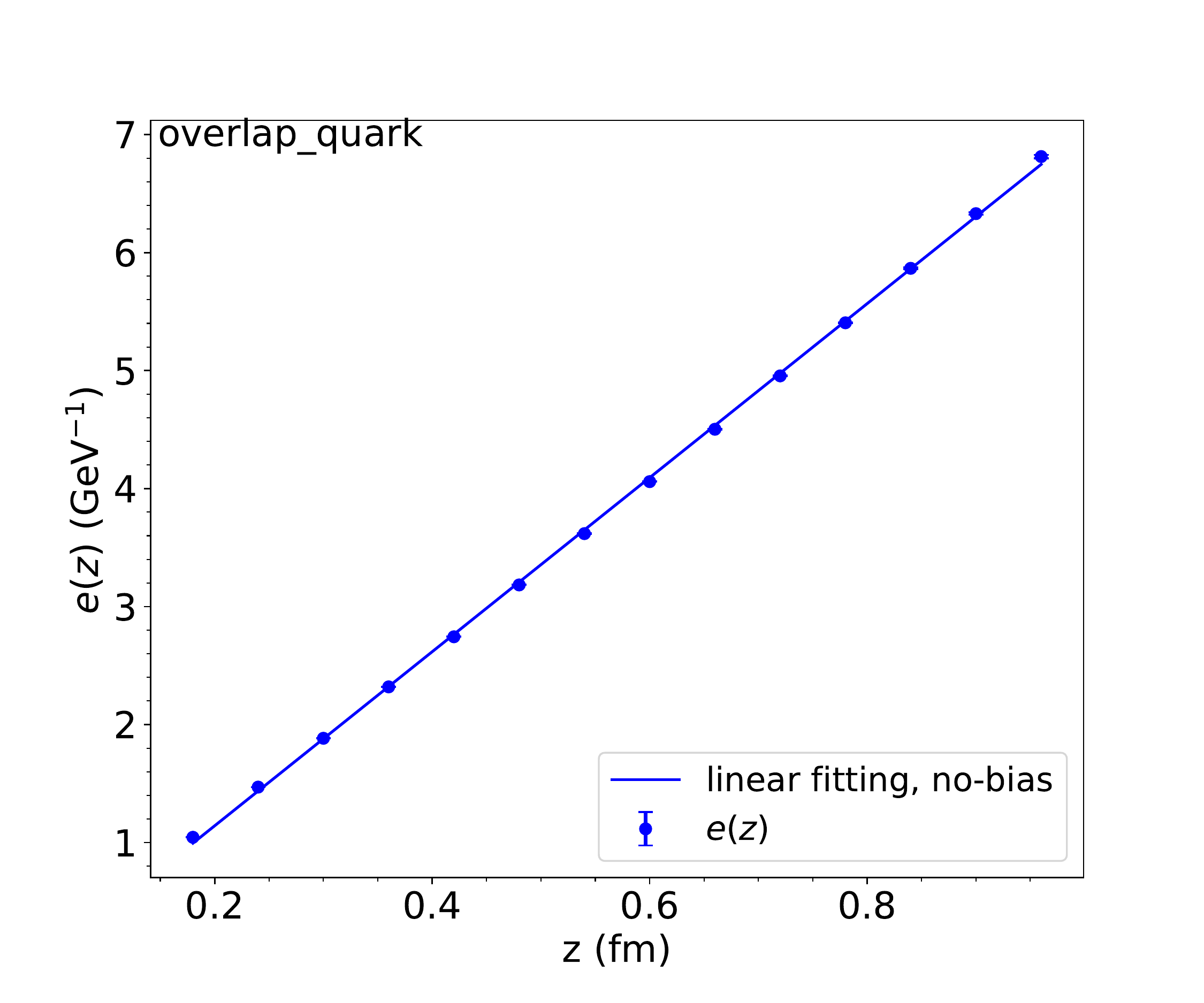}
\caption{
Upper Panel: Using Eq.~(\ref{eq:logM_s}) to fit the bare $O_{\gamma_t}(z)$ matrix element  in the off-shell quark state for overlap action without HYP smearing. Blue points are interpolated data and colorful curves are fitted curves for each $z$. $\Lambda_{\rm QCD}$ is fixed at 0.2 GeV.
Lower Panel: $e(z)$ with respect to $z$. Blue points are the fitted parameters $e(z)$ from the upper panel. The blue curve is the no-bias linear fit for $e(z)$.}
\label{fig:testlindiv_RIMOM_overlap}
\end{figure}

\begin{figure}[tbp]
\centering
\includegraphics[width=8cm]{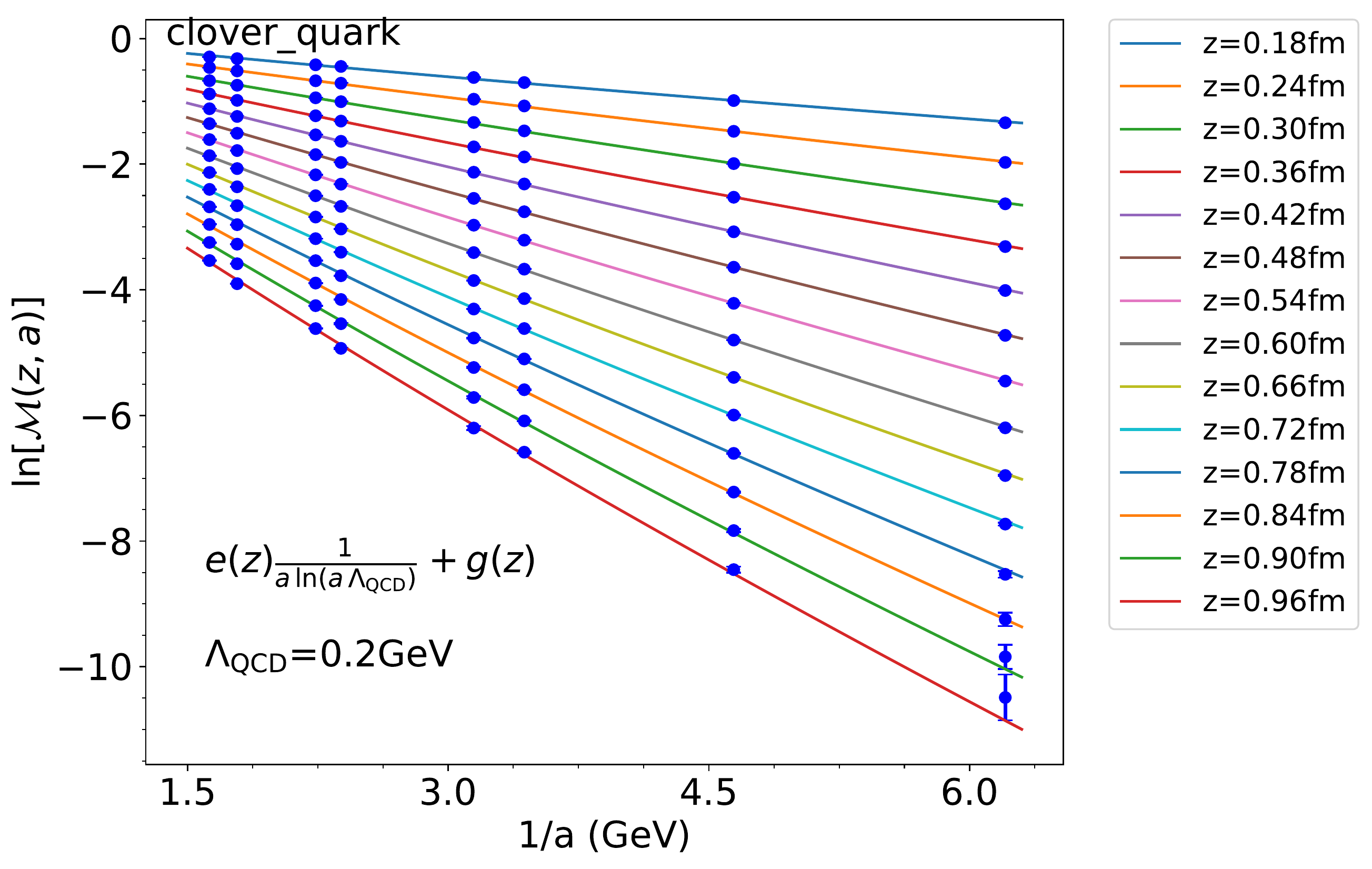}
\includegraphics[width=8cm]{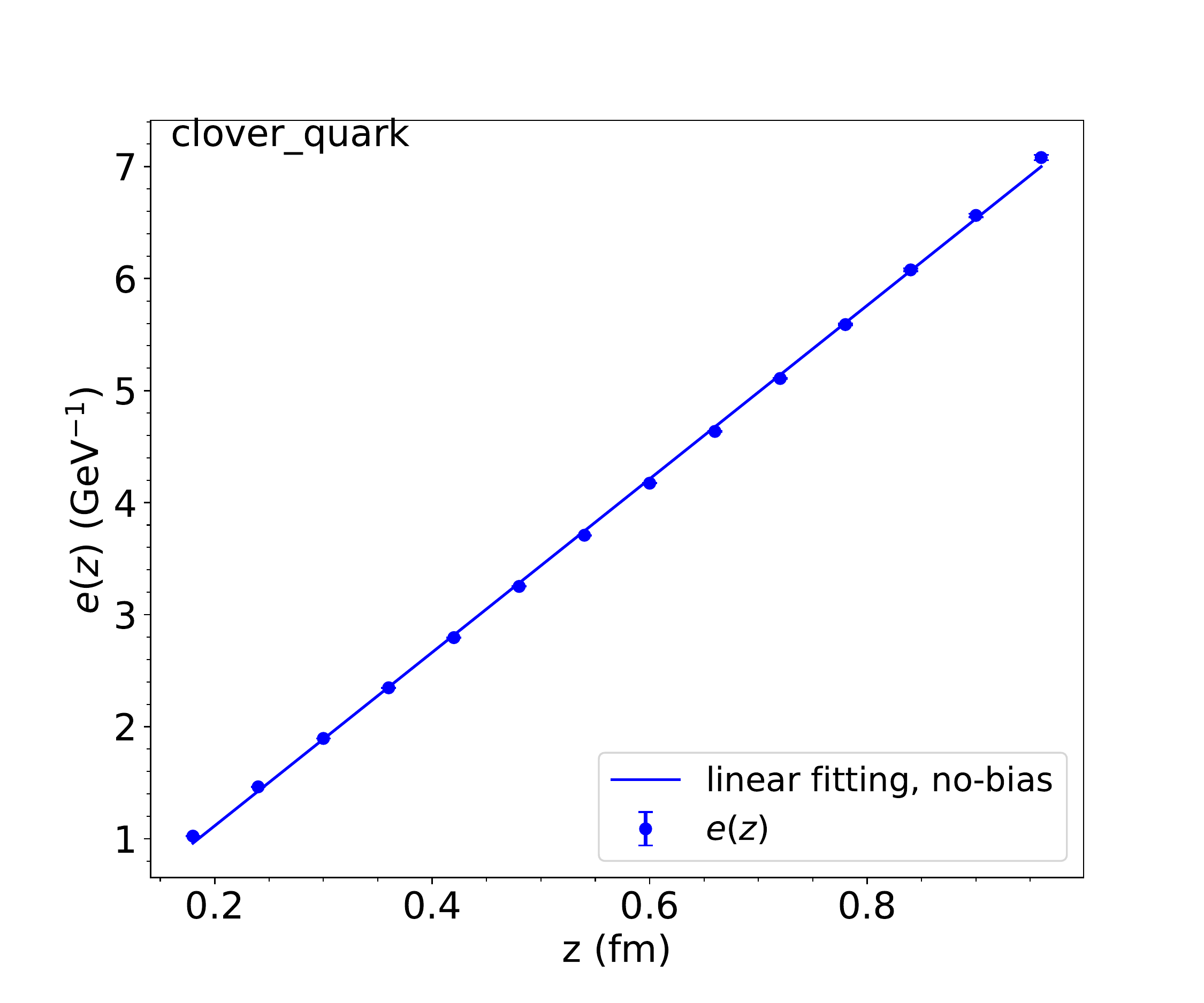}
\caption{Same as Fig.~\ref{fig:testlindiv_RIMOM_overlap}, except for the clover valence quark.}
\label{fig:testlindiv_RIMOM_clover}
\end{figure}

We use Eq.~(\ref{eq:logM_s}) to fit the bare matrix elements as a function of $a$. We treat $e(z)$ and $g(z)$ as unknown functions of $z$ so that our fitting is performed at each value of $z$ at which they are treated as free parameters. We need to do linear interpolation on the data $\ln \mathcal{M}(z,a)$ with respect to $z$. Here are the steps we perform in detail:

1. We do the linear interpolation with neighbourhood two data points to obtain the central value and uncertainty at $z=0.06\times n$ fm $(n=3,4,5,...,16)$.

2. For each $z$, we fit the dependence on $a$, treating $e(z)$, $g(z)$ as fitted parameters. For this initial test, we fix $\Lambda_{\rm QCD}$ at 0.2 GeV, which is a reasonable guess to start with~\cite{Lepage:1992xa}.

3. We plot the fitted $e(z)$ with respect to $z$ to see if $e(z)$ has a linear $z$ dependence.

If we obtain a reasonable fit in step 2, we then see approximately the $1/a$ dependence of the divergences. In step 3, we can test the linear $z$ dependence. The fitting results for the bare $O_{\gamma_t}(z)$ matrix element in the off-shell quark state for valence overlap and clover actions without HYP smearing are shown in Fig.~\ref{fig:testlindiv_RIMOM_overlap} and Fig.~\ref{fig:testlindiv_RIMOM_clover}, respectively. There
are eight different lattice spacings. The range of $z$
is taken from 0.18~fm to 0.96~fm and we analyse 14 different $z$ values in this range.

The $1/a$ dependence in both cases is fitted very well, although due to the high precision of the data, chi-square
is large. However, this is not our concern at this initial step. 
The $z$ dependence of the divergent coefficients
shows nicely the linear feature, approximately going through
zero at $z=0$. This is a strong indication that
the linear divergence follows roughly
the prediction of perturbation theory which gives
us confidence to perform more refined analyses in the next sections.

\section{Self-Renormalization of lattice matrix elements}\label{sec:StrtoRe}
\subsection{The Strategy}\label{Strategy}

The non-perturbative matrix elements contain the following important contributions: a) the linear divergence, b) a finite term coming from non-perturbative renormalon physics, and c) the residual contribution encoding intrinsic non-perturbative physics. It is part c) that is required to extract the partonic structure information. Thus, it is important to separate out the latter from the former in a systematic way and with high precision. 

Here we develop a self-renormalization method to extract the residual intrinsic non-perturbative physics and renormalization factor from a matrix element itself without using a different matrix element. The extraction of the residual is much harder than the extraction of the linear divergence term because after we take the logarithm of the matrix element (Eq.~(\ref{eq:logM_s})), the linear divergence term is dominant and the residual is very sensitive to the subtraction. We need to properly take into account the fine features of the data, including logarithmic divergences and discretization errors. 

Based on Eq.~(\ref{eq:ZO_RS_higher}), we modify the fitting function Eq.~(\ref{eq:logM_s}) to be, 
\begin{align}\label{eq:logM}
\ln \mathcal{M}(z,a) = \frac{k z}{a \ln[a \Lambda_{\rm QCD}]} 
+ g(z) + f_{1,2}(z)a \nonumber\\
+ \frac{3 C_{F}}{b_0} \ln \bigg[\frac{\ln [1 /(a \Lambda_{\rm QCD})]}{\ln [\mu / \Lambda_{\rm QCD}]}\bigg]+\ln \bigg[1+\frac{d}{\ln (a \Lambda_{\rm QCD})}\bigg],
\end{align}
where the first term is linearly divergent. $g(z)$ is the residual, which contains the non-perturbative $m_{0}$ effect and the intrinsic non-perturbative physics. $f_{1,2}(z)a$ takes into account the discretization errors, which we allow to be different for the data calculated from MILC and RBC ensembles ($f_{1}(z)$ for MILC and $f_{2}(z)$ for RBC). The last two terms come from the resummation of leading and sub-leading logarithmic divergences, which only affect the overall normalization at different lattice spacings.

We use Eq.~(\ref{eq:logM}) to fit the logarithms to the data for the bare matrix elements for each $z$. The $z$ points can be chosen in 
a large range where the lattice discretization error is small and, at the same time, the statistical error is limited. We fit the dependence on $a$, treating $g(z)$, $f_{1}(z)$, $f_{2}(z)$ as free parameters and fixing $\mu$ at 2 GeV, which is the renormalization scale we choose to use. We will discuss $k$ and $\Lambda_{\rm QCD}$ in detail in Sec.~\ref{ReUncer} and $d$ in Sec.~\ref{sec:d}. We obtain the residual $g(z)$ from the fit.

The $g(z)$ obtained this way does not correspond to what is in the perturbative $\overline{\rm MS}$ scheme because it contains a non-perturbative $m_{0}$ effect. 
We need to eliminate this effect through matching the lattice result to the continuum scheme at short distance $z$. Based on Eq.~(\ref{eq:Reoperator}) and Eq.~(\ref{eq:m0}), we use the following equation within a window $a \ll z < 1/\mu$ to extract $m_{0}$,   
\bea\label{eq:gztoMS}
g(z) - \ln[Z_{\overline{\mathrm{MS}}}(z,\mu, \Lambda_{\overline{\mathrm{MS}}})] = m_{0} z,
\eea
where $Z_{\overline{\mathrm{MS}}}$ is the perturbative matrix element in the $\overline{\mathrm{MS}}$ scheme. For the $O_{\gamma_t}(z)$ matrix elements in the pion, nucleon, and off-shell quark state, we take $Z_{\overline{\mathrm{MS}}}$ at one loop~\cite{Izubuchi:2018srq},
\bea\label{eq:ZMS}
Z_{\overline{\mathrm{MS}}}(z, \mu, \Lambda_{\overline{\mathrm{MS}}}) = 1+\frac{\alpha_{s}(\mu,\Lambda_{\overline{\mathrm{MS}}}) C_{F}}{2 \pi}\bigg[\frac{3}{2} \ln \frac{z^{2} \mu^{2}}{4 e^{-2 \gamma_{E}}}+\frac{5}{2}\bigg]\ , \nonumber\\
\eea
from operator product expansion, where we fix $\mu$ = 2 GeV and $\Lambda_{\overline{\mathrm{MS}}}$ = 0.3 GeV. 
The renormalized physical matrix element is then obtained from
\bea\label{eq:gz-m0}
\mathcal{M}(z)_{R} = \exp[g(z) - m_{0}z],
\eea
and is expected to be consistent with $Z_{\overline{\mathrm{MS}}}$ at small $z$.

One could consider higher-order $\overline{\mathrm{MS}}$ matrix elements for matching, including summing over large logarithms. We choose not to do this here as it does not affect the procedure for the subsequent analysis, and therefore does not change the main conclusions of the paper. However,
to obtain physical results at high precisions, one might need to examine this more carefully. 

We thus define a renormalization factor
\begin{align}\label{eq:ZRm0}
Z(z,a)_{R} = \exp\Big[\frac{k z}{a \ln[a \Lambda_{\rm QCD}]} 
+ m_{0}z + f_{1,2}(z)a \nonumber\\
+\frac{3 C_{F}}{b_0} \ln \bigg[\frac{\ln [1 /(a \Lambda_{\rm QCD})]}{\ln [\mu / \Lambda_{\rm QCD}]}\bigg]+\ln [1+\frac{d}{\ln (a \Lambda_{\rm QCD})}]\Big]\ ,
\end{align}
which includes the discretization error as well. 
All the parameters here are either fixed, fitted or fine-tuned through the renormalization procedure from the matrix element $\mathcal{M}(z,a)$ itself. Dividing the bare matrix element $\mathcal{M}(z,a)$ by $Z(z,a)_{R}$, we get
\bea\label{eq:M/Z}
\mathcal{\widetilde{M}}_{R}(z) = \mathcal{M}(z,a)/Z(z,a)_{R}.
\eea
If our procedure eliminates all the divergences and discretization errors, we should expect that $\mathcal{\widetilde{M}}_{R}(z)$ is $a$-independent and the same as $\mathcal{M}(z)_{R}$ (Eq.~(\ref{eq:gz-m0})). The degree to which this is the case can be used as a test of the self-renormalization procedure.

\subsection{Renormalon Uncertainty}\label{ReUncer}
Now let us turn to the parameter $k$ and $\Lambda_{\rm QCD}$ in Eq.~(\ref{eq:logM}). In Eq.~(\ref{eq:logM}), we have only considered the leading order perturbative term for the linear divergence and neglected higher-order ones. These can be included partially by a proper choice of $\Lambda_{\rm QCD}$. 
For example, the coupling constants with different choices of $\Lambda$ are related to each other by perturbation theory~\cite{Lepage:1992xa}, 
\bea\label{eq:alpha_s_series}
\alpha_{s}(Q,\Lambda_{\rm QCD}) \sim \alpha_{s}(Q,\Lambda) + c_{2} \alpha_{s}^2(Q,\Lambda) \nonumber\\ + c_{3} \alpha_{s}^3(Q,\Lambda) + ...
\eea
In principle, $\Lambda_{\rm QCD}$ depends on the specific lattice action, and one needs to perform multi-loop calculations to relate them. In our analysis, we treat $\Lambda_{\rm QCD}$ as a fitting parameter which, therefore, can effectively take into account some higher-order effects. 

\begin{figure}[tbp]
\centering
\includegraphics[width=9cm]{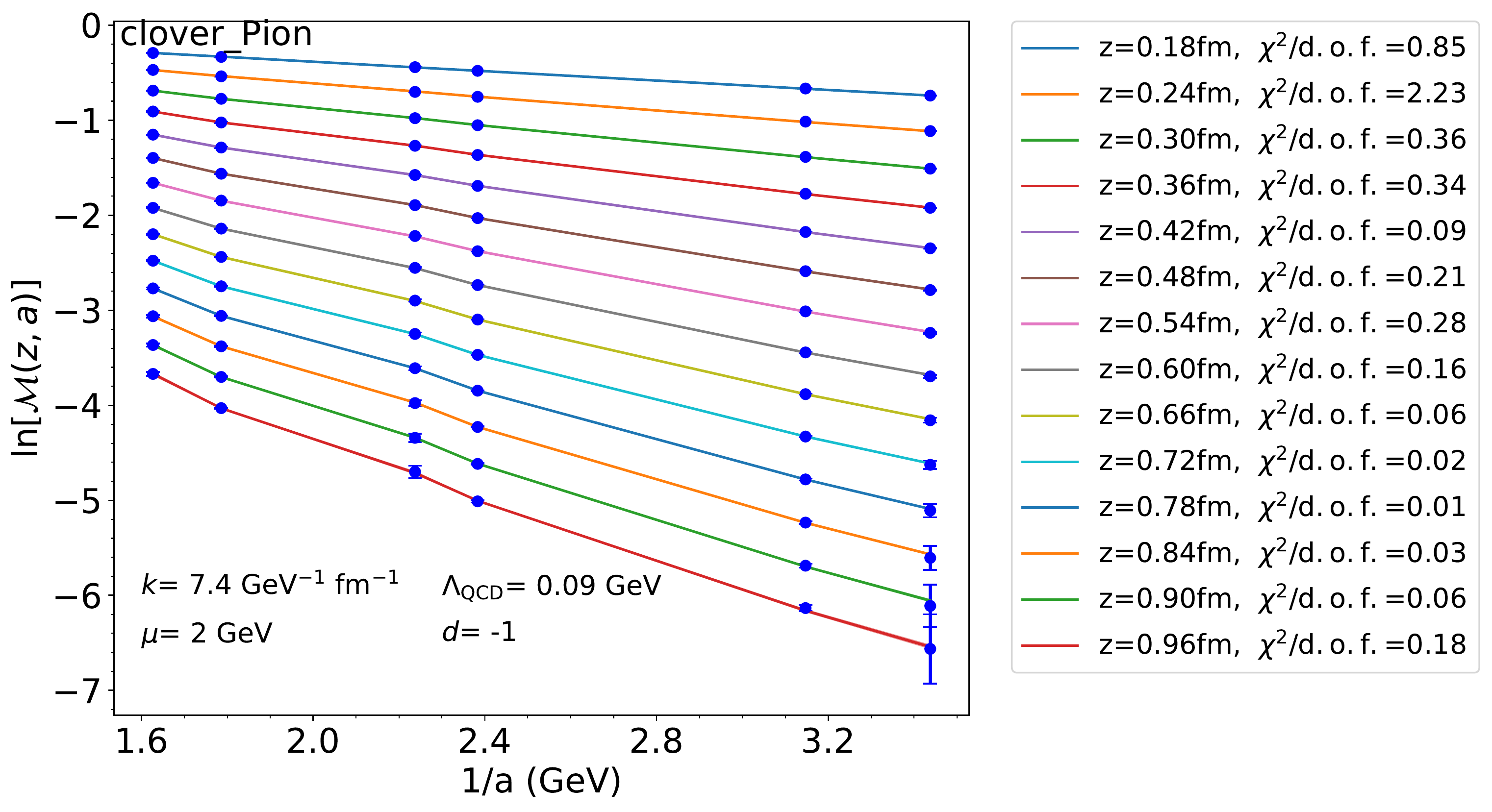}
\caption{Using Eq.~(\ref{eq:logM}) to fit the bare $O_{\gamma_t}(z)$ matrix element in the pion state for the clover action without  HYP  smearing. Blue points are interpolated data and colorful curves are fitted curves for each $z$. The fitted curves are broken lines because $f_{1}(z)$ and $f_{2}(z)$ are allowed to be different. $k$, $\Lambda_{\rm QCD}$, and $d$ are fixed at 7.4 GeV$^{-1}$fm$^{-1}$ (1.46 if dimensionless), 0.09 GeV, and $-1$ respectively. The $\chi^{2}/{\rm d.o.f.}$ of the fitting for each $z$ is listed in the plot legend.
}
\label{fig:fitting_clover_Pion}
\end{figure}

\begin{figure}[tbp]
\centering
\includegraphics[width=9cm]{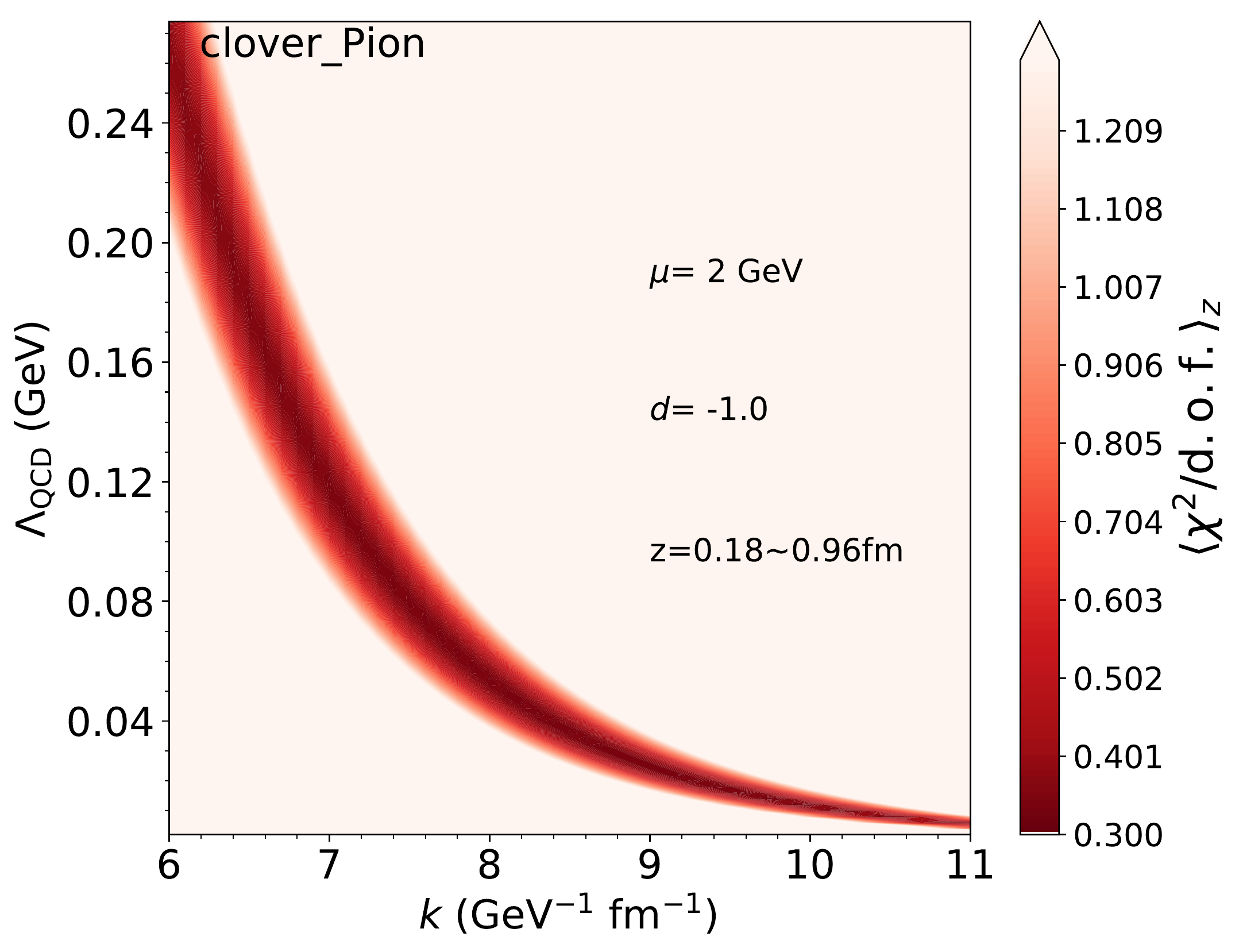}
\caption{ 
$\chi^{2}$ map with respect to $k$ and $\Lambda_{\rm QCD}$. $\langle  \chi^{2}/{\rm d.o.f.} \rangle_{z}$ is the average of $\chi^{2}/{\rm d.o.f.}$ among the fitting for each $z$. The ``small-$\chi^{2}$ band'' shows that $k$ and $\Lambda_{\rm QCD}$ are strongly correlated and uncertain for the chosen data.
}
\label{fig:chisqmap_clover_Pion}
\end{figure}

Next we want to discuss the coefficient $k$ of the linear divergence. QCD perturbation theory not only predicts the $1/a$ dependence and linear $z$ dependence (which we have tested in Sec.~\ref{sec:test}), but also the value of $k$. Comparing Eq.~(\ref{eq:delta_m}) and Eq.~(\ref{eq:logM}), we get the one-loop perturbative value of $k$:
\bea\label{eq:k}
k~=~\frac{(2 \pi)^2}{3 b_{0}} = 1.46 = 7.4 {\rm GeV^{-1} fm^{-1}}.
\eea
Next, we check whether this value is consistent with our data.

We use Eq.~(\ref{eq:logM}) to fit the bare $O_{\gamma_t}(z)$ matrix element in the pion state for the clover action without  HYP smearing, treating $g(z)$, $f_{1}(z)$, $f_{2}(z)$ as fitted parameters. If we take ($k$, $\Lambda_{\rm QCD}$) to be one set of values, e.g., (7.4 GeV$^{-1}$fm$^{-1}$, 0.09 GeV), we can perform a fitting for each $z$, as seen in Fig.~\ref{fig:fitting_clover_Pion}. There is a $\chi^{2}/{\rm d.o.f.}$ for the fit at each $z$ and we calculate the average of them, denoted as $\langle \chi^{2}/{\rm d.o.f.} \rangle_{z}$.
Varying ($k$, $\Lambda_{\rm QCD}$) generates a $\chi^{2}$ map, as seen in Fig.~\ref{fig:chisqmap_clover_Pion}. 

\begin{figure}[tbp]
\centering
\includegraphics[width=8cm]{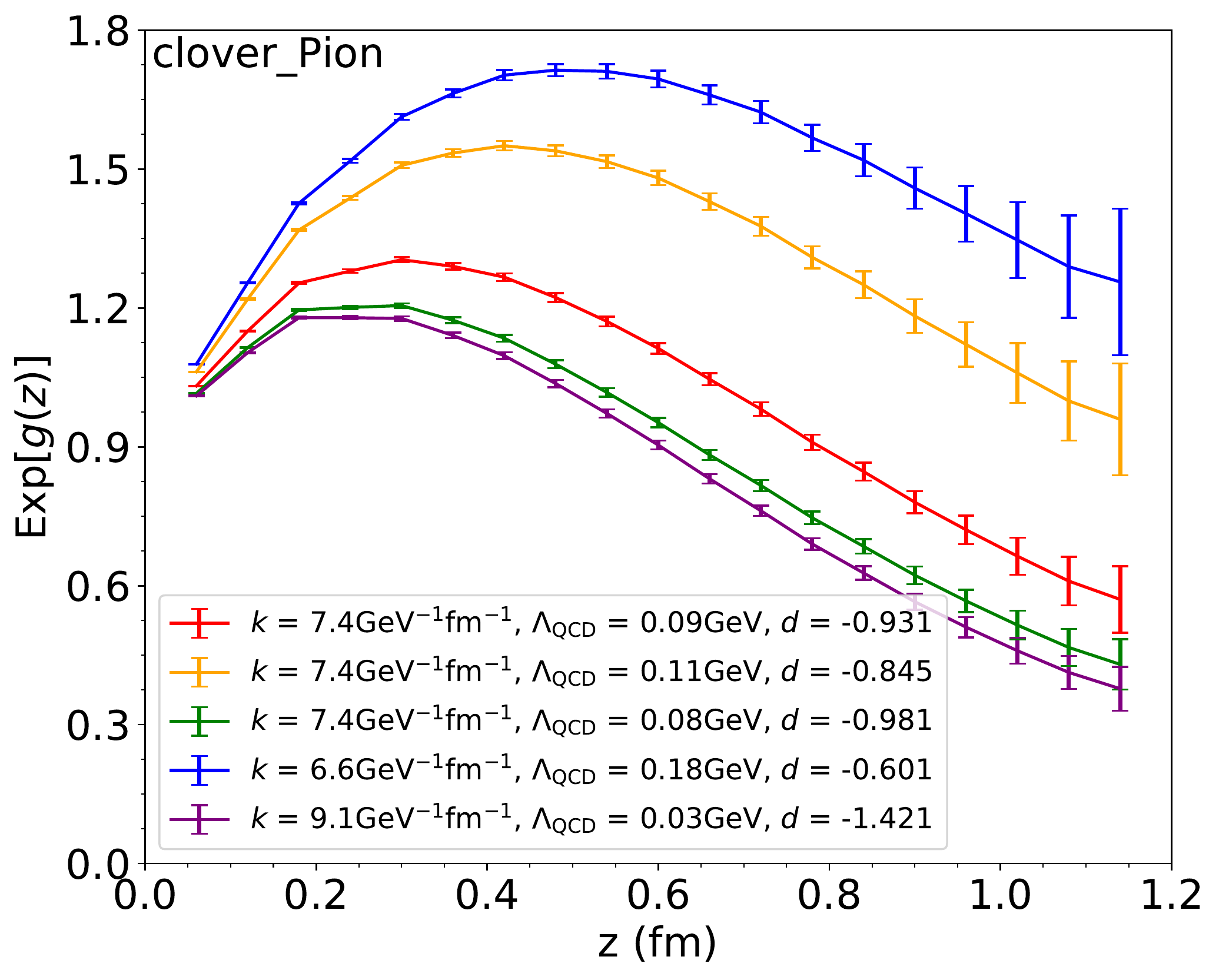}
\caption{
The fitted residual Exp[$g(z)$] for different sets of ($k$, $\Lambda_{\rm QCD}$) along the ``small-$\chi^{2}$ band''.
}
\label{fig:Egz}
\end{figure}

\begin{figure}[tbp]
\centering
\includegraphics[width=8cm]{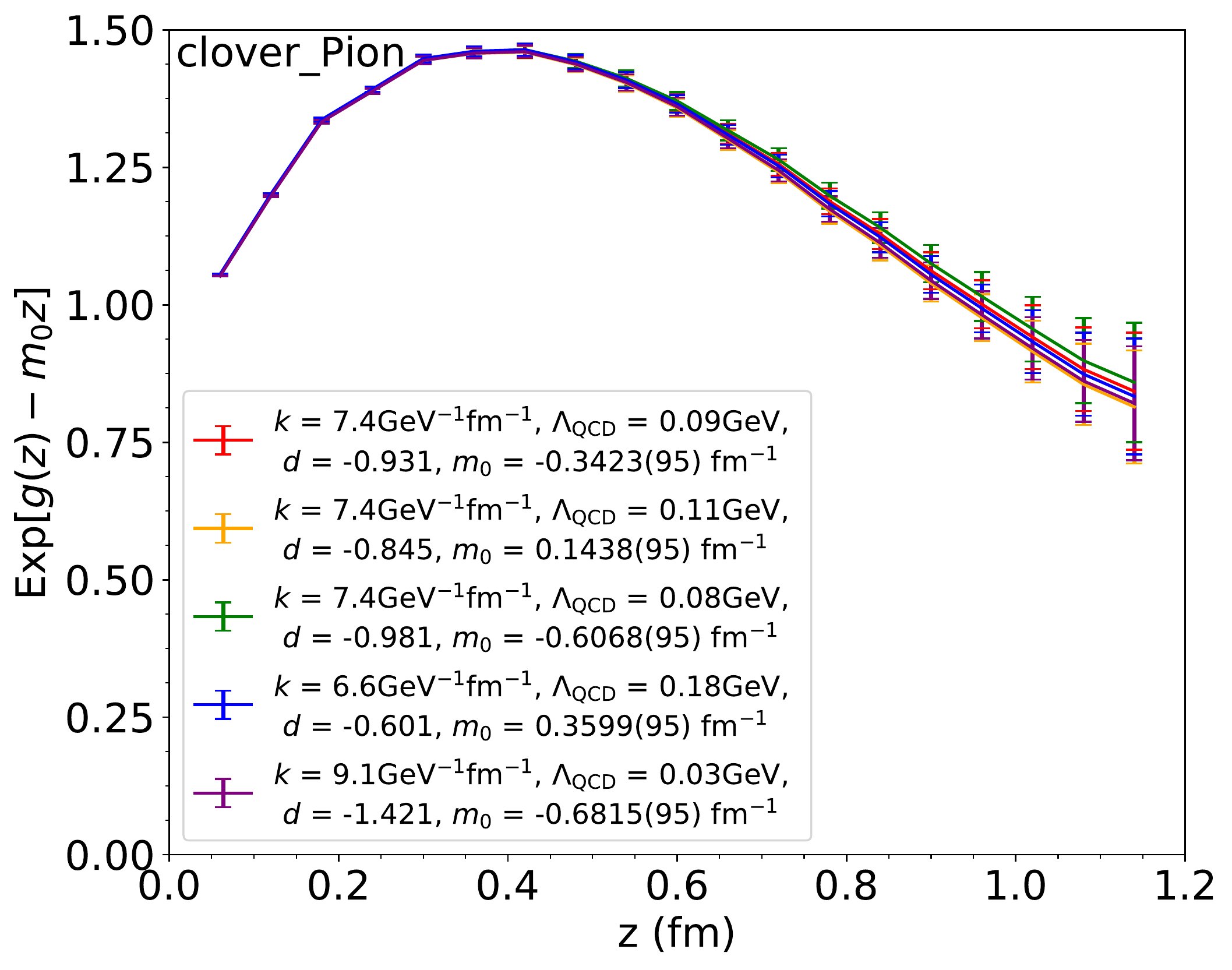}
\caption{
The renormalized matrix element Exp[$g(z)-m_{0} z$] for different sets of ($k$, $\Lambda_{\rm QCD}$) along the ``small-$\chi^{2}$ band''.
}
\label{fig:Egz_m0}
\end{figure}

The sets of ($k$, $\Lambda_{\rm QCD}$) of small $\chi^{2}$ lie in a band, which we call the ``small-$\chi^{2}$ band''. That means that $k$ and $\Lambda_{\rm QCD}$ are strongly correlated, while there is a large uncertainty for $k$ and $\Lambda_{\rm QCD}$ separately.
If we choose some sets of values from the ``small-$\chi^{2}$ band'' to do the fitting, we find the fitted residuals Exp[$g(z)$] are very different, as seen in Fig.~\ref{fig:Egz}. However, after eliminating the non-perturbative $m_{0}$ effect, the renormalized matrix elements Exp[$g(z) - m_{0} z$] are all the same (see Fig.~\ref{fig:Egz_m0}). Therefore, the uncertainty in $k$ and $\Lambda_{\rm QCD}$ will not significantly influence the final physical result.

\begin{figure}[tbp]
\centering
\includegraphics[width=8cm]{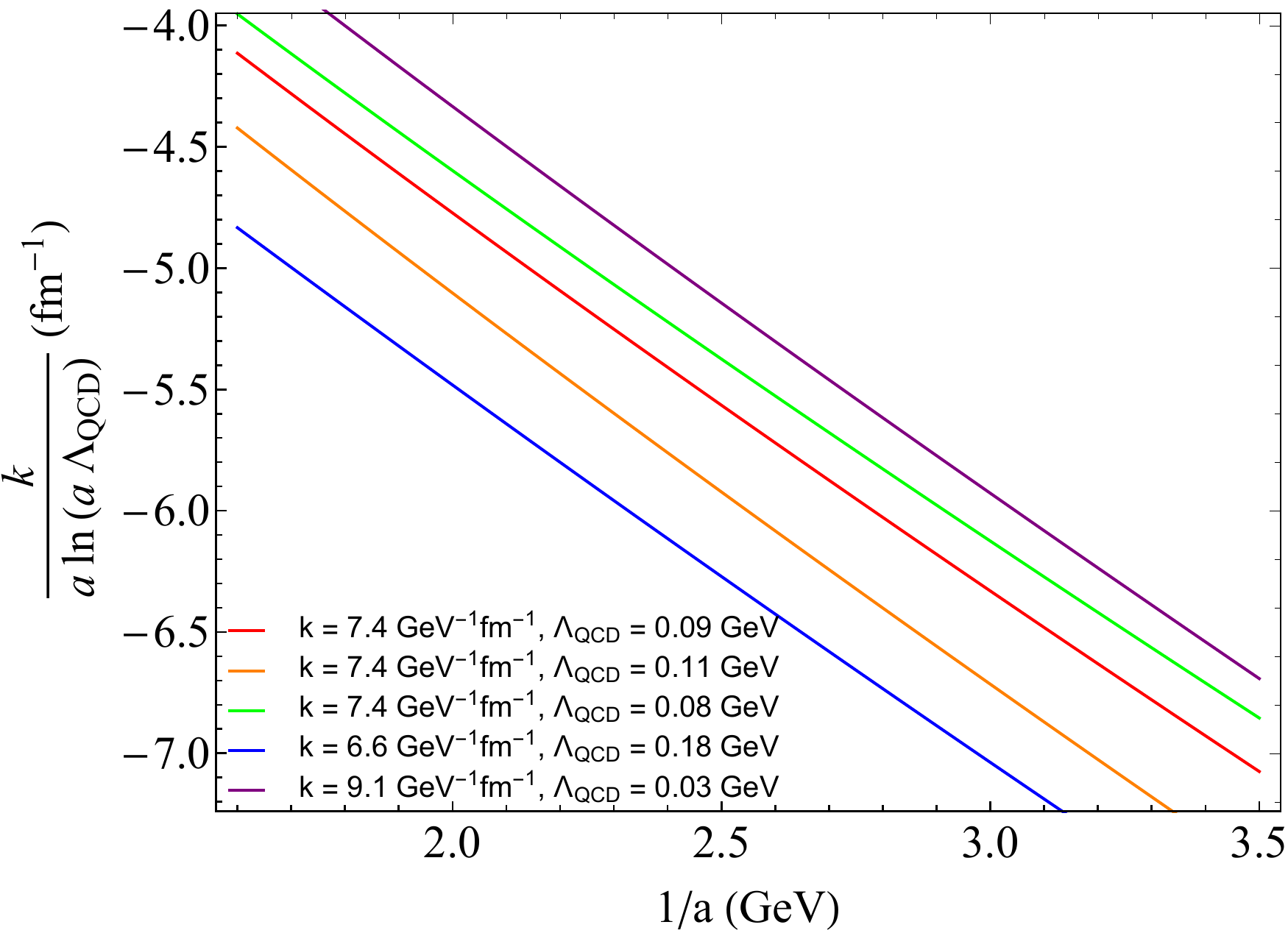}
\caption{ 
The linear divergent term in the $a$ range of our data for different sets of ($k$, $\Lambda_{\rm QCD}$) along the ``small-$\chi^{2}$ band''.
}
\label{fig:linearf}
\end{figure}

The reason for this uncertainty lies in the behavior of $\frac{k}{a \ln[a \Lambda_{\rm QCD}]}$ (the term related to the linear divergence in Eq.~(\ref{eq:logM})) in the $a$ range of our data. Varying along the "small-$\chi^2$ band'' effectively shifts $\frac{k}{a \ln[a \Lambda_{\rm QCD}]}$ by a constant $C$, see Fig.~\ref{fig:linearf}. The $Cz$ term is absorbed into $g(z)$ automatically during fitting so there is a large uncertainty for $k$ and $\Lambda_{\rm QCD}$. However, when we eliminate the $m_{0}$ effect, the $Cz$ term has been taken into account,  and the final result does not change. 

Another way to interpret this is that 
the perturbative value of $k$ is consistent with the ``small-$\chi^{2}$'' band 
and therefore is consistent with the fit. As a consequence,
we can fix $k$ at the one-loop perturbative value, 7.4 ${\rm GeV^{-1}fm^{-1}}$ (or 1.46 if dimensionless) in the subsequent analysis. Even with this
choice of $k$, $\Lambda_{\rm QCD}$ can still vary within a reasonable range of $\chi^2$ while the physical result is stabilized by the choice of $m_0$. We call this correlation between
$\Lambda_{\rm QCD}$ and $m_0$ the ``phenomenological renormalon uncertainty'', 
in the sense that $\Lambda_{\rm QCD}$ is characteristic for the different 
truncation/resummation schemes for the perturbation series, but that the non-perturbative $m_0$ compensates the effects of the differences~\cite{Ji:1995tm,Beneke:1998ui,Bauer:2011ws,Bali:2013pla}.

\subsection{Tuning Parameter-$d$ to Match the Continuum Scheme}\label{sec:d}

\begin{figure}[tbp]
\centering
\includegraphics[width=8cm]{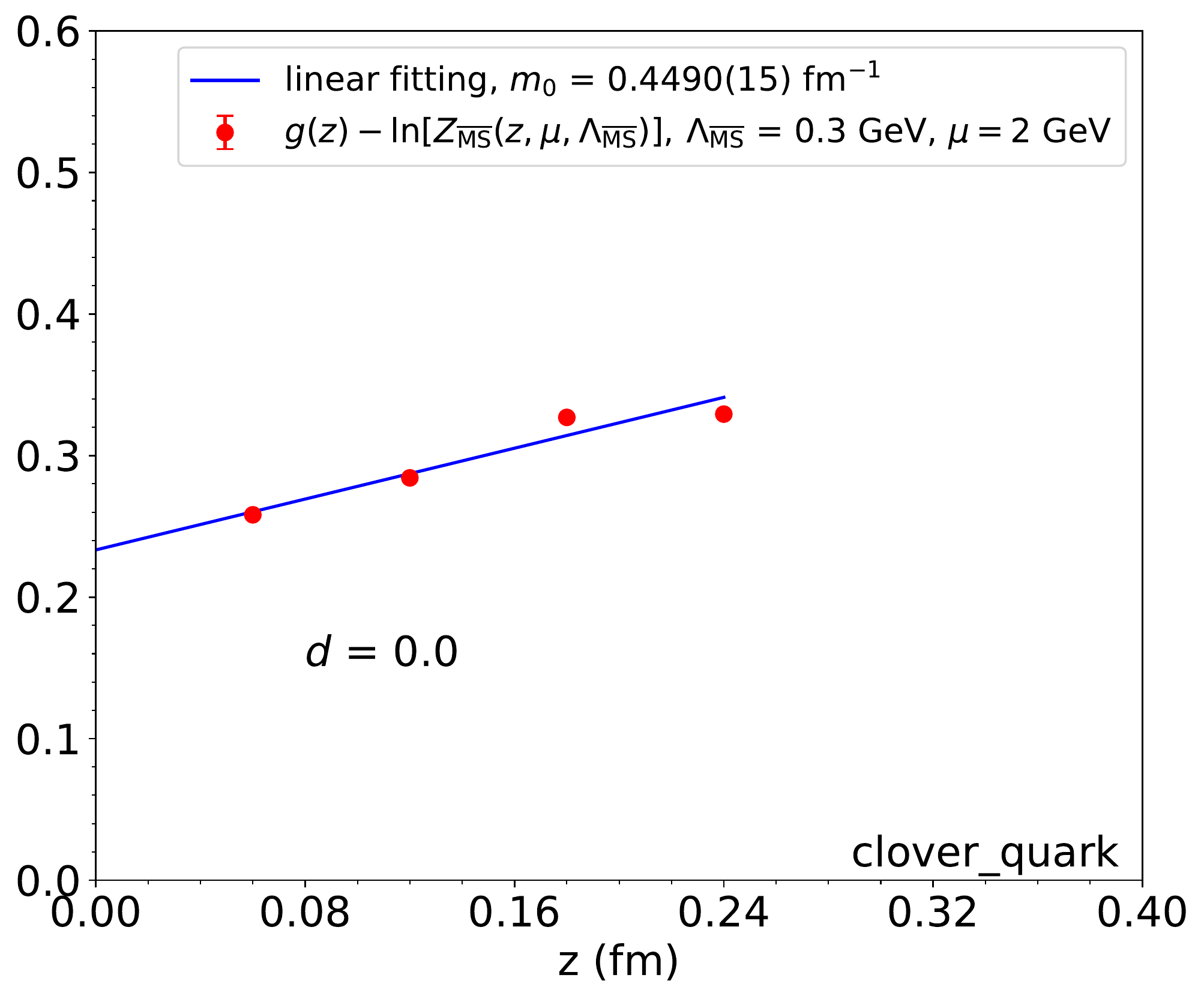}
\includegraphics[width=8cm]{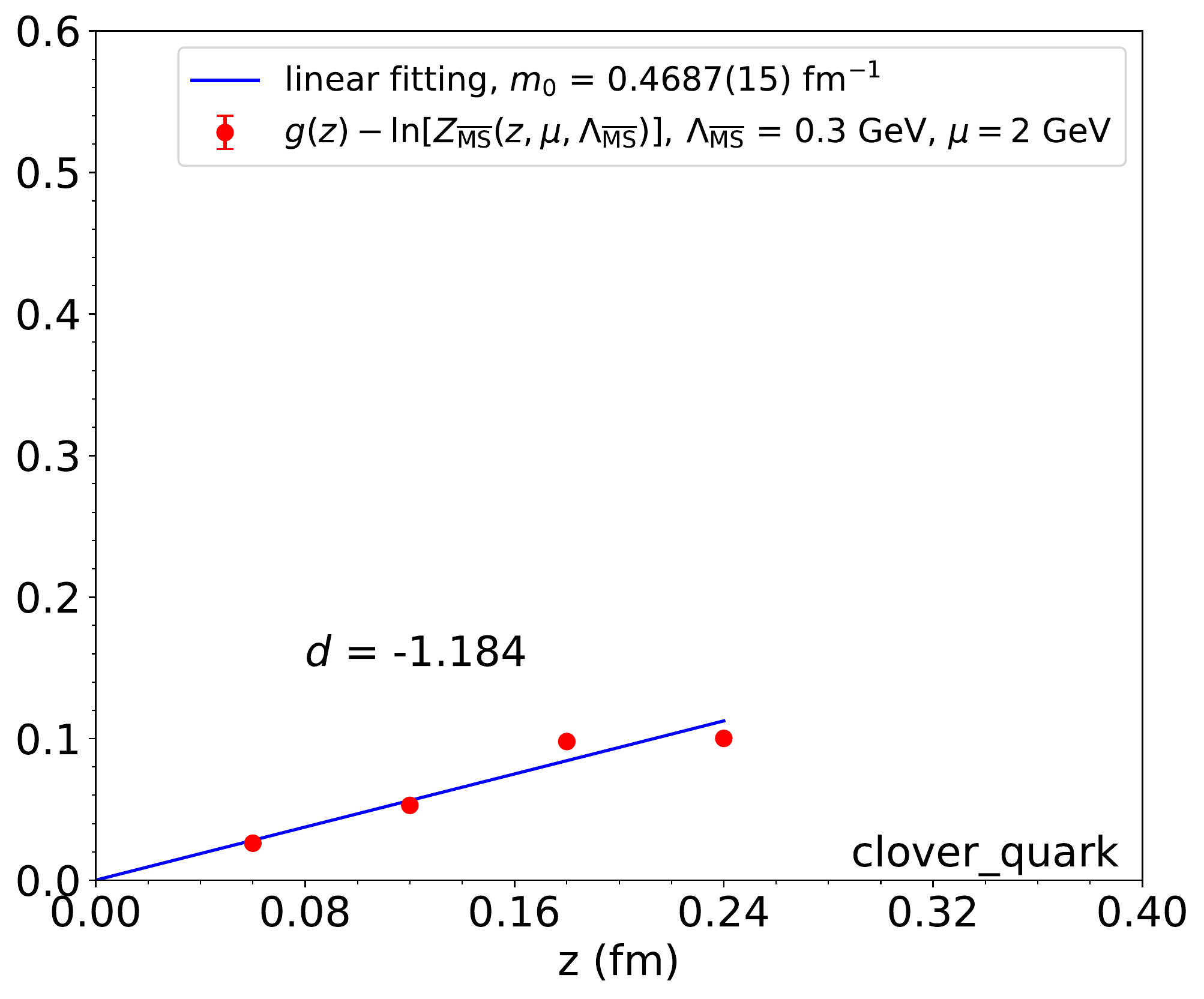}
\caption{
The fitting to extract $m_{0}$ based on Eq.~(\ref{eq:gztoMS}) for the bare $O_{\gamma_t}(z)$ matrix element in the off-shell quark state for the clover action without HYP smearing. Red points are $g(z) - \ln[Z_{\overline{\mathrm{MS}}}]$, where $g(z)$ is extracted through fitting the data for the bare matrix element using Eq.~(\ref{eq:logM}) and $Z_{\overline{\mathrm{MS}}}$ given in Eq.~(\ref{eq:ZMS}). The blue line is the linear fit of the red points (where the fitting function is $m_{0} z + C_0$). The fitted $m_{0}$ is given in the plot legend. $k$ is fixed at 7.4 GeV$^{-1}$fm$^{-1}$ (1.46 if dimensionless) and $\Lambda_{\rm QCD}$ is chosen as the best fit value 0.118 GeV.
In the upper panel, $d$ is zero and $g(z) - \ln[Z_{\overline{\mathrm{MS}}}]$ is in a linear relationship with $z$ but not proportional to $z$. (It does not go through the origin). In the lower panel, we tune $d$ to make sure that $g(z) - \ln[Z_{\overline{\mathrm{MS}}}]$ is proportional to $z$.
}
\label{fig:gz-lnZMS_d}
\end{figure}

In principle, the parameter $d$ in Eq.~(\ref{eq:logM}) is determined by the lattice action. In our data, there are two different dynamical lattice ensembles (MILC and RBC) and two different valence fermion actions (overlap and clover). Here we treat $d$ as a fine-tuning parameter to make sure that $( g(z) - \ln[Z_{\overline{\mathrm{MS}}}(z,\mu, \Lambda_{\overline{\mathrm{MS}}})] )$ in Eq.~(\ref{eq:gztoMS}) is proportional to $z$ within a window $a \ll z < 1/\mu$.
As shown in Fig.~\ref{fig:gz-lnZMS_d}, this is quite effective.
We have also tested that $d$ has little influence on $\chi^2$ since the effect of tuning $d$ shifts $g(z)$ simply by a constant.

\begin{figure}[tbp]
\centering
\includegraphics[width=8cm]{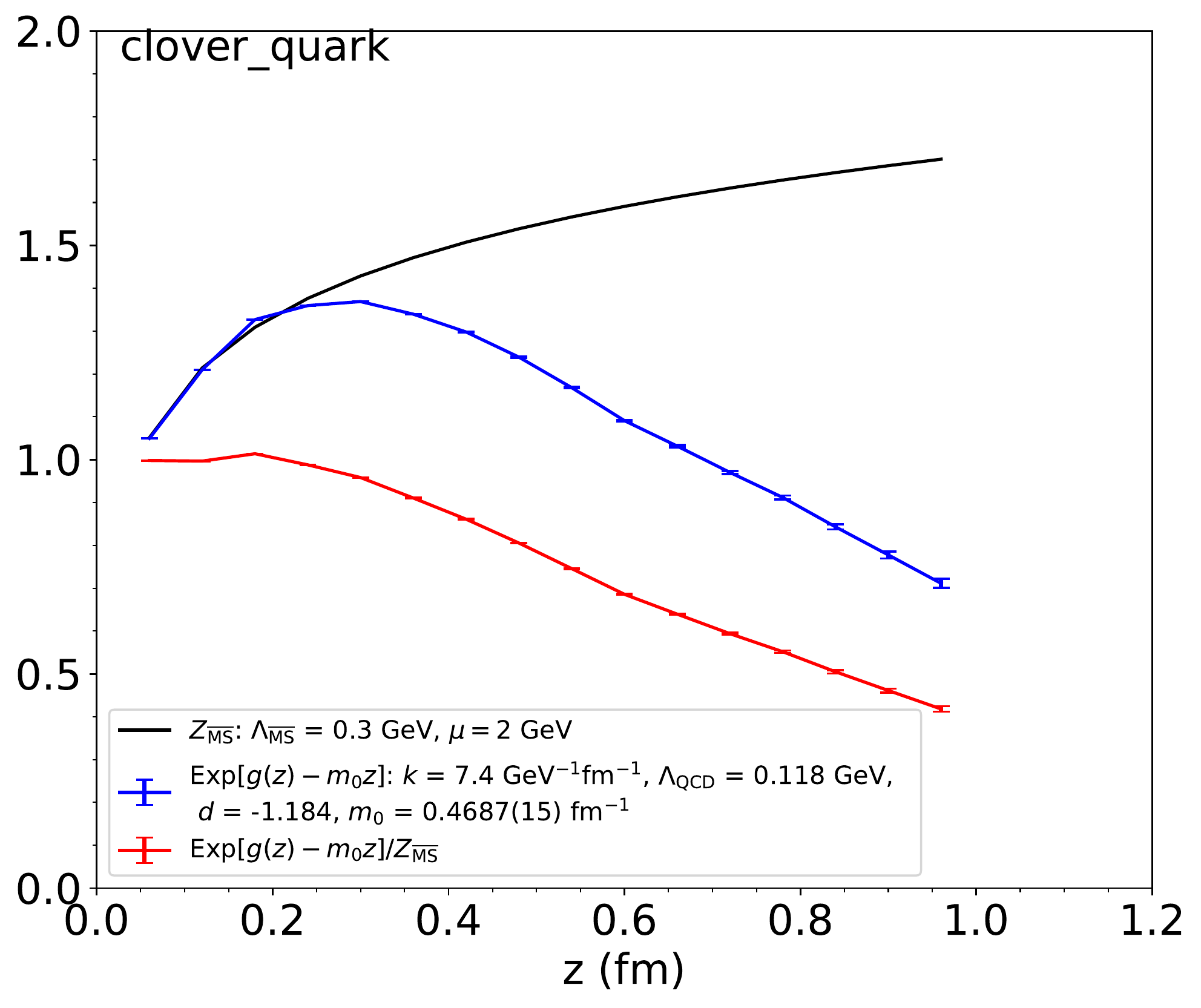}
\caption{
Renormalized $O_{\gamma_t}(z)$ matrix element in the off-shell quark state for the clover action without HYP smearing. Blue points are the renormalized matrix element Exp[$g(z)-m_{0}z$]. The black curve denotes $Z_{\overline{\mathrm{MS}}}$ and the red points are their ratio. $k$ is fixed at 7.4 GeV$^{-1}$fm$^{-1}$ (1.46 if dimensionless) and $\Lambda_{\rm QCD}$ is chosen as the best fitted value 0.118 GeV. The parameter $d$ is fine-tuned to be $-1.184$.
}
\label{fig:nonpert_clover_quark}
\end{figure}

Fig.~\ref{fig:nonpert_clover_quark} shows that the renormalized matrix element Exp[$g(z)-m_{0}z$] (blue) for clover quarks is consistent with $Z_{\overline{\mathrm{MS}}}$ (black) at small $z$. However, at large $z$, there is a significant discrepancy between Exp[$g(z)-m_{0}z$] and $Z_{\overline{\mathrm{MS}}}$. That means that there is a significant non-perturbative effect at large $z$ in the popular RI/MOM and ratio renormalization schemes used previously.
To our knowledge, this is the first time that this difference has been studied 
in the literature. This finding supports the usage of the hybrid renormalization procedure proposed recently~\cite{Ji:2020brr}.

\subsection{Self-Renormalization Procedure}\label{sec:reofpro}

The procedure developed in the previous subsections 
forms our self-renormalization strategy. This procedure can overcome
some of the problems encountered in the renormalization
of the linear divergence due to the numerical uncertainties associated
with exponentially-amplified small errors. 
There are three steps in this process:

1. Use Eq.~(\ref{eq:logM}) to fit the data of the bare matrix elements $\mathcal{M}$. For each $z$ (here $z$ can be chosen in a large range and linear interpolation with respect to $z$ might be needed 
for the data in ln$\mathcal{M}(z,a)$ ), fit the dependence on $a$, treating $\Lambda_{\rm QCD}$ as a global parameter, namely the same for all $z$, and $g(z)$, $f_{1}(z)$, $f_{2}(z)$ as fit parameters, with fixing $k$=7.4 GeV$^{-1}$fm$^{-1}$ (or 1.46 if dimensionless), and $\mu$=2 GeV. Fine-tune $d$ to make sure that $g(z) - \ln[Z_{\overline{\mathrm{MS}}}]$ is proportional to $z$ within a window $a \ll z < 1/\mu$. We obtain the residual $g(z)$ through fitting;

2. Use Eq.~(\ref{eq:gztoMS}) to fit the dependence on $z$ (within a window $a \ll  z < 1/\mu$) to extract $m_{0}$. Then calculate the renormalized matrix element Exp[$g(z)-m_{0}z$] (Eq.~(\ref{eq:gz-m0})), which is considered valid for a large range of $z$;

3. Calculate $\mathcal{M}(z,a)/Z(z,a)_{R}$ (Eq.~(\ref{eq:M/Z})) to see if showing any dependence on the lattice constant $a$ and compare with Exp[$g(z)-m_{0}z$]. 

In step 1 and 2, one can use another way to get $d$ and $m_{0}$: Modify Eq.~(\ref{eq:logM}) through replacing $g(z)$ with ($\ln[Z_{\overline{\mathrm{MS}}}(z,\mu, \Lambda_{\overline{\mathrm{MS}}})] + m_{0} z$), and then use the modified fitting function to fit the bare matrix element at small $z$ to extract $d$ and $m_{0}$. This way should be equivalent to the procedure outlined above and we will not follow it here.

In step 3, we need to calculate $Z(z,a)_{R}$ as well as estimate its error. We shall not propagate the fitting error of $\Lambda_{\rm QCD}$ to $Z(z,a)_{R}$ because the effect of slight variation of $\Lambda_{\rm QCD}$ can be compensated by $m_{0}$ and will not influence our result as we have discussed in Sec.~\ref{ReUncer}.

Since the lattice data we use do not provide systematic errors, we will add a dummy systematic error $\delta_{\rm sys}$ to the data during our fitting~\cite{Zhang:2020rsx}
\bea\label{eq:error}
(\sigma_{\ln \mathcal{M}})_{\rm new} = \sqrt{(\sigma_{\ln \mathcal{M}})_{\rm old}^{2} + (\delta_{\rm sys} a \mu)^{2}},
\eea
where we assume that there is an $a\mu$ dependence as well.
This choice is intended to give more weight to the small lattice spacing 
data in the fitting.

\subsection{Comparing Results from Different Actions and Matrix Elements}

In this subsection, we apply the self-renormalization method to the $O_{\gamma_t}(z)$ matrix 
elements in the pion, nucleon, and off-shell quark state with valence clover and overlap actions 
without HYP smearing. We compare the renormalization factors and physics results. 
We find that apart from the case of the RI/MOM matrix element with clover valence fermion, all other renormalization factors are similar. Despite this difference in renormalization factor, the physical
results are independent of fermion actions.

\begin{figure*}[tbp]
\centering
\subfigure[]{\includegraphics[width=8cm]{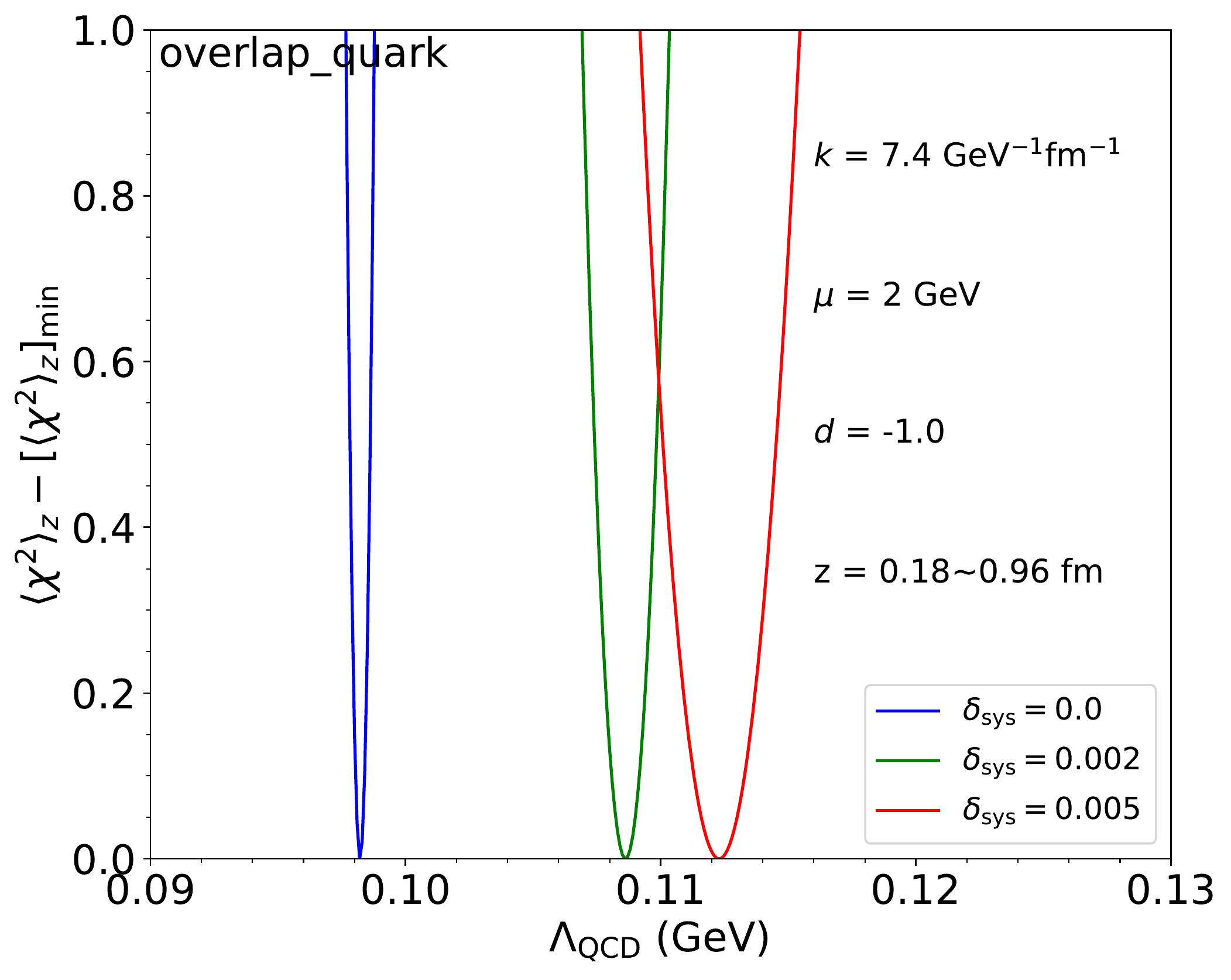}}
\subfigure[]{\includegraphics[height=5.5cm,width=8cm]{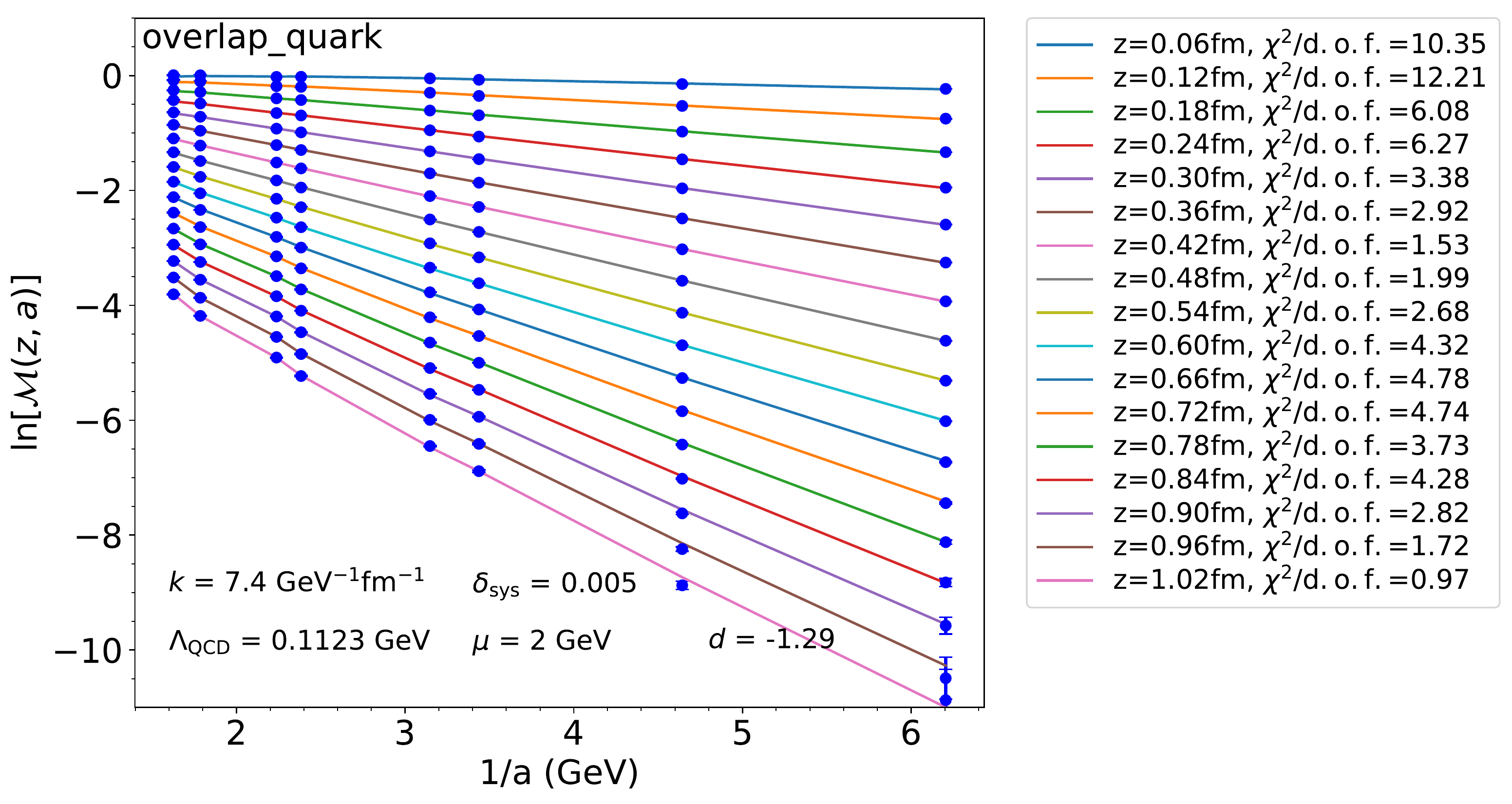}}
\subfigure[]{\includegraphics[width=8cm]{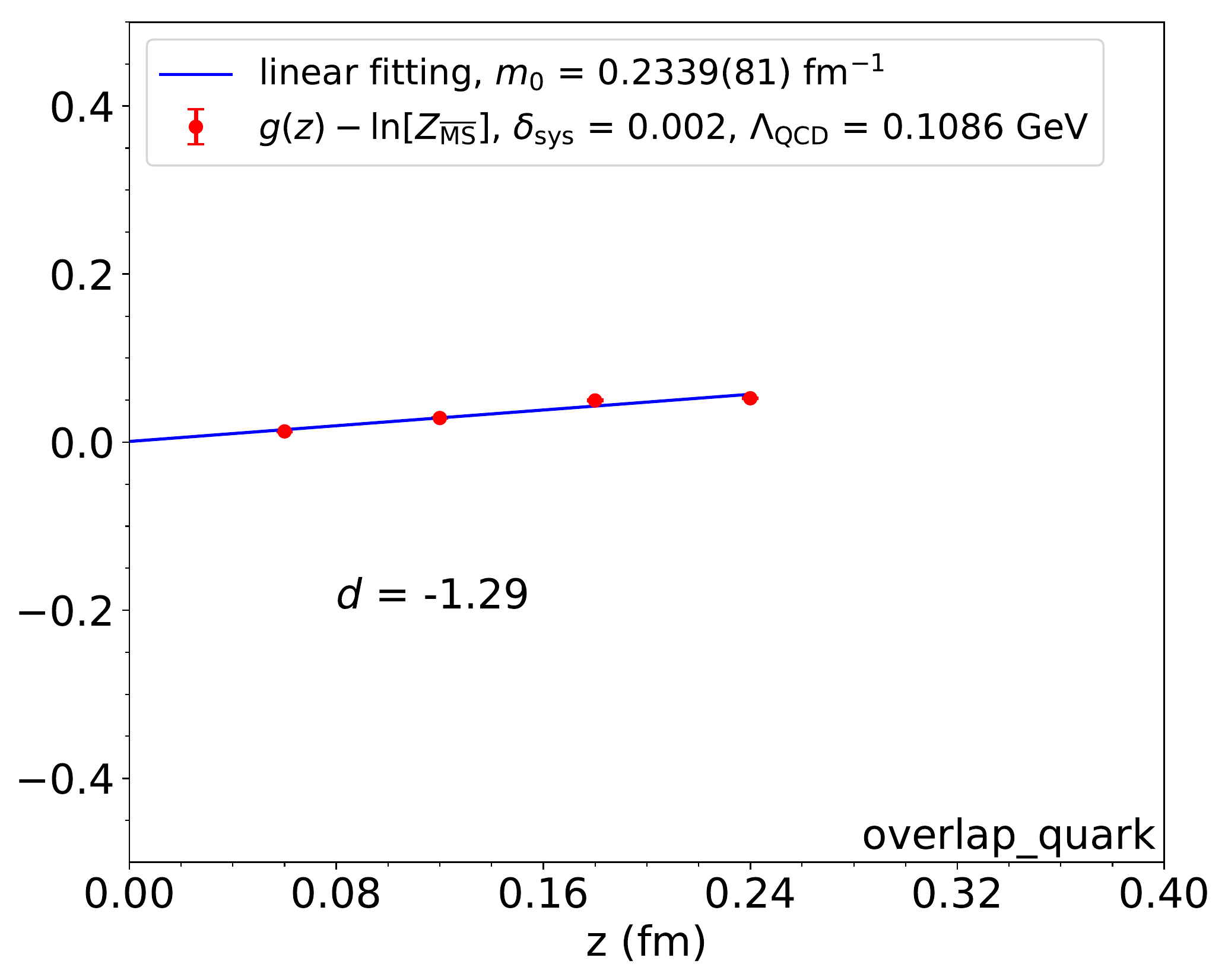}}
\subfigure[]{\includegraphics[width=7.5cm]{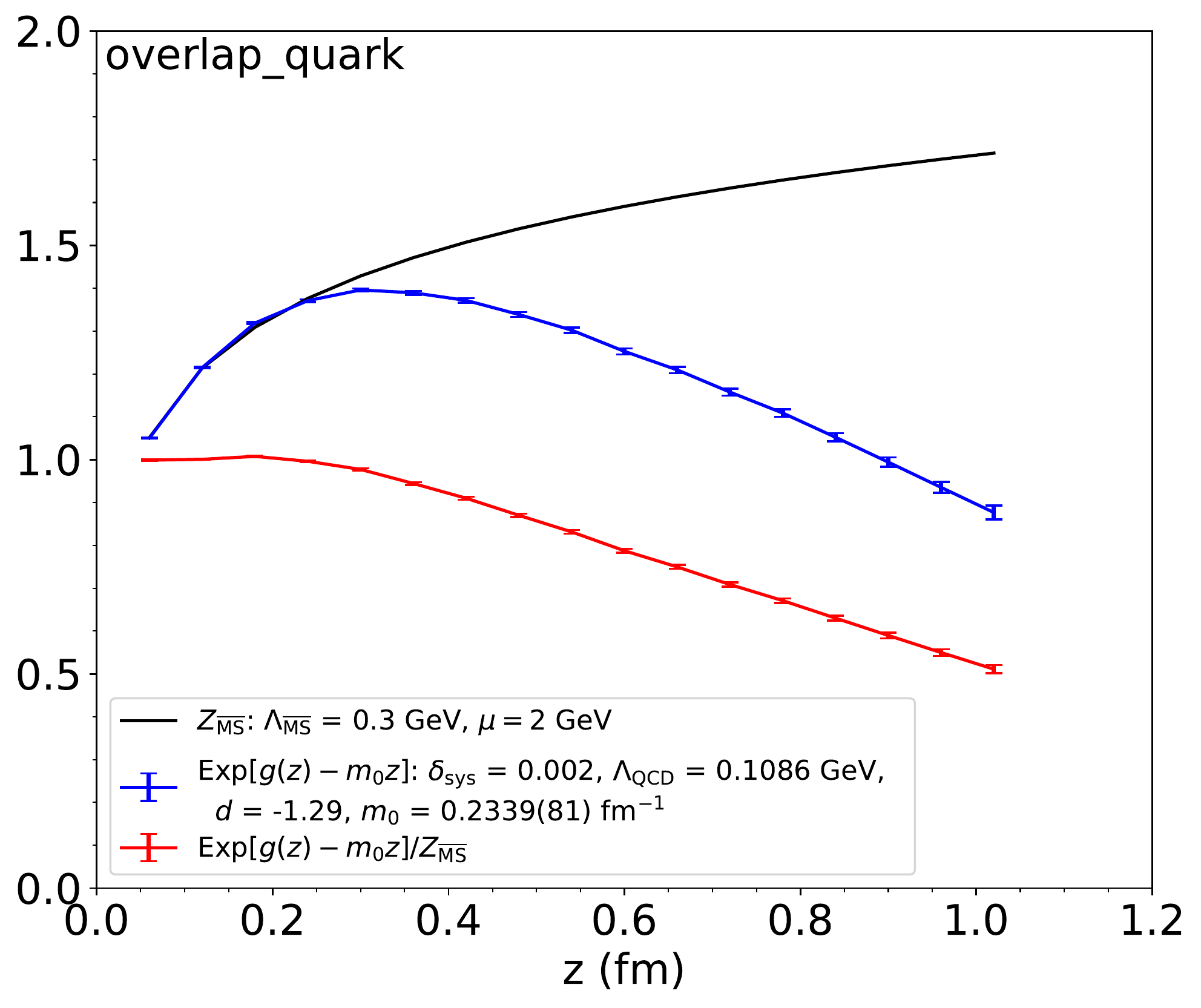}}
\subfigure[]{\includegraphics[width=8cm]{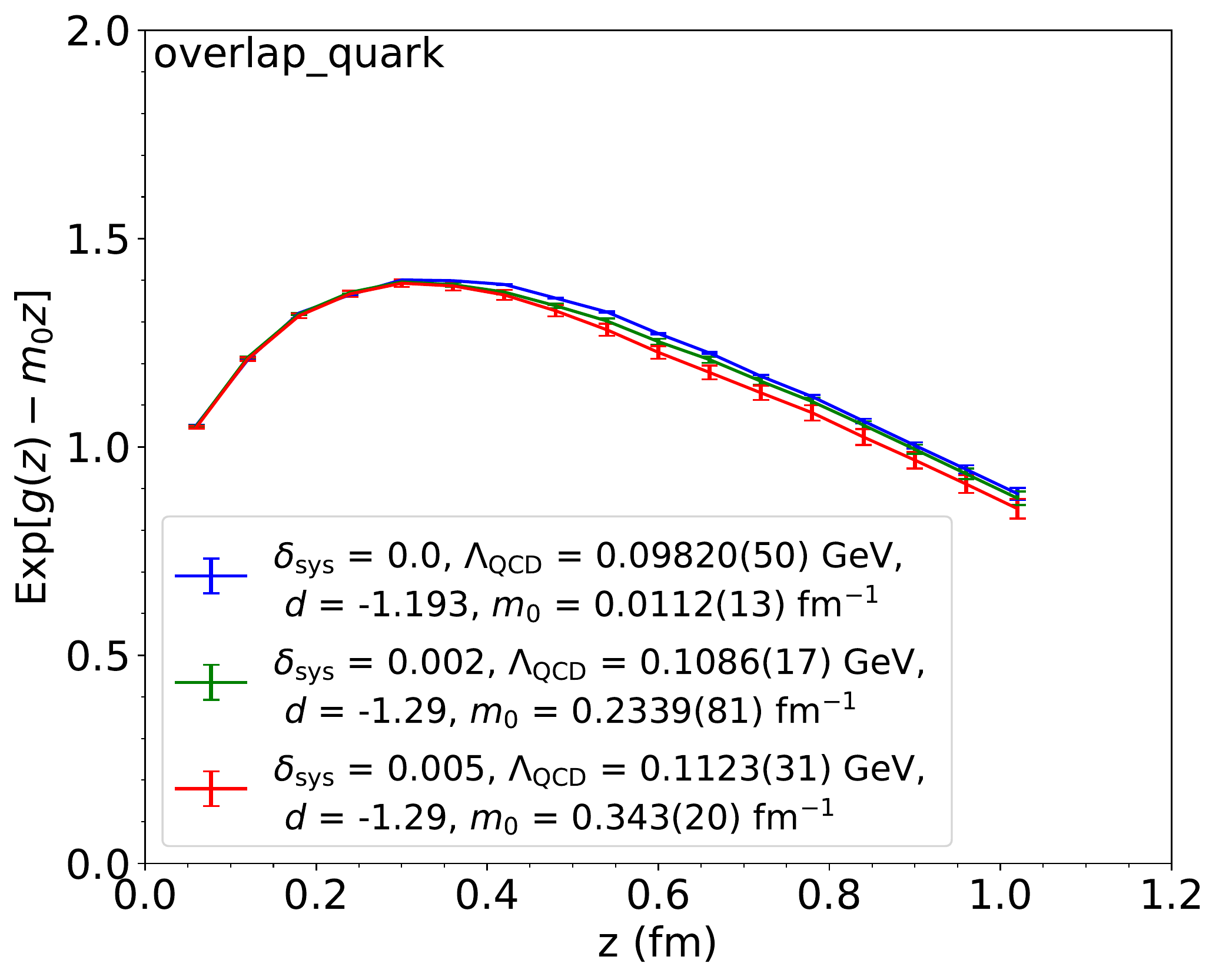}}
\subfigure[]{\includegraphics[width=8cm]{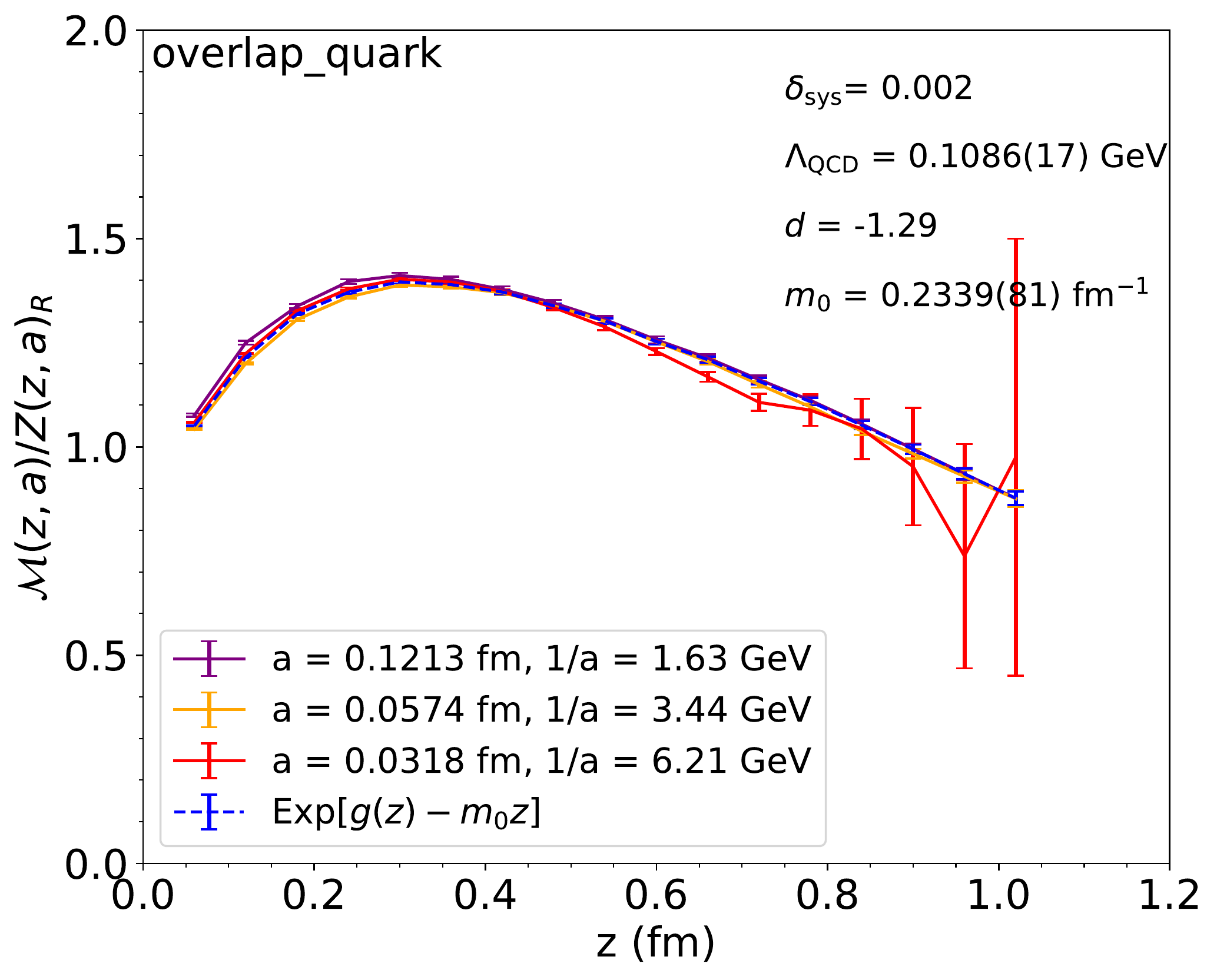}}
\caption{
Self-renormalizing the $O_{\gamma_t}(z)$ correlator in the off-shell quark state calculated for the overlap action without HYP smearing, using Eq.~(\ref{eq:logM}). (a) shows $\langle  \chi^{2} \rangle_{z} - [\langle  \chi^{2} \rangle_{z}]_{\rm min}$ with respect to $\Lambda_{\rm QCD}$. The best fitted $\Lambda_{\rm QCD}$ with its error can be estimated here. (b), (c) and (d) show the procedure to get the renormalized matrix element Exp[$g(z)-m_{0}z$]. (b) shows the fitting of the bare matrix element to extract $g(z)$. Blue points are interpolated data and colorful curves are fitted curves. (c) shows the fitting to extract $m_{0}$. (d) shows Exp[$g(z)-m_{0}z$] (blue), compared with $Z_{\overline{\mathrm{MS}}}$ (black). (e) is the comparison of Exp[$g(z)-m_{0}z$] for different $\delta_{\rm sys}$. (f) is the ratio of bare matrix element and renormalization factor (connected with solid lines), compared with Exp[$g(z)-m_{0}z$] (connected with dashed lines).
}
\label{fig:Re_overlap_quark}
\end{figure*}

\begin{figure*}[tbp]
\centering
\subfigure[]{\includegraphics[width=8cm]{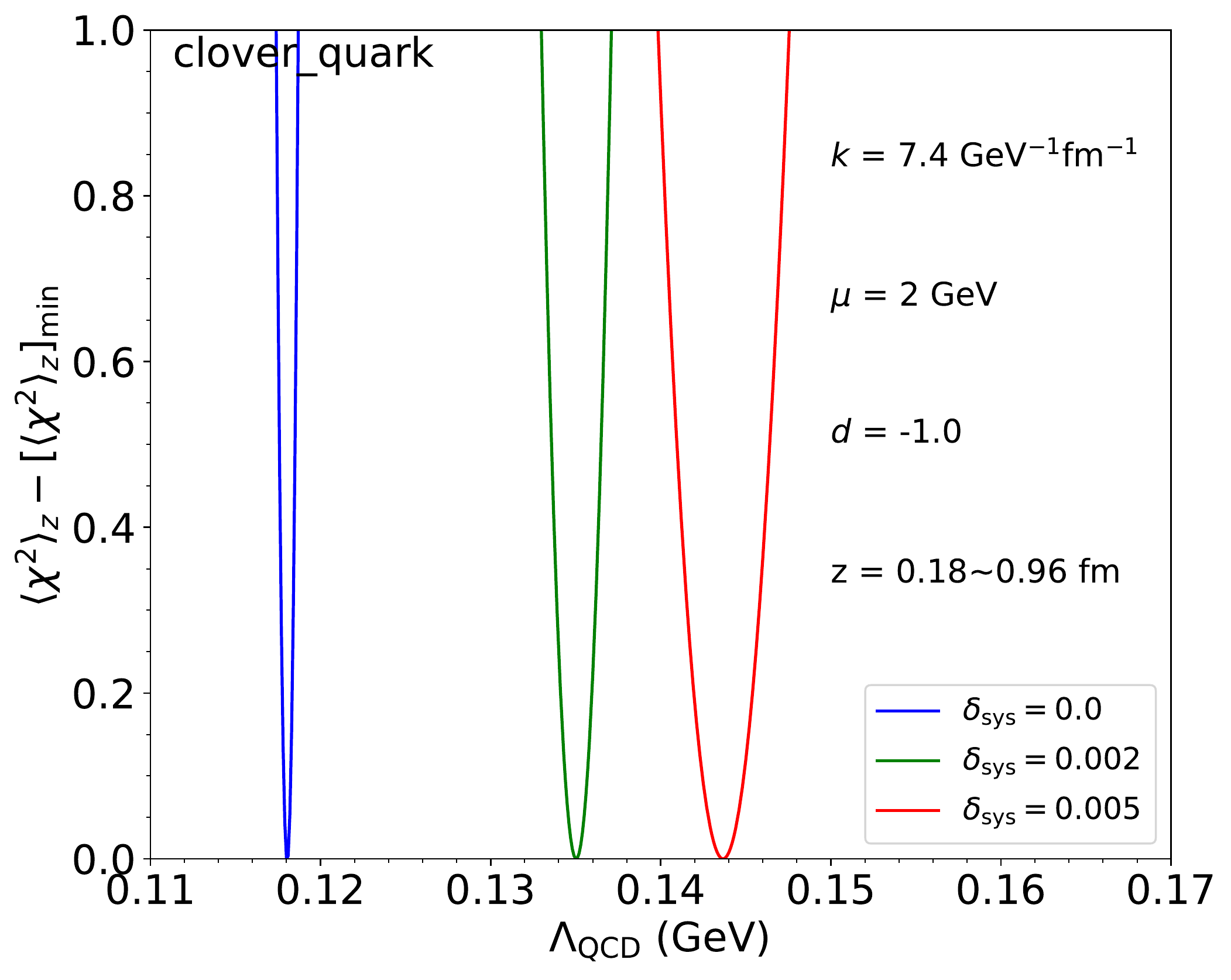}}
\subfigure[]{\includegraphics[height=5.5cm,width=8cm]{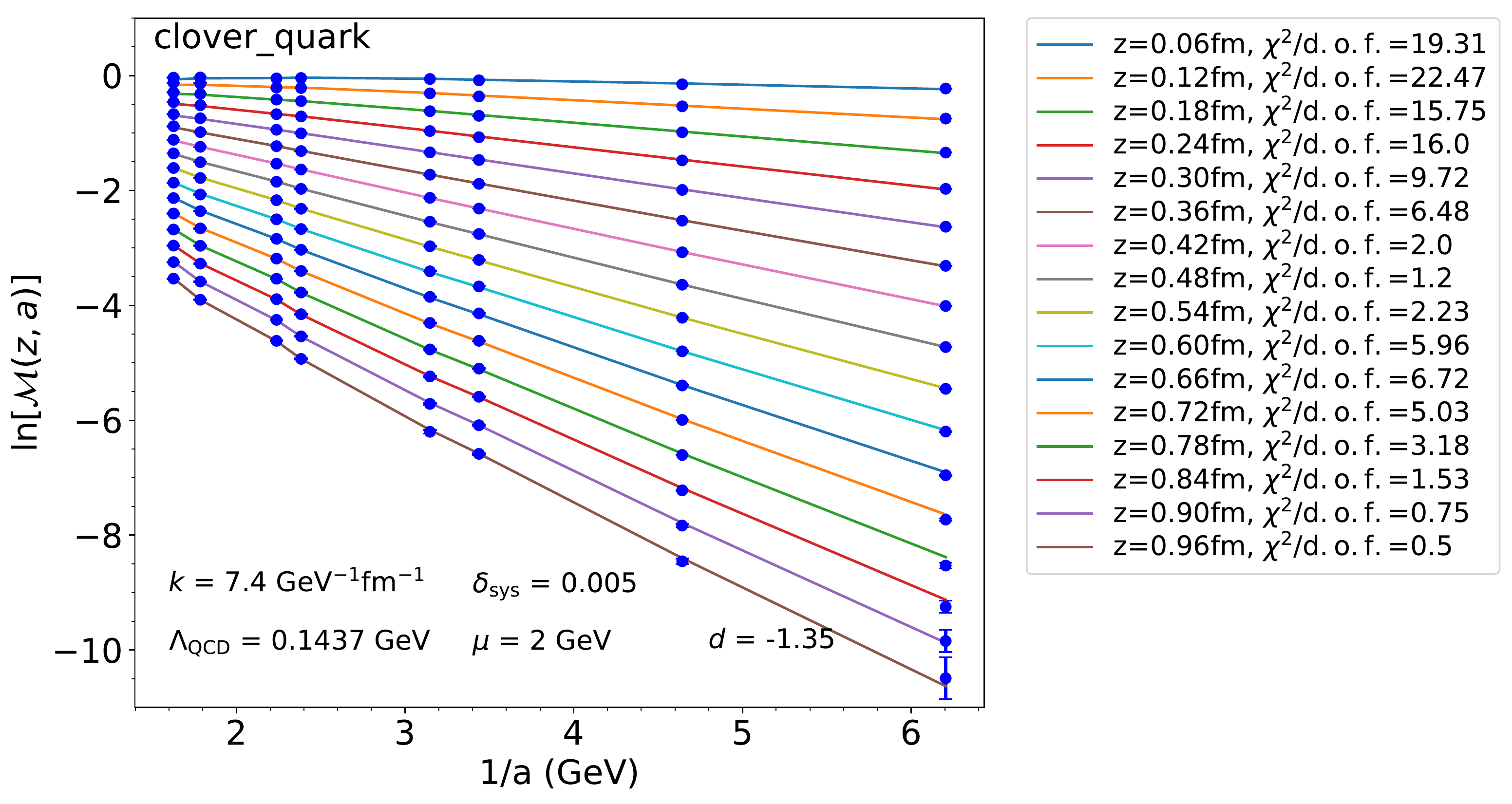}}
\subfigure[]{\includegraphics[width=8cm]{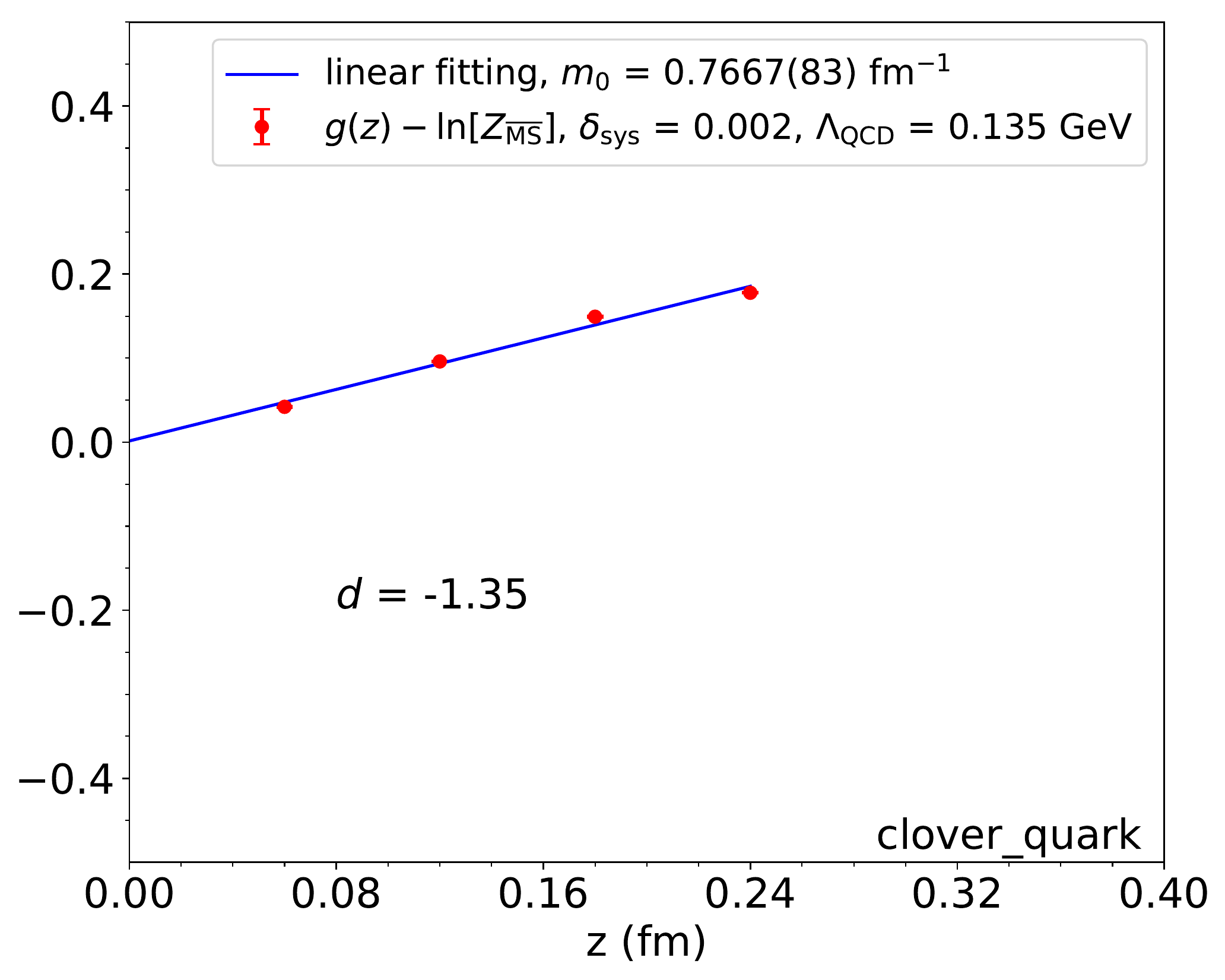}}
\subfigure[]{\includegraphics[width=7.5cm]{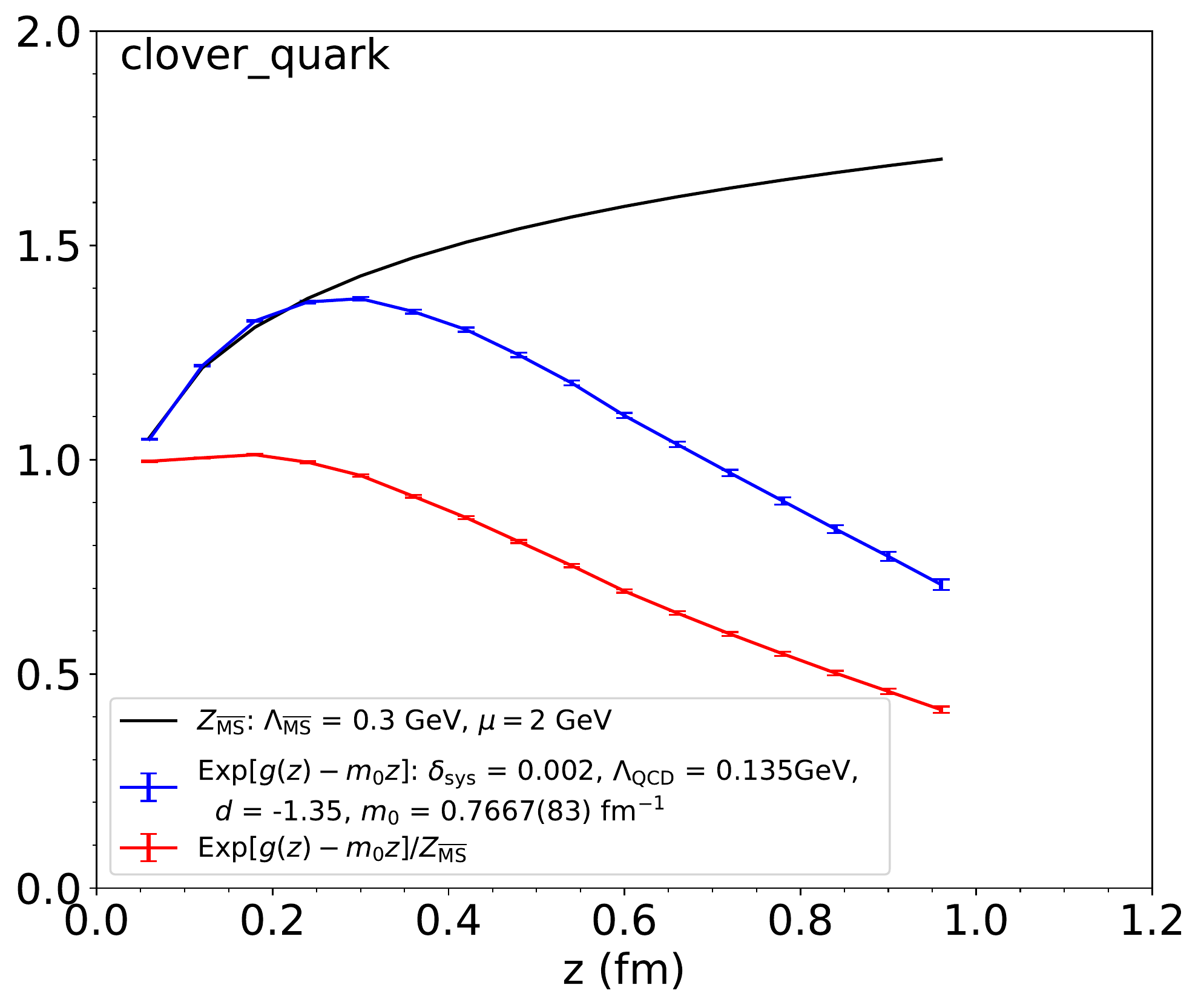}}
\subfigure[]{\includegraphics[width=8cm]{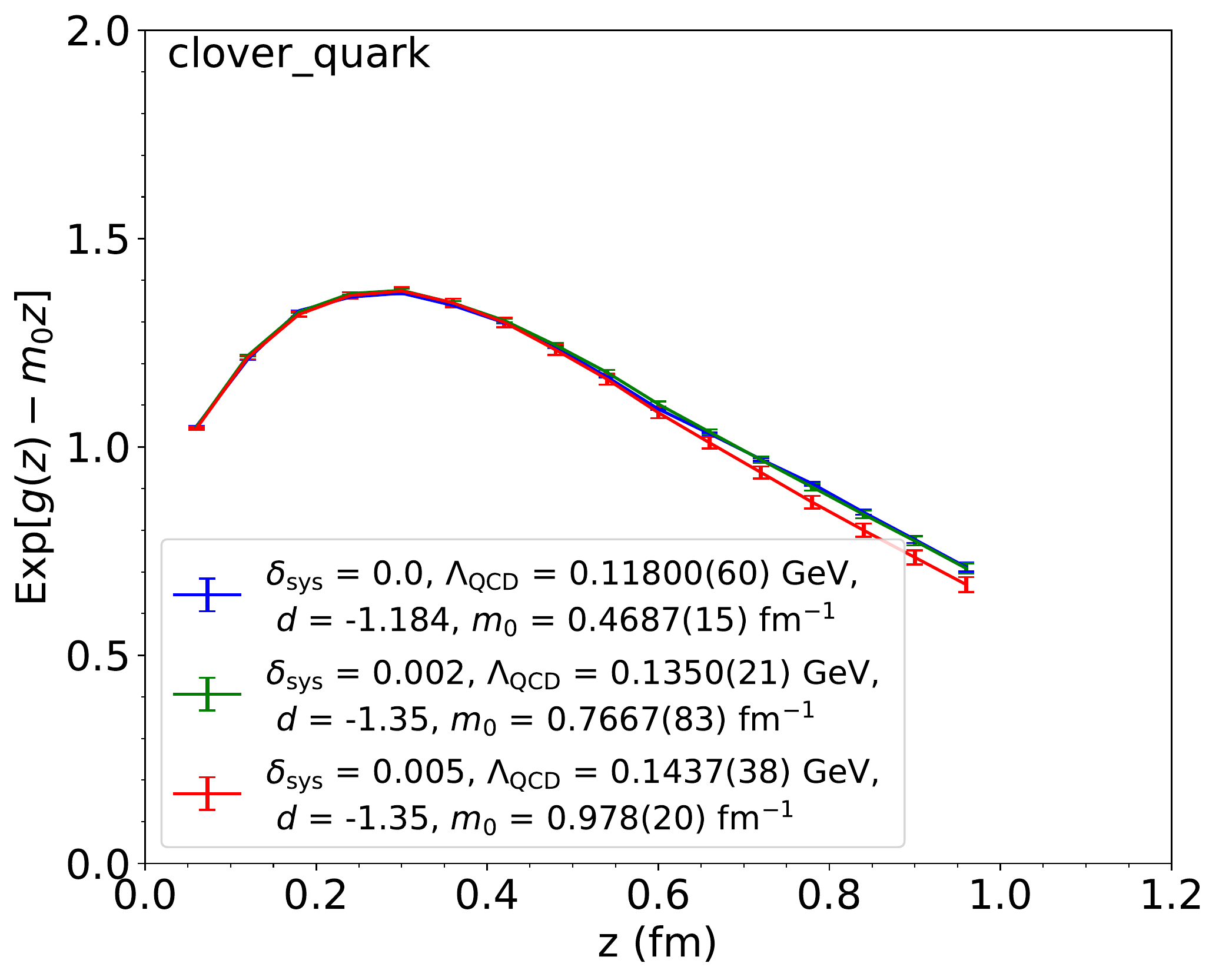}}
\subfigure[]{\includegraphics[width=8cm]{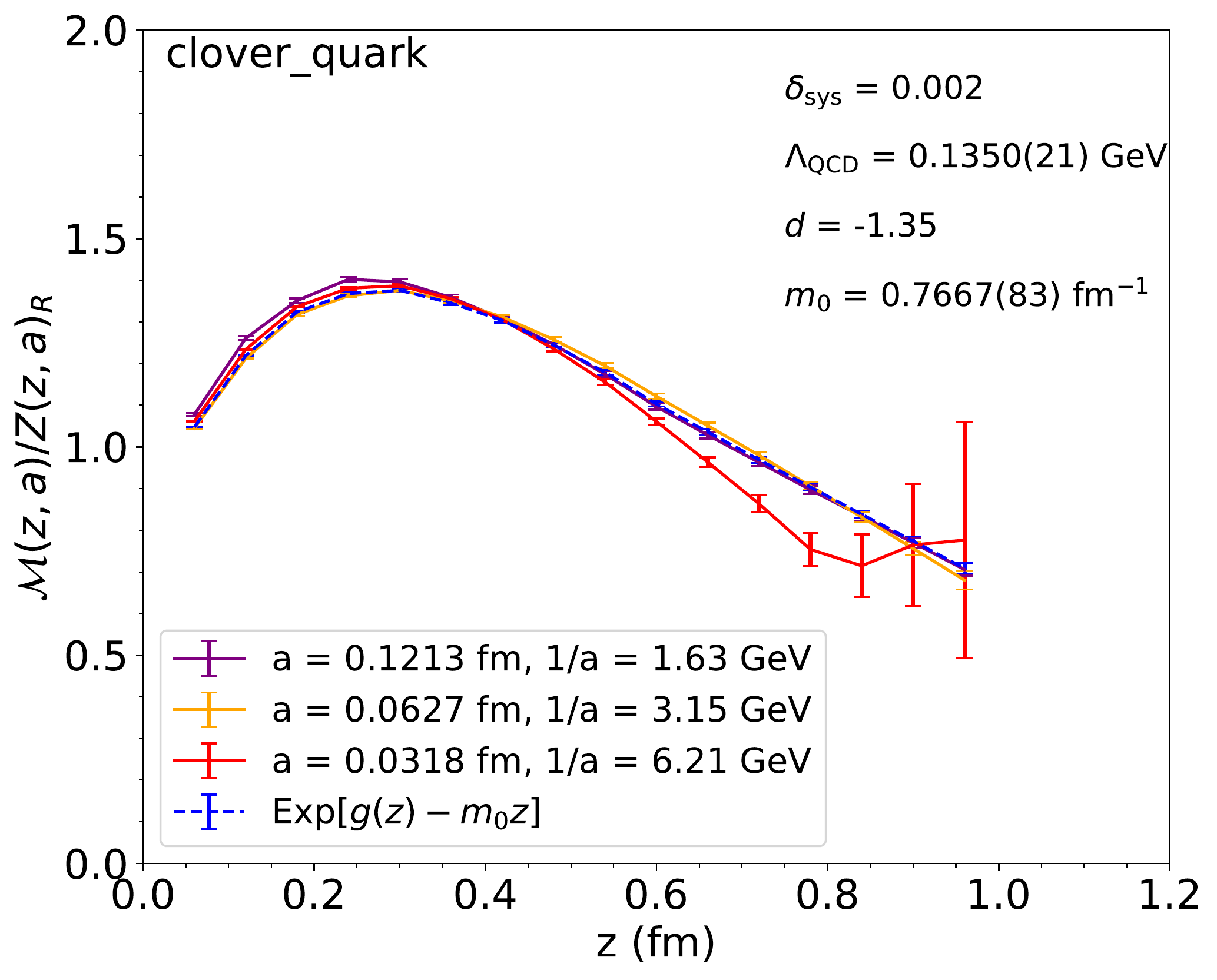}}
\caption{
Same as Fig.~\ref{fig:Re_overlap_quark}, with the $O_{\gamma_t}(z)$ correlator in the off-shell quark state calculated for the clover action.
}
\label{fig:Re_clover_quark}
\end{figure*}

\begin{figure*}[tbp]
\centering
\subfigure[]{\includegraphics[width=8cm]{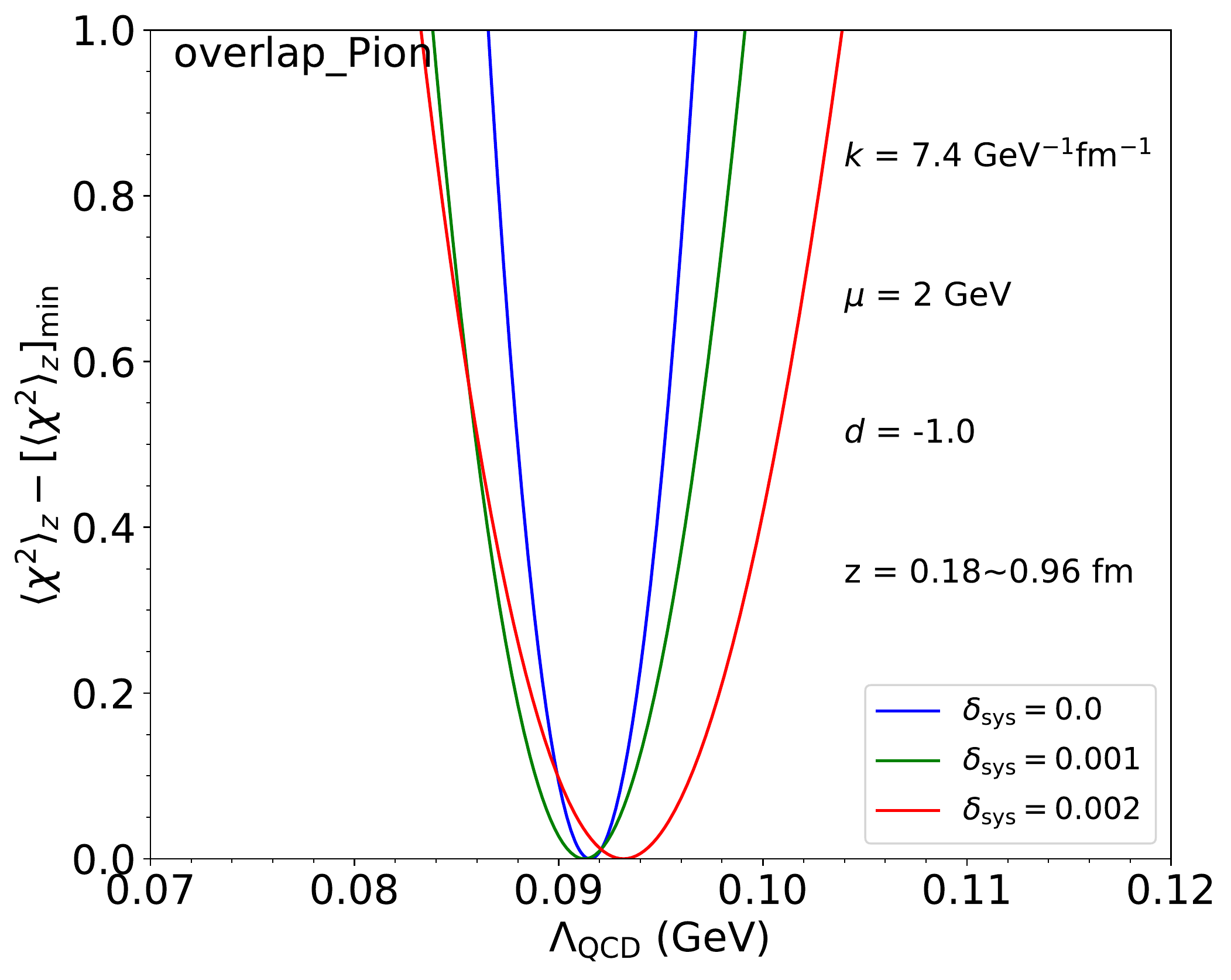}}
\subfigure[]{\includegraphics[height=5.5cm,width=8cm]{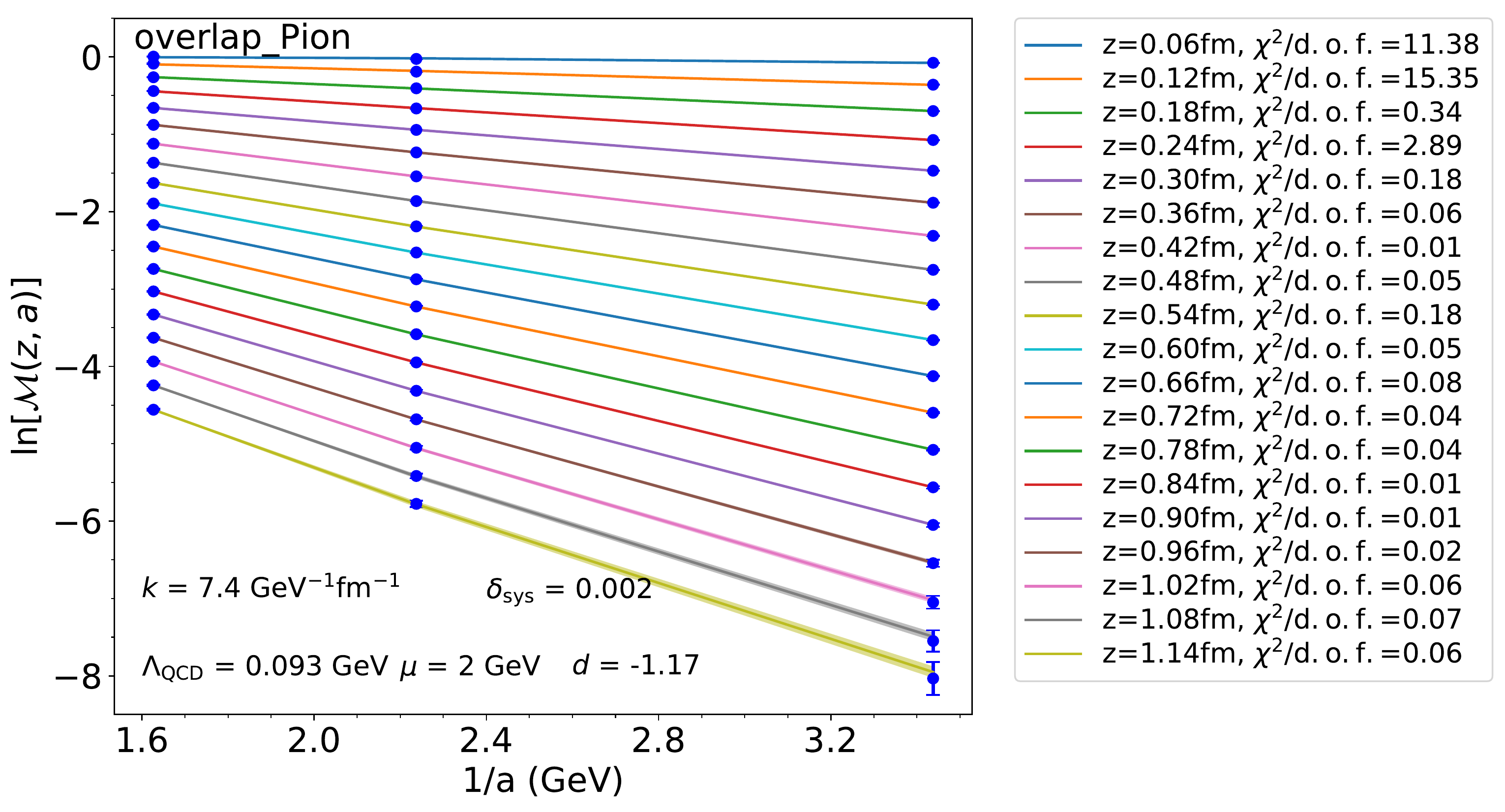}}
\subfigure[]{\includegraphics[width=8cm]{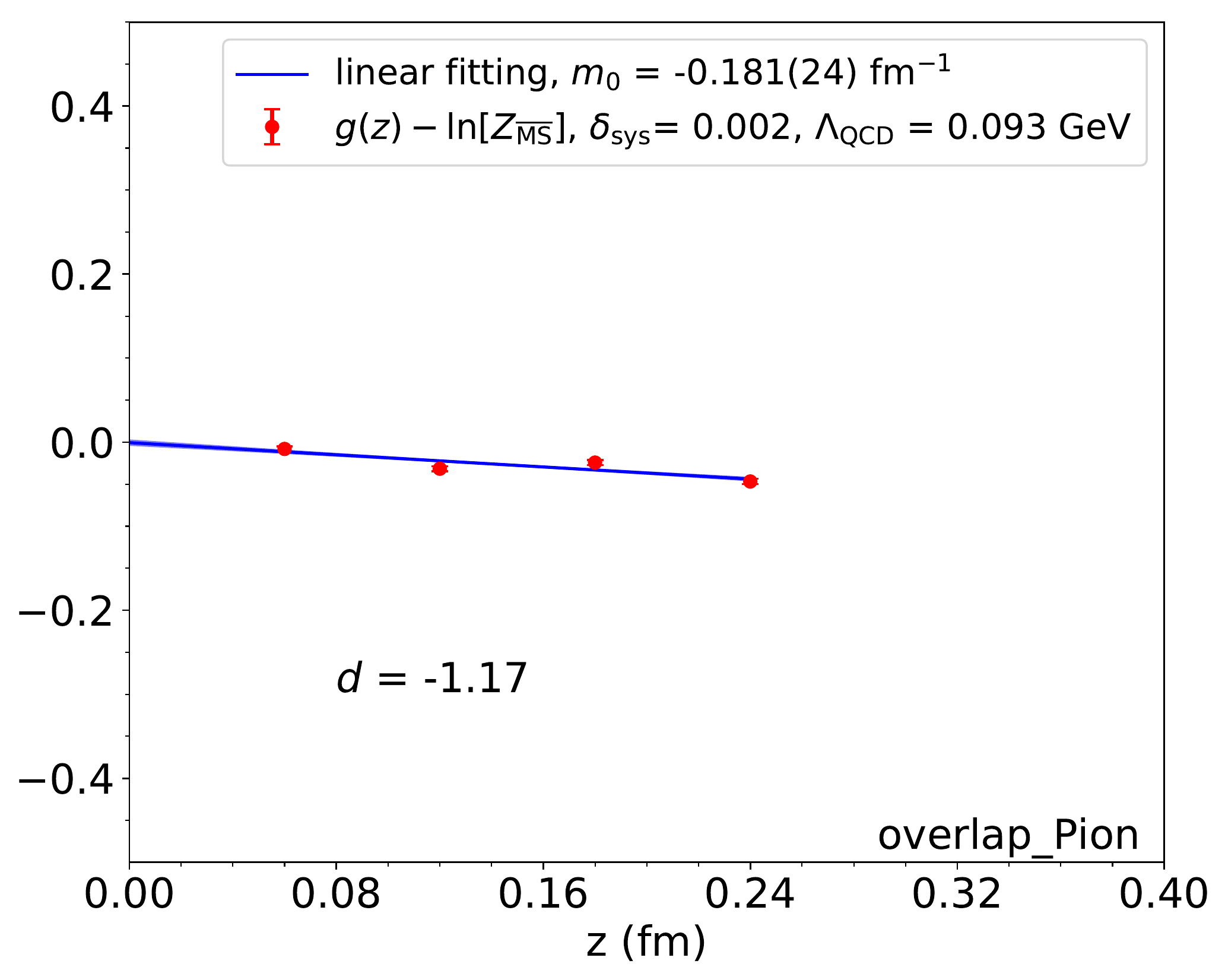}}
\subfigure[]{\includegraphics[width=7.5cm]{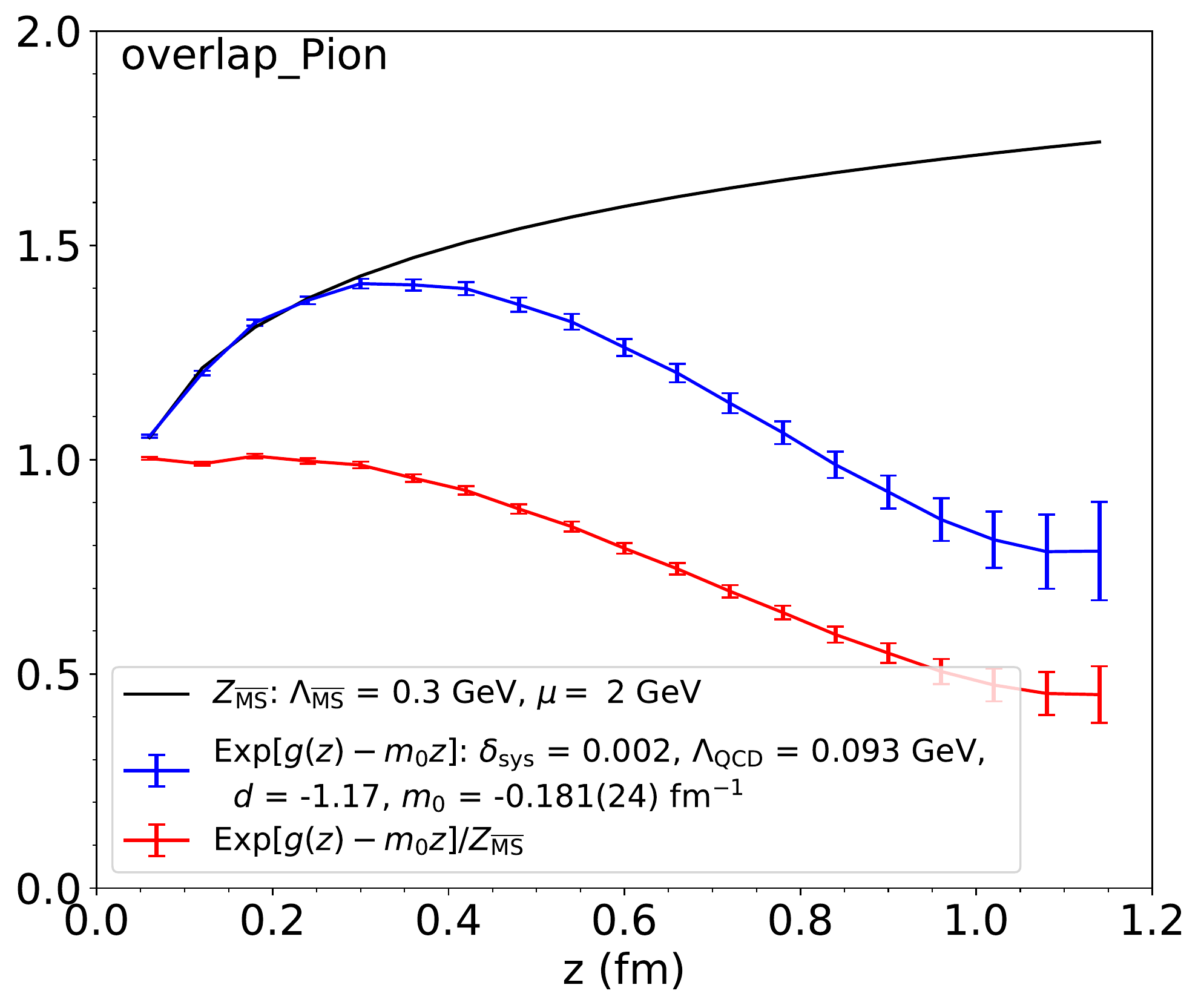}}
\subfigure[]{\includegraphics[width=8cm]{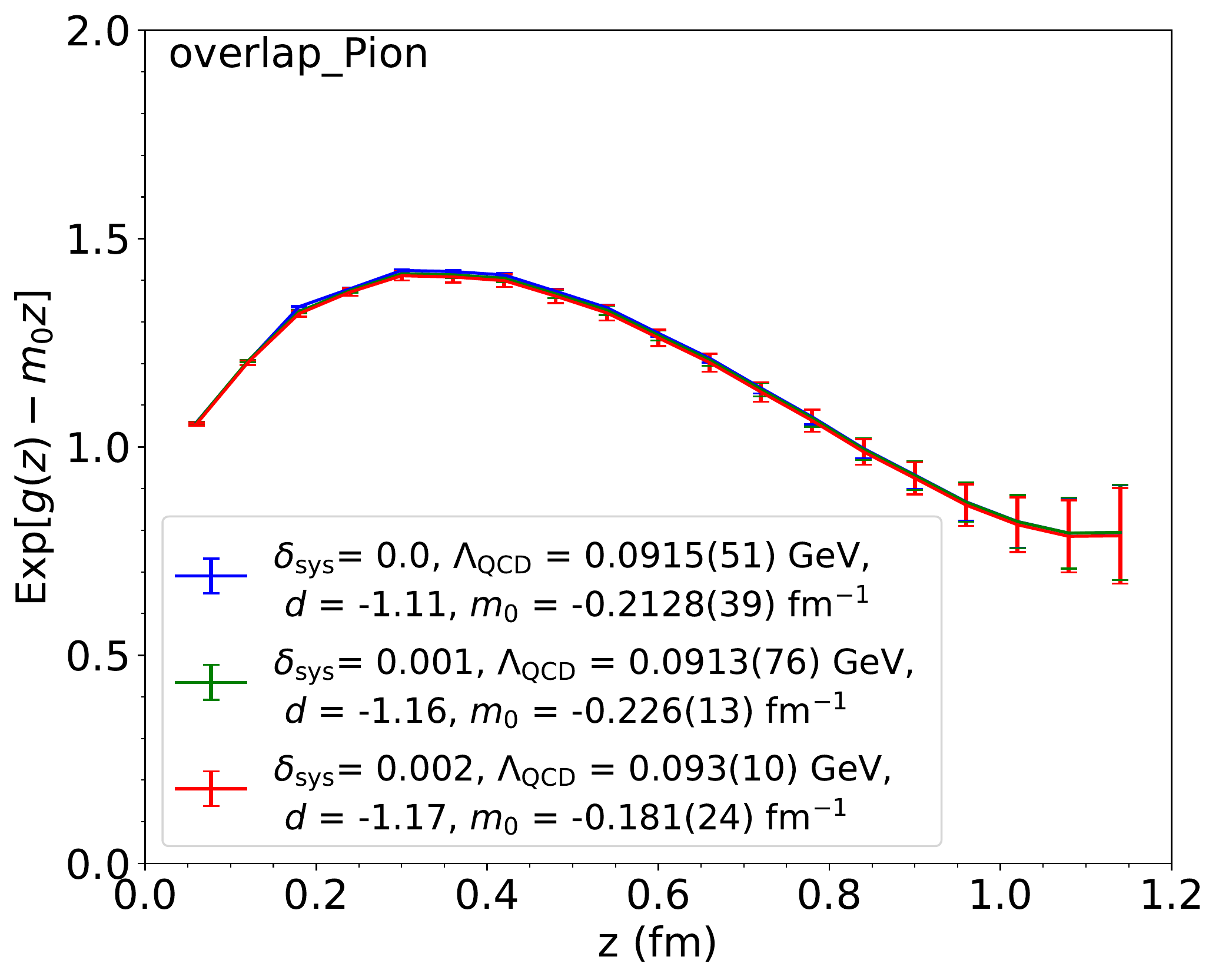}}
\subfigure[]{\includegraphics[width=8cm]{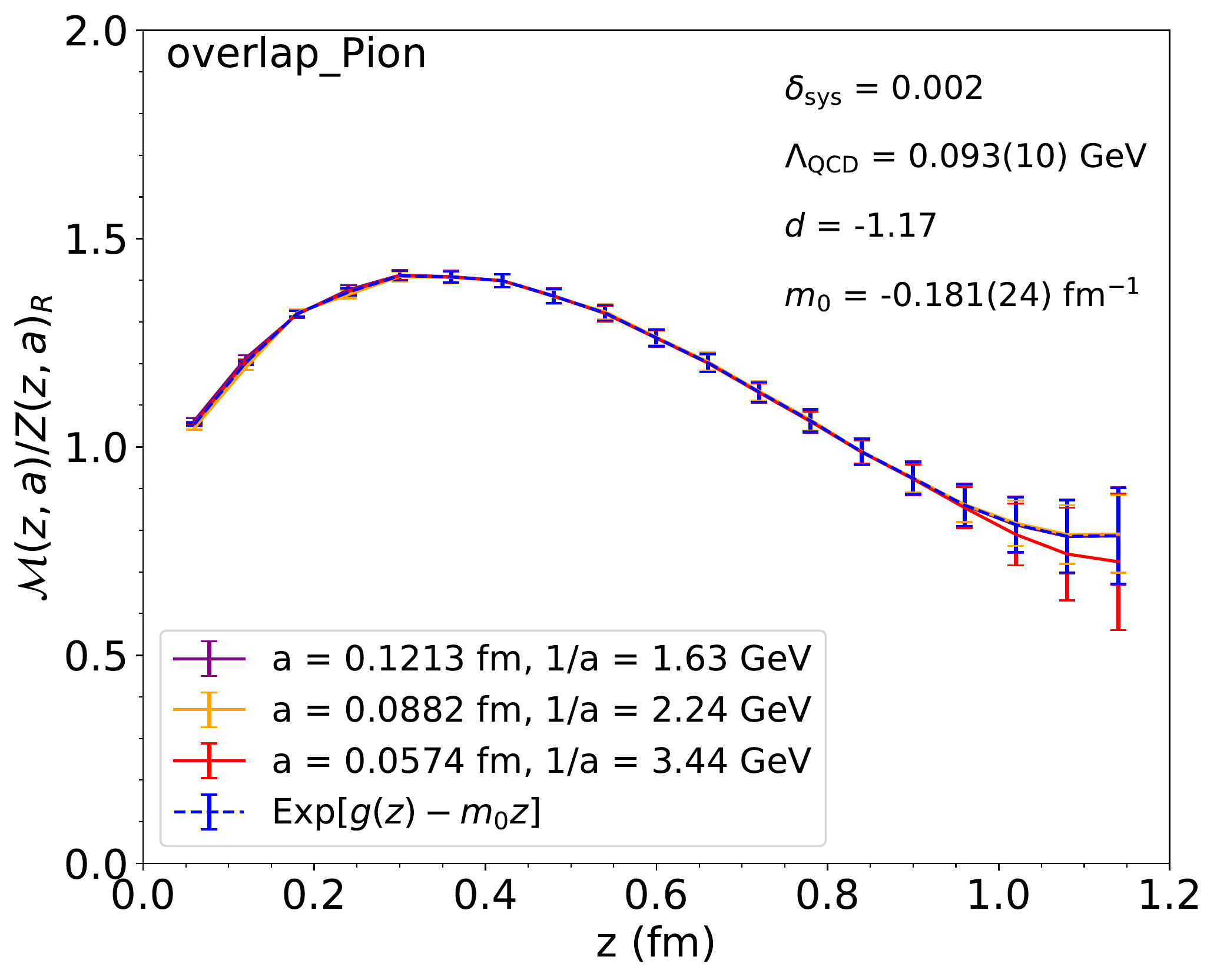}}
\caption{Same as Fig. \ref{fig:Re_overlap_quark}, with quasi-LF correlation in zero-momentum pion state calculated with the overlap fermion action.
}
\label{fig:Re_overlap_Pion}
\end{figure*}

\begin{figure*}[tbp]
\centering
\subfigure[]{\includegraphics[width=8cm]{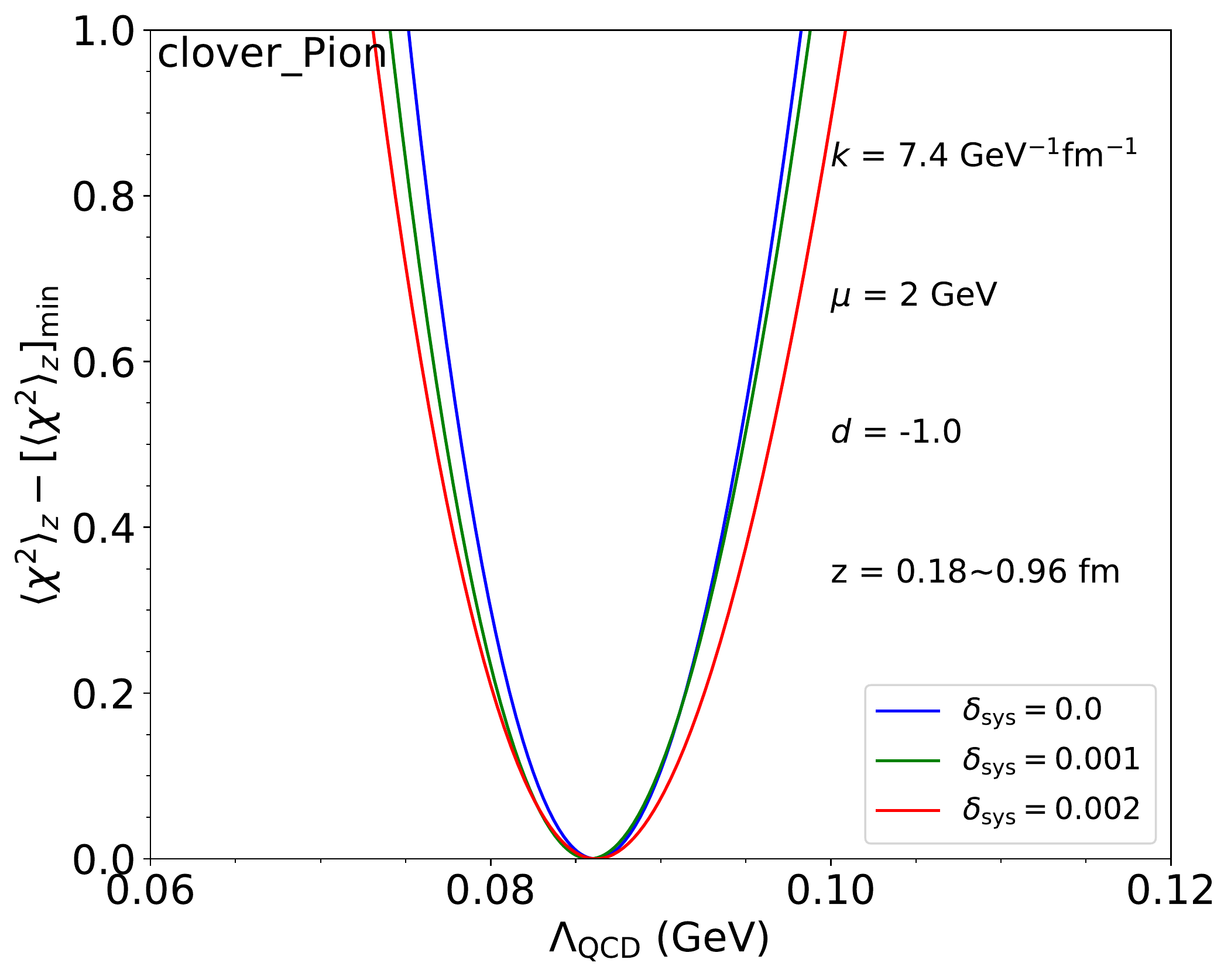}}
\subfigure[]{\includegraphics[height=5.5cm,width=8cm]{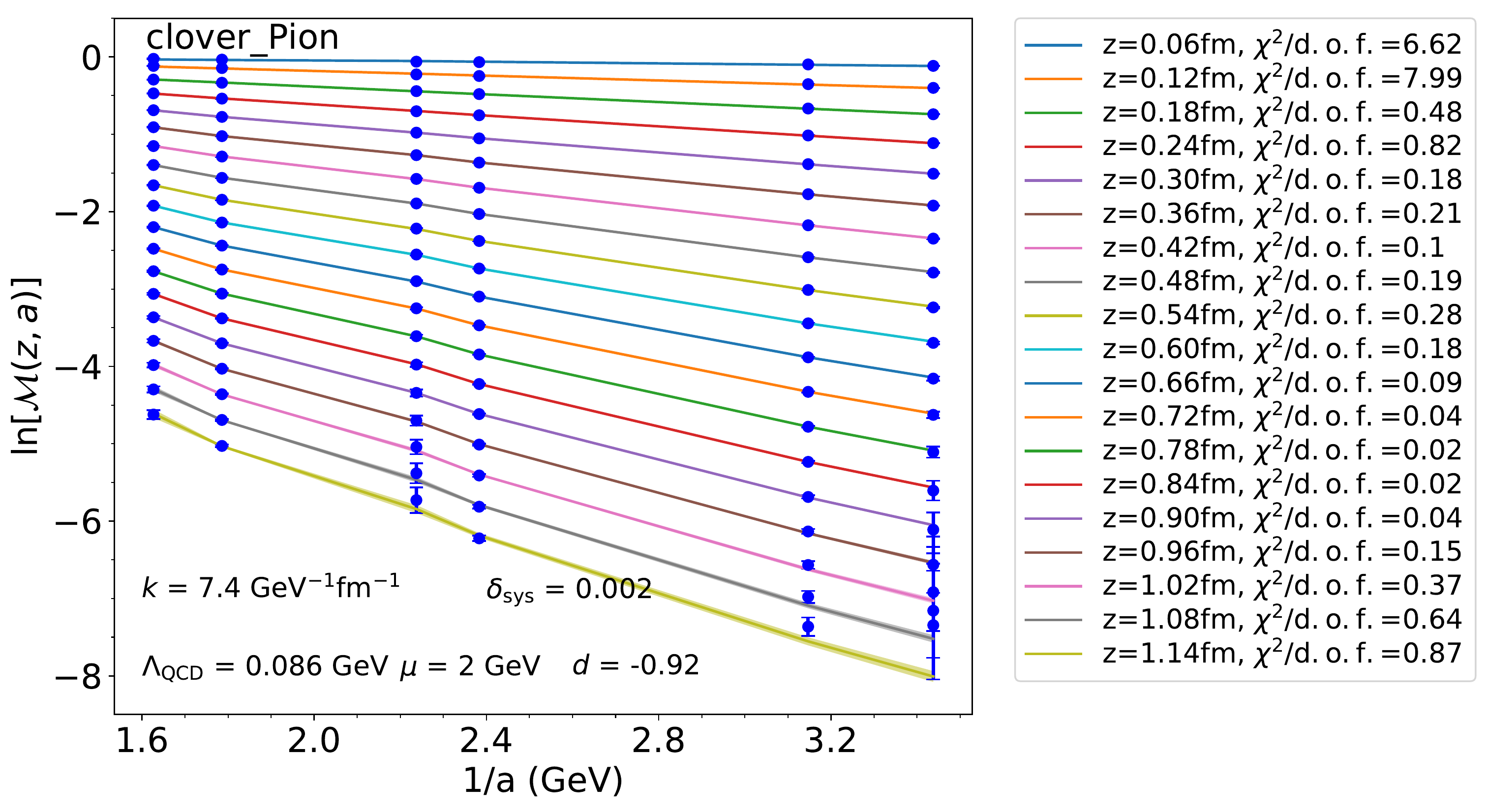}}
\subfigure[]{\includegraphics[width=8cm]{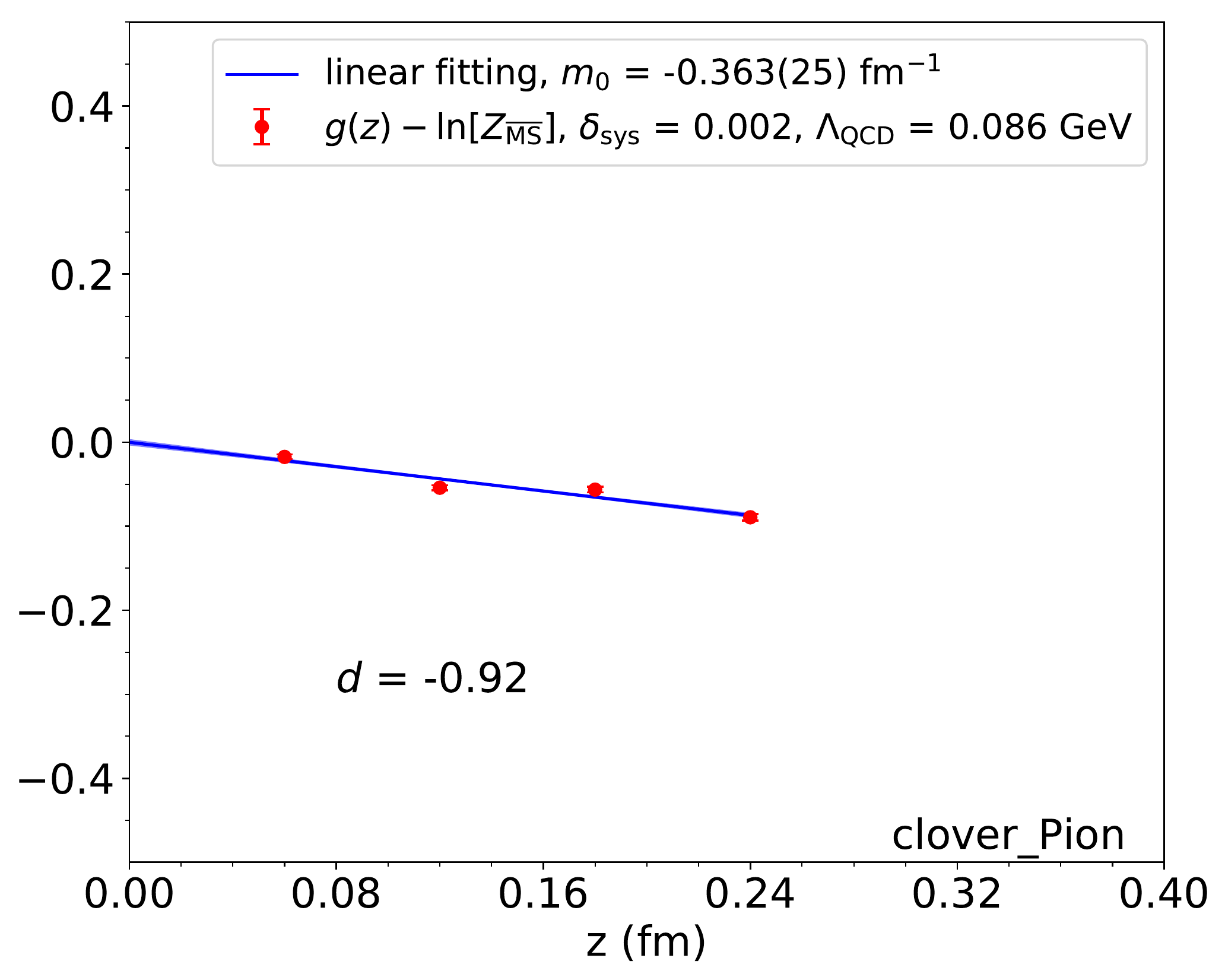}}
\subfigure[]{\includegraphics[width=7.5cm]{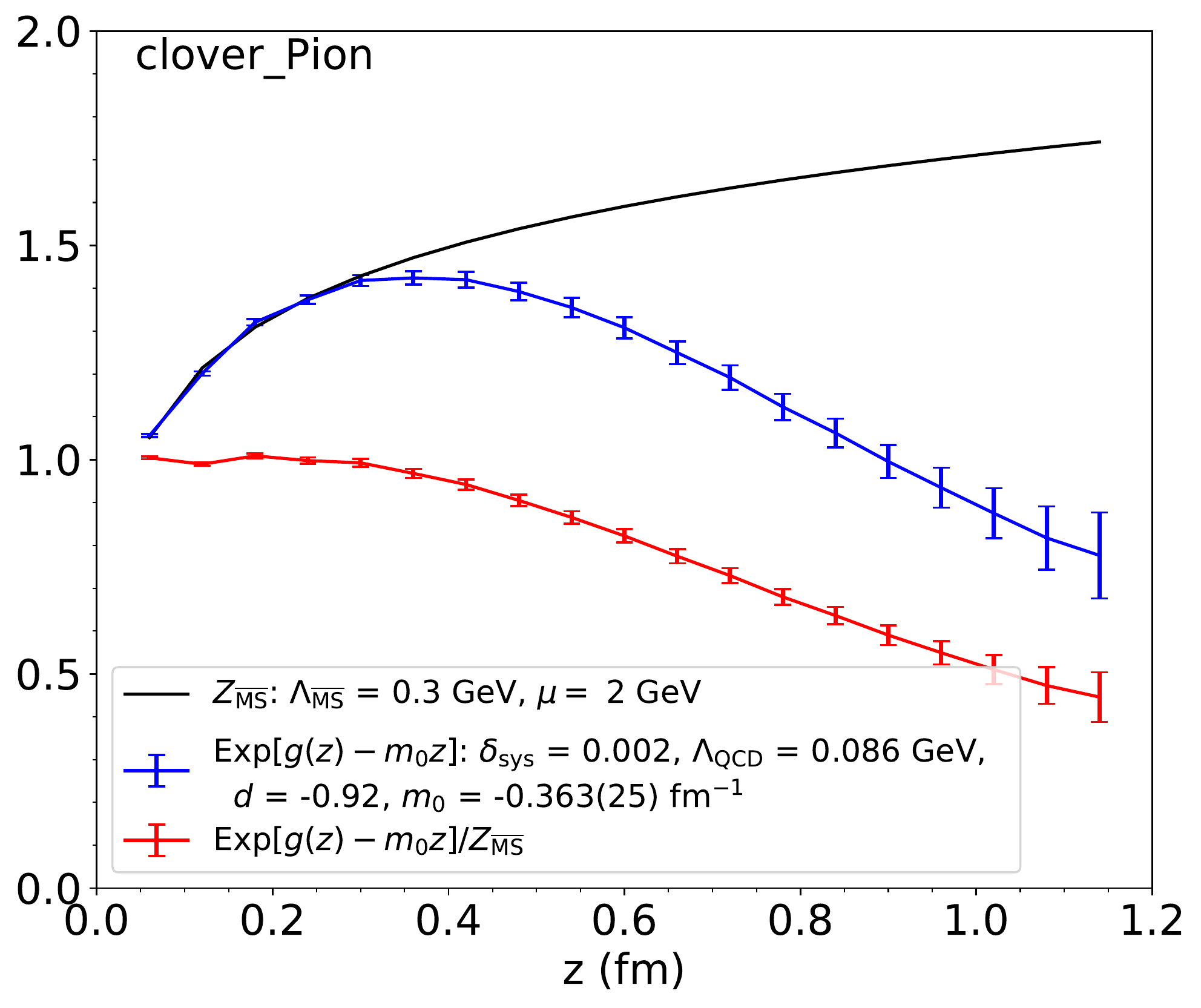}}
\subfigure[]{\includegraphics[width=8cm]{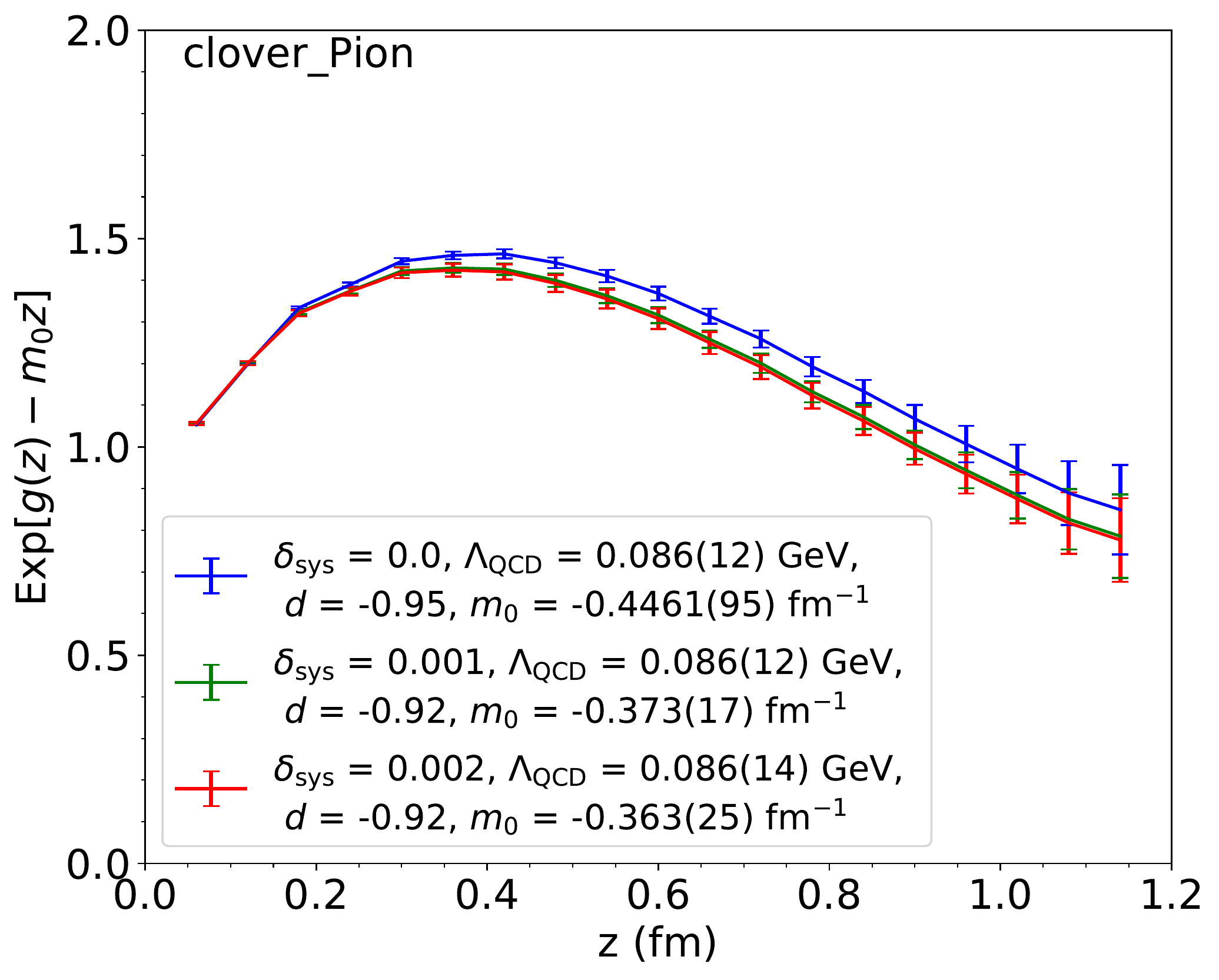}}
\subfigure[]{\includegraphics[width=8cm]{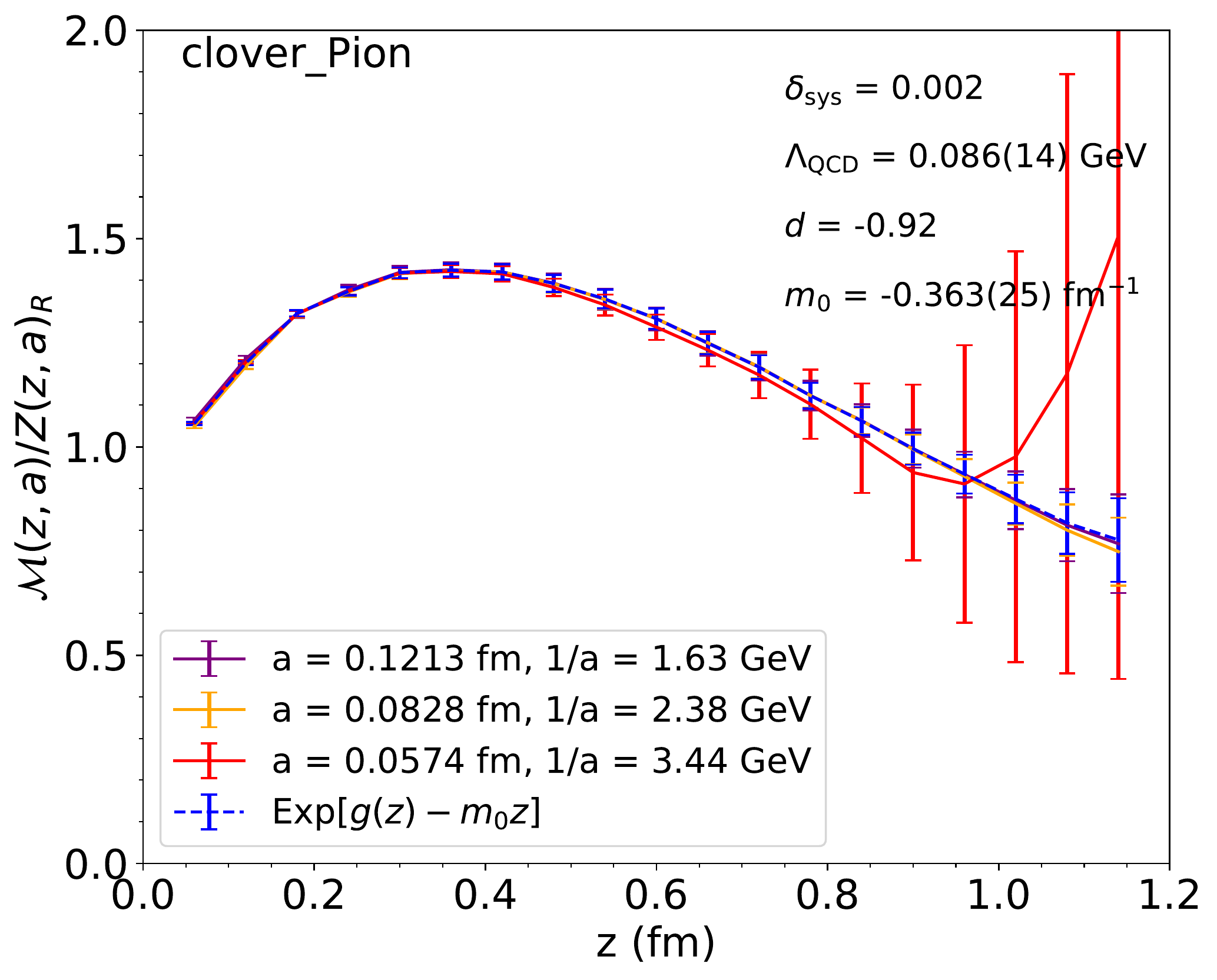}}
\caption{Same as Fig. \ref{fig:Re_overlap_quark}, with quasi-LF correlation in zero-momentum pion state calculated with the clover fermion action.
}
\label{fig:Re_clover_Pion}
\end{figure*}

\begin{figure*}[tbp]
\centering
\subfigure[]{\includegraphics[width=8cm]{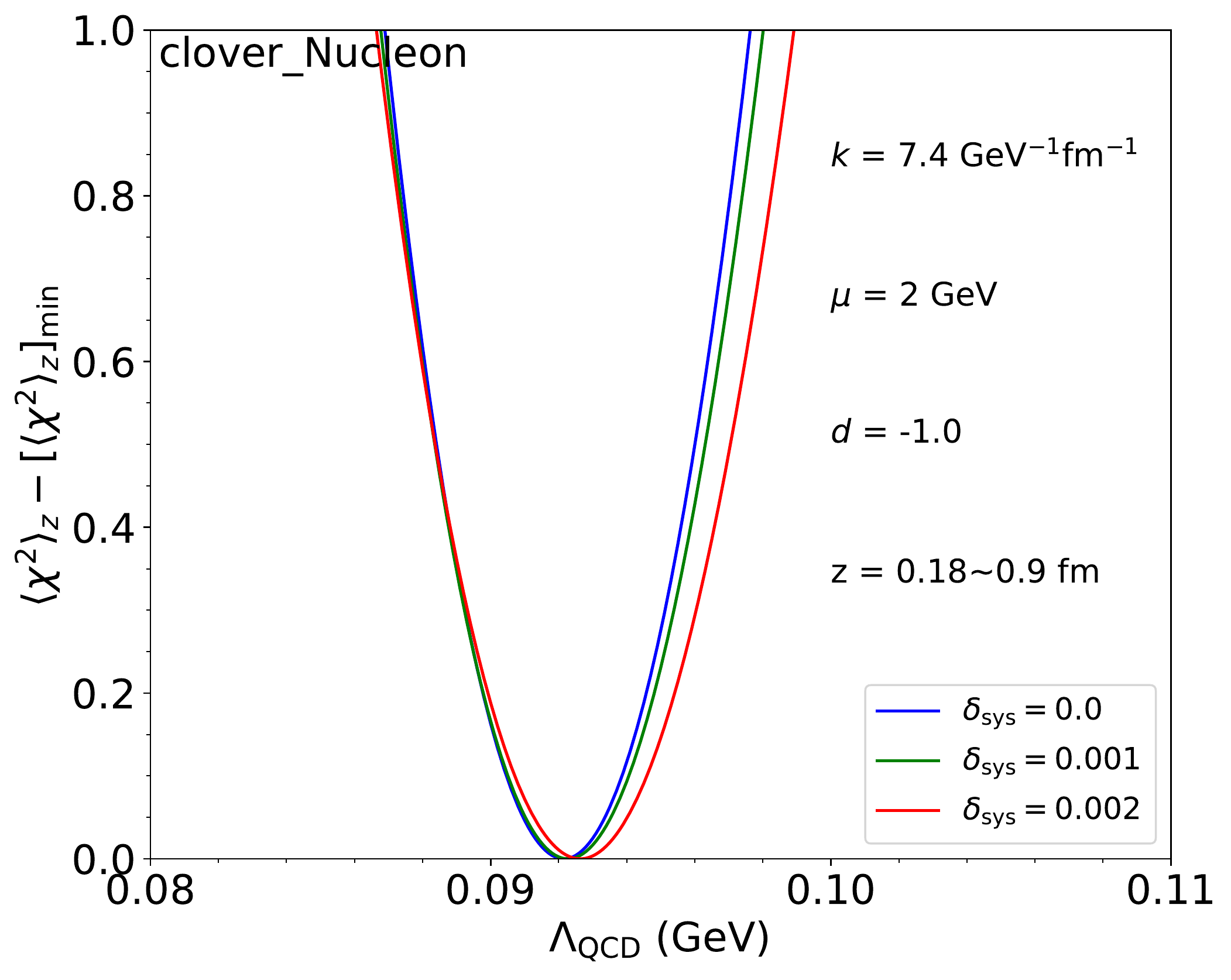}}
\subfigure[]{\includegraphics[height=5.5cm,width=8cm]{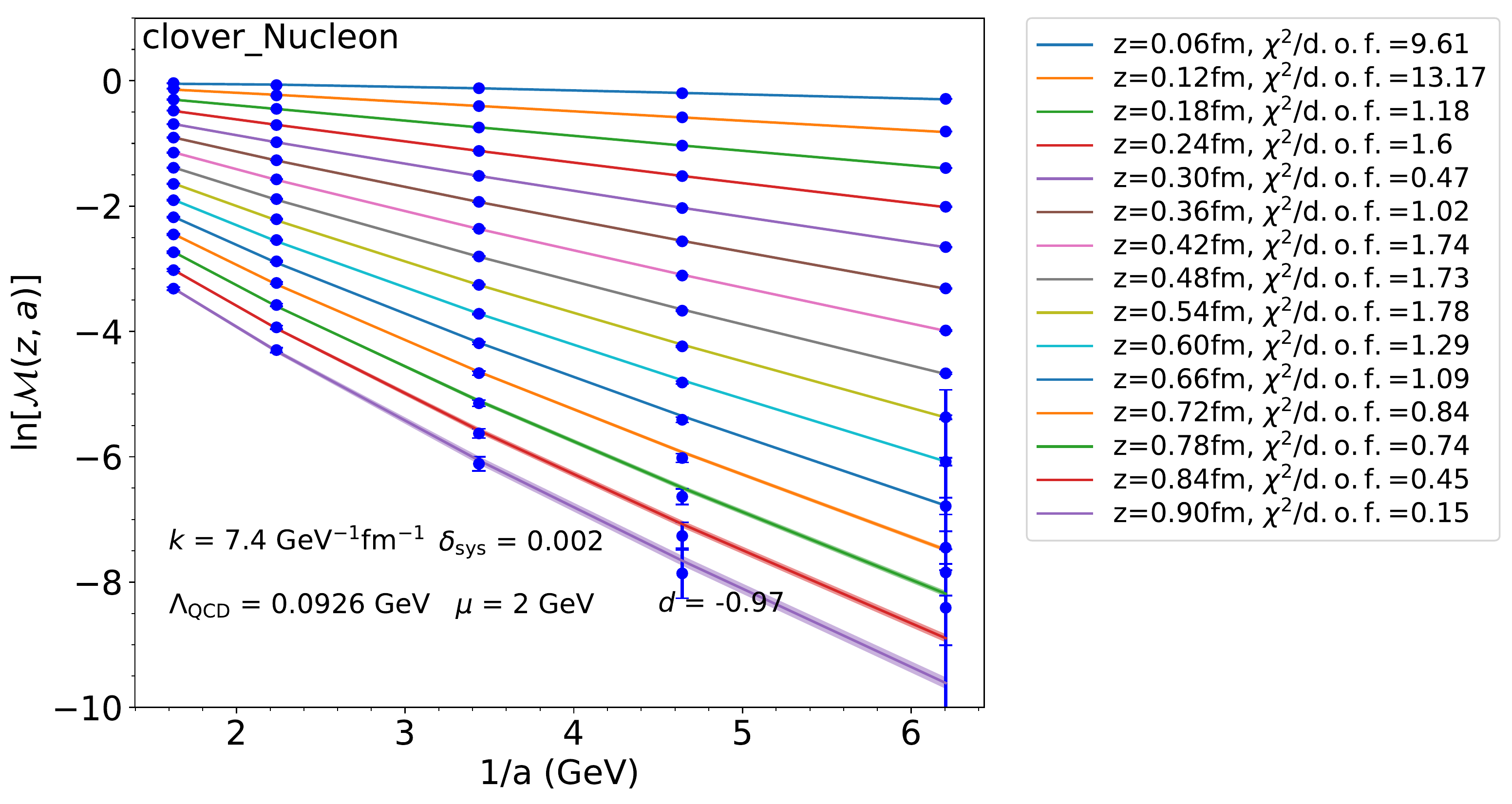}}
\subfigure[]{\includegraphics[width=8cm]{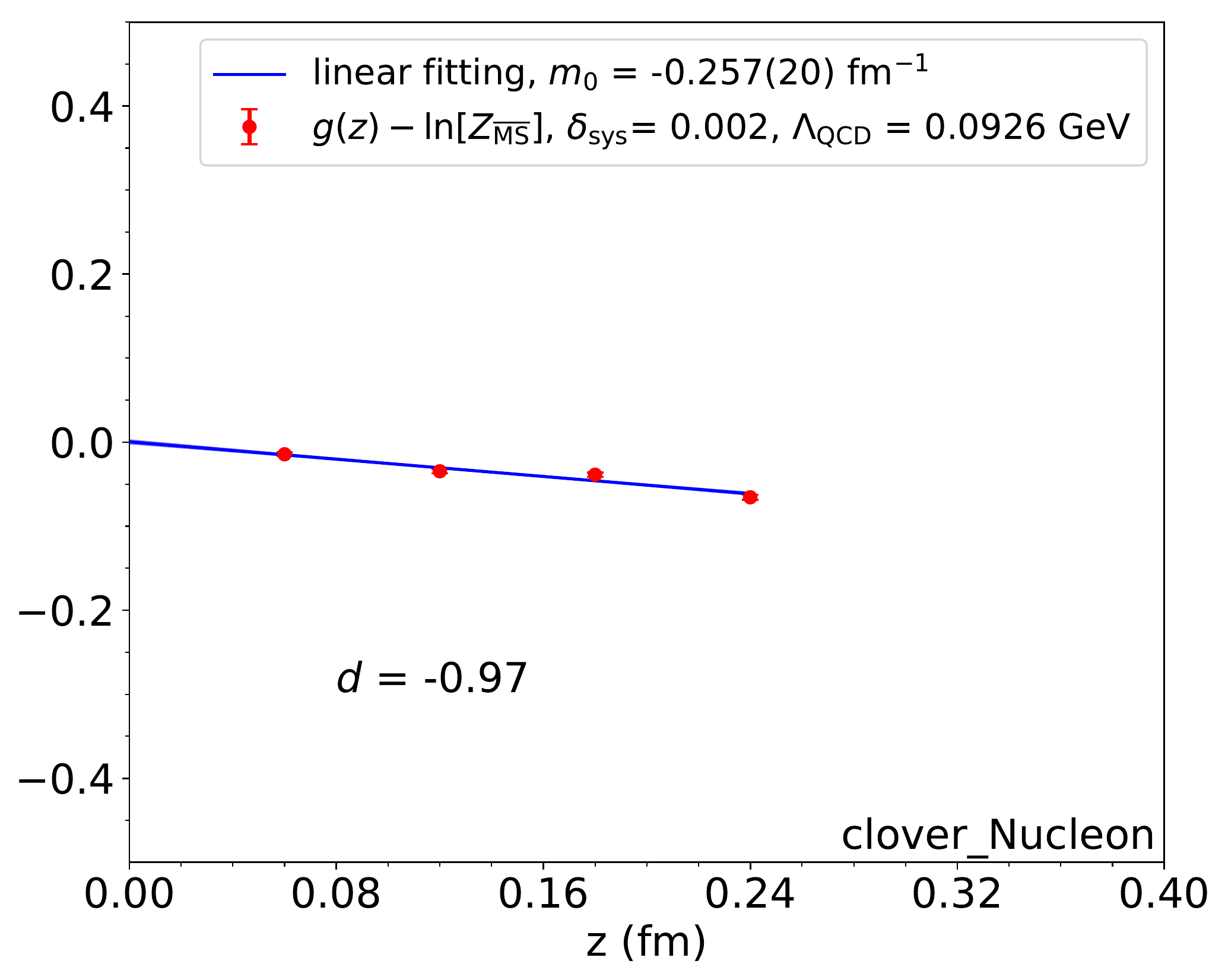}}
\subfigure[]{\includegraphics[width=7.5cm]{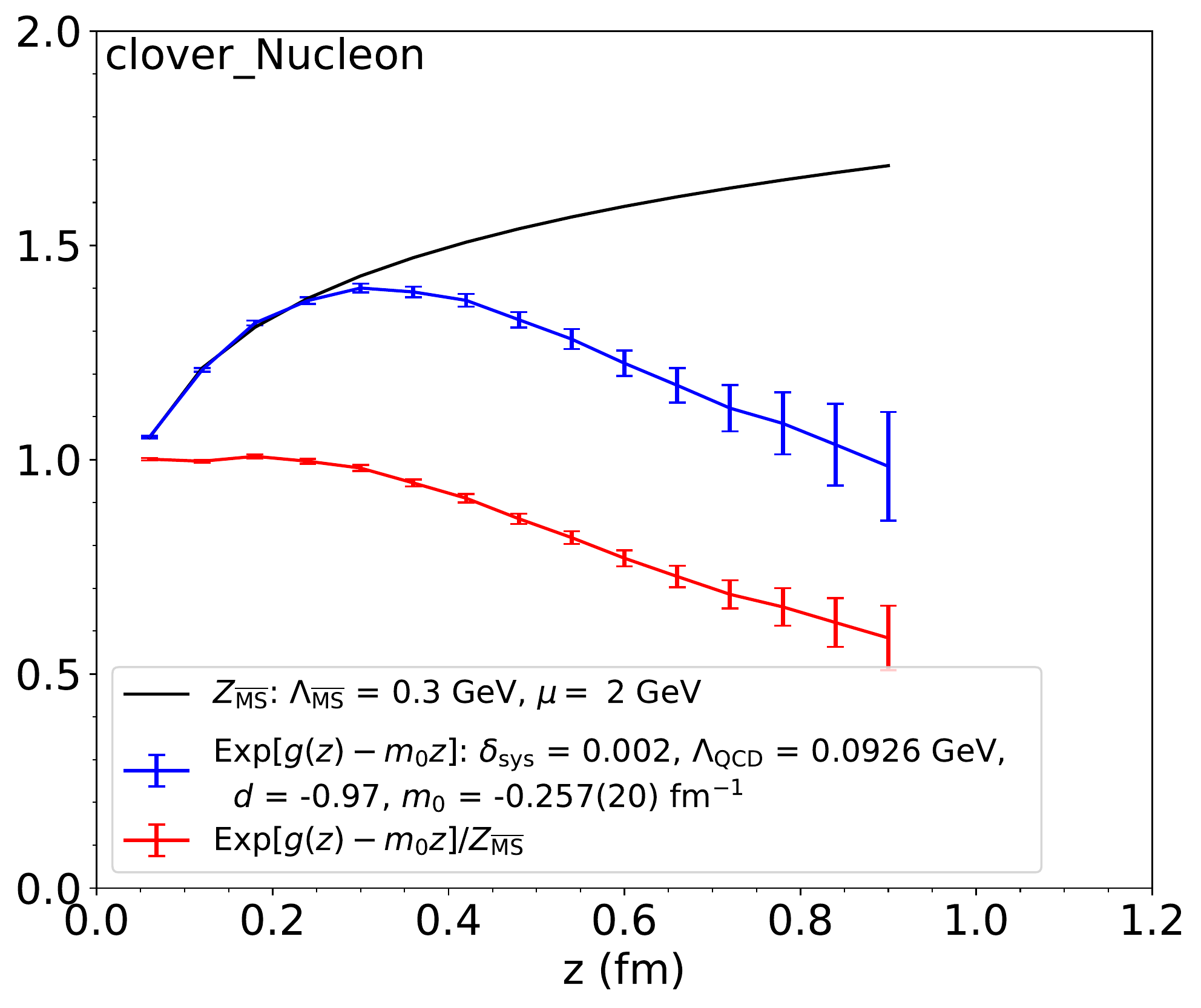}}
\subfigure[]{\includegraphics[width=8cm]{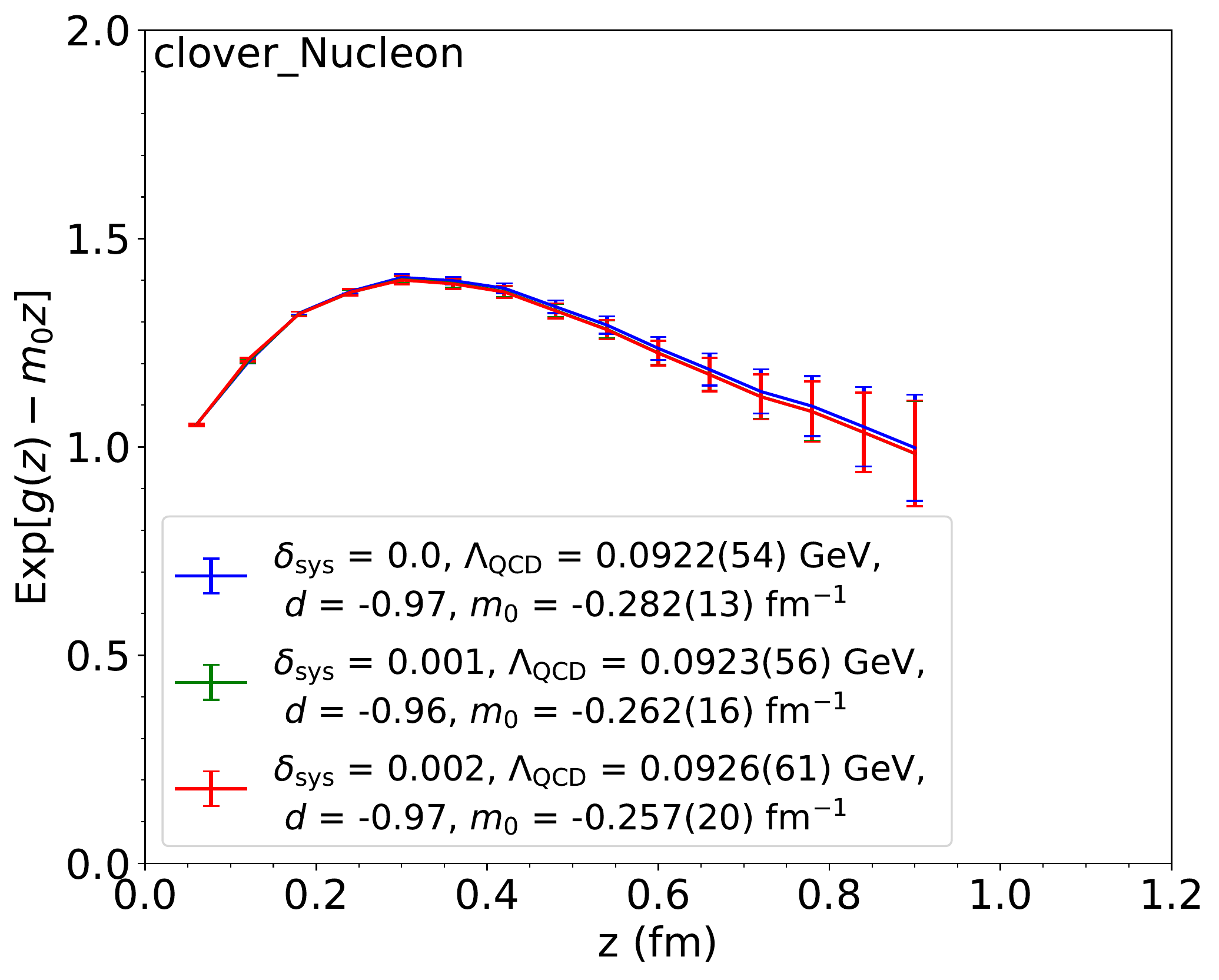}}
\subfigure[]{\includegraphics[width=8cm]{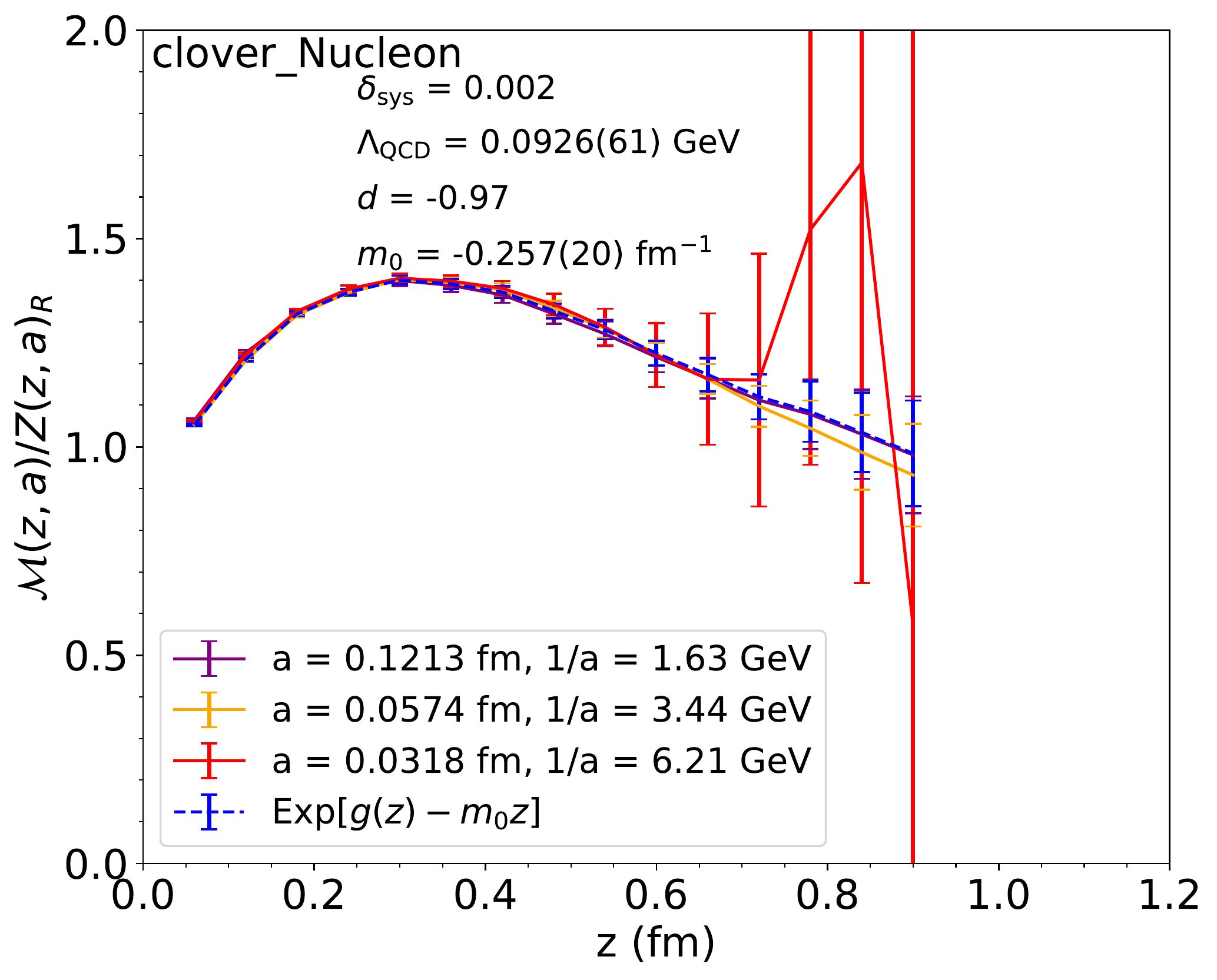}}
\caption{Same as Fig. \ref{fig:Re_overlap_quark}, with quasi-LF correlation in zero-momentum nucleon state calculated with the clover fermion action.
}
\label{fig:Re_clover_Nucleon}
\end{figure*}

Figs.~\ref{fig:Re_overlap_quark} through \ref{fig:Re_clover_Nucleon} show the detailed results of applying the
self-renormalization method for different cases. For the RI/MOM factor with overlap and clover fermions, we analyse eight different lattice spacings from two different types of gauge ensembles. For the pion matrix element calculated with the overlap fermion action, we analyse three different lattice spacings from MILC ensembles. In the clover pion case, we use six different lattice spacings from two different types of gauge ensembles. For the nucleon matrix element with clover fermion action, we analyse five different lattice spacings from MILC ensembles. 

Subfigure (a) in each figure is used to estimate the best fitted value of the global parameter $\Lambda_{\rm QCD}$ as well as its error. For each chosen $\Lambda_{\rm QCD}$, we can use Eq.~(\ref{eq:logM}) to fit the bare matrix element. The fitting for each $z$ gives us a separate $\chi^{2}$. We can calculate the averaged $\chi^{2}$ for different $z$, denoted as $\langle  \chi^{2} \rangle_{z}$. Subfigure (a) shows $\langle  \chi^{2} \rangle_{z} - [\langle  \chi^{2} \rangle_{z}]_{\rm min}$ with respect to $\Lambda_{\rm QCD}$. $\langle  \chi^{2} \rangle_{z} - [\langle  \chi^{2} \rangle_{z}]_{\rm min}$=0 gives us the best fitted $\Lambda_{\rm QCD}$ and $\langle  \chi^{2} \rangle_{z} - [\langle  \chi^{2} \rangle_{z}]_{\rm min}$=1 gives us its error. If the data points for different $z$ were independent, we should use the total $\chi^{2}$ for the fits at different $z$ to estimate the error of $\Lambda_{\rm QCD}$. However, since we have done linear interpolations of the original data points, some of the points are correlated to each other. So here we just use $\langle  \chi^{2} \rangle_{z}$ to estimate the error of $\Lambda_{\rm QCD}$. The range of $z$ starts from 0.18 fm here but not 0.06 fm because the $\chi^{2}$ for $z$ = 0.06 or 0.12 fm is much larger than others. $d$ is fixed to be $-1$ since it has little influence on $\chi^{2}$. 

For $\delta_{\rm sys}=0.002$, the best fitted values of $\Lambda_{\rm QCD}$ are 0.1086(17), 0.1350(21), 0.093(10), 0.086(14), 0.0926(61) GeV for the correlations in the overlap quark, clover quark, overlap pion, clover pion, and clover nucleon, respectively. As one can see, they are quite similar except for the correlation in the clover quark state. While the size of the systematic
error has some effects for the correlation in the quark case, it has little influence on the correlation in the physical 
states. 

In subfigure (b) in each figure, after taking $\Lambda_{\rm QCD}$ to be the best fitted value, we use Eq.~(\ref{eq:logM}) to fit the bare matrix elements to extract $g(z)$. The range of $z$ is taken from 0.06 fm to 1.02 fm with 17 different $z$ values for correlations in the overlap quark case, from 0.06 fm to 1.14 fm, with 19 different $z$ values for correlations in the overlap pion and clover pion cases, from 0.06 fm to 0.96 fm, with 16 different $z$ in the clover quark case, from 0.06 fm to 0.9 fm, and  with 15 different $z$ values in the clover nucleon case. We fine-tune $d$ to make sure that $g(z) - \ln[Z_{\overline{\mathrm{MS}}}]$ is proportional to $z$ within the window 0.06 fm $\leq z \leq$ 0.24 fm. 
The fine-tuned $d$ values are $-1.29, -1.35, -1.17, -0.92, -0.97$ in the five cases, respectively. They are all close to $-1$. 

Subfigure (c) in each figure shows the fitting to extract $m_{0}$. For $\delta_{\rm sys}=0.002$, the fitted values of $m_{0}$ are 0.2339(81), 0.7667(83), $-$0.181(24), $-0.363(25)$, and $-0.257(20)$ fm$^{-1}$ in the five cases, respectively. They are all about the order of $\Lambda_{\rm QCD} \sim 0.1$ GeV (about 0.5 fm$^{-1}$), which supports the argument in Sec.~\ref{sec:theory} that $m_{0}$ originates from the non-perturbative renormalon effect. 

In subfigure (d) in each figure, we show our result for the renormalized matrix element Exp[$g(z)-m_{0}z$] in blue points connected by lines. As $z$ increases, Exp[$g(z)-m_{0}z$] increases first, following the perturbative result, 
and then decreases from about $z$ = 0.25 fm on. In all the cases, Exp[$g(z)-m_{0}z$] is consistent with $Z_{\overline{\mathrm{MS}}}$ at small $z$. However, at large $z$, there is a significant discrepancy between Exp[$g(z)-m_{0}z$] and $Z_{\overline{\mathrm{MS}}}$. This means that there is a large non-perturbative effect at large $z$ in the popular RI/MOM and ratio renormalization scheme used previously, which supports the hybrid renormalization procedure proposed recently~\cite{Ji:2020brr}. We have already 
briefly discussed this in Sec.~\ref{sec:d}. 

Subfigure (e) shows the renormalized matrix element Exp[$g(z)-m_{0}z$] for different $\delta_{\rm sys}$. A change in $\delta_{\rm sys}$ does not influence the central values of Exp[$g(z)-m_{0}z$] for most cases except for the clover pion. 
In the clover pion case, the increase of $\delta_{\rm sys}$ can slightly decrease the renormalized matrix element Exp[$g(z)-m_{0}z$]. 

Finally, subfigure (f) shows the ratio of bare matrix element and renormalization factor $\mathcal{M}(z,a)/Z(z,a)_{R}$, compared with the renormalized matrix element Exp[$g(z)-m_{0}z$]. $Z(z,a)_{R}$ here is not extracted from a different matrix elements but from $\mathcal{M}(z,a)$ itself. This subfigure is a consistency check of our method. In all cases, $\mathcal{M}(z,a)/Z(z,a)_{R}$ has little dependence on $a$ within error and is consistent with Exp[$g(z)-m_{0}z$]. 
Therefore, our self-renormalization method can isolate the residual intrinsic physics from the linear divergence, 
renormalon uncertainty, log divergence and discretization error with a reasonable precision. 

\begin{figure}[tbp]
\centering
\includegraphics[width=8.5cm]{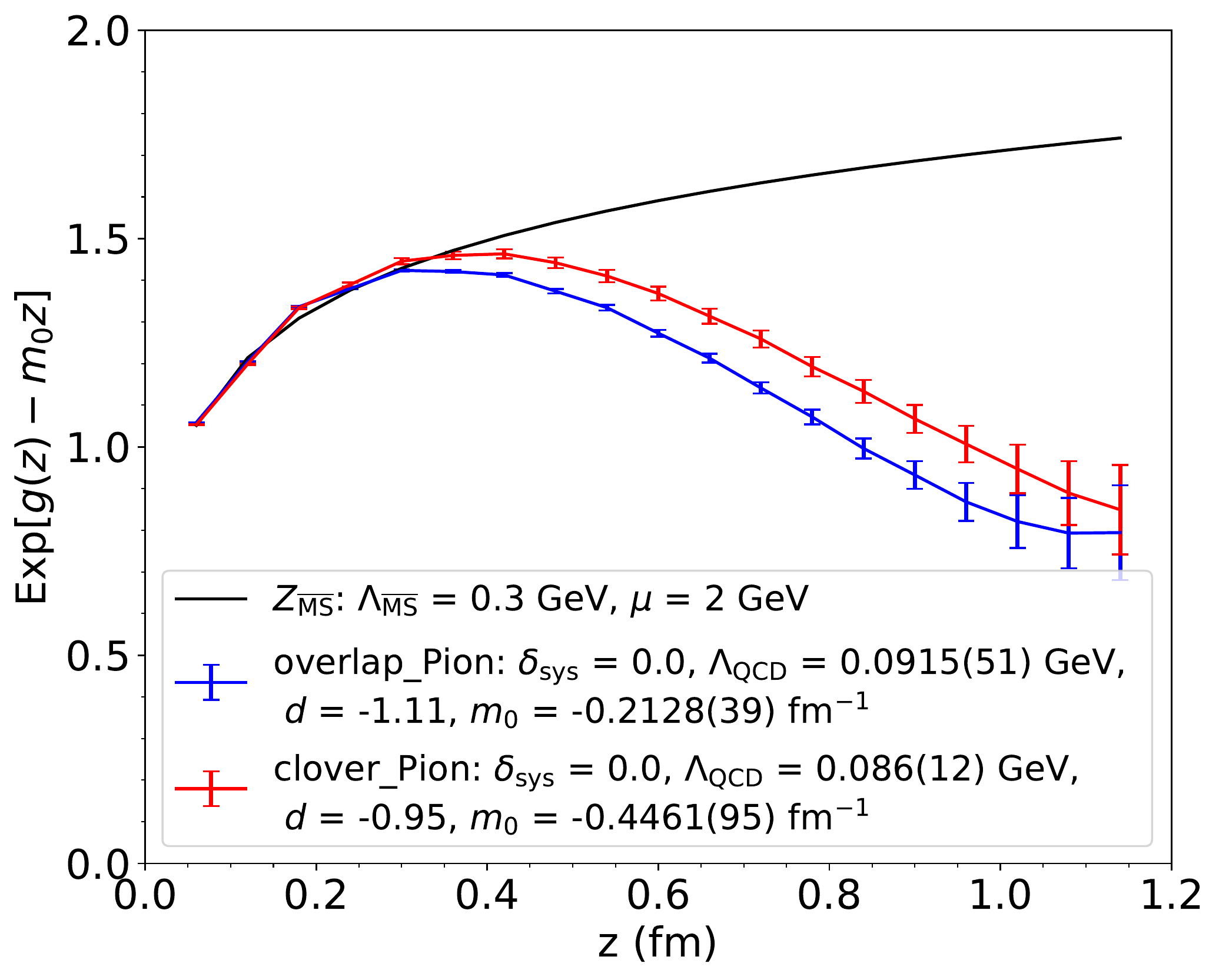}
\includegraphics[width=8.5cm]{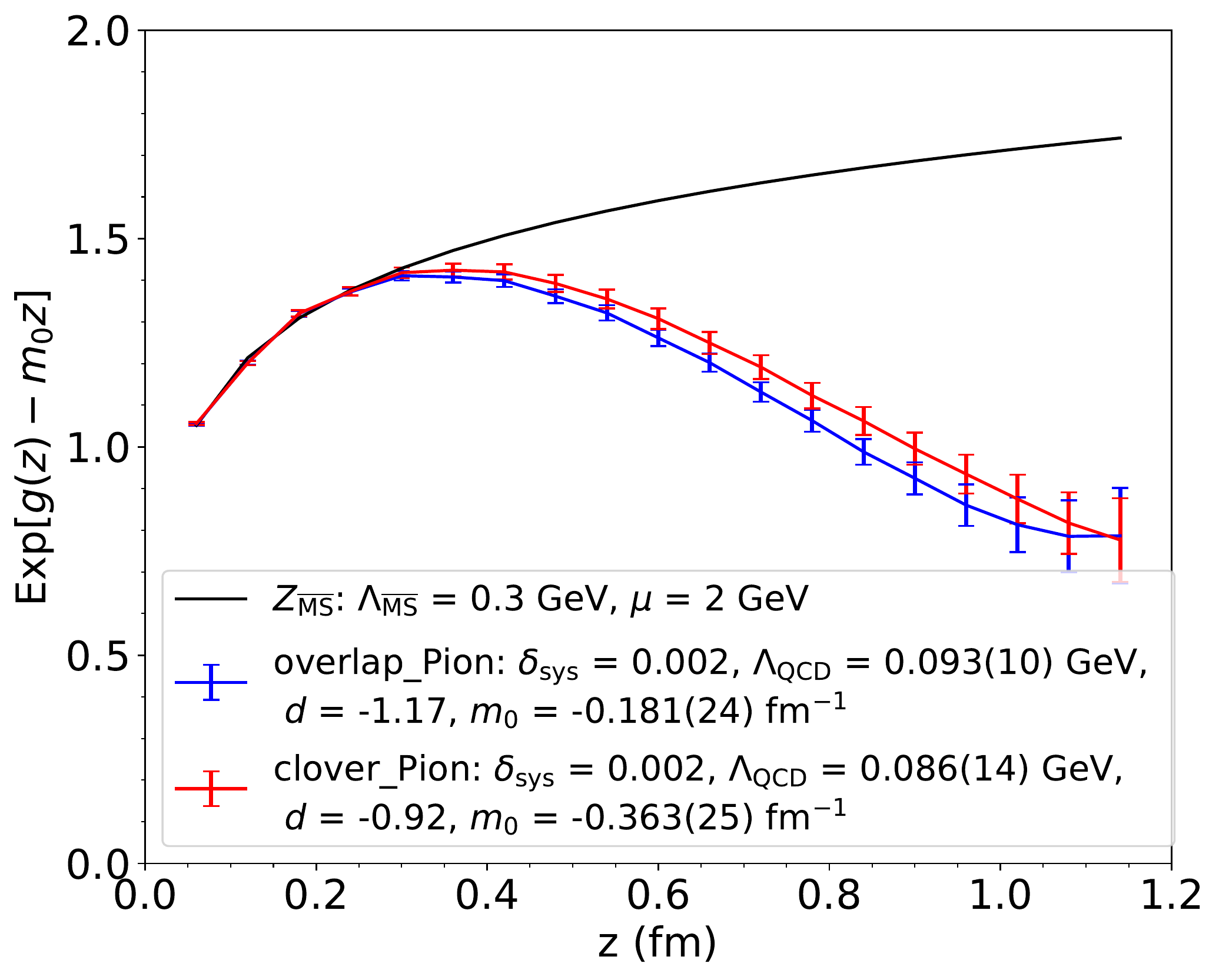}
\caption{
Comparison of renormalized $O_{\gamma_t}(z)$ correlator in the zero-momentum pion state between overlap and clover 
actions without HYP smearing. Points connected by solid lines are renormalized matrix elements Exp[$g(z)-m_{0}z$]. Black curve is $Z_{\overline{\mathrm{MS}}}$. In the upper panel, $\delta_{\rm sys}=0$. In the lower panel, $\delta_{\rm sys}=0.002$.
}
\label{fig:Re_overlap_clover_Pion}
\end{figure}

Although the ratio $\mathcal{M}(z,a)/Z(z,a)_{R}$ has a good behavior for most of the lattice spacings, it may have large errors for small lattice spacings like 0.0318 fm or 0.0574 fm. Therefore we do not recommend using $\mathcal{M}(z,a)/Z(z,a)_{R}$ as the renormalized matrix element. Since Exp[$g(z)-m_{0}z$] is achieved through our fitting, the
large-error data points have little influence on it. We take this as 
the renormalized matrix element, while using $\mathcal{M}(z,a)/Z(z,a)_{R}$ only 
to test consistency.

We show in Fig.~\ref{fig:Re_overlap_clover_Pion} a comparison of renormalized $O_{\gamma_t}(z)$ correlation in the pion state between overlap and clover actions. Before we add the systematic error, both results show appreciable difference, which
may indicate that two different valence fermions will lead to different results. After we add a small dummy systematic error during the fitting, the correlations in both cases 
lead to similar results. That means that the systematic error is important, and the residual intrinsic non-perturbative physics is independent of valence quark formulations if it is properly included.

\begin{figure}[tbp]
\centering
\includegraphics[width=8.5cm]{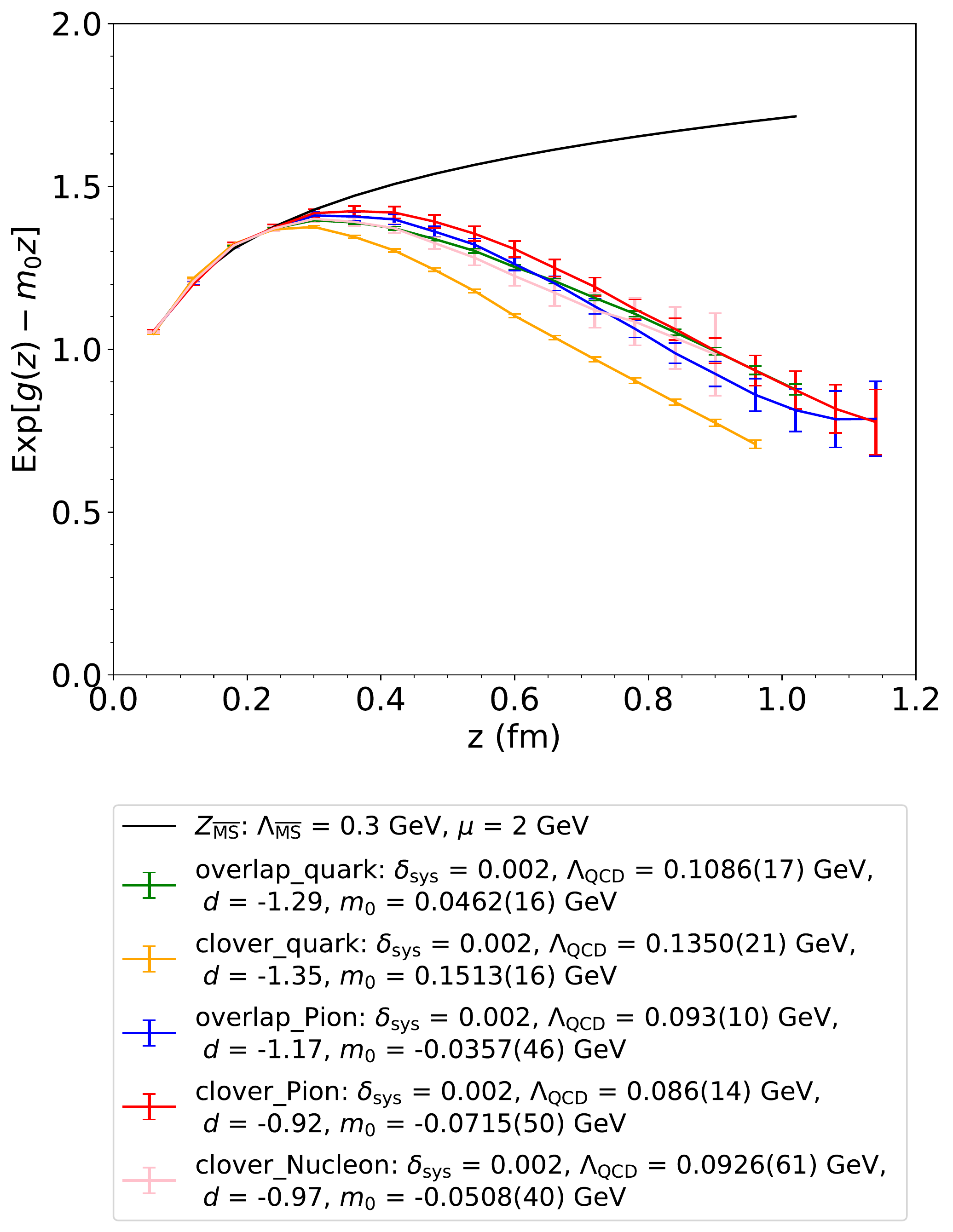}
\caption{
Comparison of renormalized $O_{\gamma_t}(z)$ correlations for all the cases without HYP smearing. Points connected by solid lines are renormalized matrix elements Exp[$g(z)-m_{0}z$]. Black curve is $Z_{\overline{\mathrm{MS}}}$.
}
\label{fig:Re_all}
\end{figure}

Finally, for curiosity, we show in Fig.~\ref{fig:Re_all} the renormalized correlations as well as parameters related to the divergence for all cases we studied. As mentioned before, the fitted $\Lambda_{\rm QCD}$ for overlap quark, overlap pion, clover pion and clover nucleon are similar to each other. But for the clover quark, it is markedly different. We do not yet understand the reason, but it could be due to the combination of chiral symmetry breaking and the linearly-divergent quark mass. We plan to investigate this in the future, but for the time being, at least we understand why we cannot eliminate the linear divergence using RI/MOM for the clover action~\cite{Zhang:2020rsx}. Quite surprisingly, the residual intrinsic non-perturbative 
correlations for overlap quark, pion and nucleon are very similar to each other,
which we also do not fully understand.

\section{Fits with other options and stability }\label{sec:highorder}

In a phenomenological analysis, the amount of information and the accuracy one can obtain depends, obviously, on the quality of the lattice data. Given the data we have, we would like to push the limit of the analysis by including more physics in the fit until the analysis no longer yields useful information. 

\begin{figure}[tbp]
\centering
\includegraphics[width=9cm]{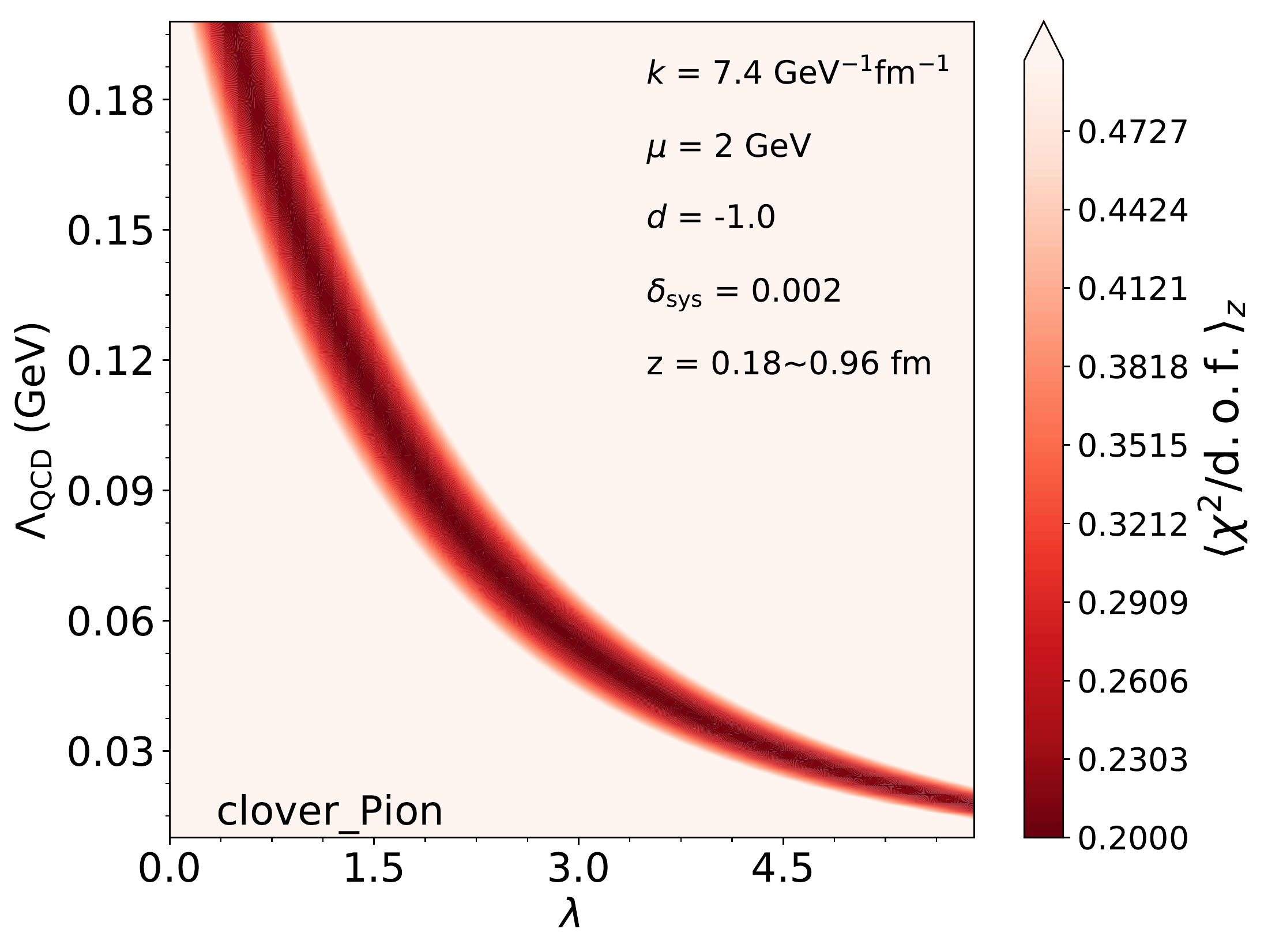}
\includegraphics[width=8cm]{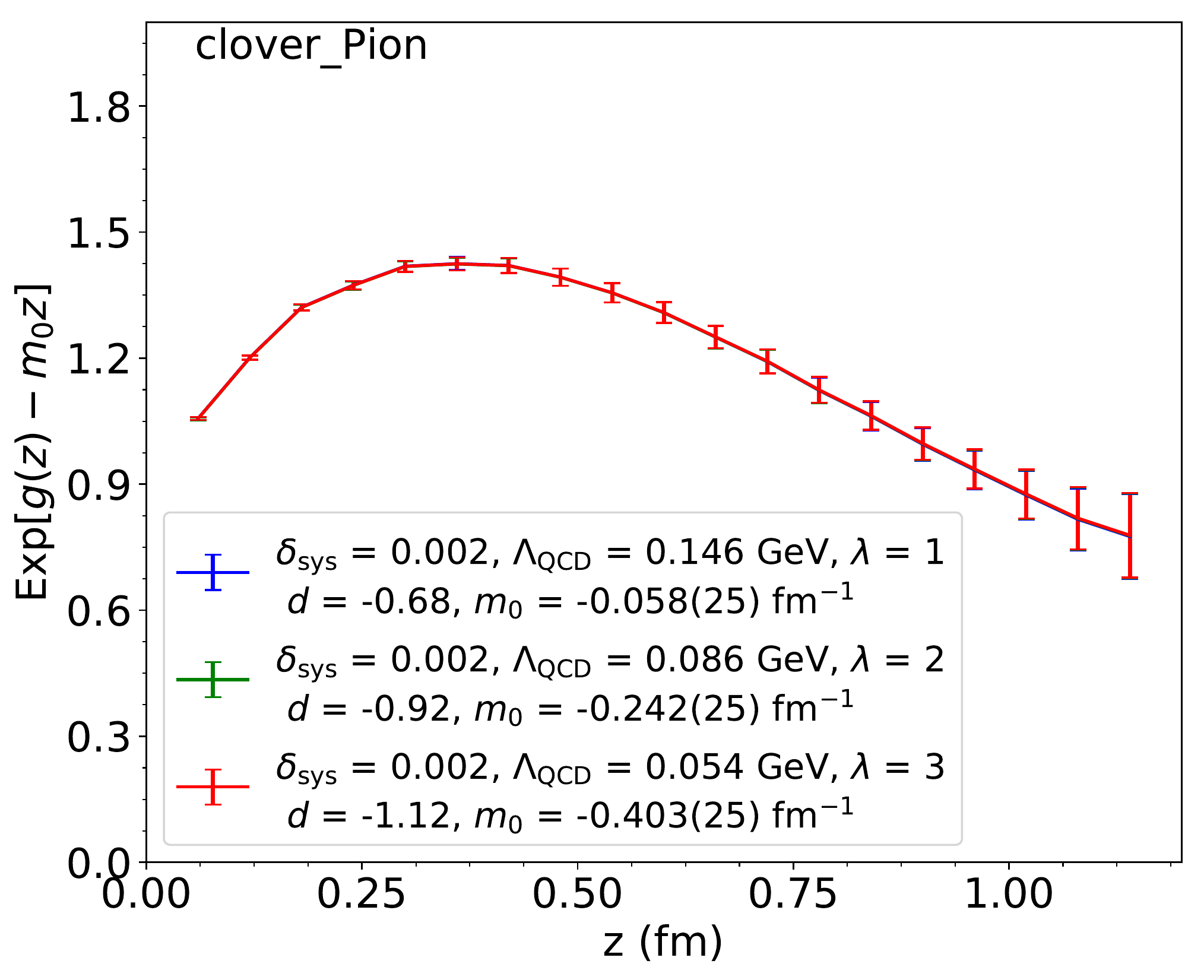}
\caption{
Using Eq.~(\ref{eq:logM_2}) to renormalize the $O_{\gamma_t}(z)$ matrix element in the pion state for the clover action without HYP smearing. Upper Panel: $\chi^2$ map with respect to $\lambda$ and $\Lambda_{\rm QCD}$. $\langle  \chi^{2}/{\rm d.o.f.} \rangle_{z}$ is the average of $\chi^{2}/{\rm d.o.f.}$ from fitting for each $z$. Lower Panel: renormalized matrix element for several sets of ($\Lambda_{\rm QCD}$, $\lambda$) along the 'small-$\chi^2$ band'. 
}
\label{fig:Re_overlap_quark_2}
\end{figure}

\begin{figure}[tbp]
\centering
\includegraphics[width=9cm]{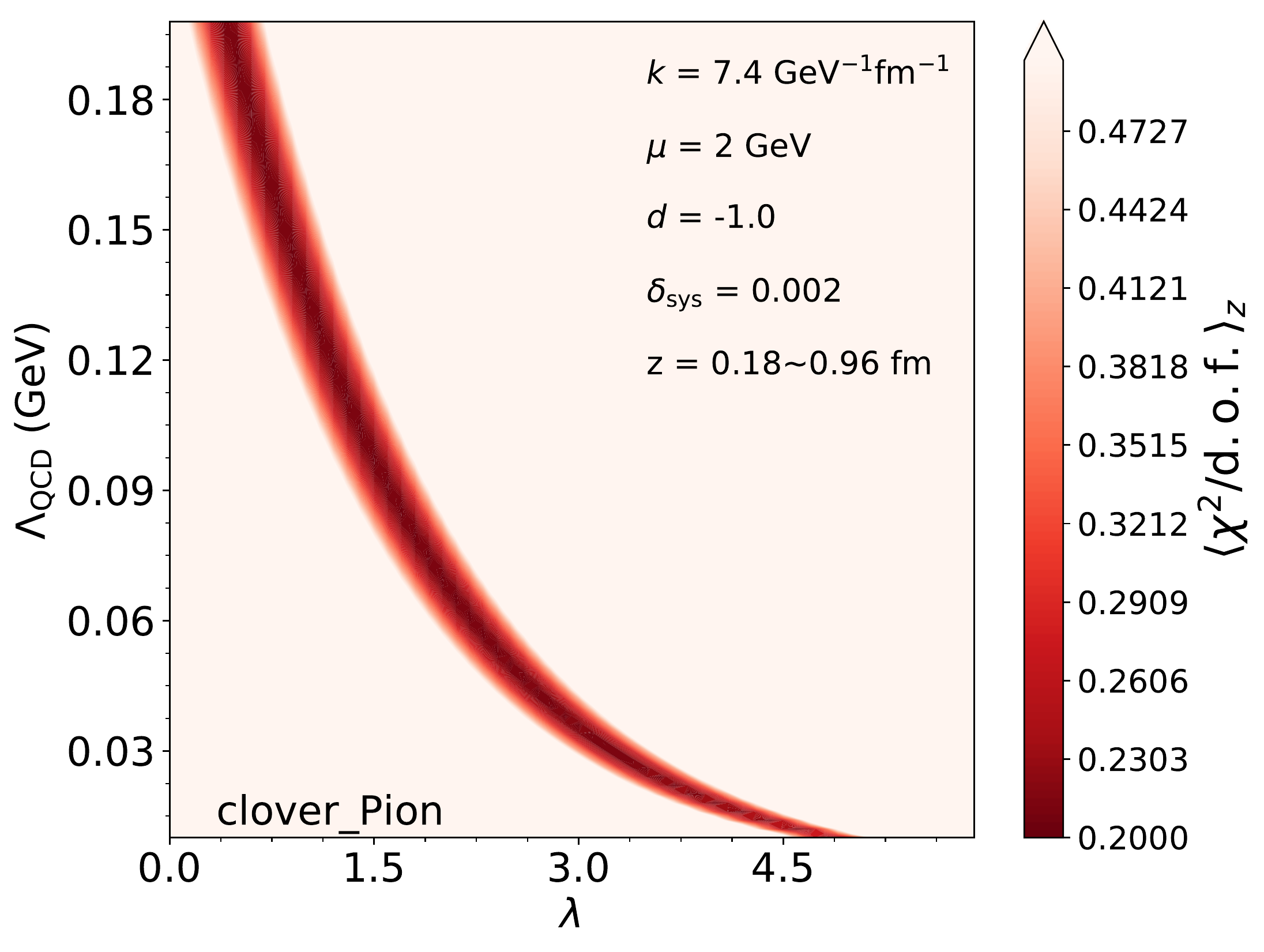}
\includegraphics[width=8cm]{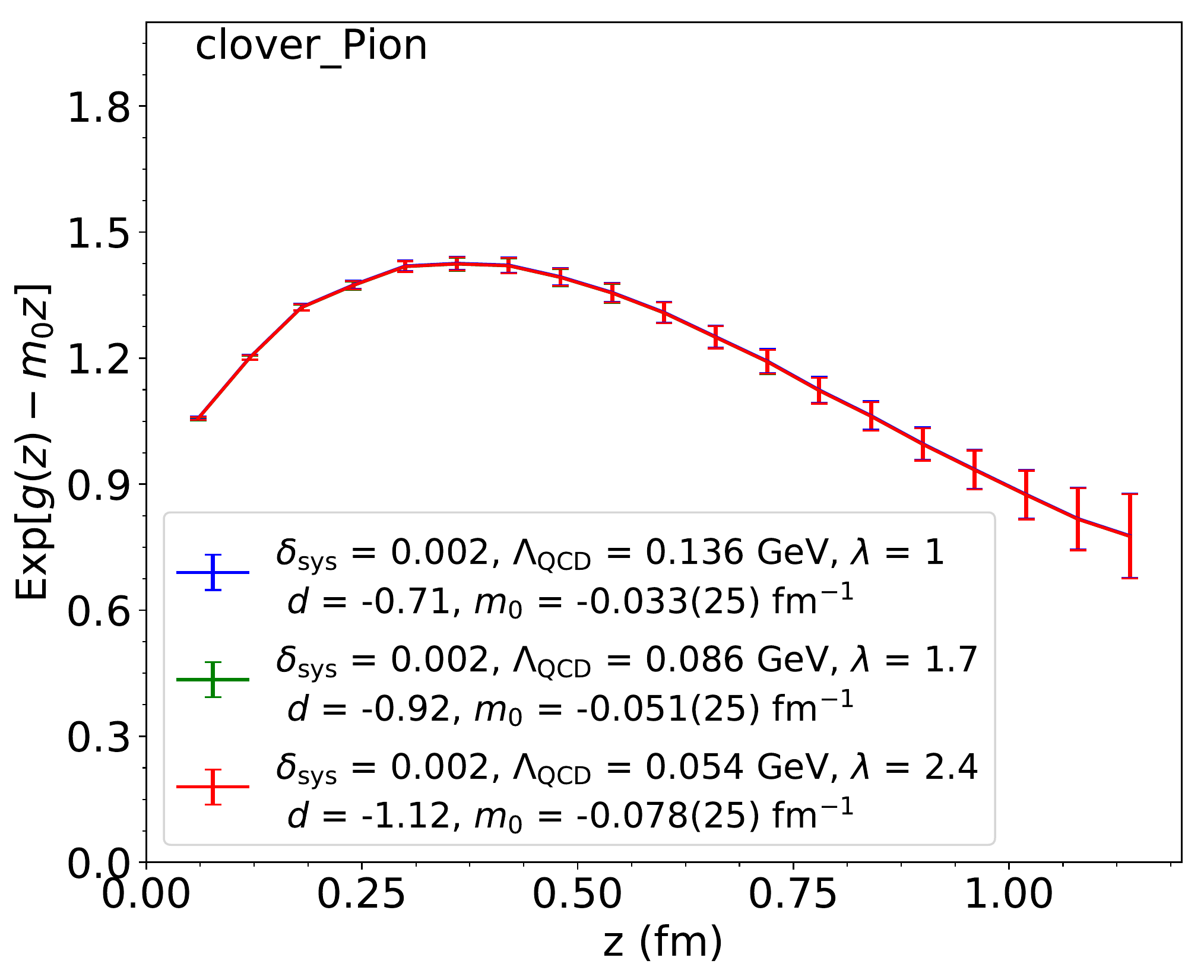}
\caption{Same as Fig. \ref{fig:Re_overlap_quark_2}, fit using Eq. (\ref{eq:logM_3}).
}
\label{fig:Re_overlap_quark_3}
\end{figure}

The first question we would like to address is whether one can get more accurate information
about the linear divergence. We have used so far the one-loop perturbative
QCD prediction, but treated $\Lambda_{\rm QCD}$ as a free parameter to partially take 
into account higher order corrections. However, one might consider
directly including at least the second-order corrections. Our strategy here 
is to fix $k$ to its perturbative value, and to vary $\Lambda_{\rm QCD}$
and the explicit second order correction characterized by
a new parameter $\lambda$
\begin{align}\label{eq:logM_2}
\ln \mathcal{M}(z,a) = \frac{k' z}{a} \tilde{\alpha_{s}}(1+\lambda \tilde{\alpha_{s}})
+ g(z) + ...
\end{align}
where the terms omitted are the same as those we had before in Eq.~(\ref{eq:logM}). $k'$ is related to $k$ by $-2\pi/b_0$ (since $-k'2\pi/b_0=k$). On the other hand, one might use the $\lambda$ parameter to include
some higher order corrections~\cite{Bauer:2011ws}, 

\begin{align}\label{eq:logM_3}
\ln \mathcal{M}(z,a) = \frac{k' z}{a} \frac{\tilde{\alpha_{s}}}{1-\lambda \tilde{\alpha_{s}}}
+ g(z) + ...
\end{align}
To be consistent, we now have to use the QCD running coupling constant $\tilde{\alpha_{s}}$ up to the second order $\beta$ function $b_{1}=102-\frac{38}{3} n_{f}$,
\bea\label{eq:alpha_s2}
\tilde{\alpha_{s}}=\frac{4 \pi}{b_0 \ln (\frac{1}{a^{2} \Lambda_{\rm QCD}^{2}})}\left(1-\frac{b_1}{b_0^{2}} \times \frac{\ln[\ln (\frac{1}{a^{2} \Lambda_{\rm QCD}^{2}})]}{\ln(\frac{1}{a^{2} \Lambda_{\rm QCD}^{2}})}\right).
\eea
Therefore, again we have two global parameters $\lambda$ and $\Lambda_{\rm QCD}$ in the fit.

\begin{figure}[tbp]
\centering
\includegraphics[width=8cm]{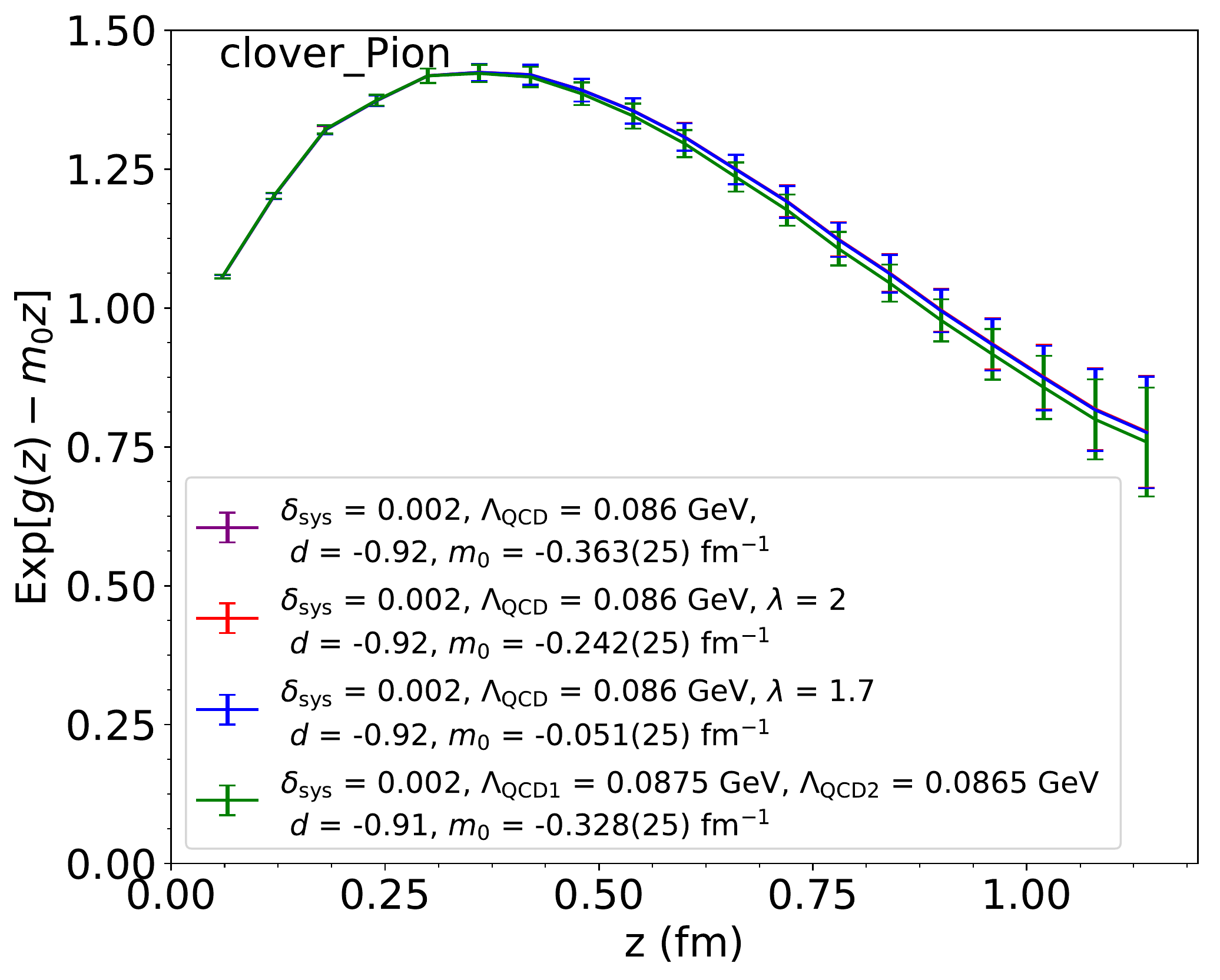}
\caption{
Renormalized $O_{\gamma_t}(z)$ correlations in the pion state for the clover action without HYP smearing based on different fitting functions: Eq.~(\ref{eq:logM}) (purple), Eq.~(\ref{eq:logM_2}) (red), Eq.~(\ref{eq:logM_3}) (blue), Eq.~(\ref{eq:logM_4}) (green).
}
\label{fig:Re_overlap_quark_dff}
\end{figure}

The fitting process with the above parametrization is similar to what is described in Sec.~\ref{sec:reofpro}. The results are shown in Figs.~\ref{fig:Re_overlap_quark_2} and \ref{fig:Re_overlap_quark_3}. 
As one can see, the range of $\lambda$ and $\Lambda_{\rm QCD}$ are rather large, and strongly correlated. After picking several correlated values of the pair, matching to the perturbative result, the resulting residual matrix elements are shown in the lower panel. One can hardly see any difference between the different choices. Moreover, the results are essentially the same as the fit
without the extra $\lambda$ parameter (Fig.~\ref{fig:Re_overlap_quark_dff}). Therefore, we conclude
that with the data at hand, our result is stable with respect to the higher order corrections. Only with more accurate data, one might be able to see the difference of a higher-order fit. 

\begin{figure}[tbp]
\centering
\includegraphics[width=9cm]{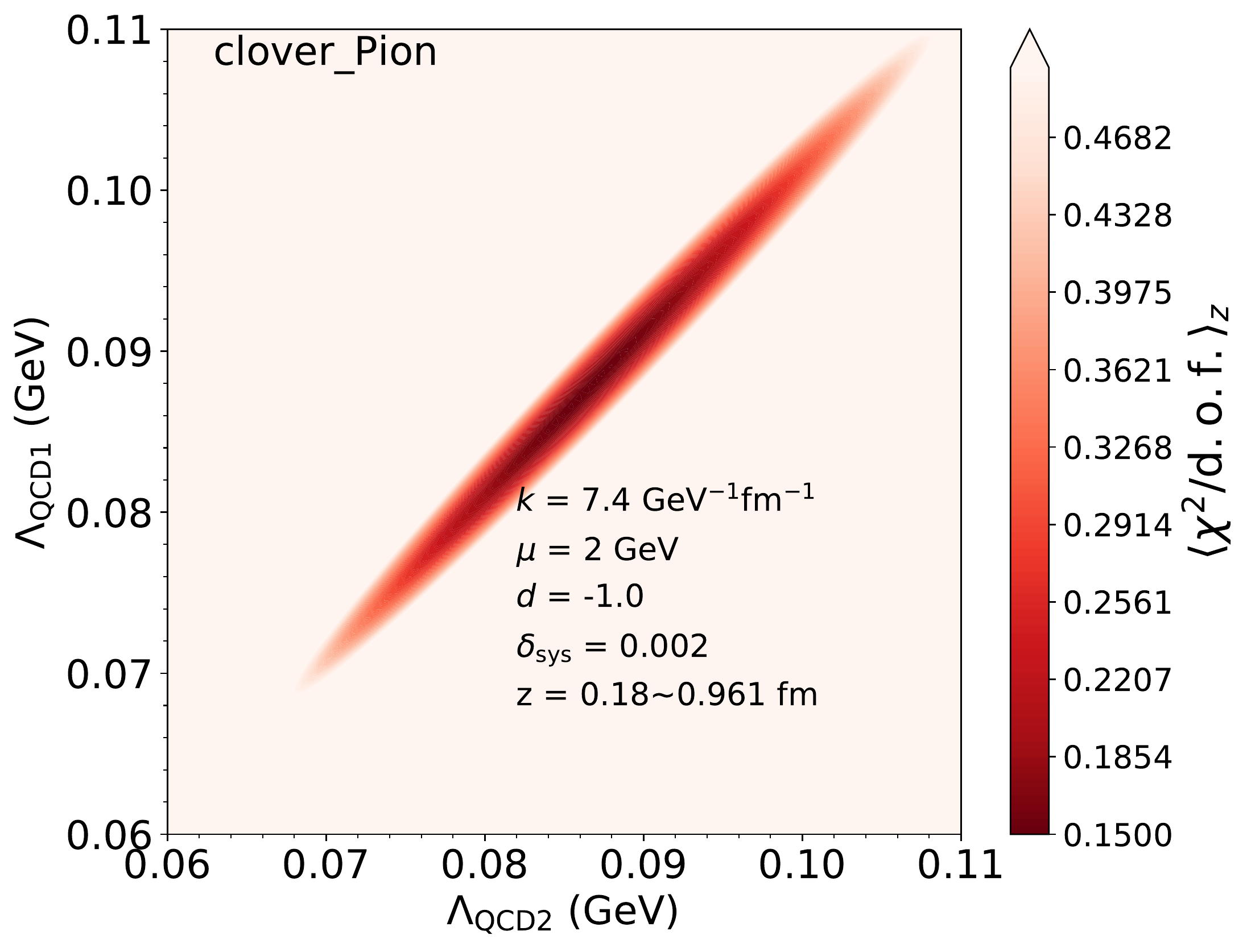}
\includegraphics[width=8cm]{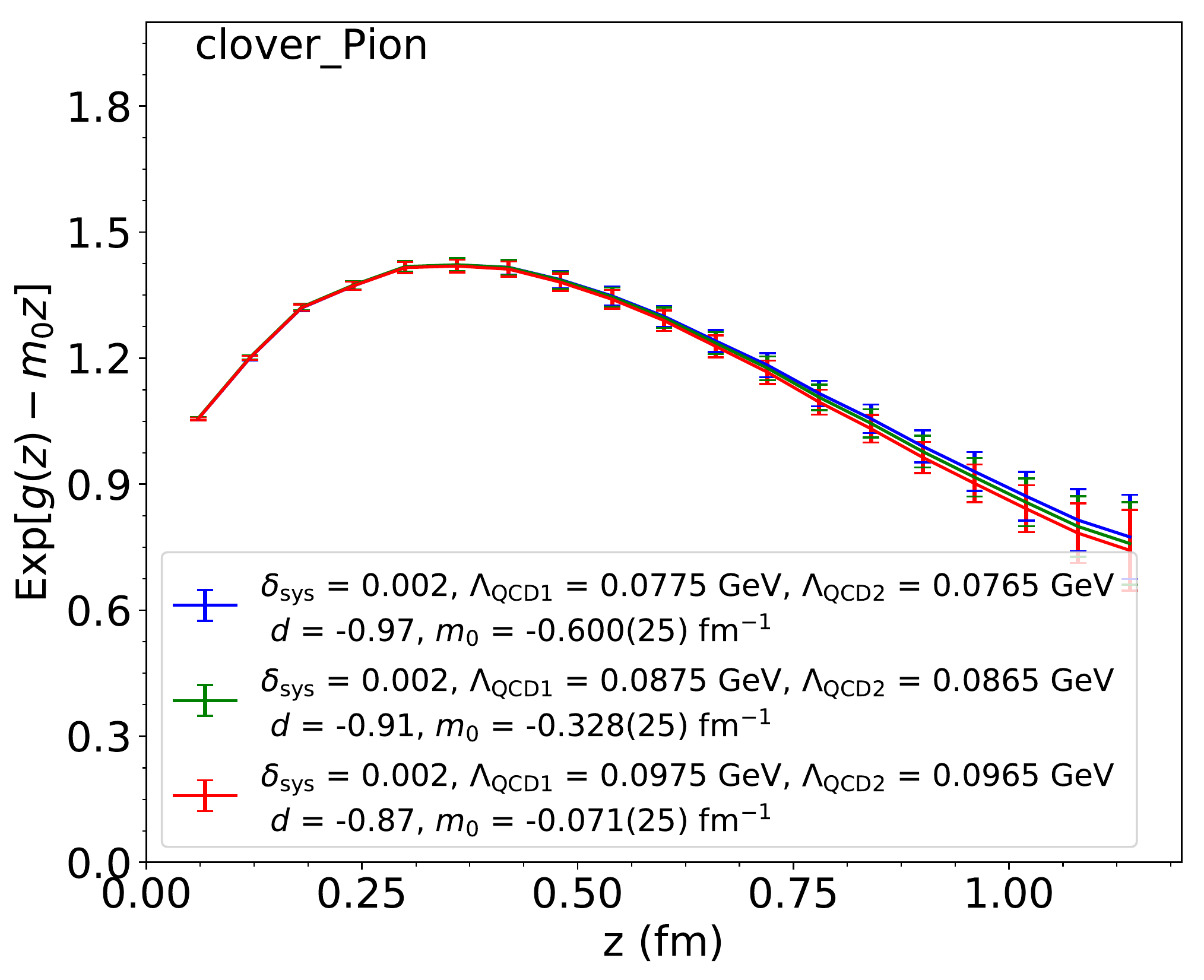}
\caption{
Using Eq.~(\ref{eq:logM_4}) to renormalize the $O_{\gamma_t}(z)$ matrix element in the pion state for the clover action without HYP smearing. Upper Panel: $\chi^2$ map with respect to $\Lambda_{\rm QCD1}$ and $\Lambda_{\rm QCD2}$. $\langle  \chi^{2}/{\rm d.o.f.} \rangle_{z}$ is the average of $\chi^{2}/{\rm d.o.f.}$ from fitting for each $z$. Lower Panel: renormalized matrix element for several sets of ($\Lambda_{\rm QCD1}$, $\Lambda_{\rm QCD2}$) along the 'small-$\chi^2$ band'.
}
\label{fig:Re_overlap_quark_4}
\end{figure}

Another possibility one can explore is that the higher-order corrections
represented by $\Lambda_{\rm QCD}$ might be different for different ensembles. Therefore, one can discuss different  
$\Lambda_{\rm QCD}$ values for MILC and RBC data,
\begin{align}\label{eq:logM_4}
\ln \mathcal{M}(z,a) = \frac{k z}{a \ln[a \Lambda_{\rm QCD1,2}]} + g(z) + f_{1,2} (z) a \nonumber\\
+\frac{3 C_{F}}{b_0} \ln [\frac{\ln [1 /(a \Lambda_{\rm QCD1,2})]}{\ln [\mu / \Lambda_{\rm QCD1,2}]}] + \ln [1+\frac{d}{\ln (a \Lambda_{\rm QCD1,2})}],
\end{align}
where we fix the $k$ parameter and obtain a two parameter fit, $\Lambda_{\rm QCD1}$ and $\Lambda_{\rm QCD2}$. 
The results of using the above equation is shown in Fig.~\ref{fig:Re_overlap_quark_4}. As shown, the two parameters 
are strongly correlated and proportional to each other. The difference 
of the two is limited to a very small range. Furthermore, with different
choices of the correlated pair ($\Lambda_{\rm QCD1}$, $\Lambda_{\rm QCD2}$), the final renormalized results do not change. 

In Fig.~\ref{fig:Re_overlap_quark_dff}, we show the results of renormalized matrix elements based on different fitting functions. All of these fitting functions lead to the same result. Thus, given the data, it is 
unnecessary to consider other high-order terms in the fitting function and Eq.~(\ref{eq:logM}) is good enough. This also supports the argument made in Sec.~\ref{ReUncer} that we can use the fitted $\Lambda_{\rm QCD}$ in leading order to partially represent higher order corrections.

Finally, we consider $O(a^{2})$ corrections to our fitting formula. These fits were unstable, which indicates that we introduce too many fit parameters for the given data quality. The conclusion is that higher precision data is needed to constrain $O(a^2)$ terms. One can also make fits
by including  $O(a^{2})$  instead of the linear correction. The physics for doing it is
weak in our cases.
  
\section{Renormalization of lattice smeared matrix elements}\label{sec:smearing}
  
In many lattice calculations, the data can be very noisy
which calls for more statistics. However, 
large statistics means more resources which are hard to 
come by sometimes. Therefore, lattice practitioners have
invented phenomenological approaches to quench the short-distance
fluctuations: smearing. However, smearing might lead 
to configurations that are not connected with fundamental theory. On the other hand, smearing in some sense just makes the effective UV cutoff (the inverse of the lattice spacing) smaller, and in the limit $a\to 0$, all smearing becomes local effects which can be taken into account properly through some kind of renormalization. 
In this section, we check if the procedure we discussed
in the previous section still works for smeared matrix elements. 

\begin{figure*}[tbp]
\centering
\includegraphics[width=8.8cm]{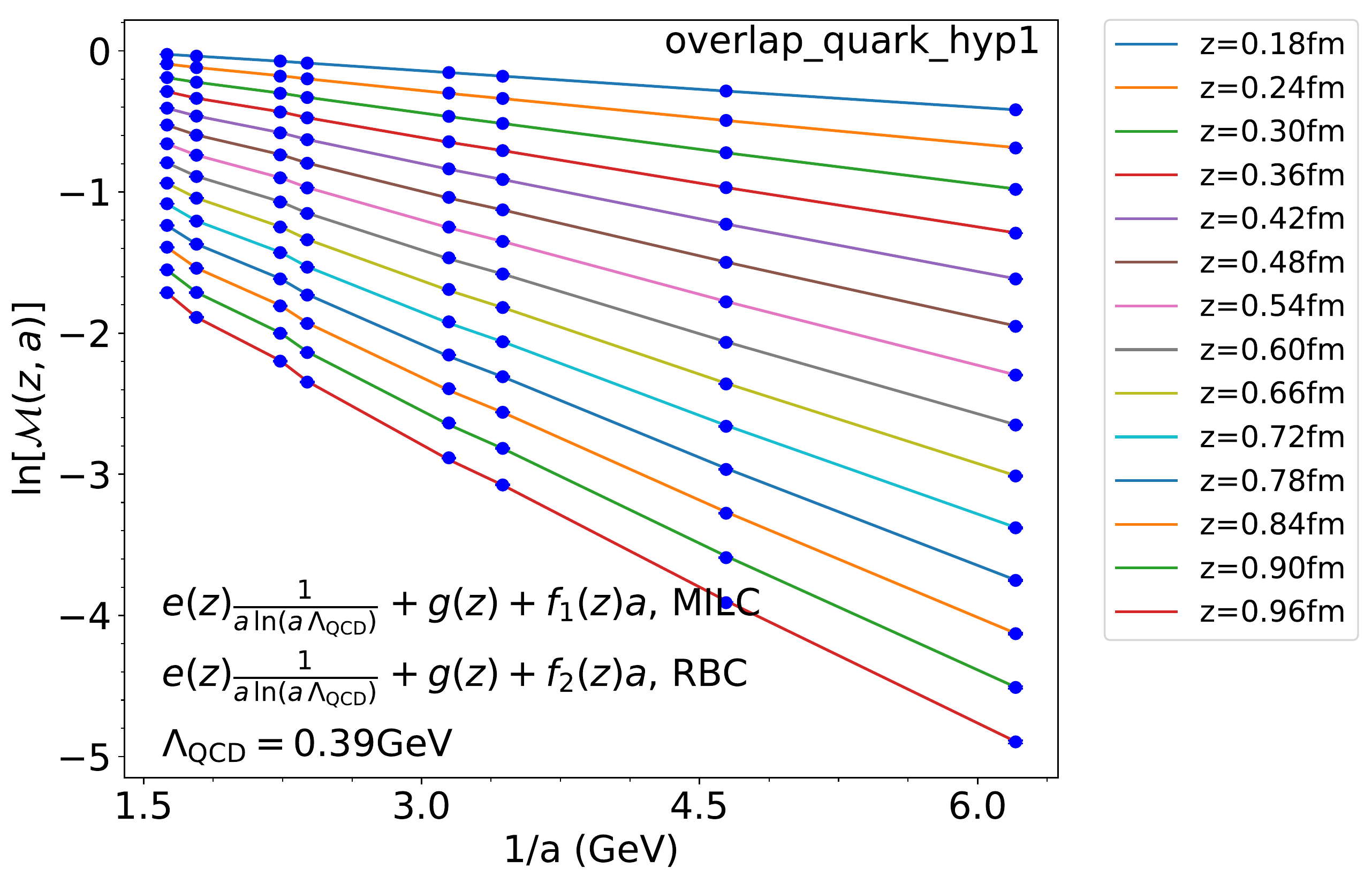}
\includegraphics[width=8.8cm]{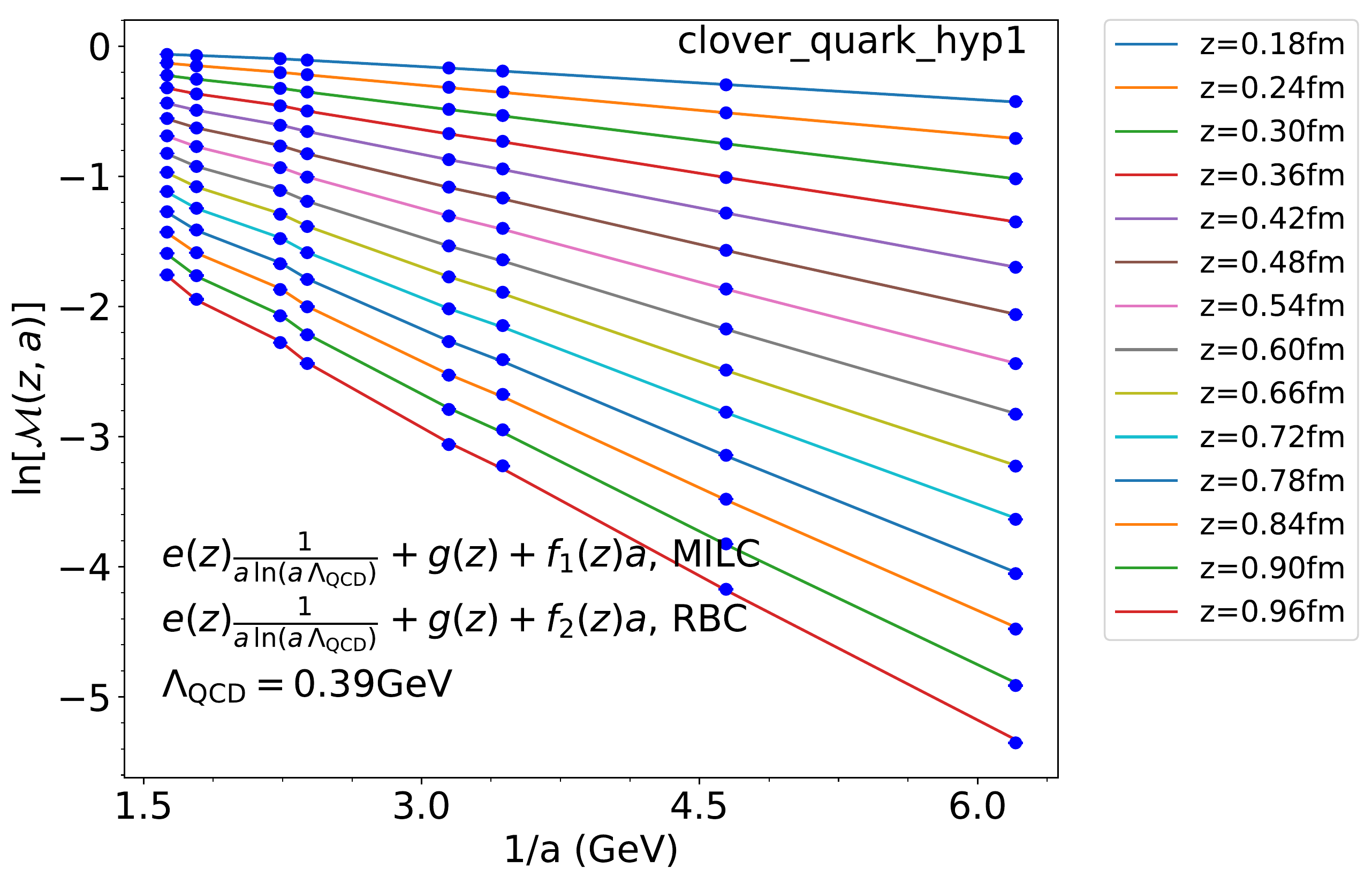}
\includegraphics[width=8.8cm]{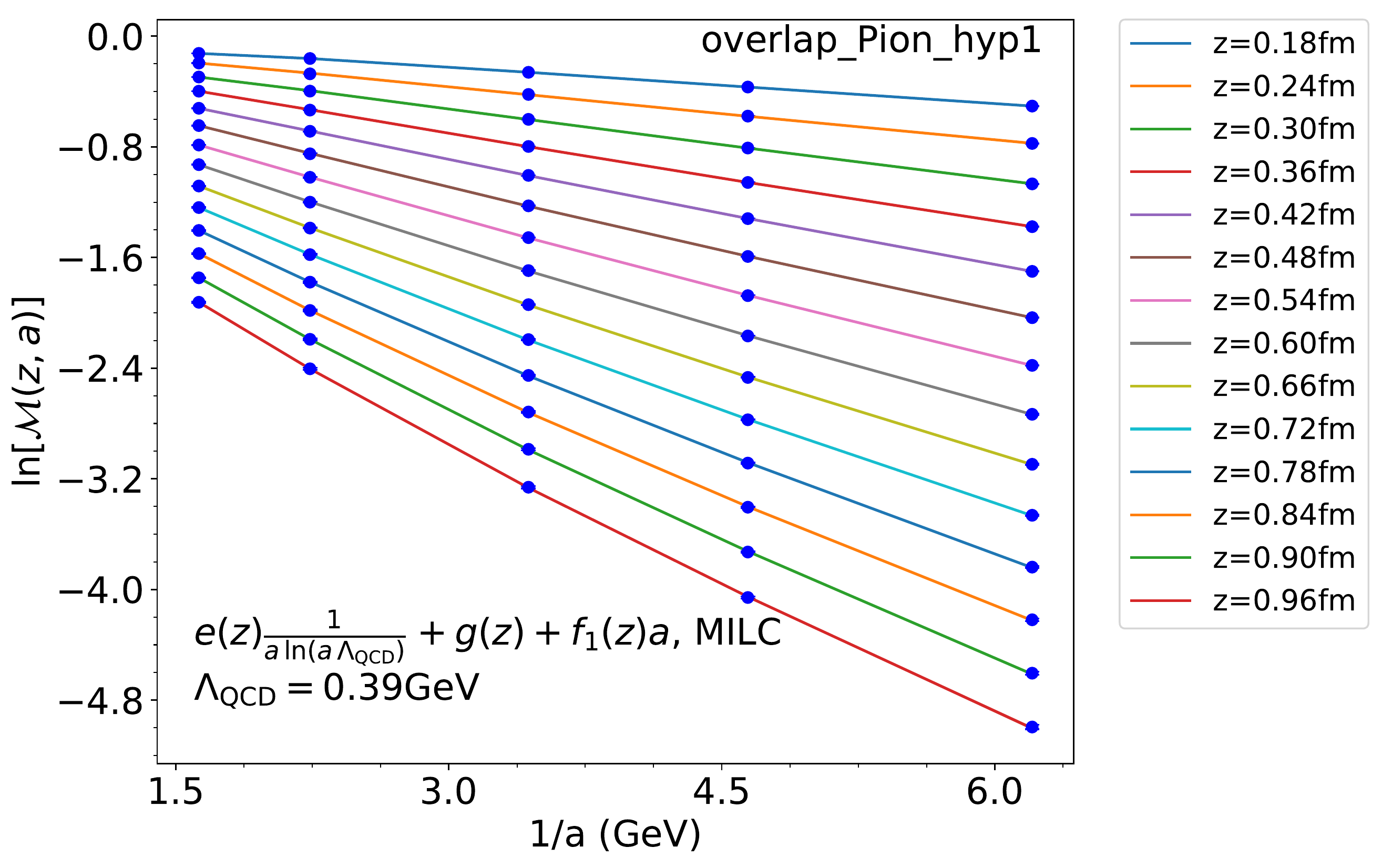}
\includegraphics[width=8.8cm]{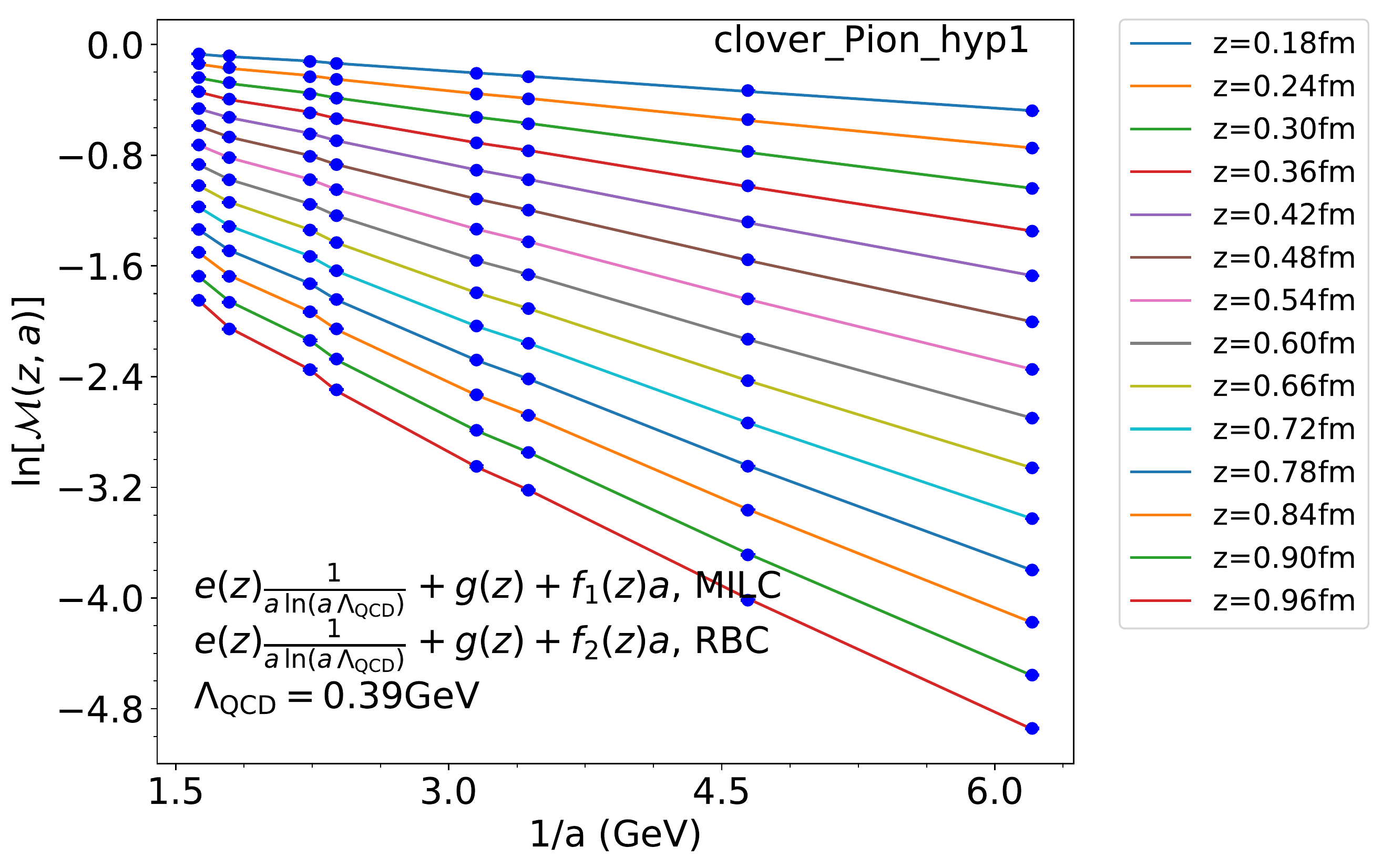}
\includegraphics[width=8.8cm]{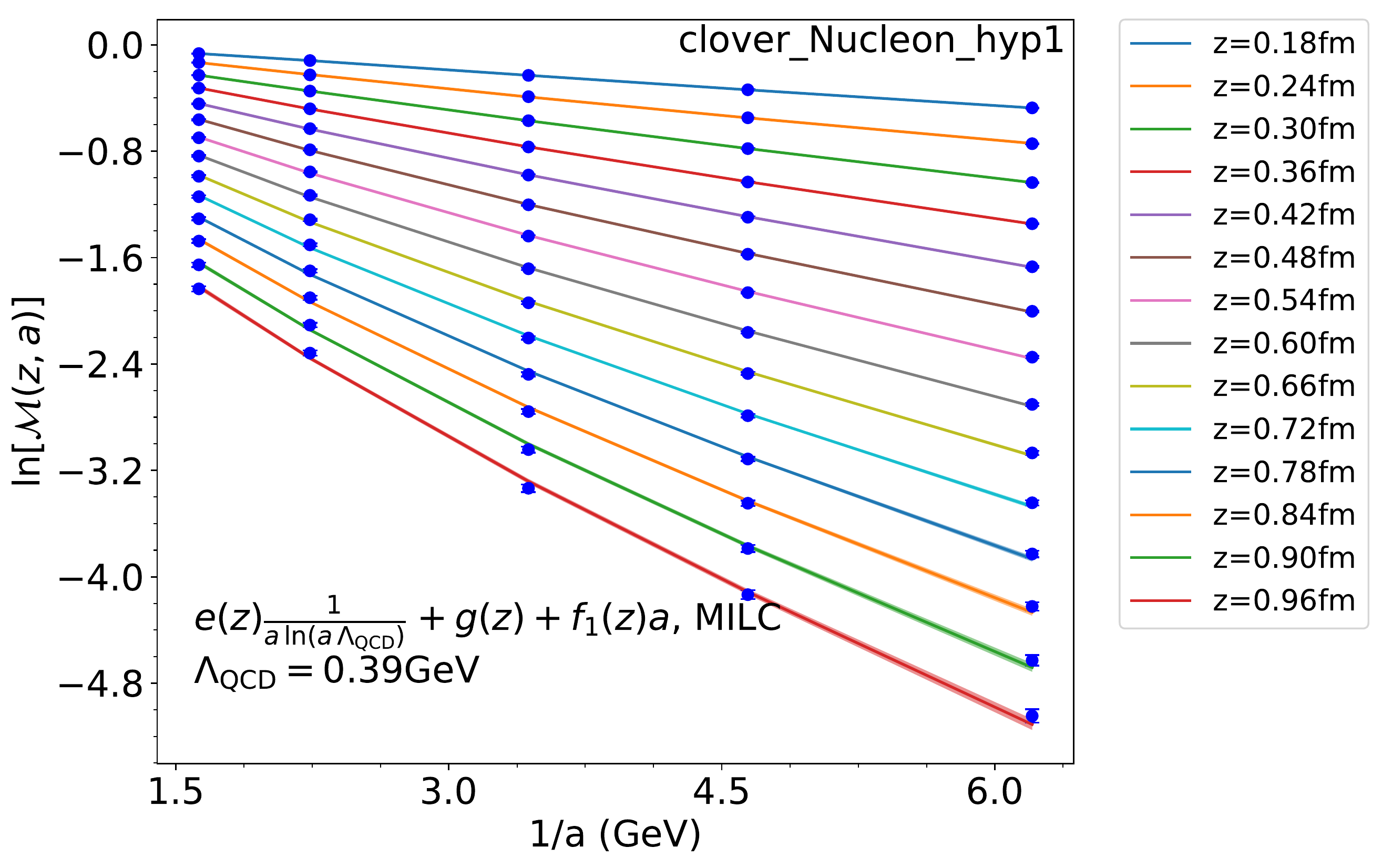}
\caption{Using Eq.~(\ref{eq:logM_ss}) to fit $O_{\gamma_t}(z)$ matrix element for HYP smearing cases. $\Lambda_{\rm QCD}$ is fixed at 0.39 GeV.
}
\label{fig:fitting_hyp}
\end{figure*}

\begin{figure*}[tbp]
\centering
\includegraphics[width=8.5cm]{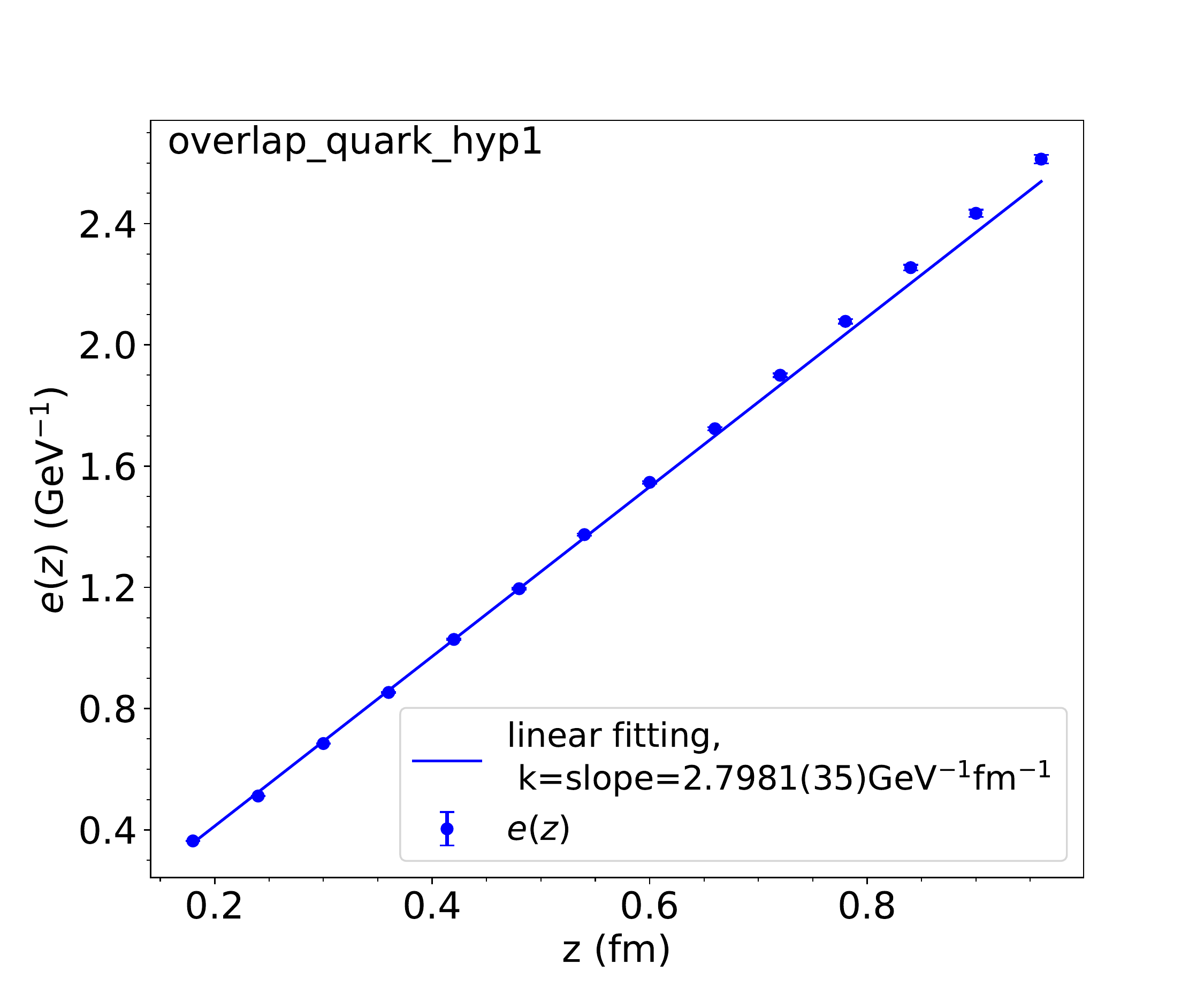}
\includegraphics[width=8.5cm]{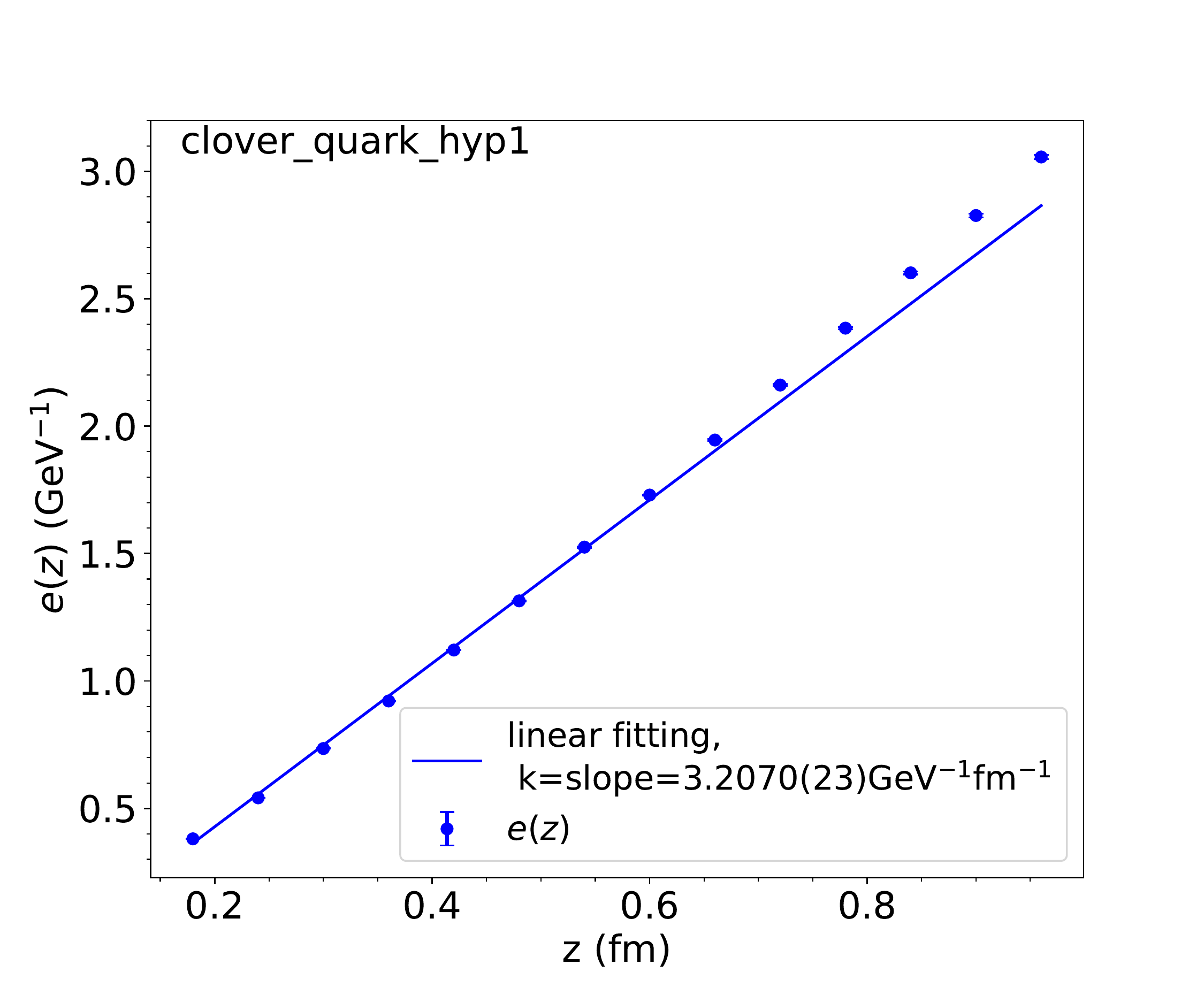}
\includegraphics[width=8.5cm]{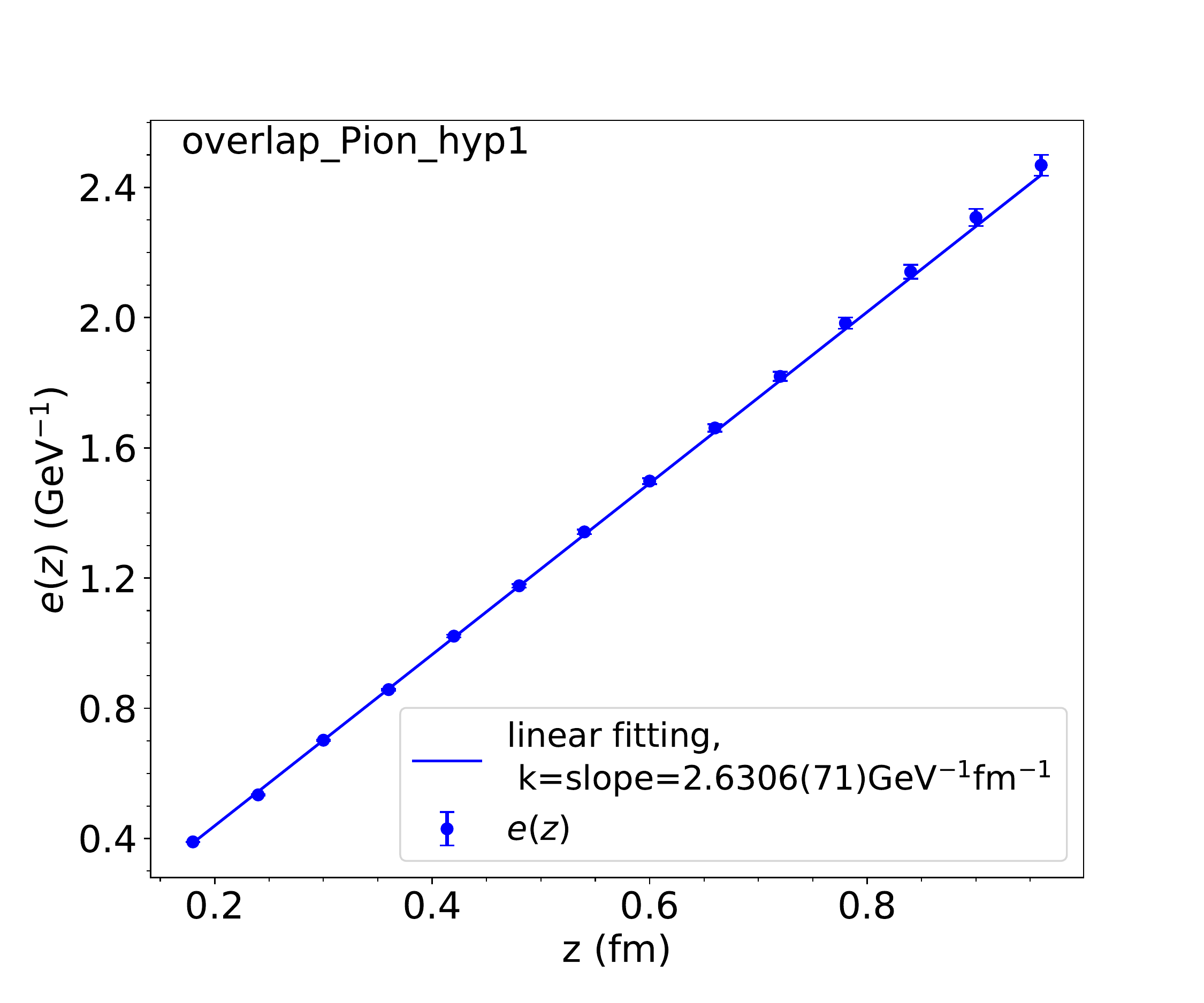}
\includegraphics[width=8.5cm]{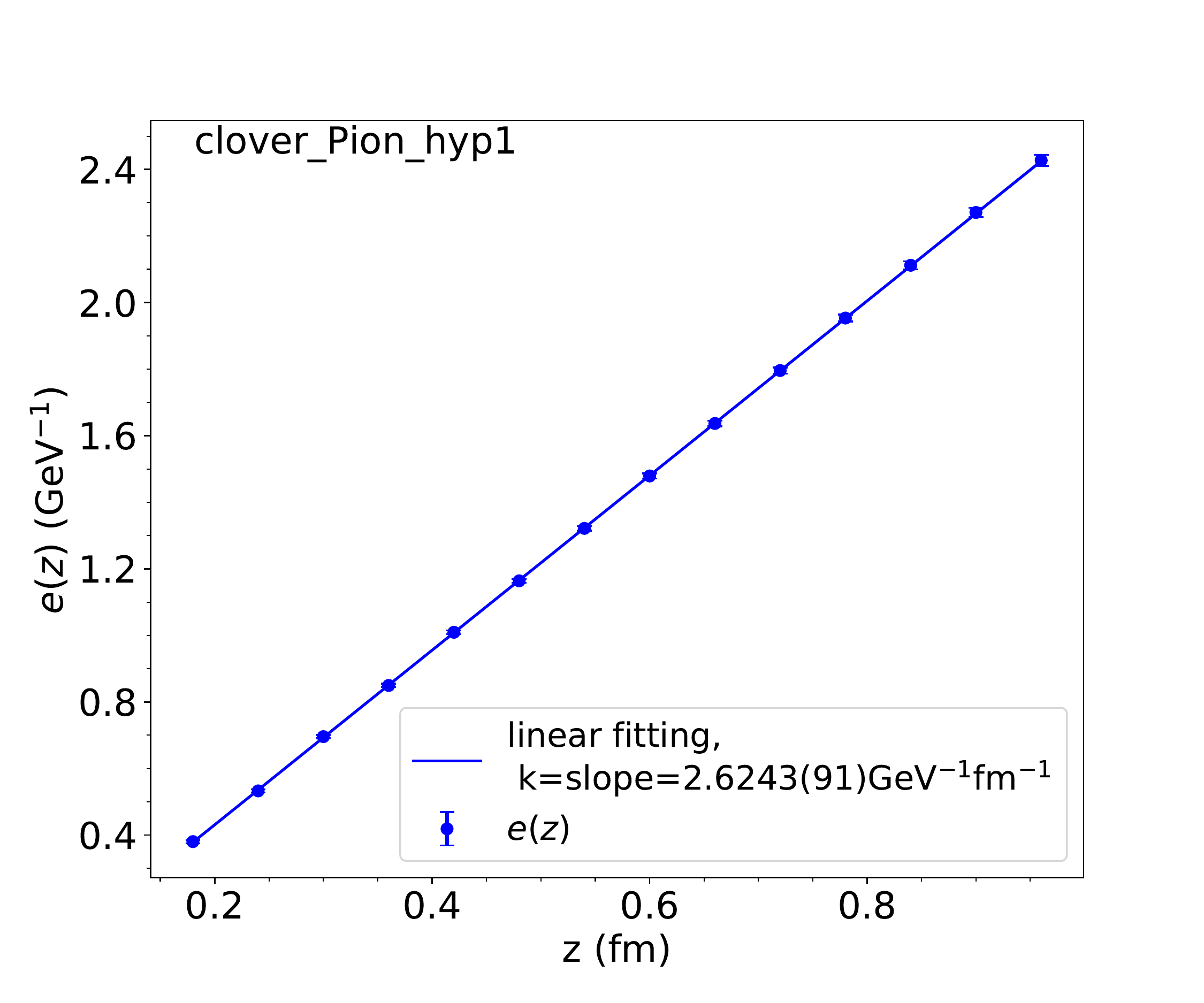}
\includegraphics[width=8.5cm]{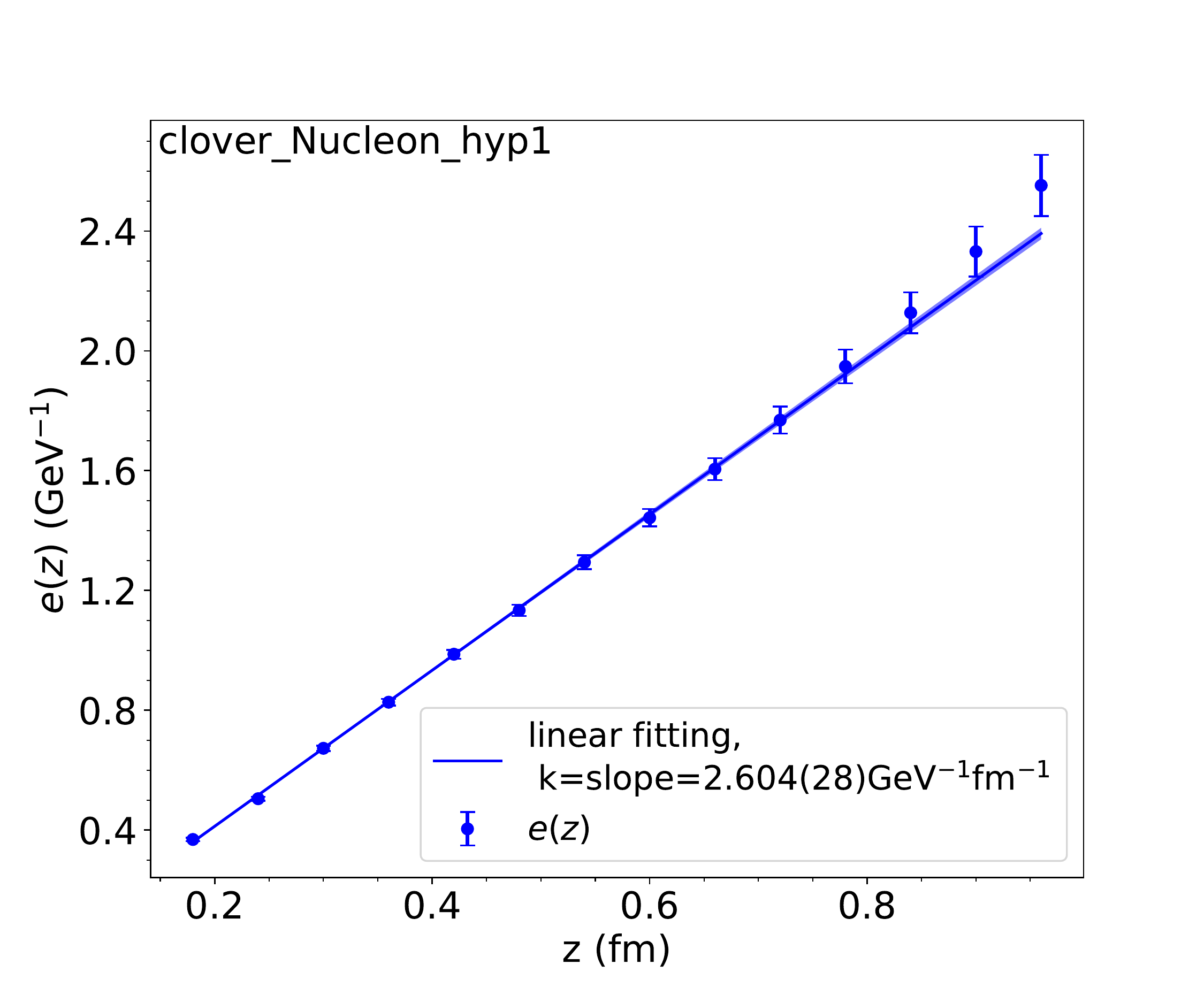}
\caption{$e(z)$ with respect to $z$. $e(z)$ is extracted from the fitting in Fig.~\ref{fig:fitting_hyp}. The slopes are labeled in the plot legend.
}
\label{fig:e(z)_hyp}
\end{figure*}

To ensure that the HYP smearing data are useful for refined analysis, we can do a simple test on whether the HYP smearing data show the properties of the linear divergence predicted by perturbation theory, as we have done in Sec.~\ref{sec:test}. Here we use the following function to fit the bare matrix element $\mathcal{M}$ to extract the linear divergence factor,
\begin{align}\label{eq:logM_ss}
\ln \mathcal{M}(z,a) = \frac{e(z)}{a \ln[a \Lambda_{\rm QCD}]} 
+ g(z) + f_{1,2}(z)a,
\end{align}
where we allow for a discretization error to get a better fit. We fix $\Lambda_{\rm QCD}$ at 0.39 GeV in this simple test. Fig.~\ref{fig:fitting_hyp} shows the fits for different actions and states which all look reasonable. 
We have not shown $\chi^2$ which will be considered in the extraction
of the renormalization factor.

Fig.~\ref{fig:e(z)_hyp} shows the extracted $e(z)$ with respect to $z$. The linear $z$ dependence is approximately reproduced but some deviations are seen at large $z$. We can perform a linear fitting of $e(z)$ and compare the fitted slopes.
Overall, the quality of the fits is not as good as in the unsmeared cases, although it is still rather impressive. 
The slope extracted for the linear divergence for the clover quark is totally different from the others, which explains why we cannot use the RI/MOM factor to eliminate it in the clover case for HYP smearing data~\cite{Zhang:2020rsx}. Although the slopes for the overlap quark and pion, clover pion and nucleon are similar to each other, there are some small discrepancies. Using the RI/MOM factor for these cases may leave small residual linear divergences, and the self-renormalization method provides a better way to renormalize the matrix elements. 

Next, we extend our self-renormalization method to the HYP smeared data. Our fitting function, Eq.~(\ref{eq:logM}) may have no clear physical meanings for the HYP smeared date, 
but it may be regarded as a phenomenological model.  

\begin{figure*}[tbp]
\centering
\includegraphics[width=8.5cm]{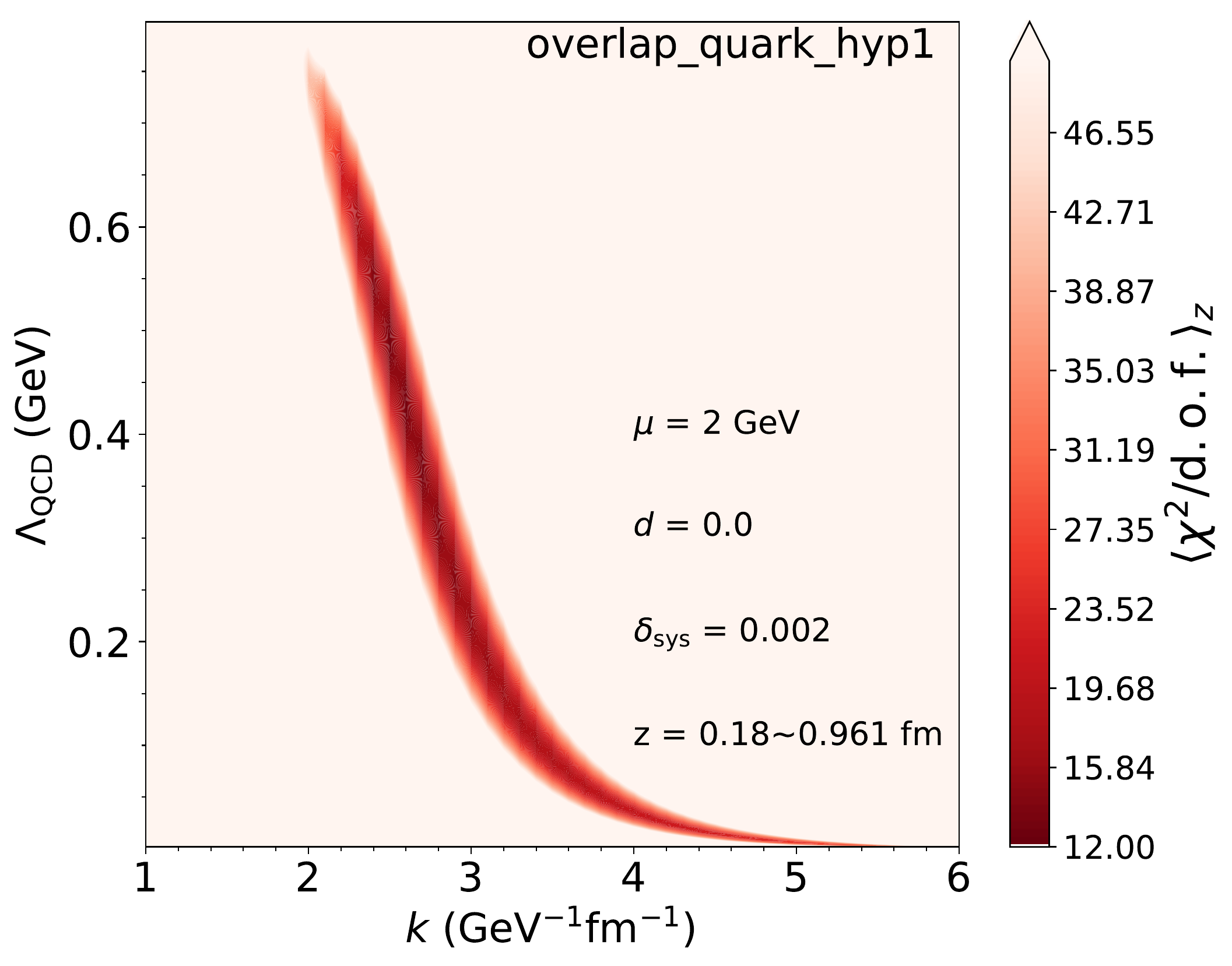}
\includegraphics[width=8.5cm]{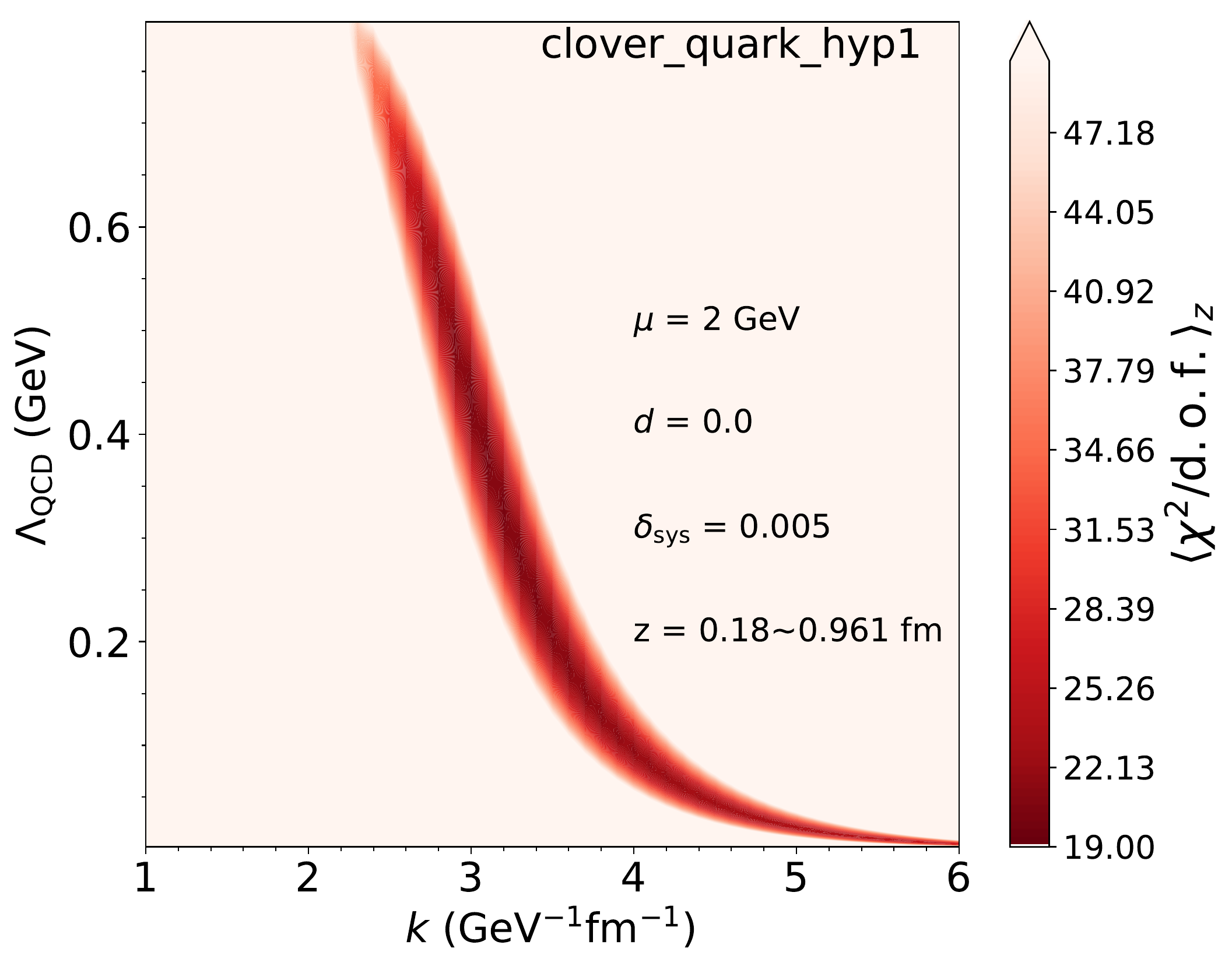}
\includegraphics[width=8.5cm]{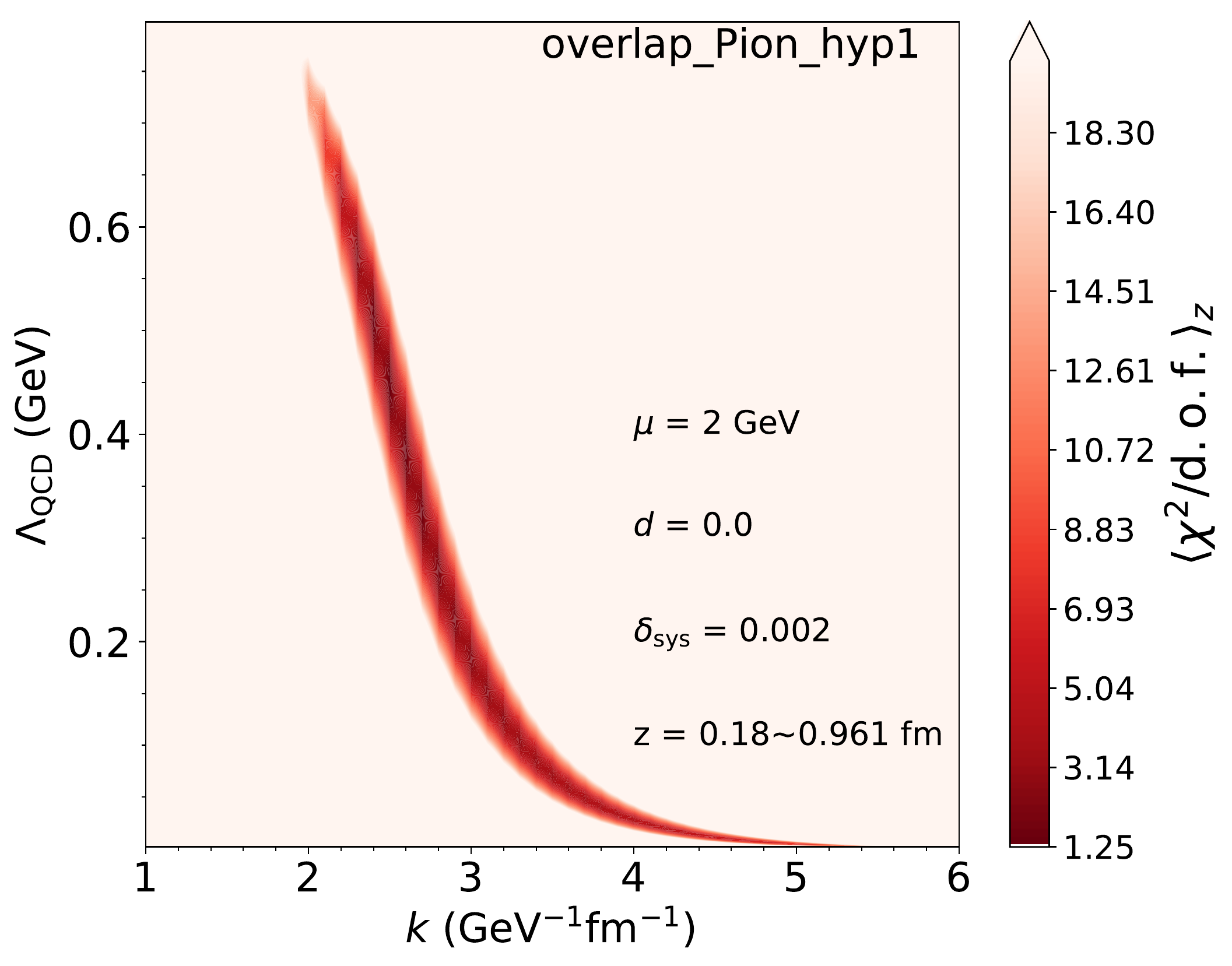}
\includegraphics[width=8.5cm]{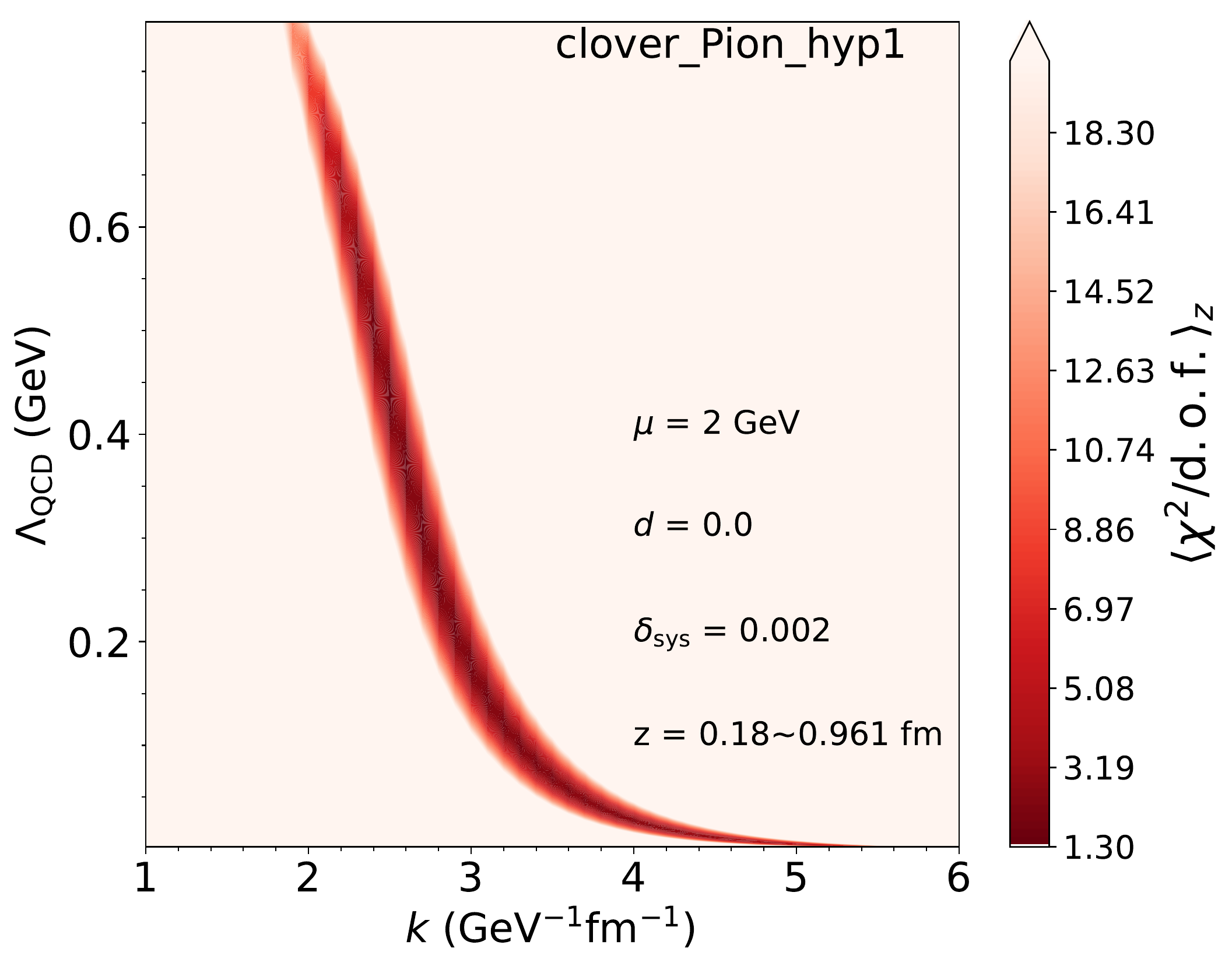}
\includegraphics[width=8.5cm]{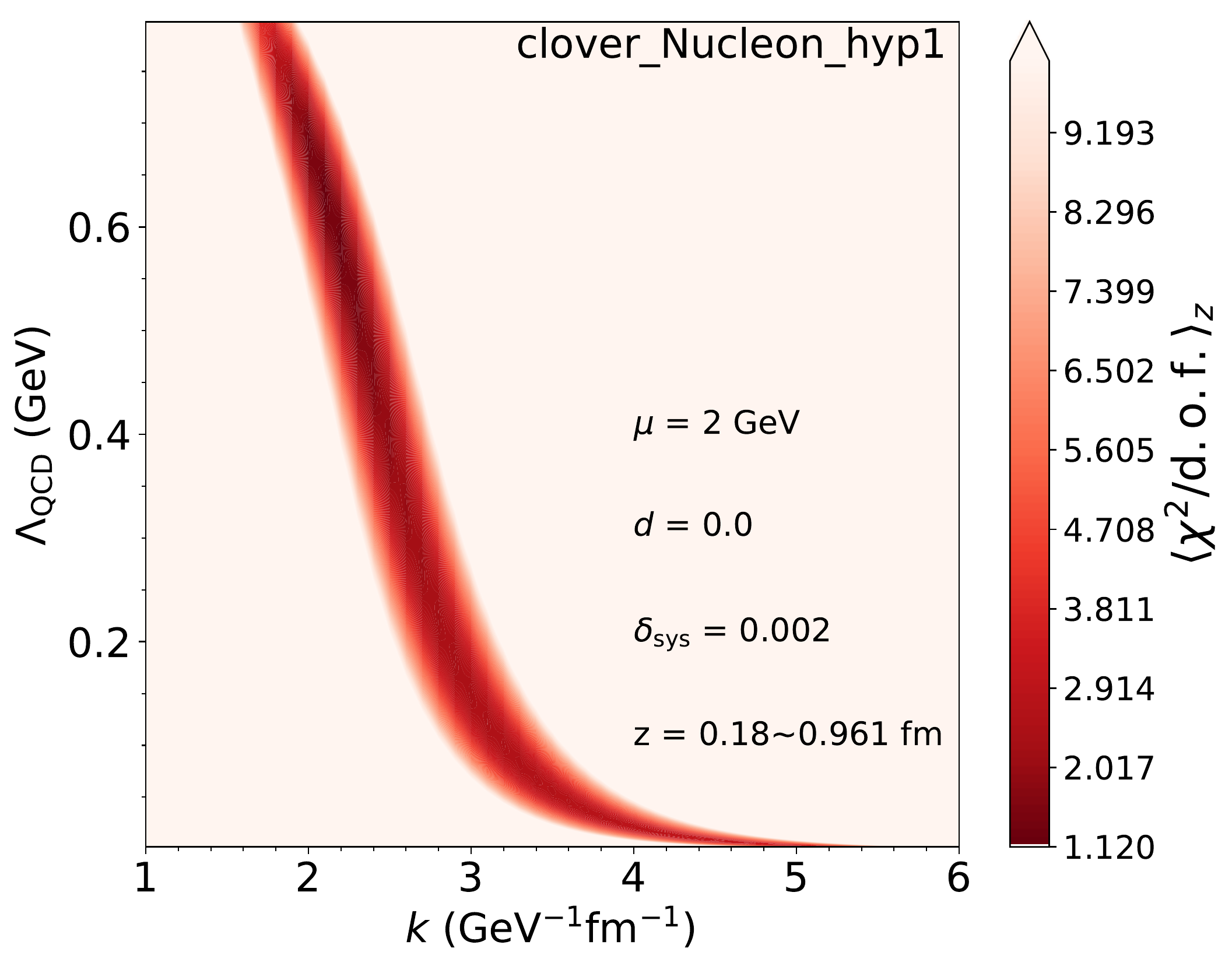}
\caption{Using Eq.~(\ref{eq:logM}) to fit the $O_{\gamma_t}(z)$ matrix element for HYP smearing. The plot shows the $\chi^{2}$ map with respect to $k$ and $\Lambda_{\rm QCD}$ for different actions and states. $\langle  \chi^{2}/{\rm d.o.f.} \rangle_{z}$ is the average of $\chi^{2}/{\rm d.o.f.}$ from fitting for each $z$.
}
\label{fig:chisqmap_hyp}
\end{figure*}

\begin{figure*}[tbp]
\centering
\includegraphics[width=8.5cm]{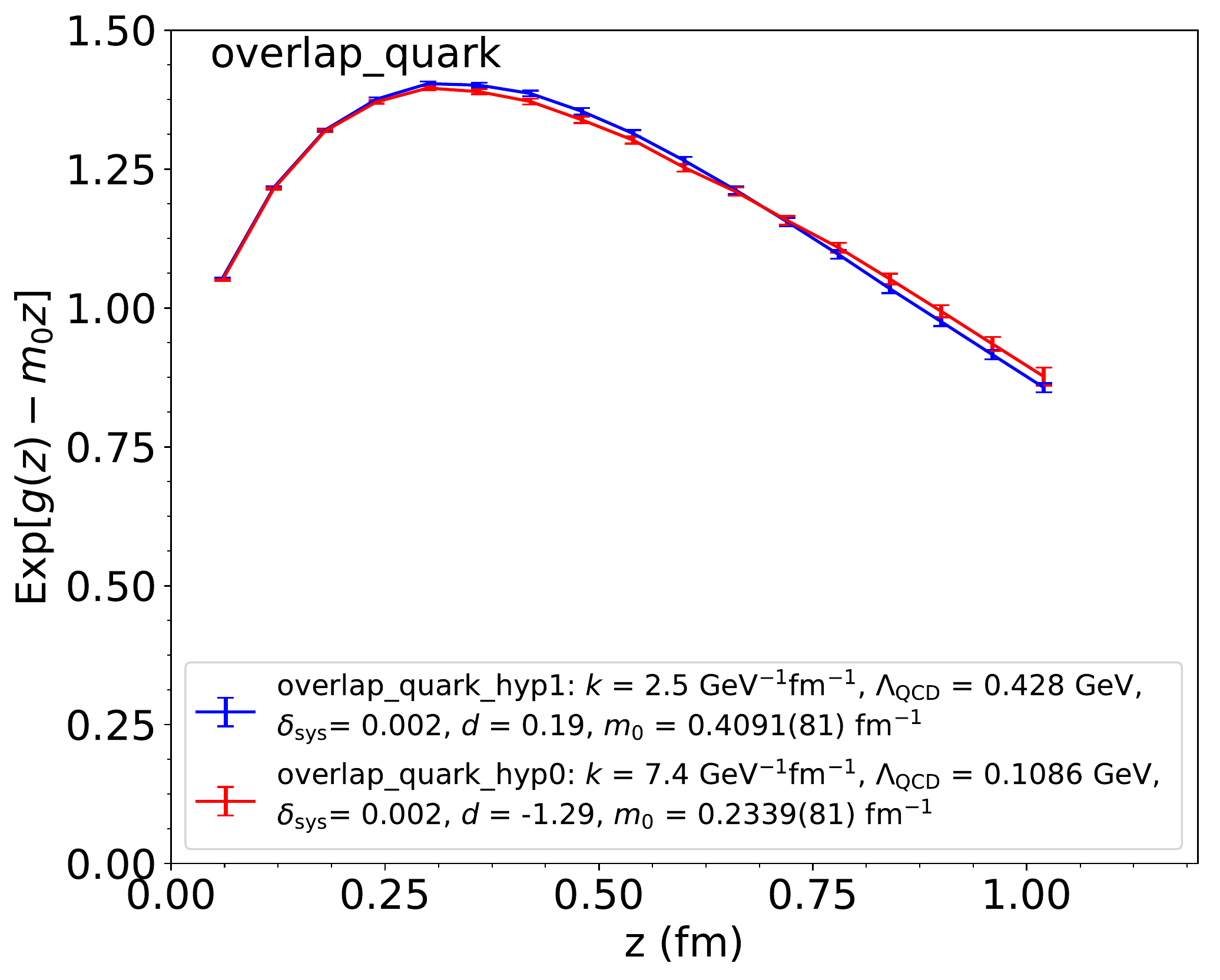}
\includegraphics[width=8.5cm]{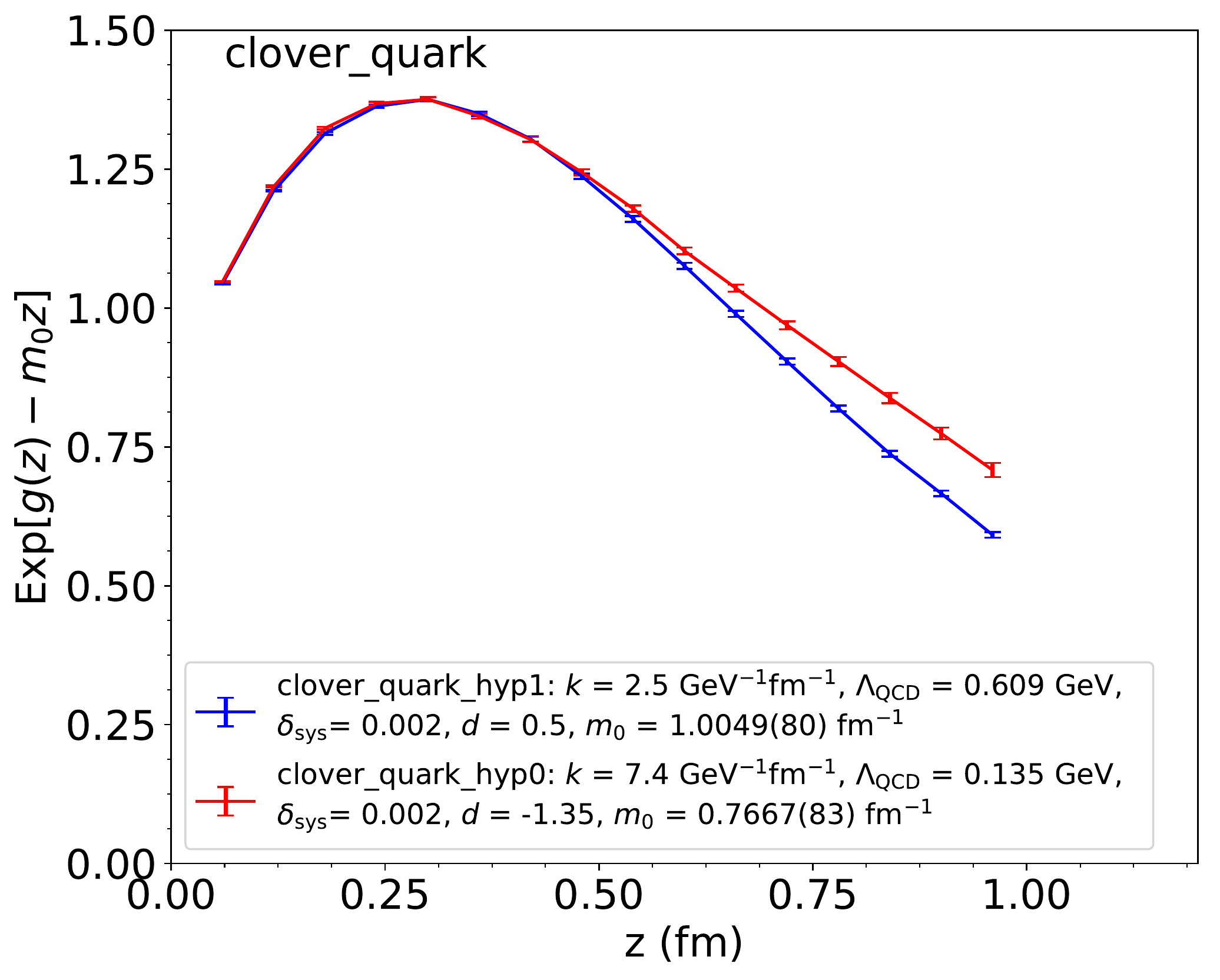}
\includegraphics[width=8.5cm]{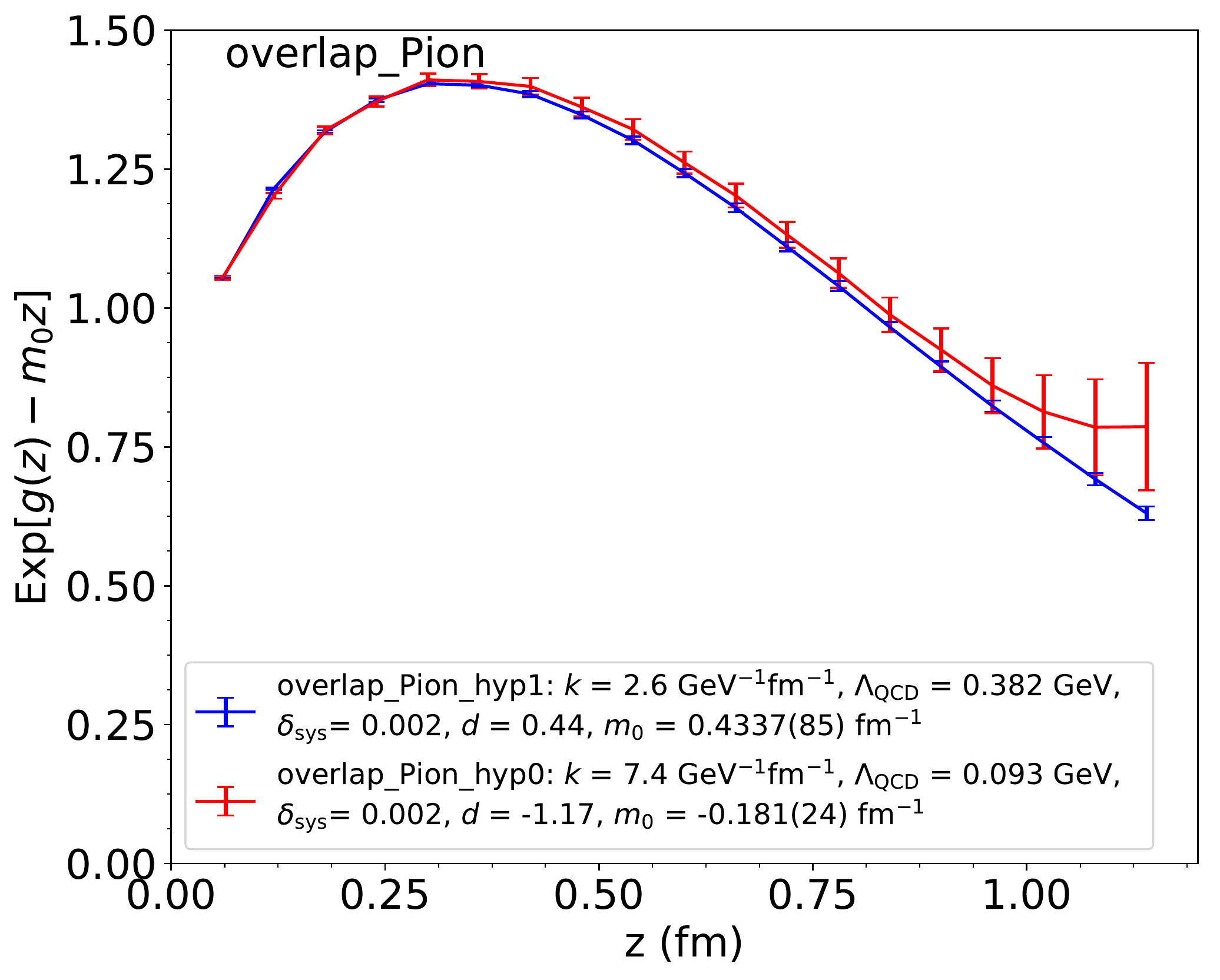}
\includegraphics[width=8.5cm]{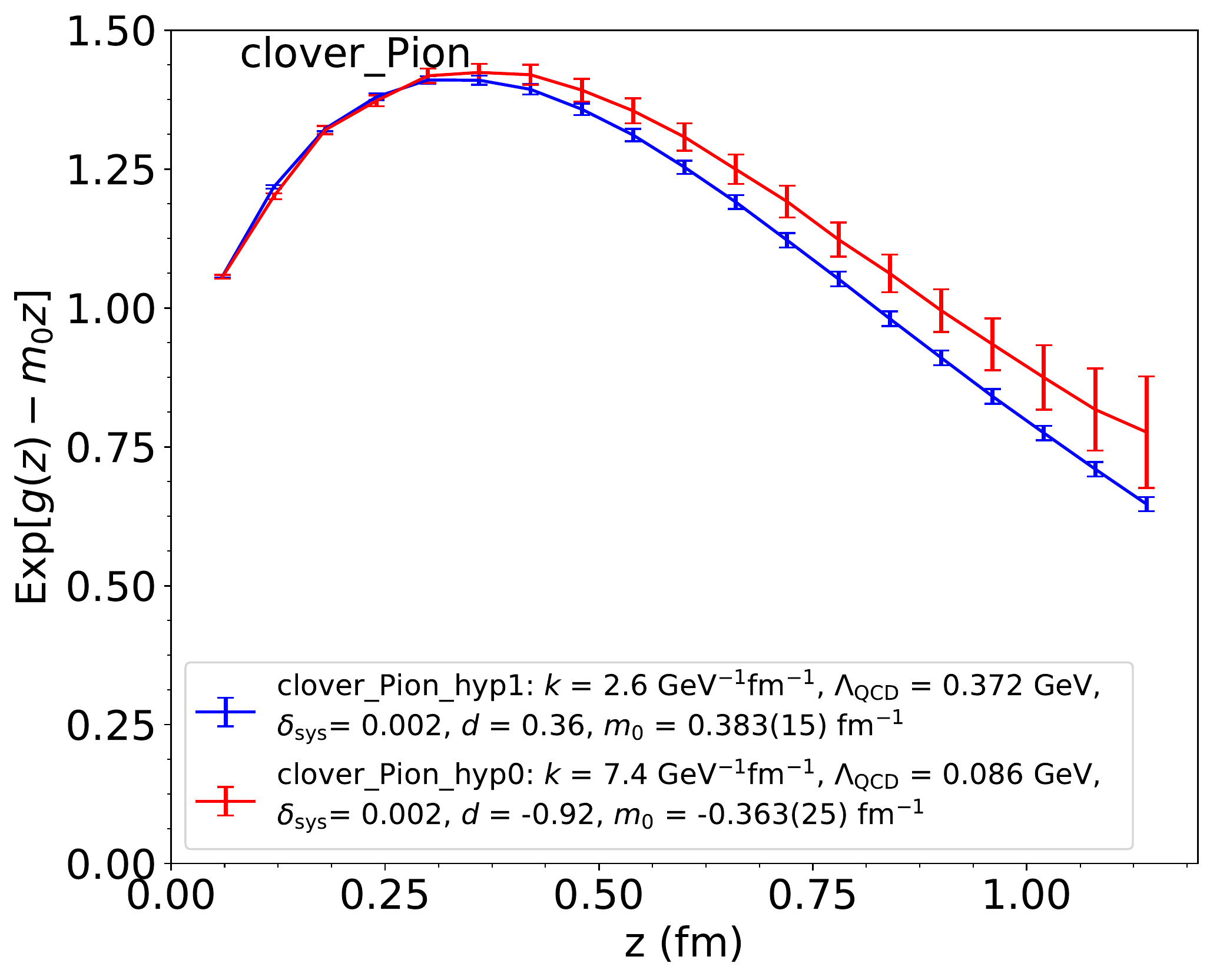}
\includegraphics[width=8.5cm]{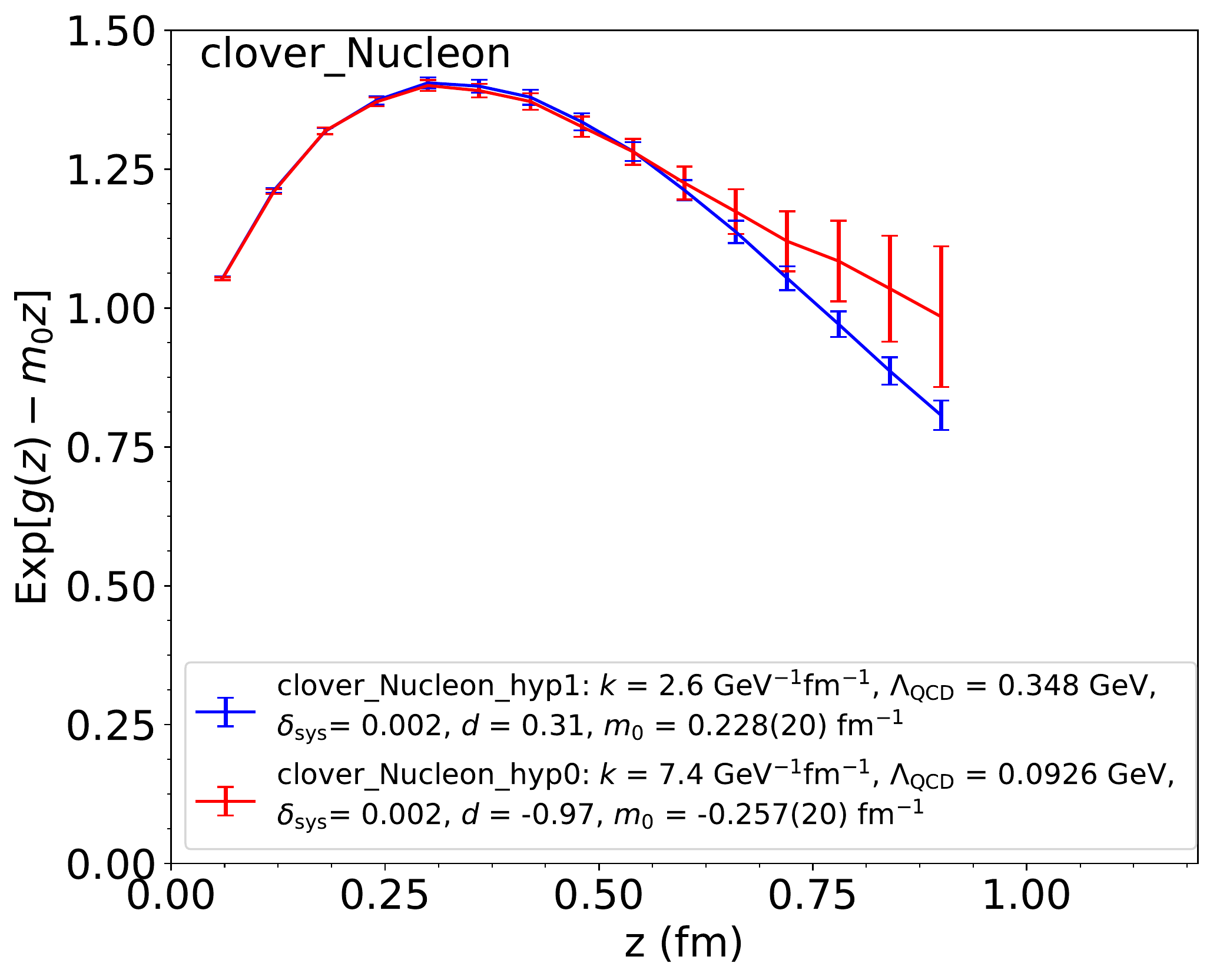}
\caption{Comparison between renormalized matrix elements Exp[$g(z)-m_{0}z$] for unsmearing (hyp0, red) and HYP smearing (hyp1, blue) cases.
}
\label{fig:Re_hyp}
\end{figure*}

We start with a fit function with two parameters in the linearly divergent part, $k$ and $\Lambda_{\rm QCD}$. 
Fig.~\ref{fig:chisqmap_hyp} shows the $\chi^{2}$ map with respect to them. It is clear
from the plots that in the ``small-$\chi^2$ band'', we cannot find a solution for $k$ with the  one-loop perturbative value 7.4 GeV$^{-1}$fm$^{-1}$ (1.46 if dimensionless, see Eq.~(\ref{eq:k})) any more. 
This is expected because the lattice space is enlarged by a factor of two, and therefore the values of $k$ in the ``small-$\chi^2$ band'' are expected to be at most
half of this, named 3.7 GeV$^{-1}$fm$^{-1}$ (0.73 if dimensionless). In fact, most probable $k$ is now around 
3.0 GeV$^{-1}$fm$^{-1}$ (0.59 if dimensionless) or below. Another way of saying this is HYP smearing smooths out the short-range behaviors such as the linear divergence. The value of $\Lambda_{\rm QCD}$, on the other hand, has a larger range,
possibly corresponding again to a larger effective lattice spacing.

So for the HYP smearing cases, we choose a set of ($k$, $\Lambda_{\rm QCD}$) near the center of the ``small-$\chi^{2}$ band''. We choose $k=2.5$ GeV$^{-1}$fm$^{-1}$ (0.49 if dimensionless) for overlap quark and clover quark, and $k=2.6$ GeV$^{-1}$fm$^{-1}$ (0.51 if dimensionless) for overlap pion, clover pion and clover nucleon. The corresponding $\Lambda_{\rm QCD}$ shows a larger dispersion, 
with $\Lambda_{\rm QCD} = 0.43$ and $0.38$ GeV for overlap quark and pion correlations respectively, and $\Lambda_{\rm QCD} = 0.61$, $0.37$ and $0.35$ GeV for clover 
quark, pion and nucleon correlations, respectively.

Once the ($k$, $\Lambda_{\rm QCD}$) parameters are chosen, we 
subtract the linear divergences, and match the result
to the perturbative matrix elements to determine possible
further subtractions for the non-perturbative mass
effect. Fig.~\ref{fig:Re_hyp} shows that the renormalized matrix elements for HYP smearing in blue ticks and lines, 
in comparison with the cases without HYP smearing. 

The large difference between smeared and unsmeared matrix
elements is seen in the case of clover quark correlation 
at large $z$. 
Again, we speculate that due to the chiral symmetry breaking effects, the linear divergence has some curious behavior there. However, the difference is smaller in the pion matrix element where the error bars are also bigger. For the overlap case, 
the correlations for quark and pion show no substantial change 
due to smearing effects. On the other hand, the matrix
element in the nucleon case shows a much smaller 
error bar in the smeared case, which demonstrates the power 
of smearing. 

To conclude, our analysis method to renormalize
the linear divergent matrix elements can successfully be used
for smeared matrix elements as well. Smearing does not seem to change the behavior of renormalization qualitatively, but only quantitatively. 
The intrinsic physics does not seem to change under (moderate) smearing,
consistent with the expectation that smearing is an alternative method to reduce the statistical noise in lattice calculations.

\section{Linear Divergence of Correlations in QCD Vacuum  }\label{sec:otherresult}

In this section, we analyze the linear divergence of 
the Wilson line in the matrix elements taken in the QCD
vacuum. We try to test
if the divergence is universal and if the divergent mass can be extracted successfully from these matrix elements. We consider
three new types of matrix elements: 1) large-size Wilson loop
that has been used to extract heavy-quark potential, 2) the
vacuum matrix element of the quasi-PDF operator, 3) the vacuum matrix element of Wilson link operator in a fixed gauge. 
The data are for MILC ensembles and, when applicable, with the clover action.
According to the last section, smearing does not change
the form of renormalization, and therefore, we consider
here the higher-precision data with one step smearing. Since we just want to test the linear divergence, we can use the simple function Eq.~(\ref{eq:logM_ss}) in our fitting. 

\subsection{Wilson Loop}

To extract the linear divergence from lattice data, 
a Wilson loop is a good choice because it is easy to calculate, 
gauge invariant, and has high precision. It has been extensively studied in
the literature~\cite{Chen:2016fxx,Zhang:2017bzy,Musch:2010ka,Green:2017xeu,Chen:2017gck}, particularly in relation to the heavy quark potential. 

\begin{figure}[tbp]
\centering
\includegraphics[width=8cm]{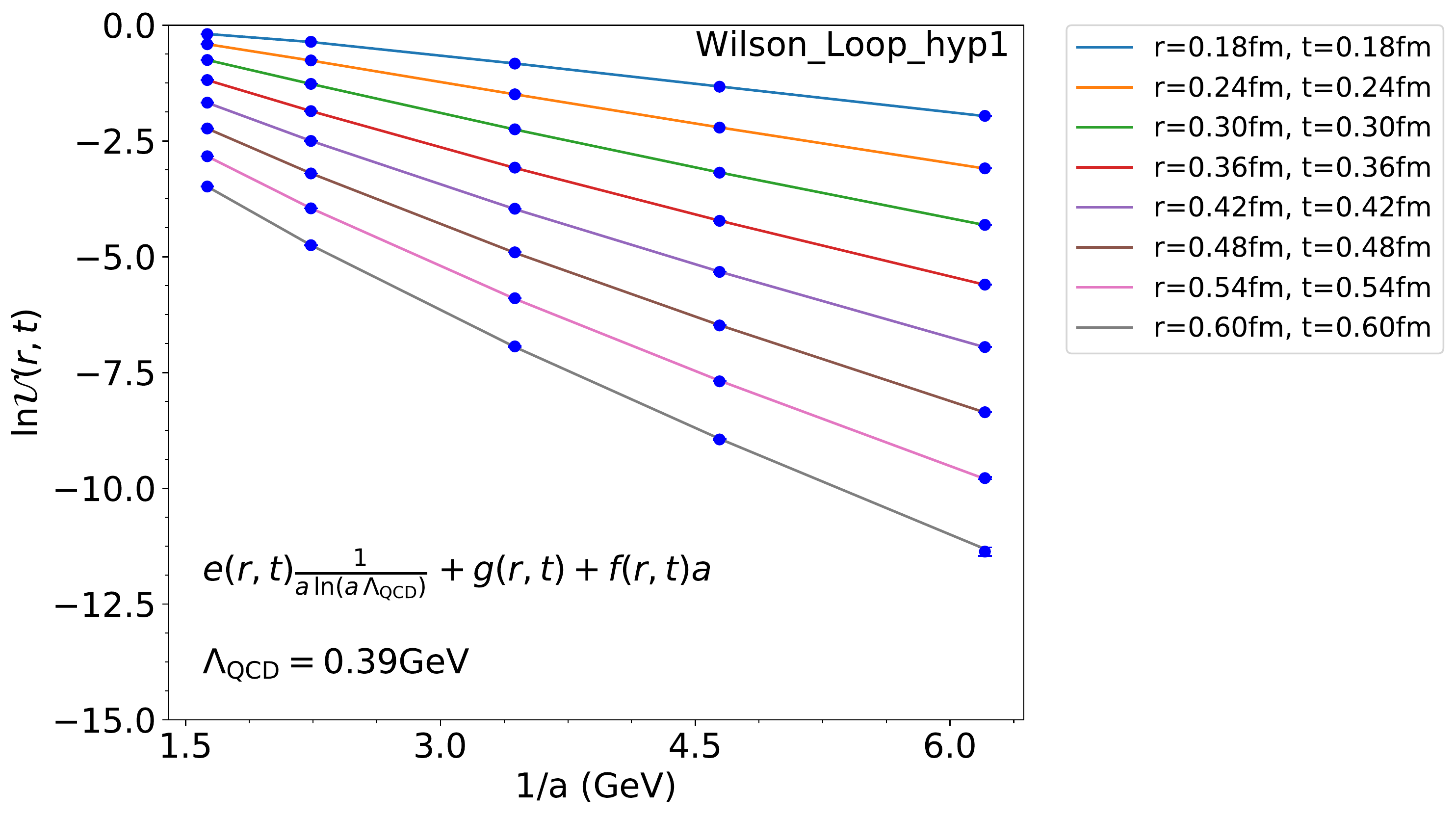}
\includegraphics[width=8cm]{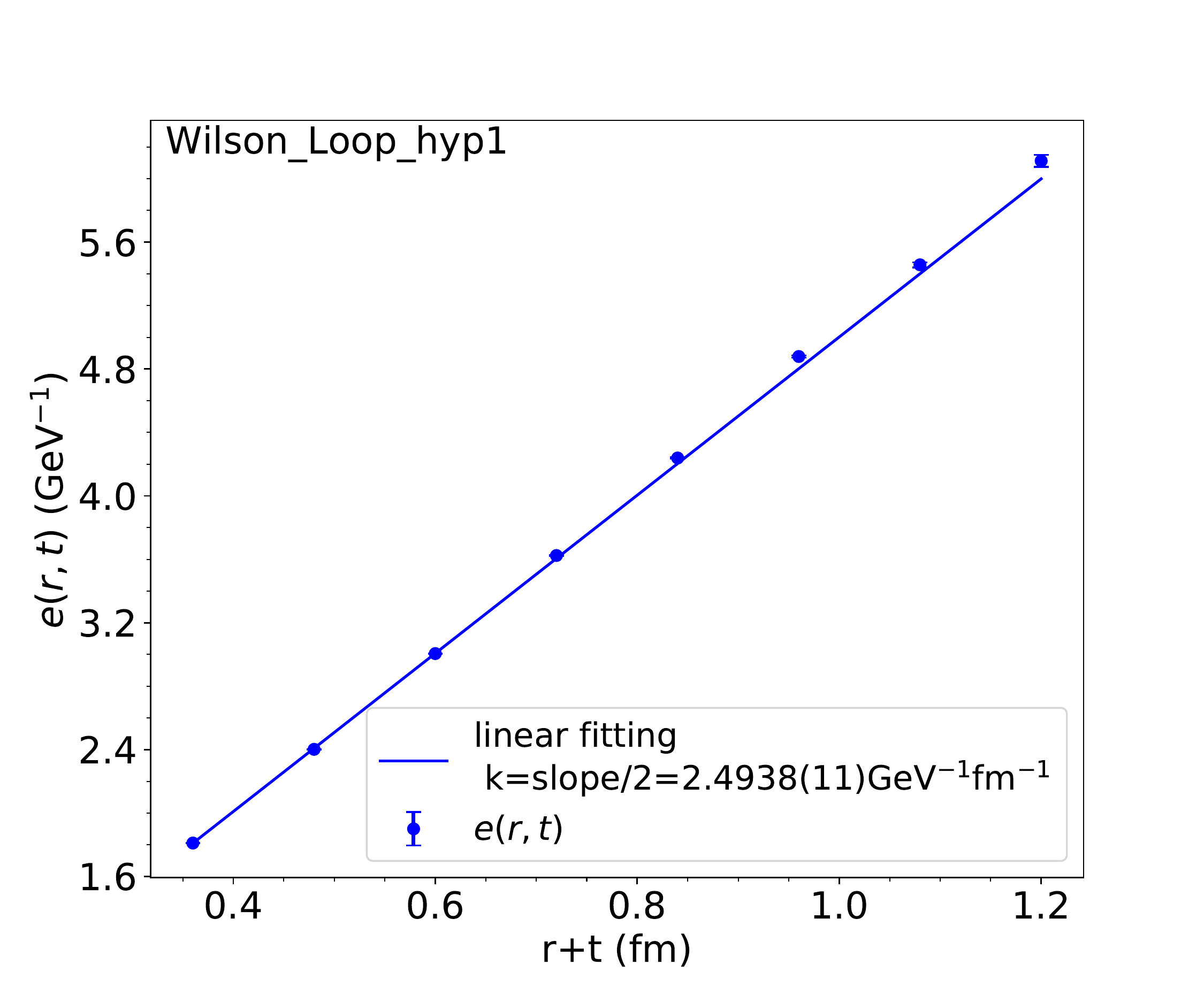}
\caption{Test of linear divergence for a Wilson Loop for HYP smearing cases.
The slope is consistent approximately with hadronic matrix elements from the previous section.}
\label{fig:LP_test}
\end{figure}

In our analysis, we will not consider 
the heavy quark potential interpretation but simply consider
a Wilson loop as a matrix element and fit
its $1/a$ divergence by looking at the logarithm
of the matrix element, as shown in 
Fig.~\ref{fig:LP_test}. The lattice
spacing dependence has been fitted well
with our formula with a choice of $\Lambda_{\rm QCD}=
0.39$ GeV, which is a value favored by
our fitting in the previous section. After isolating the linear
divergent coefficient, we plot it as a function
of $r+t$, for which half of the slope gives us
the divergent coefficient. Our value is
about 2.5 GeV$^{-1}$fm$^{-1}$ (0.49 if dimensionless). This value is very similar
to the value we found in the previous section (Fig.~\ref{fig:e(z)_hyp}),
demonstrating that the linear divergence 
can be very well extracted from the Wilson loop.
 
\subsection{Vacuum Matrix Element of Quasi-PDF Operator}

\begin{figure}[tbp]
\centering
\includegraphics[width=8cm]{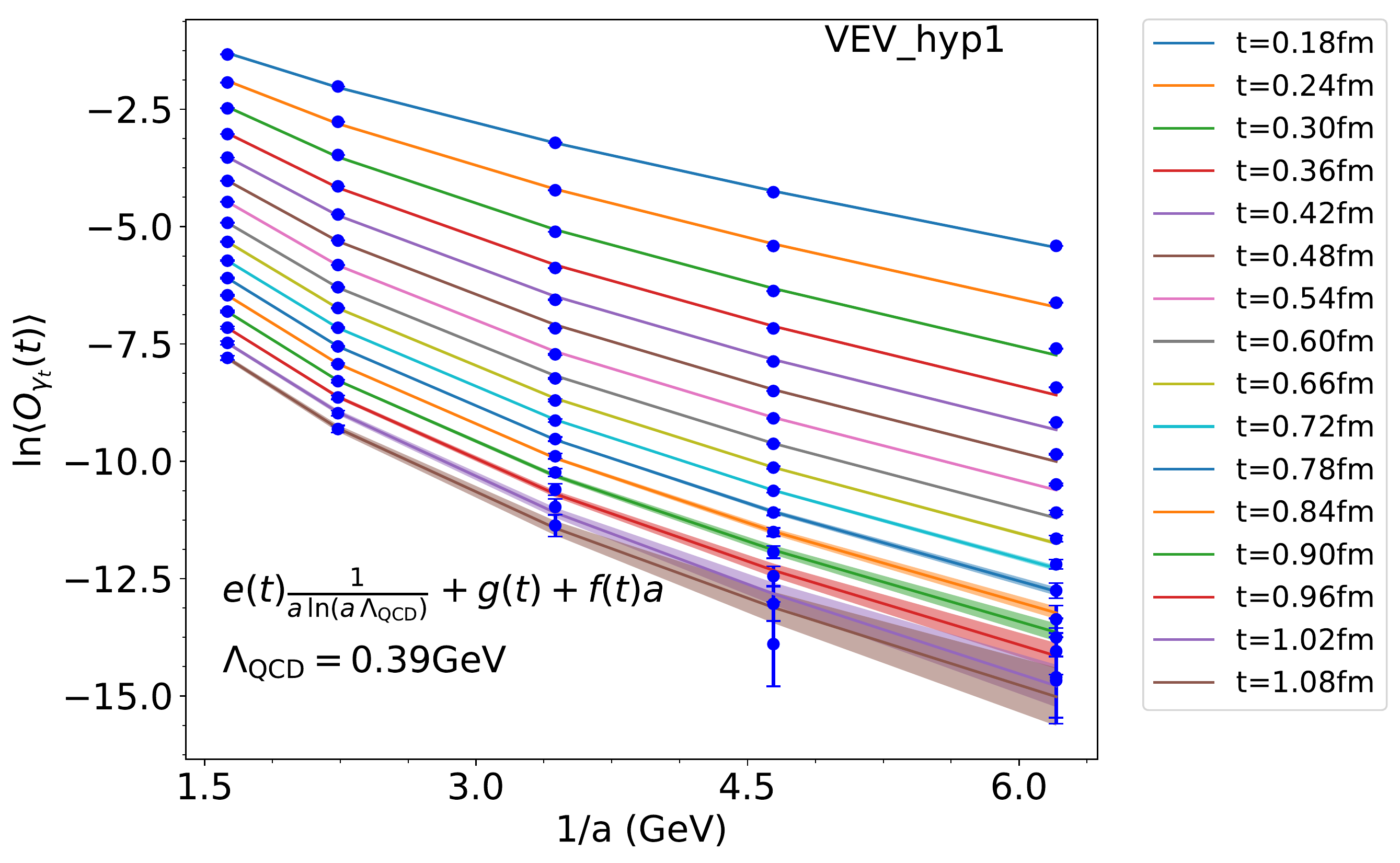}
\includegraphics[width=8cm]{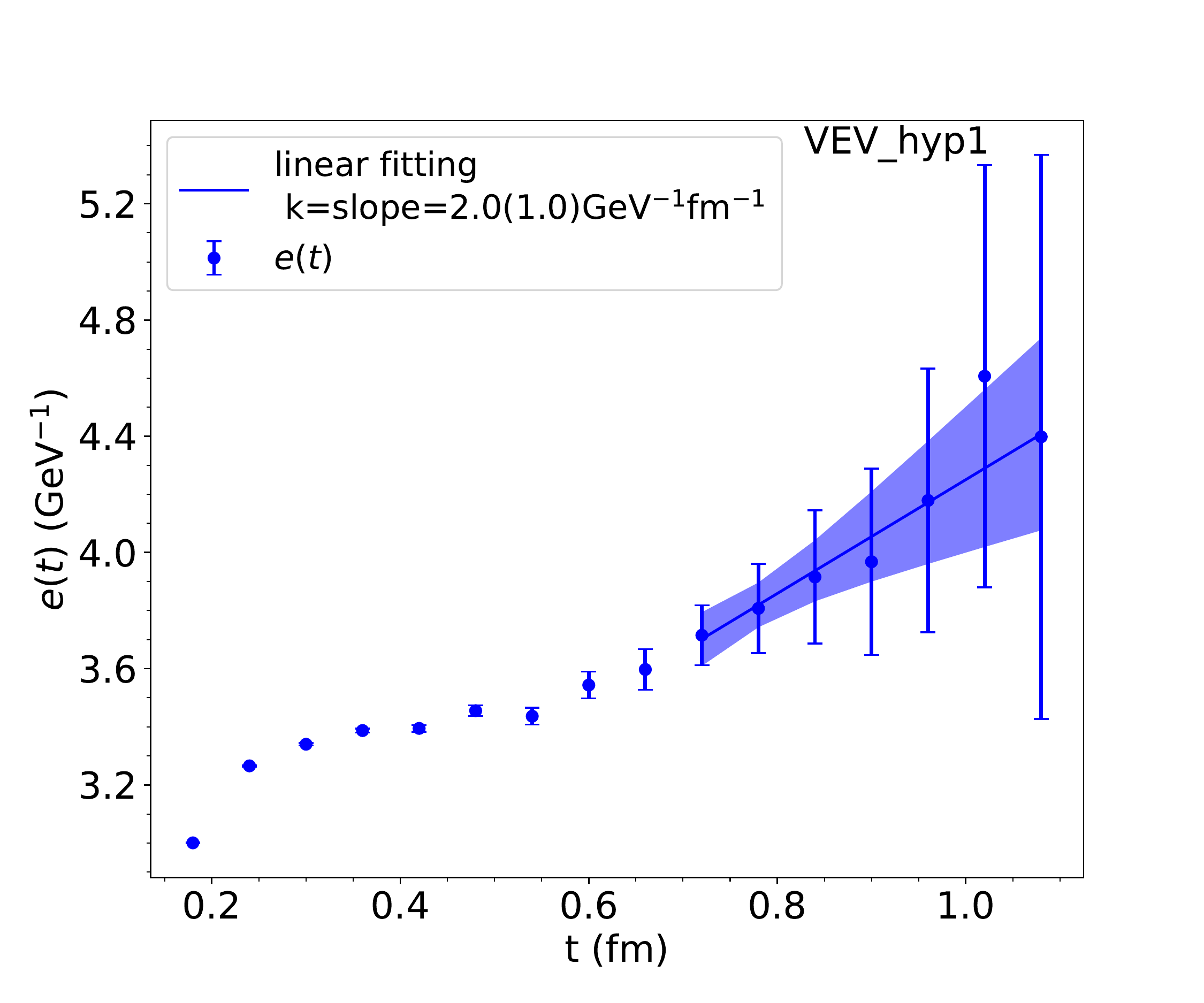}
\caption{Test of linear divergence for the vacuum matrix of the quasi-PDF operator. The matrix
element is much suppressed and there is no leading twist contribution.}
\label{fig:vev12}
\end{figure}

Here we consider the vacuum matrix elements of a quasi-PDF operator
which were suggested as choices for the renormalization
factor in Ref.~\cite{Braun:2018brg}. 

One can consider the vacuum expectation value (VEV) of $O_{\gamma_t}(t)$. As a gauge invariant choice proposed in Ref.~\cite{Braun:2018brg}. $\langle O_{\Gamma}(t)\rangle$
can be obtained through the following stochastic estimation:
\bea\label{eq:vev1}
&&\langle O_{\Gamma}(t)\rangle =\frac{1}{L_s^3}\langle \sum_{\vec{x}}\textrm{Tr}[U(\vec{x},0;\vec{x},t)\Gamma S_w(\vec{x},t)]\rangle\nonumber\\
&=&\frac{1}{L_s^3}\langle \sum_{\vec{x}}\textrm{Tr}[U(\vec{x},0;\vec{x},t)\Gamma S(\vec{x},t;\vec{x},0)]\rangle\nonumber\\
&&+\frac{1}{L_s^3}\sum_{\vec{x},\vec{y},\vec{y}\neq\vec{x}}\langle \textrm{Tr}[U(\vec{x},0;\vec{x},t)\Gamma S(\vec{x},t;\vec{y},0)]\rangle
\eea
where the second term on the right hand side of the second line vanishes as it is not gauge invariant, and $S_w(\vec{x},t)=\sum_{\vec{y}}S(\vec{x},t;\vec{y},0)$ is a wall source quark propagator without gauge fixing. Such a proposal can only be applied for the non-vanishing cases with $\Gamma=\gamma_t$ or ${\cal I}$. We will apply it to $O_{\gamma_t}$.  

However, it is simple to see that the $O_{\gamma_t}$ matrix element in the vacuum vanishes at small $t$ using operator product expansion.
Therefore the matrix element is susceptible 
to large O($a$) correction. Only at very large $t$,
one might see the correct slope. 

Results of a test of the linear divergence for the vacuum matrix element of the PDF operator is shown in Fig.~\ref{fig:vev12}. 
The slope at large $t$ appears to be consistent with that from the Wilson loop case but only within uninterestingly large errors.

\subsection{Landau-gauge-fixed Wilson link}

For the gauge-dependent matrix elements, we consider 
the Wilson line in Landau gauge as the simplest choice. 

\begin{figure}[tbp]
\centering
\includegraphics[width=8cm]{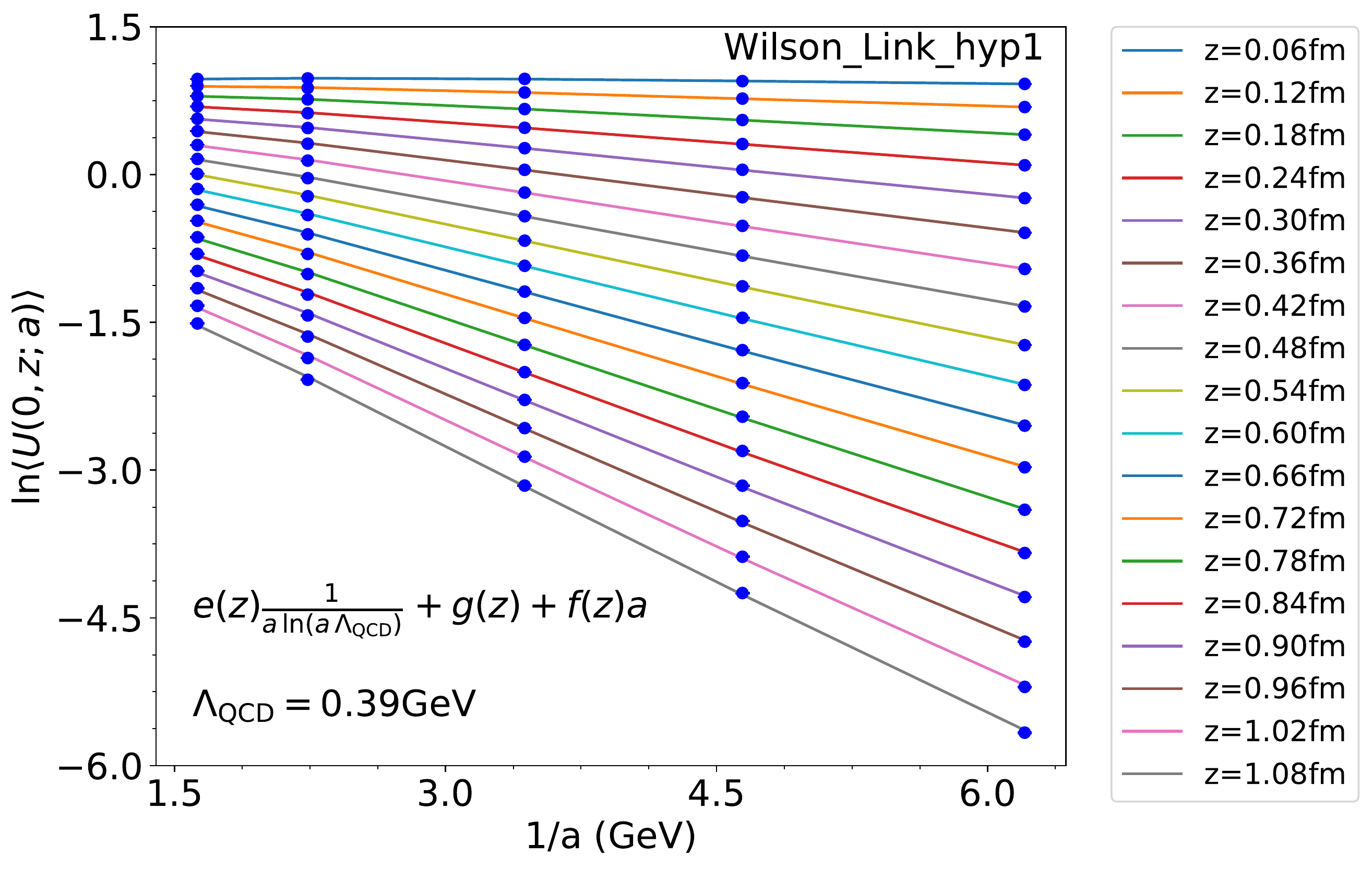}
\includegraphics[width=8cm]{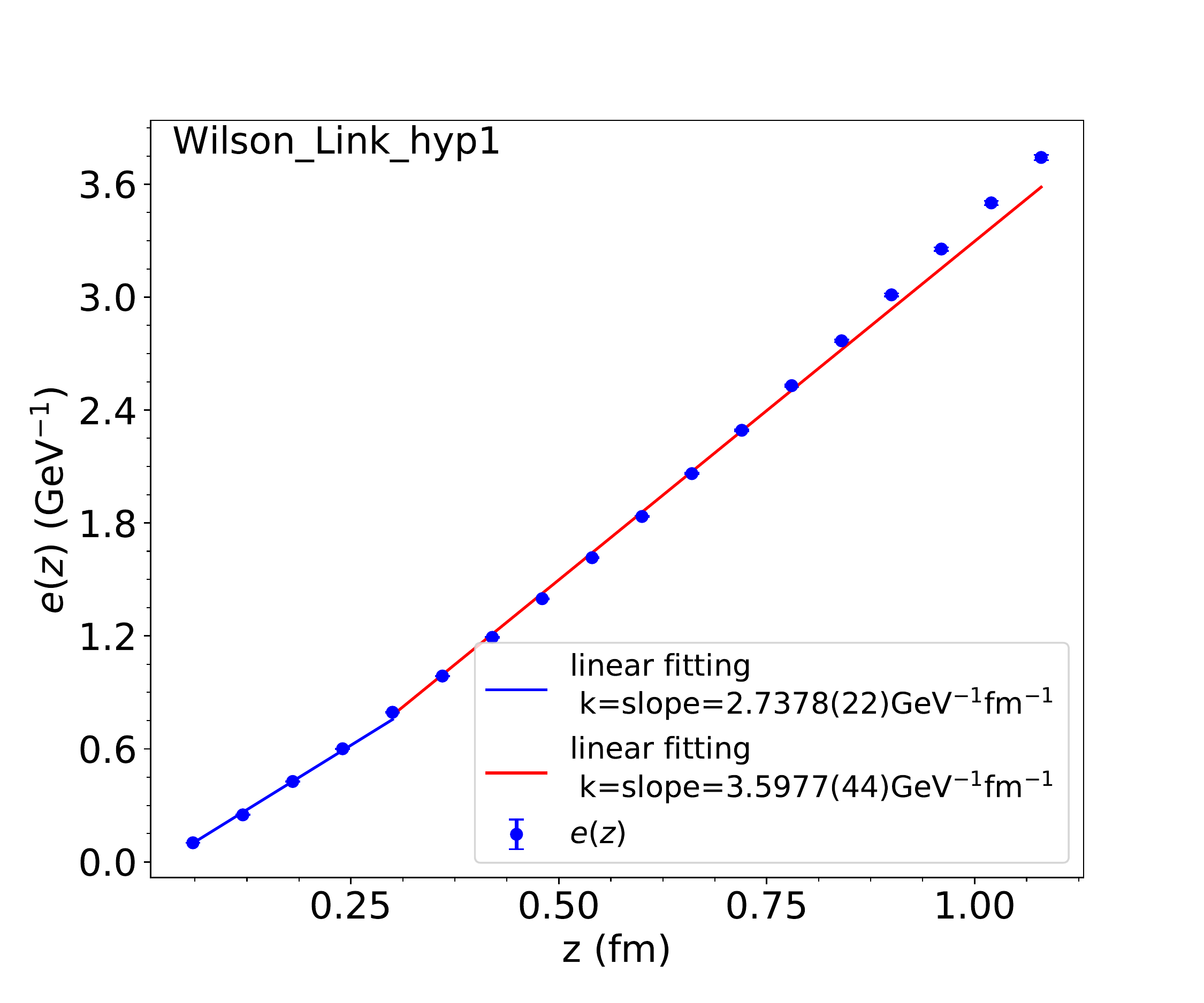}
\caption{Test of linear divergence for Landau gauge fixed Wilson link for HYP smearing cases. At small-$z$, it can be calculated in perturbation theory where the linear divergence works well. For large-$z$, the matrix element does not have a transfer-matrix 
interpretation in this gauge.}
\label{fig:link_test}
\end{figure}

The result of our test of the linear divergence is shown in Fig.~\ref{fig:link_test}.
At small-$z$, where perturbation theory works well, the slope roughly agrees with 
the prediction from  perturbation theory. For large-$z$, the matrix element does not
have a transfer-matrix interpretation in this gauge.

\begin{figure}[tbp]
\centering
\includegraphics[width=8cm]{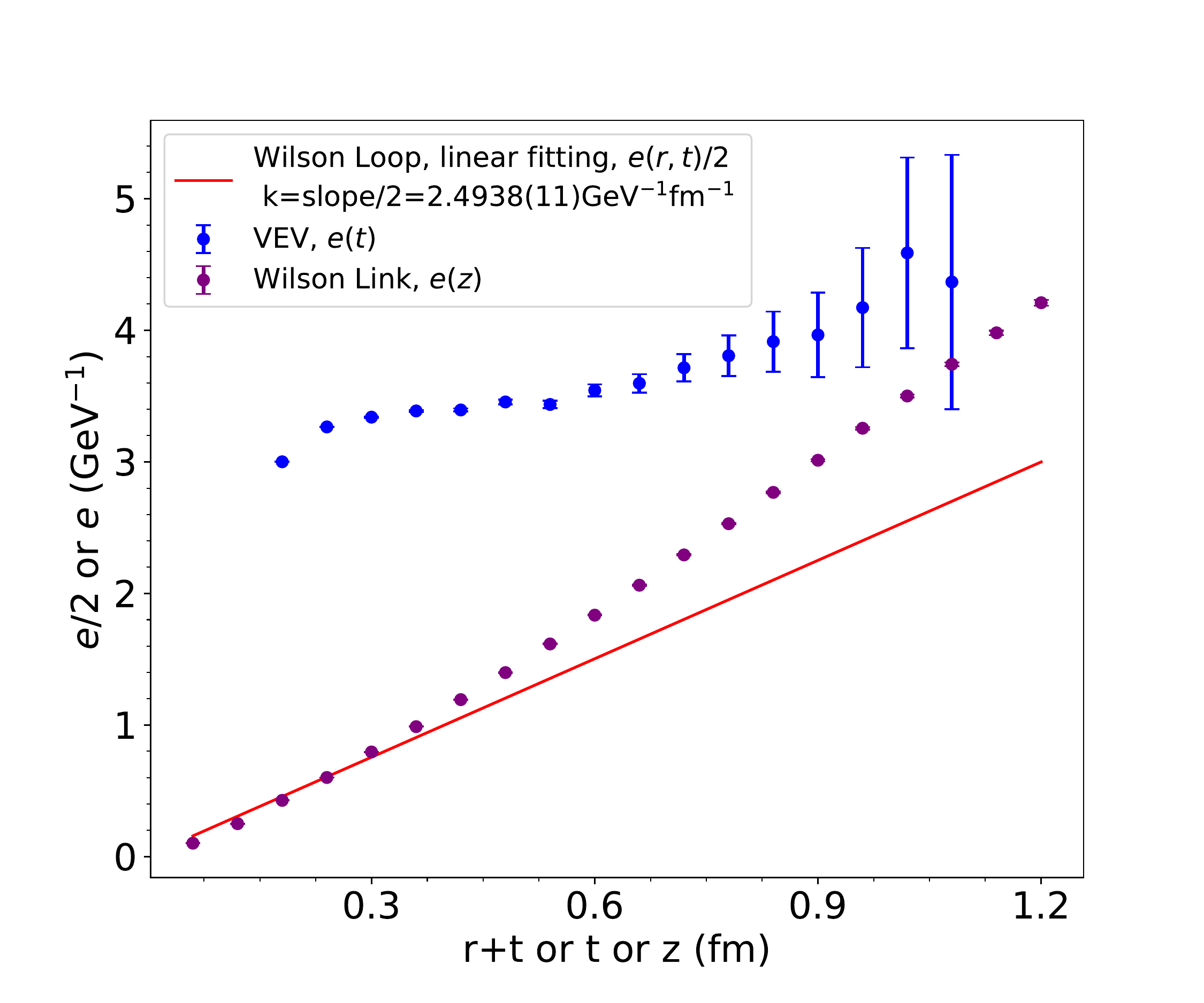}
\caption{Comparison of linear divergences in different vacuum matrix elements.}
\label{fig:com_test}
\end{figure}

Finally, Fig.~\ref{fig:com_test} is a comparison of the linearly divergent term for Wilson Loop, VEV and Wilson link. At small $z$, the linear divergence terms of Wilson Loop and Wilson link are consistent while at large $z$ they disagree strongly. The linear divergence term of VEV is different from the others because the large discretization errors 
make the extraction of the linear divergence very imprecise.

\section{Summary}\label{sec:summary}

The proper renormalization of quasi-LF correlations calculated on the lattice has been a main challenge for the applications of LaMET to studying parton physics. Such correlations contain linear divergences associated with Wilson lines which have to be eliminated with high precision to extract the desired physics information. In this work, we propose a self-renormalization method to eliminate both the linear and the logarithmic divergence, as well as the discretization error, from a quasi-LF matrix element which can be matched to a continuum scheme at short distance. As a paradigmatic example, we show that our method works well for $O_{\gamma_t}$ matrix elements in the pion, nucleon, and off-shell quark state. Our analysis shows that the renormalization factors are universal in the hadron state considered. Moreover, the renormalized correlations in the pion state for clover and overlap fermions are similar to each other. Besides, We find a large non-perturbative effect in the popular RI/MOM and ratio renormalization scheme used previously, which has to and can be avoided in the hybrid renormalization scheme proposed recently. It is certainly interesting to see that our method works also for HYP smeared matrix elements. This paves the way to do a phenomenological renormalization for smeared high-precision lattice data. 

For the convenience of the reader, we collect the parameters related to renormalization for various matrix elements discussed in the previous sections, see Tables~\ref{tab:para_hyp0} and~\ref{tab:para_hyp1}. For HYP smearing cases, there is much freedom for the choice of $\Lambda_{\rm QCD}$. We choose $\Lambda_{\rm QCD}=0.39$ GeV in Table~\ref{tab:para_hyp1}. But we also find that $\Lambda_{\rm QCD}=0.1$ GeV gives us smaller fitted discretization errors.

\begin{table}[htbp]
  \centering
  \begin{tabular}{c|cccc}
  \toprule
 Cases & $k$ & $\Lambda_{\rm QCD}$(GeV) & $d$ & $m_{0}$(GeV) \\
\hline
overlap quark & 1.46 & 0.109(02) & -1.29 & 0.0462(16)\\
\hline
clover quark & 1.46 & 0.135(02) & -1.35 & 0.1513(16)\\
\hline
overlap pion & 1.46 & 0.093(10) & -1.17 & -0.0357(46)\\
\hline
clover pion & 1.46 & 0.086(14) & -0.92 &  -0.0715(50)\\
\hline
clover nucleon & 1.46 & 0.093(06) & -0.97 &  -0.0508(40)\\
\hline
  \end{tabular}
  \caption{Renormalization parameters based on the fitting functions Eq.~(\ref{eq:logM}) and Eq.~(\ref{eq:gztoMS}) for the cases without HYP smearing for MILC and RBC ensembles, where $\delta_{\rm sys}=0.002$. $k$ is dimensionless here.}
  \label{tab:para_hyp0}
\end{table}

\begin{table}[htbp]
  \centering
  \begin{tabular}{c|cccc}
  \toprule
 Cases & $k$ & $\Lambda_{\rm QCD}$(GeV)  \\
\hline
overlap quark & 0.5521(07) & 0.39 \\
\hline
clover quark & 0.6328(05) & 0.39 \\
\hline
overlap pion & 0.5191(14) & 0.39 \\
\hline
clover pion & 0.5178(18) & 0.39 \\
\hline
clover nucleon & 0.5139(54) & 0.39 \\
\hline
Wilson Loop   & 0.4921(02) &  0.39 \\
\hline
VEV           & 0.39(21)  &  0.39 \\
\hline
Wilson Link   & \multicolumn{1}{p{1cm}<{\centering}}{0.5402(04), 0.7099(09)} & 0.39  &     &  \\
\hline
  \end{tabular}
  \caption{Renormalization parameters based on the fitting functions Eq.~(\ref{eq:logM_ss}) for the cases with HYP smearing for MILC and RBC ensembles. $k$ is the slope of $e(z)$ (half of the slope for Wilson Loop) and is dimensionless here.}
  \label{tab:para_hyp1}
\end{table}

For comparison purposes, we also show results based on 2+1 flavor clover quark and Luescher-Weisz (equivalent to Symanzik) gauge ensembles from CLS collaboration~\cite{Bruno:2014jqa}. We choose valence fermion action to be clover action, the same as sea fermion action. The renormalization parameters for CLS ensembles are shown in Table~\ref{tab:CLS}. The parameters $k$ and $\Lambda_{\rm QCD}$ for pion matrix elements for CLS ensembles are similar to those for MILC and RBC ensembles, suggesting consistency between results using mixed actions or not. 

\begin{table}[htbp]
  \centering
  \begin{tabular}{c|cccc}
  \toprule
 Cases & $k$ & $\Lambda_{\rm QCD}$(GeV) & $d$ & $m_{0}$(GeV) \\
\hline
clover quark hyp0 & 1.46 & 0.170(06)  &  -0.68 &  0.1936(26) \\
\hline
clover pion hyp0 & 1.46 & 0.113(09)  &  -2.24  &  -0.0941(49) \\
\hline
clover quark hyp1 & 0.573(05)  & 0.39 & 0.53  &  0.2298(26) \\
\hline
clover pion hyp1 & 0.486(10) & 0.39 & -0.21 &  0.0312(29) \\
\hline
  \end{tabular}
  \caption{Renormalization parameters based on the fitting functions Eq.~(\ref{eq:logM}) and Eq.~(\ref{eq:gztoMS}) for CLS ensembles, where $\delta_{\rm sys}=0.002$. “hyp0” is the unsmearing cases and “hyp1” is the HYP smearing cases. $k$ is dimensionless here.}
  \label{tab:CLS}
\end{table}

We would like to stress that the application of our method is not limited to the matrix elements discussed in this paper. Instead, it can be applied to any matrix element of an operator in the form of Eq.~(\ref{eq:operator}). 
It can also be generalized, in principle, to any matrix element with a Wilson line
in LaMET applications~\cite{Ji:2020ect}. Of course, in these cases one also needs to consider the physical interpretations of the matrix elements and the details related to lattice calculations carefully.

~\\
~\\
~\\
~\\
~\\
~\\
~\\
~\\

\section*{Acknowledgement}

We thank A. Kronfeld, P. Petreczky, O. Philipsen, Y. Zhao for valuable comments and discussions related to linear divergence.
We thank the MILC and RBC/UKQCD collaborations for providing us their gauge configurations.
The numerical calculation is supported by Strategic Priority Research Program of Chinese Academy of Sciences, Grant No. XDC01040100, HPC Cluster of ITP-CAS, and also Jiangsu Key Lab for NSLSCS. 
L.C. Gui is supported by Natural Science Foundation of Hunan Province under Grants No. 2020JJ5343, No. 20A310. 
P. Sun is supported by Natural Science Foundation of China under grant No. 11975127. 
W. Wang is supported by  Natural Science Foundation of China under grant Nos. 11735010, and U2032102. 
Y.-B. Yang is supported by Strategic Priority Research Program of Chinese Academy of Sciences, Grant No.XDC01040100 and XDB34030303. J.-H. Zhang is supported in part by National Natural Science Foundation of China under Grant No. 11975051, and by the Fundamental Research Funds for the Central Universities. P. Sun, A. Sch\"afer, W. Wang, Y.-B. Yang and J.-H. Zhang are also supported by a NSFC-DFG joint grant under grant No. 12061131006 and SCHA~~458/22. X.-D. Ji  is supported by the U.S. Department of Energy, Office of Science, Office of Nuclear Physics, under contract number DE-SC0020682.

\bibliographystyle{IEEEtran}

\begin{thebibliography}{71}%
\makeatletter
\providecommand \@ifxundefined [1]{%
 \@ifx{#1\undefined}
}%
\providecommand \@ifnum [1]{%
 \ifnum #1\expandafter \@firstoftwo
 \else \expandafter \@secondoftwo
 \fi
}%
\providecommand \@ifx [1]{%
 \ifx #1\expandafter \@firstoftwo
 \else \expandafter \@secondoftwo
 \fi
}%
\providecommand \natexlab [1]{#1}%
\providecommand \enquote  [1]{``#1''}%
\providecommand \bibnamefont  [1]{#1}%
\providecommand \bibfnamefont [1]{#1}%
\providecommand \citenamefont [1]{#1}%
\providecommand \href@noop [0]{\@secondoftwo}%
\providecommand \href [0]{\begingroup \@sanitize@url \@href}%
\providecommand \@href[1]{\@@startlink{#1}\@@href}%
\providecommand \@@href[1]{\endgroup#1\@@endlink}%
\providecommand \@sanitize@url [0]{\catcode `\\12\catcode `\$12\catcode
  `\&12\catcode `\#12\catcode `\^12\catcode `\_12\catcode `\%12\relax}%
\providecommand \@@startlink[1]{}%
\providecommand \@@endlink[0]{}%
\providecommand \url  [0]{\begingroup\@sanitize@url \@url }%
\providecommand \@url [1]{\endgroup\@href {#1}{\urlprefix }}%
\providecommand \urlprefix  [0]{URL }%
\providecommand \Eprint [0]{\href }%
\providecommand \doibase [0]{http://dx.doi.org/}%
\providecommand \selectlanguage [0]{\@gobble}%
\providecommand \bibinfo  [0]{\@secondoftwo}%
\providecommand \bibfield  [0]{\@secondoftwo}%
\providecommand \translation [1]{[#1]}%
\providecommand \BibitemOpen [0]{}%
\providecommand \bibitemStop [0]{}%
\providecommand \bibitemNoStop [0]{.\EOS\space}%
\providecommand \EOS [0]{\spacefactor3000\relax}%
\providecommand \BibitemShut  [1]{\csname bibitem#1\endcsname}%
\let\auto@bib@innerbib\@empty
\bibitem [{\citenamefont {Ellis}\ \emph {et~al.}(2011)\citenamefont {Ellis},
  \citenamefont {Stirling},\ and\ \citenamefont {Webber}}]{Ellis:1991qj}%
  \BibitemOpen
  \bibfield  {author} {\bibinfo {author} {\bibfnamefont {R.}~\bibnamefont
  {Ellis}}, \bibinfo {author} {\bibfnamefont {W.}~\bibnamefont {Stirling}}, \
  and\ \bibinfo {author} {\bibfnamefont {B.}~\bibnamefont {Webber}},\
  }\href@noop {} {\emph {\bibinfo {title} {{QCD and collider physics}}}},\
  Vol.~\bibinfo {volume} {8}\ (\bibinfo  {publisher} {Cambridge University
  Press},\ \bibinfo {year} {2011})\BibitemShut {NoStop}%
\bibitem [{\citenamefont {Thomas}\ and\ \citenamefont
  {Weise}(2001)}]{Thomas:2001kw}%
  \BibitemOpen
  \bibfield  {author} {\bibinfo {author} {\bibfnamefont {A.~W.}\ \bibnamefont
  {Thomas}}\ and\ \bibinfo {author} {\bibfnamefont {W.}~\bibnamefont {Weise}},\
  }\href {\doibase 10.1002/352760314X} {\emph {\bibinfo {title} {{The Structure
  of the Nucleon}}}}\ (\bibinfo  {publisher} {Wiley},\ \bibinfo {address}
  {Germany},\ \bibinfo {year} {2001})\BibitemShut {NoStop}%
\bibitem [{\citenamefont {Gao}\ \emph {et~al.}(2018)\citenamefont {Gao},
  \citenamefont {Harland-Lang},\ and\ \citenamefont {Rojo}}]{Gao:2017yyd}%
  \BibitemOpen
  \bibfield  {author} {\bibinfo {author} {\bibfnamefont {J.}~\bibnamefont
  {Gao}}, \bibinfo {author} {\bibfnamefont {L.}~\bibnamefont {Harland-Lang}}, \
  and\ \bibinfo {author} {\bibfnamefont {J.}~\bibnamefont {Rojo}},\ }\href
  {\doibase 10.1016/j.physrep.2018.03.002} {\bibfield  {journal} {\bibinfo
  {journal} {Phys. Rept.}\ }\textbf {\bibinfo {volume} {742}},\ \bibinfo
  {pages} {1} (\bibinfo {year} {2018})},\ \Eprint
  {http://arxiv.org/abs/1709.04922} {arXiv:1709.04922 [hep-ph]} \BibitemShut
  {NoStop}%
\bibitem [{\citenamefont {Accardi}\ \emph {et~al.}(2016)\citenamefont {Accardi}
  \emph {et~al.}}]{Accardi:2012qut}%
  \BibitemOpen
  \bibfield  {author} {\bibinfo {author} {\bibfnamefont {A.}~\bibnamefont
  {Accardi}} \emph {et~al.},\ }\href {\doibase 10.1140/epja/i2016-16268-9}
  {\bibfield  {journal} {\bibinfo  {journal} {Eur. Phys. J.}\ }\textbf
  {\bibinfo {volume} {A52}},\ \bibinfo {pages} {268} (\bibinfo {year}
  {2016})},\ \Eprint {http://arxiv.org/abs/1212.1701} {arXiv:1212.1701
  [nucl-ex]} \BibitemShut {NoStop}%
\bibitem [{\citenamefont {Cichy}\ and\ \citenamefont
  {Constantinou}(2019)}]{Cichy:2018mum}%
  \BibitemOpen
  \bibfield  {author} {\bibinfo {author} {\bibfnamefont {K.}~\bibnamefont
  {Cichy}}\ and\ \bibinfo {author} {\bibfnamefont {M.}~\bibnamefont
  {Constantinou}},\ }\href {\doibase 10.1155/2019/3036904} {\bibfield
  {journal} {\bibinfo  {journal} {Adv. High Energy Phys.}\ }\textbf {\bibinfo
  {volume} {2019}},\ \bibinfo {pages} {3036904} (\bibinfo {year} {2019})},\
  \Eprint {http://arxiv.org/abs/1811.07248} {arXiv:1811.07248 [hep-lat]}
  \BibitemShut {NoStop}%
\bibitem [{\citenamefont {Ji}(2013)}]{Ji:2013dva}%
  \BibitemOpen
  \bibfield  {author} {\bibinfo {author} {\bibfnamefont {X.}~\bibnamefont
  {Ji}},\ }\href {\doibase 10.1103/PhysRevLett.110.262002} {\bibfield
  {journal} {\bibinfo  {journal} {Phys. Rev. Lett.}\ }\textbf {\bibinfo
  {volume} {110}},\ \bibinfo {pages} {262002} (\bibinfo {year} {2013})},\
  \Eprint {http://arxiv.org/abs/1305.1539} {arXiv:1305.1539 [hep-ph]}
  \BibitemShut {NoStop}%
\bibitem [{\citenamefont {Ji}\ \emph {et~al.}(2015)\citenamefont {Ji},
  \citenamefont {Sun}, \citenamefont {Xiong},\ and\ \citenamefont
  {Yuan}}]{Ji:2014hxa}%
  \BibitemOpen
  \bibfield  {author} {\bibinfo {author} {\bibfnamefont {X.}~\bibnamefont
  {Ji}}, \bibinfo {author} {\bibfnamefont {P.}~\bibnamefont {Sun}}, \bibinfo
  {author} {\bibfnamefont {X.}~\bibnamefont {Xiong}}, \ and\ \bibinfo {author}
  {\bibfnamefont {F.}~\bibnamefont {Yuan}},\ }\href {\doibase
  10.1103/PhysRevD.91.074009} {\bibfield  {journal} {\bibinfo  {journal} {Phys.
  Rev.}\ }\textbf {\bibinfo {volume} {D91}},\ \bibinfo {pages} {074009}
  (\bibinfo {year} {2015})},\ \Eprint {http://arxiv.org/abs/1405.7640}
  {arXiv:1405.7640 [hep-ph]} \BibitemShut {NoStop}%
\bibitem [{\citenamefont {Ji}\ \emph {et~al.}(2020{\natexlab{a}})\citenamefont
  {Ji}, \citenamefont {Liu}, \citenamefont {Liu}, \citenamefont {Zhang},\ and\
  \citenamefont {Zhao}}]{Ji:2020ect}%
  \BibitemOpen
  \bibfield  {author} {\bibinfo {author} {\bibfnamefont {X.}~\bibnamefont
  {Ji}}, \bibinfo {author} {\bibfnamefont {Y.-S.}\ \bibnamefont {Liu}},
  \bibinfo {author} {\bibfnamefont {Y.}~\bibnamefont {Liu}}, \bibinfo {author}
  {\bibfnamefont {J.-H.}\ \bibnamefont {Zhang}}, \ and\ \bibinfo {author}
  {\bibfnamefont {Y.}~\bibnamefont {Zhao}},\ }\href@noop {} {\  (\bibinfo
  {year} {2020}{\natexlab{a}})},\ \Eprint {http://arxiv.org/abs/2004.03543}
  {arXiv:2004.03543 [hep-ph]} \BibitemShut {NoStop}%
\bibitem [{\citenamefont {Lin}\ \emph {et~al.}(2015)\citenamefont {Lin},
  \citenamefont {Chen}, \citenamefont {Cohen},\ and\ \citenamefont
  {Ji}}]{Lin:2014zya}%
  \BibitemOpen
  \bibfield  {author} {\bibinfo {author} {\bibfnamefont {H.-W.}\ \bibnamefont
  {Lin}}, \bibinfo {author} {\bibfnamefont {J.-W.}\ \bibnamefont {Chen}},
  \bibinfo {author} {\bibfnamefont {S.~D.}\ \bibnamefont {Cohen}}, \ and\
  \bibinfo {author} {\bibfnamefont {X.}~\bibnamefont {Ji}},\ }\href {\doibase
  10.1103/PhysRevD.91.054510} {\bibfield  {journal} {\bibinfo  {journal} {Phys.
  Rev. D}\ }\textbf {\bibinfo {volume} {91}},\ \bibinfo {pages} {054510}
  (\bibinfo {year} {2015})},\ \Eprint {http://arxiv.org/abs/1402.1462}
  {arXiv:1402.1462 [hep-ph]} \BibitemShut {NoStop}%
\bibitem [{\citenamefont {Alexandrou}\ \emph {et~al.}(2015)\citenamefont
  {Alexandrou}, \citenamefont {Cichy}, \citenamefont {Drach}, \citenamefont
  {Garcia-Ramos}, \citenamefont {Hadjiyiannakou}, \citenamefont {Jansen},
  \citenamefont {Steffens},\ and\ \citenamefont {Wiese}}]{Alexandrou:2015rja}%
  \BibitemOpen
  \bibfield  {author} {\bibinfo {author} {\bibfnamefont {C.}~\bibnamefont
  {Alexandrou}}, \bibinfo {author} {\bibfnamefont {K.}~\bibnamefont {Cichy}},
  \bibinfo {author} {\bibfnamefont {V.}~\bibnamefont {Drach}}, \bibinfo
  {author} {\bibfnamefont {E.}~\bibnamefont {Garcia-Ramos}}, \bibinfo {author}
  {\bibfnamefont {K.}~\bibnamefont {Hadjiyiannakou}}, \bibinfo {author}
  {\bibfnamefont {K.}~\bibnamefont {Jansen}}, \bibinfo {author} {\bibfnamefont
  {F.}~\bibnamefont {Steffens}}, \ and\ \bibinfo {author} {\bibfnamefont
  {C.}~\bibnamefont {Wiese}},\ }\href {\doibase 10.1103/PhysRevD.92.014502}
  {\bibfield  {journal} {\bibinfo  {journal} {Phys. Rev. D}\ }\textbf {\bibinfo
  {volume} {92}},\ \bibinfo {pages} {014502} (\bibinfo {year} {2015})},\
  \Eprint {http://arxiv.org/abs/1504.07455} {arXiv:1504.07455 [hep-lat]}
  \BibitemShut {NoStop}%
\bibitem [{\citenamefont {Chen}\ \emph {et~al.}(2016)\citenamefont {Chen},
  \citenamefont {Cohen}, \citenamefont {Ji}, \citenamefont {Lin},\ and\
  \citenamefont {Zhang}}]{Chen:2016utp}%
  \BibitemOpen
  \bibfield  {author} {\bibinfo {author} {\bibfnamefont {J.-W.}\ \bibnamefont
  {Chen}}, \bibinfo {author} {\bibfnamefont {S.~D.}\ \bibnamefont {Cohen}},
  \bibinfo {author} {\bibfnamefont {X.}~\bibnamefont {Ji}}, \bibinfo {author}
  {\bibfnamefont {H.-W.}\ \bibnamefont {Lin}}, \ and\ \bibinfo {author}
  {\bibfnamefont {J.-H.}\ \bibnamefont {Zhang}},\ }\href {\doibase
  10.1016/j.nuclphysb.2016.07.033} {\bibfield  {journal} {\bibinfo  {journal}
  {Nucl. Phys. B}\ }\textbf {\bibinfo {volume} {911}},\ \bibinfo {pages} {246}
  (\bibinfo {year} {2016})},\ \Eprint {http://arxiv.org/abs/1603.06664}
  {arXiv:1603.06664 [hep-ph]} \BibitemShut {NoStop}%
\bibitem [{\citenamefont {Alexandrou}\ \emph
  {et~al.}(2017{\natexlab{a}})\citenamefont {Alexandrou}, \citenamefont
  {Cichy}, \citenamefont {Constantinou}, \citenamefont {Hadjiyiannakou},
  \citenamefont {Jansen}, \citenamefont {Steffens},\ and\ \citenamefont
  {Wiese}}]{Alexandrou:2016jqi}%
  \BibitemOpen
  \bibfield  {author} {\bibinfo {author} {\bibfnamefont {C.}~\bibnamefont
  {Alexandrou}}, \bibinfo {author} {\bibfnamefont {K.}~\bibnamefont {Cichy}},
  \bibinfo {author} {\bibfnamefont {M.}~\bibnamefont {Constantinou}}, \bibinfo
  {author} {\bibfnamefont {K.}~\bibnamefont {Hadjiyiannakou}}, \bibinfo
  {author} {\bibfnamefont {K.}~\bibnamefont {Jansen}}, \bibinfo {author}
  {\bibfnamefont {F.}~\bibnamefont {Steffens}}, \ and\ \bibinfo {author}
  {\bibfnamefont {C.}~\bibnamefont {Wiese}},\ }\href {\doibase
  10.1103/PhysRevD.96.014513} {\bibfield  {journal} {\bibinfo  {journal} {Phys.
  Rev. D}\ }\textbf {\bibinfo {volume} {96}},\ \bibinfo {pages} {014513}
  (\bibinfo {year} {2017}{\natexlab{a}})},\ \Eprint
  {http://arxiv.org/abs/1610.03689} {arXiv:1610.03689 [hep-lat]} \BibitemShut
  {NoStop}%
\bibitem [{\citenamefont {Alexandrou}\ \emph
  {et~al.}(2018{\natexlab{a}})\citenamefont {Alexandrou}, \citenamefont
  {Cichy}, \citenamefont {Constantinou}, \citenamefont {Jansen}, \citenamefont
  {Scapellato},\ and\ \citenamefont {Steffens}}]{Alexandrou:2018pbm}%
  \BibitemOpen
  \bibfield  {author} {\bibinfo {author} {\bibfnamefont {C.}~\bibnamefont
  {Alexandrou}}, \bibinfo {author} {\bibfnamefont {K.}~\bibnamefont {Cichy}},
  \bibinfo {author} {\bibfnamefont {M.}~\bibnamefont {Constantinou}}, \bibinfo
  {author} {\bibfnamefont {K.}~\bibnamefont {Jansen}}, \bibinfo {author}
  {\bibfnamefont {A.}~\bibnamefont {Scapellato}}, \ and\ \bibinfo {author}
  {\bibfnamefont {F.}~\bibnamefont {Steffens}},\ }\href {\doibase
  10.1103/PhysRevLett.121.112001} {\bibfield  {journal} {\bibinfo  {journal}
  {Phys. Rev. Lett.}\ }\textbf {\bibinfo {volume} {121}},\ \bibinfo {pages}
  {112001} (\bibinfo {year} {2018}{\natexlab{a}})},\ \Eprint
  {http://arxiv.org/abs/1803.02685} {arXiv:1803.02685 [hep-lat]} \BibitemShut
  {NoStop}%
\bibitem [{\citenamefont {Chen}\ \emph
  {et~al.}(2018{\natexlab{a}})\citenamefont {Chen}, \citenamefont {Jin},
  \citenamefont {Lin}, \citenamefont {Liu}, \citenamefont {Yang}, \citenamefont
  {Zhang},\ and\ \citenamefont {Zhao}}]{Chen:2018xof}%
  \BibitemOpen
  \bibfield  {author} {\bibinfo {author} {\bibfnamefont {J.-W.}\ \bibnamefont
  {Chen}}, \bibinfo {author} {\bibfnamefont {L.}~\bibnamefont {Jin}}, \bibinfo
  {author} {\bibfnamefont {H.-W.}\ \bibnamefont {Lin}}, \bibinfo {author}
  {\bibfnamefont {Y.-S.}\ \bibnamefont {Liu}}, \bibinfo {author} {\bibfnamefont
  {Y.-B.}\ \bibnamefont {Yang}}, \bibinfo {author} {\bibfnamefont {J.-H.}\
  \bibnamefont {Zhang}}, \ and\ \bibinfo {author} {\bibfnamefont
  {Y.}~\bibnamefont {Zhao}},\ }\href@noop {} {\  (\bibinfo {year}
  {2018}{\natexlab{a}})},\ \Eprint {http://arxiv.org/abs/1803.04393}
  {arXiv:1803.04393 [hep-lat]} \BibitemShut {NoStop}%
\bibitem [{\citenamefont {Lin}\ \emph {et~al.}(2018)\citenamefont {Lin},
  \citenamefont {Chen}, \citenamefont {Ji}, \citenamefont {Jin}, \citenamefont
  {Li}, \citenamefont {Liu}, \citenamefont {Yang}, \citenamefont {Zhang},\ and\
  \citenamefont {Zhao}}]{Lin:2018pvv}%
  \BibitemOpen
  \bibfield  {author} {\bibinfo {author} {\bibfnamefont {H.-W.}\ \bibnamefont
  {Lin}}, \bibinfo {author} {\bibfnamefont {J.-W.}\ \bibnamefont {Chen}},
  \bibinfo {author} {\bibfnamefont {X.}~\bibnamefont {Ji}}, \bibinfo {author}
  {\bibfnamefont {L.}~\bibnamefont {Jin}}, \bibinfo {author} {\bibfnamefont
  {R.}~\bibnamefont {Li}}, \bibinfo {author} {\bibfnamefont {Y.-S.}\
  \bibnamefont {Liu}}, \bibinfo {author} {\bibfnamefont {Y.-B.}\ \bibnamefont
  {Yang}}, \bibinfo {author} {\bibfnamefont {J.-H.}\ \bibnamefont {Zhang}}, \
  and\ \bibinfo {author} {\bibfnamefont {Y.}~\bibnamefont {Zhao}},\ }\href
  {\doibase 10.1103/PhysRevLett.121.242003} {\bibfield  {journal} {\bibinfo
  {journal} {Phys. Rev. Lett.}\ }\textbf {\bibinfo {volume} {121}},\ \bibinfo
  {pages} {242003} (\bibinfo {year} {2018})},\ \Eprint
  {http://arxiv.org/abs/1807.07431} {arXiv:1807.07431 [hep-lat]} \BibitemShut
  {NoStop}%
\bibitem [{\citenamefont {Liu}\ \emph {et~al.}(2020)\citenamefont {Liu} \emph
  {et~al.}}]{Liu:2018uuj}%
  \BibitemOpen
  \bibfield  {author} {\bibinfo {author} {\bibfnamefont {Y.-S.}\ \bibnamefont
  {Liu}} \emph {et~al.} (\bibinfo {collaboration} {Lattice Parton}),\ }\href
  {\doibase 10.1103/PhysRevD.101.034020} {\bibfield  {journal} {\bibinfo
  {journal} {Phys. Rev.}\ }\textbf {\bibinfo {volume} {D101}},\ \bibinfo
  {pages} {034020} (\bibinfo {year} {2020})},\ \Eprint
  {http://arxiv.org/abs/1807.06566} {arXiv:1807.06566 [hep-lat]} \BibitemShut
  {NoStop}%
\bibitem [{\citenamefont {Alexandrou}\ \emph
  {et~al.}(2018{\natexlab{b}})\citenamefont {Alexandrou}, \citenamefont
  {Cichy}, \citenamefont {Constantinou}, \citenamefont {Jansen}, \citenamefont
  {Scapellato},\ and\ \citenamefont {Steffens}}]{Alexandrou:2018eet}%
  \BibitemOpen
  \bibfield  {author} {\bibinfo {author} {\bibfnamefont {C.}~\bibnamefont
  {Alexandrou}}, \bibinfo {author} {\bibfnamefont {K.}~\bibnamefont {Cichy}},
  \bibinfo {author} {\bibfnamefont {M.}~\bibnamefont {Constantinou}}, \bibinfo
  {author} {\bibfnamefont {K.}~\bibnamefont {Jansen}}, \bibinfo {author}
  {\bibfnamefont {A.}~\bibnamefont {Scapellato}}, \ and\ \bibinfo {author}
  {\bibfnamefont {F.}~\bibnamefont {Steffens}},\ }\href {\doibase
  10.1103/PhysRevD.98.091503} {\bibfield  {journal} {\bibinfo  {journal} {Phys.
  Rev. D}\ }\textbf {\bibinfo {volume} {98}},\ \bibinfo {pages} {091503}
  (\bibinfo {year} {2018}{\natexlab{b}})},\ \Eprint
  {http://arxiv.org/abs/1807.00232} {arXiv:1807.00232 [hep-lat]} \BibitemShut
  {NoStop}%
\bibitem [{\citenamefont {Liu}\ \emph {et~al.}(2018)\citenamefont {Liu},
  \citenamefont {Chen}, \citenamefont {Jin}, \citenamefont {Li}, \citenamefont
  {Lin}, \citenamefont {Yang}, \citenamefont {Zhang},\ and\ \citenamefont
  {Zhao}}]{Liu:2018hxv}%
  \BibitemOpen
  \bibfield  {author} {\bibinfo {author} {\bibfnamefont {Y.-S.}\ \bibnamefont
  {Liu}}, \bibinfo {author} {\bibfnamefont {J.-W.}\ \bibnamefont {Chen}},
  \bibinfo {author} {\bibfnamefont {L.}~\bibnamefont {Jin}}, \bibinfo {author}
  {\bibfnamefont {R.}~\bibnamefont {Li}}, \bibinfo {author} {\bibfnamefont
  {H.-W.}\ \bibnamefont {Lin}}, \bibinfo {author} {\bibfnamefont {Y.-B.}\
  \bibnamefont {Yang}}, \bibinfo {author} {\bibfnamefont {J.-H.}\ \bibnamefont
  {Zhang}}, \ and\ \bibinfo {author} {\bibfnamefont {Y.}~\bibnamefont {Zhao}},\
  }\href@noop {} {\  (\bibinfo {year} {2018})},\ \Eprint
  {http://arxiv.org/abs/1810.05043} {arXiv:1810.05043 [hep-lat]} \BibitemShut
  {NoStop}%
\bibitem [{\citenamefont {Zhang}\ \emph
  {et~al.}(2019{\natexlab{a}})\citenamefont {Zhang}, \citenamefont {Chen},
  \citenamefont {Jin}, \citenamefont {Lin}, \citenamefont {Sch\"afer},\ and\
  \citenamefont {Zhao}}]{Chen:2018fwa}%
  \BibitemOpen
  \bibfield  {author} {\bibinfo {author} {\bibfnamefont {J.-H.}\ \bibnamefont
  {Zhang}}, \bibinfo {author} {\bibfnamefont {J.-W.}\ \bibnamefont {Chen}},
  \bibinfo {author} {\bibfnamefont {L.}~\bibnamefont {Jin}}, \bibinfo {author}
  {\bibfnamefont {H.-W.}\ \bibnamefont {Lin}}, \bibinfo {author} {\bibfnamefont
  {A.}~\bibnamefont {Sch\"afer}}, \ and\ \bibinfo {author} {\bibfnamefont
  {Y.}~\bibnamefont {Zhao}},\ }\href {\doibase 10.1103/PhysRevD.100.034505}
  {\bibfield  {journal} {\bibinfo  {journal} {Phys. Rev. D}\ }\textbf {\bibinfo
  {volume} {100}},\ \bibinfo {pages} {034505} (\bibinfo {year}
  {2019}{\natexlab{a}})},\ \Eprint {http://arxiv.org/abs/1804.01483}
  {arXiv:1804.01483 [hep-lat]} \BibitemShut {NoStop}%
\bibitem [{\citenamefont {Izubuchi}\ \emph {et~al.}(2019)\citenamefont
  {Izubuchi}, \citenamefont {Jin}, \citenamefont {Kallidonis}, \citenamefont
  {Karthik}, \citenamefont {Mukherjee}, \citenamefont {Petreczky},
  \citenamefont {Shugert},\ and\ \citenamefont {Syritsyn}}]{Izubuchi:2019lyk}%
  \BibitemOpen
  \bibfield  {author} {\bibinfo {author} {\bibfnamefont {T.}~\bibnamefont
  {Izubuchi}}, \bibinfo {author} {\bibfnamefont {L.}~\bibnamefont {Jin}},
  \bibinfo {author} {\bibfnamefont {C.}~\bibnamefont {Kallidonis}}, \bibinfo
  {author} {\bibfnamefont {N.}~\bibnamefont {Karthik}}, \bibinfo {author}
  {\bibfnamefont {S.}~\bibnamefont {Mukherjee}}, \bibinfo {author}
  {\bibfnamefont {P.}~\bibnamefont {Petreczky}}, \bibinfo {author}
  {\bibfnamefont {C.}~\bibnamefont {Shugert}}, \ and\ \bibinfo {author}
  {\bibfnamefont {S.}~\bibnamefont {Syritsyn}},\ }\href {\doibase
  10.1103/PhysRevD.100.034516} {\bibfield  {journal} {\bibinfo  {journal}
  {Phys. Rev. D}\ }\textbf {\bibinfo {volume} {100}},\ \bibinfo {pages}
  {034516} (\bibinfo {year} {2019})},\ \Eprint
  {http://arxiv.org/abs/1905.06349} {arXiv:1905.06349 [hep-lat]} \BibitemShut
  {NoStop}%
\bibitem [{\citenamefont {Shugert}\ \emph {et~al.}(2020)\citenamefont
  {Shugert}, \citenamefont {Gao}, \citenamefont {Izubichi}, \citenamefont
  {Jin}, \citenamefont {Kallidonis}, \citenamefont {Karthik}, \citenamefont
  {Mukherjee}, \citenamefont {Petreczky}, \citenamefont {Syritsyn},\ and\
  \citenamefont {Zhao}}]{Shugert:2020tgq}%
  \BibitemOpen
  \bibfield  {author} {\bibinfo {author} {\bibfnamefont {C.}~\bibnamefont
  {Shugert}}, \bibinfo {author} {\bibfnamefont {X.}~\bibnamefont {Gao}},
  \bibinfo {author} {\bibfnamefont {T.}~\bibnamefont {Izubichi}}, \bibinfo
  {author} {\bibfnamefont {L.}~\bibnamefont {Jin}}, \bibinfo {author}
  {\bibfnamefont {C.}~\bibnamefont {Kallidonis}}, \bibinfo {author}
  {\bibfnamefont {N.}~\bibnamefont {Karthik}}, \bibinfo {author} {\bibfnamefont
  {S.}~\bibnamefont {Mukherjee}}, \bibinfo {author} {\bibfnamefont
  {P.}~\bibnamefont {Petreczky}}, \bibinfo {author} {\bibfnamefont
  {S.}~\bibnamefont {Syritsyn}}, \ and\ \bibinfo {author} {\bibfnamefont
  {Y.}~\bibnamefont {Zhao}},\ }in\ \href@noop {} {\emph {\bibinfo {booktitle}
  {{37th International Symposium on Lattice Field Theory}}}}\ (\bibinfo {year}
  {2020})\ \Eprint {http://arxiv.org/abs/2001.11650} {arXiv:2001.11650
  [hep-lat]} \BibitemShut {NoStop}%
\bibitem [{\citenamefont {Chai}\ \emph {et~al.}(2020)\citenamefont {Chai} \emph
  {et~al.}}]{Chai:2020nxw}%
  \BibitemOpen
  \bibfield  {author} {\bibinfo {author} {\bibfnamefont {Y.}~\bibnamefont
  {Chai}} \emph {et~al.},\ }\href {\doibase 10.1103/PhysRevD.102.014508}
  {\bibfield  {journal} {\bibinfo  {journal} {Phys. Rev. D}\ }\textbf {\bibinfo
  {volume} {102}},\ \bibinfo {pages} {014508} (\bibinfo {year} {2020})},\
  \Eprint {http://arxiv.org/abs/2002.12044} {arXiv:2002.12044 [hep-lat]}
  \BibitemShut {NoStop}%
\bibitem [{\citenamefont {Lin}\ \emph {et~al.}(2020)\citenamefont {Lin},
  \citenamefont {Chen}, \citenamefont {Fan}, \citenamefont {Zhang},\ and\
  \citenamefont {Zhang}}]{Lin:2020ssv}%
  \BibitemOpen
  \bibfield  {author} {\bibinfo {author} {\bibfnamefont {H.-W.}\ \bibnamefont
  {Lin}}, \bibinfo {author} {\bibfnamefont {J.-W.}\ \bibnamefont {Chen}},
  \bibinfo {author} {\bibfnamefont {Z.}~\bibnamefont {Fan}}, \bibinfo {author}
  {\bibfnamefont {J.-H.}\ \bibnamefont {Zhang}}, \ and\ \bibinfo {author}
  {\bibfnamefont {R.}~\bibnamefont {Zhang}},\ }\href@noop {} {\  (\bibinfo
  {year} {2020})},\ \Eprint {http://arxiv.org/abs/2003.14128} {arXiv:2003.14128
  [hep-lat]} \BibitemShut {NoStop}%
\bibitem [{\citenamefont {Fan}\ \emph {et~al.}(2020)\citenamefont {Fan},
  \citenamefont {Gao}, \citenamefont {Li}, \citenamefont {Lin}, \citenamefont
  {Karthik}, \citenamefont {Mukherjee}, \citenamefont {Petreczky},
  \citenamefont {Syritsyn}, \citenamefont {Yang},\ and\ \citenamefont
  {Zhang}}]{Fan:2020nzz}%
  \BibitemOpen
  \bibfield  {author} {\bibinfo {author} {\bibfnamefont {Z.}~\bibnamefont
  {Fan}}, \bibinfo {author} {\bibfnamefont {X.}~\bibnamefont {Gao}}, \bibinfo
  {author} {\bibfnamefont {R.}~\bibnamefont {Li}}, \bibinfo {author}
  {\bibfnamefont {H.-W.}\ \bibnamefont {Lin}}, \bibinfo {author} {\bibfnamefont
  {N.}~\bibnamefont {Karthik}}, \bibinfo {author} {\bibfnamefont
  {S.}~\bibnamefont {Mukherjee}}, \bibinfo {author} {\bibfnamefont
  {P.}~\bibnamefont {Petreczky}}, \bibinfo {author} {\bibfnamefont
  {S.}~\bibnamefont {Syritsyn}}, \bibinfo {author} {\bibfnamefont {Y.-B.}\
  \bibnamefont {Yang}}, \ and\ \bibinfo {author} {\bibfnamefont
  {R.}~\bibnamefont {Zhang}},\ }\href {\doibase 10.1103/PhysRevD.102.074504}
  {\bibfield  {journal} {\bibinfo  {journal} {Phys. Rev. D}\ }\textbf {\bibinfo
  {volume} {102}},\ \bibinfo {pages} {074504} (\bibinfo {year} {2020})},\
  \Eprint {http://arxiv.org/abs/2005.12015} {arXiv:2005.12015 [hep-lat]}
  \BibitemShut {NoStop}%
\bibitem [{\citenamefont {Chen}\ \emph
  {et~al.}(2020{\natexlab{a}})\citenamefont {Chen}, \citenamefont {Lin},\ and\
  \citenamefont {Zhang}}]{Chen:2019lcm}%
  \BibitemOpen
  \bibfield  {author} {\bibinfo {author} {\bibfnamefont {J.-W.}\ \bibnamefont
  {Chen}}, \bibinfo {author} {\bibfnamefont {H.-W.}\ \bibnamefont {Lin}}, \
  and\ \bibinfo {author} {\bibfnamefont {J.-H.}\ \bibnamefont {Zhang}},\ }\href
  {\doibase 10.1016/j.nuclphysb.2020.114940} {\bibfield  {journal} {\bibinfo
  {journal} {Nucl. Phys. B}\ }\textbf {\bibinfo {volume} {952}},\ \bibinfo
  {pages} {114940} (\bibinfo {year} {2020}{\natexlab{a}})},\ \Eprint
  {http://arxiv.org/abs/1904.12376} {arXiv:1904.12376 [hep-lat]} \BibitemShut
  {NoStop}%
\bibitem [{\citenamefont {Alexandrou}\ \emph {et~al.}(2019)\citenamefont
  {Alexandrou}, \citenamefont {Cichy}, \citenamefont {Constantinou},
  \citenamefont {Hadjiyiannakou}, \citenamefont {Jansen}, \citenamefont
  {Scapellato},\ and\ \citenamefont {Steffens}}]{Alexandrou:2019dax}%
  \BibitemOpen
  \bibfield  {author} {\bibinfo {author} {\bibfnamefont {C.}~\bibnamefont
  {Alexandrou}}, \bibinfo {author} {\bibfnamefont {K.}~\bibnamefont {Cichy}},
  \bibinfo {author} {\bibfnamefont {M.}~\bibnamefont {Constantinou}}, \bibinfo
  {author} {\bibfnamefont {K.}~\bibnamefont {Hadjiyiannakou}}, \bibinfo
  {author} {\bibfnamefont {K.}~\bibnamefont {Jansen}}, \bibinfo {author}
  {\bibfnamefont {A.}~\bibnamefont {Scapellato}}, \ and\ \bibinfo {author}
  {\bibfnamefont {F.}~\bibnamefont {Steffens}},\ }\href {\doibase
  10.22323/1.363.0036} {\bibfield  {journal} {\bibinfo  {journal} {PoS}\
  }\textbf {\bibinfo {volume} {LATTICE2019}},\ \bibinfo {pages} {036} (\bibinfo
  {year} {2019})},\ \Eprint {http://arxiv.org/abs/1910.13229} {arXiv:1910.13229
  [hep-lat]} \BibitemShut {NoStop}%
\bibitem [{\citenamefont {Zhang}\ \emph {et~al.}(2017)\citenamefont {Zhang},
  \citenamefont {Chen}, \citenamefont {Ji}, \citenamefont {Jin},\ and\
  \citenamefont {Lin}}]{Zhang:2017bzy}%
  \BibitemOpen
  \bibfield  {author} {\bibinfo {author} {\bibfnamefont {J.-H.}\ \bibnamefont
  {Zhang}}, \bibinfo {author} {\bibfnamefont {J.-W.}\ \bibnamefont {Chen}},
  \bibinfo {author} {\bibfnamefont {X.}~\bibnamefont {Ji}}, \bibinfo {author}
  {\bibfnamefont {L.}~\bibnamefont {Jin}}, \ and\ \bibinfo {author}
  {\bibfnamefont {H.-W.}\ \bibnamefont {Lin}},\ }\href {\doibase
  10.1103/PhysRevD.95.094514} {\bibfield  {journal} {\bibinfo  {journal} {Phys.
  Rev.}\ }\textbf {\bibinfo {volume} {D95}},\ \bibinfo {pages} {094514}
  (\bibinfo {year} {2017})},\ \Eprint {http://arxiv.org/abs/1702.00008}
  {arXiv:1702.00008 [hep-lat]} \BibitemShut {NoStop}%
\bibitem [{\citenamefont {Zhang}\ \emph
  {et~al.}(2019{\natexlab{b}})\citenamefont {Zhang}, \citenamefont {Jin},
  \citenamefont {Lin}, \citenamefont {Sch\"afer}, \citenamefont {Sun},
  \citenamefont {Yang}, \citenamefont {Zhang}, \citenamefont {Zhao},\ and\
  \citenamefont {Chen}}]{Chen:2017gck}%
  \BibitemOpen
  \bibfield  {author} {\bibinfo {author} {\bibfnamefont {J.-H.}\ \bibnamefont
  {Zhang}}, \bibinfo {author} {\bibfnamefont {L.}~\bibnamefont {Jin}}, \bibinfo
  {author} {\bibfnamefont {H.-W.}\ \bibnamefont {Lin}}, \bibinfo {author}
  {\bibfnamefont {A.}~\bibnamefont {Sch\"afer}}, \bibinfo {author}
  {\bibfnamefont {P.}~\bibnamefont {Sun}}, \bibinfo {author} {\bibfnamefont
  {Y.-B.}\ \bibnamefont {Yang}}, \bibinfo {author} {\bibfnamefont
  {R.}~\bibnamefont {Zhang}}, \bibinfo {author} {\bibfnamefont
  {Y.}~\bibnamefont {Zhao}}, \ and\ \bibinfo {author} {\bibfnamefont {J.-W.}\
  \bibnamefont {Chen}} (\bibinfo {collaboration} {LP3}),\ }\href {\doibase
  10.1016/j.nuclphysb.2018.12.020} {\bibfield  {journal} {\bibinfo  {journal}
  {Nucl. Phys.}\ }\textbf {\bibinfo {volume} {B939}},\ \bibinfo {pages} {429}
  (\bibinfo {year} {2019}{\natexlab{b}})},\ \Eprint
  {http://arxiv.org/abs/1712.10025} {arXiv:1712.10025 [hep-ph]} \BibitemShut
  {NoStop}%
\bibitem [{\citenamefont {Zhang}\ \emph
  {et~al.}(2020{\natexlab{a}})\citenamefont {Zhang}, \citenamefont {Honkala},
  \citenamefont {Lin},\ and\ \citenamefont {Chen}}]{Zhang:2020gaj}%
  \BibitemOpen
  \bibfield  {author} {\bibinfo {author} {\bibfnamefont {R.}~\bibnamefont
  {Zhang}}, \bibinfo {author} {\bibfnamefont {C.}~\bibnamefont {Honkala}},
  \bibinfo {author} {\bibfnamefont {H.-W.}\ \bibnamefont {Lin}}, \ and\
  \bibinfo {author} {\bibfnamefont {J.-W.}\ \bibnamefont {Chen}},\ }\href@noop
  {} {\  (\bibinfo {year} {2020}{\natexlab{a}})},\ \Eprint
  {http://arxiv.org/abs/2005.13955} {arXiv:2005.13955 [hep-lat]} \BibitemShut
  {NoStop}%
\bibitem [{\citenamefont {Shanahan}\ \emph
  {et~al.}(2020{\natexlab{a}})\citenamefont {Shanahan}, \citenamefont
  {Wagman},\ and\ \citenamefont {Zhao}}]{Shanahan:2019zcq}%
  \BibitemOpen
  \bibfield  {author} {\bibinfo {author} {\bibfnamefont {P.}~\bibnamefont
  {Shanahan}}, \bibinfo {author} {\bibfnamefont {M.~L.}\ \bibnamefont
  {Wagman}}, \ and\ \bibinfo {author} {\bibfnamefont {Y.}~\bibnamefont
  {Zhao}},\ }\href {\doibase 10.1103/PhysRevD.101.074505} {\bibfield  {journal}
  {\bibinfo  {journal} {Phys. Rev. D}\ }\textbf {\bibinfo {volume} {101}},\
  \bibinfo {pages} {074505} (\bibinfo {year} {2020}{\natexlab{a}})},\ \Eprint
  {http://arxiv.org/abs/1911.00800} {arXiv:1911.00800 [hep-lat]} \BibitemShut
  {NoStop}%
\bibitem [{\citenamefont {Shanahan}\ \emph
  {et~al.}(2020{\natexlab{b}})\citenamefont {Shanahan}, \citenamefont
  {Wagman},\ and\ \citenamefont {Zhao}}]{Shanahan:2020zxr}%
  \BibitemOpen
  \bibfield  {author} {\bibinfo {author} {\bibfnamefont {P.}~\bibnamefont
  {Shanahan}}, \bibinfo {author} {\bibfnamefont {M.}~\bibnamefont {Wagman}}, \
  and\ \bibinfo {author} {\bibfnamefont {Y.}~\bibnamefont {Zhao}},\ }\href
  {\doibase 10.1103/PhysRevD.102.014511} {\bibfield  {journal} {\bibinfo
  {journal} {Phys. Rev. D}\ }\textbf {\bibinfo {volume} {102}},\ \bibinfo
  {pages} {014511} (\bibinfo {year} {2020}{\natexlab{b}})},\ \Eprint
  {http://arxiv.org/abs/2003.06063} {arXiv:2003.06063 [hep-lat]} \BibitemShut
  {NoStop}%
\bibitem [{\citenamefont {Zhang}\ \emph
  {et~al.}(2020{\natexlab{b}})\citenamefont {Zhang} \emph
  {et~al.}}]{Zhang:2020dbb}%
  \BibitemOpen
  \bibfield  {author} {\bibinfo {author} {\bibfnamefont {Q.-A.}\ \bibnamefont
  {Zhang}} \emph {et~al.} (\bibinfo {collaboration} {Lattice Parton}),\ }\href
  {\doibase 10.1103/PhysRevLett.125.192001} {\bibfield  {journal} {\bibinfo
  {journal} {Phys. Rev. Lett.}\ }\textbf {\bibinfo {volume} {125}},\ \bibinfo
  {pages} {192001} (\bibinfo {year} {2020}{\natexlab{b}})},\ \Eprint
  {http://arxiv.org/abs/2005.14572} {arXiv:2005.14572 [hep-lat]} \BibitemShut
  {NoStop}%
\bibitem [{\citenamefont {Ji}(2020)}]{Ji:2020byp}%
  \BibitemOpen
  \bibfield  {author} {\bibinfo {author} {\bibfnamefont {X.}~\bibnamefont
  {Ji}},\ }\href@noop {} {\  (\bibinfo {year} {2020})},\ \Eprint
  {http://arxiv.org/abs/2007.06613} {arXiv:2007.06613 [hep-ph]} \BibitemShut
  {NoStop}%
\bibitem [{\citenamefont {Alexandrou}\ \emph
  {et~al.}(2017{\natexlab{b}})\citenamefont {Alexandrou}, \citenamefont
  {Cichy}, \citenamefont {Constantinou}, \citenamefont {Hadjiyiannakou},
  \citenamefont {Jansen}, \citenamefont {Panagopoulos},\ and\ \citenamefont
  {Steffens}}]{Alexandrou:2017huk}%
  \BibitemOpen
  \bibfield  {author} {\bibinfo {author} {\bibfnamefont {C.}~\bibnamefont
  {Alexandrou}}, \bibinfo {author} {\bibfnamefont {K.}~\bibnamefont {Cichy}},
  \bibinfo {author} {\bibfnamefont {M.}~\bibnamefont {Constantinou}}, \bibinfo
  {author} {\bibfnamefont {K.}~\bibnamefont {Hadjiyiannakou}}, \bibinfo
  {author} {\bibfnamefont {K.}~\bibnamefont {Jansen}}, \bibinfo {author}
  {\bibfnamefont {H.}~\bibnamefont {Panagopoulos}}, \ and\ \bibinfo {author}
  {\bibfnamefont {F.}~\bibnamefont {Steffens}},\ }\href {\doibase
  10.1016/j.nuclphysb.2017.08.012} {\bibfield  {journal} {\bibinfo  {journal}
  {Nucl. Phys.}\ }\textbf {\bibinfo {volume} {B923}},\ \bibinfo {pages} {394}
  (\bibinfo {year} {2017}{\natexlab{b}})},\ \Eprint
  {http://arxiv.org/abs/1706.00265} {arXiv:1706.00265 [hep-lat]} \BibitemShut
  {NoStop}%
\bibitem [{\citenamefont {Ji}\ and\ \citenamefont {Zhang}(2015)}]{Ji:2015jwa}%
  \BibitemOpen
  \bibfield  {author} {\bibinfo {author} {\bibfnamefont {X.}~\bibnamefont
  {Ji}}\ and\ \bibinfo {author} {\bibfnamefont {J.-H.}\ \bibnamefont {Zhang}},\
  }\href {\doibase 10.1103/PhysRevD.92.034006} {\bibfield  {journal} {\bibinfo
  {journal} {Phys. Rev.}\ }\textbf {\bibinfo {volume} {D92}},\ \bibinfo {pages}
  {034006} (\bibinfo {year} {2015})},\ \Eprint
  {http://arxiv.org/abs/1505.07699} {arXiv:1505.07699 [hep-ph]} \BibitemShut
  {NoStop}%
\bibitem [{\citenamefont {Ji}\ \emph {et~al.}(2018)\citenamefont {Ji},
  \citenamefont {Zhang},\ and\ \citenamefont {Zhao}}]{Ji:2017oey}%
  \BibitemOpen
  \bibfield  {author} {\bibinfo {author} {\bibfnamefont {X.}~\bibnamefont
  {Ji}}, \bibinfo {author} {\bibfnamefont {J.-H.}\ \bibnamefont {Zhang}}, \
  and\ \bibinfo {author} {\bibfnamefont {Y.}~\bibnamefont {Zhao}},\ }\href
  {\doibase 10.1103/PhysRevLett.120.112001} {\bibfield  {journal} {\bibinfo
  {journal} {Phys. Rev. Lett.}\ }\textbf {\bibinfo {volume} {120}},\ \bibinfo
  {pages} {112001} (\bibinfo {year} {2018})},\ \Eprint
  {http://arxiv.org/abs/1706.08962} {arXiv:1706.08962 [hep-ph]} \BibitemShut
  {NoStop}%
\bibitem [{\citenamefont {Ishikawa}\ \emph {et~al.}(2017)\citenamefont
  {Ishikawa}, \citenamefont {Ma}, \citenamefont {Qiu},\ and\ \citenamefont
  {Yoshida}}]{Ishikawa:2017faj}%
  \BibitemOpen
  \bibfield  {author} {\bibinfo {author} {\bibfnamefont {T.}~\bibnamefont
  {Ishikawa}}, \bibinfo {author} {\bibfnamefont {Y.-Q.}\ \bibnamefont {Ma}},
  \bibinfo {author} {\bibfnamefont {J.-W.}\ \bibnamefont {Qiu}}, \ and\
  \bibinfo {author} {\bibfnamefont {S.}~\bibnamefont {Yoshida}},\ }\href
  {\doibase 10.1103/PhysRevD.96.094019} {\bibfield  {journal} {\bibinfo
  {journal} {Phys. Rev.}\ }\textbf {\bibinfo {volume} {D96}},\ \bibinfo {pages}
  {094019} (\bibinfo {year} {2017})},\ \Eprint
  {http://arxiv.org/abs/1707.03107} {arXiv:1707.03107 [hep-ph]} \BibitemShut
  {NoStop}%
\bibitem [{\citenamefont {Green}\ \emph {et~al.}(2018)\citenamefont {Green},
  \citenamefont {Jansen},\ and\ \citenamefont {Steffens}}]{Green:2017xeu}%
  \BibitemOpen
  \bibfield  {author} {\bibinfo {author} {\bibfnamefont {J.}~\bibnamefont
  {Green}}, \bibinfo {author} {\bibfnamefont {K.}~\bibnamefont {Jansen}}, \
  and\ \bibinfo {author} {\bibfnamefont {F.}~\bibnamefont {Steffens}},\ }\href
  {\doibase 10.1103/PhysRevLett.121.022004} {\bibfield  {journal} {\bibinfo
  {journal} {Phys. Rev. Lett.}\ }\textbf {\bibinfo {volume} {121}},\ \bibinfo
  {pages} {022004} (\bibinfo {year} {2018})},\ \Eprint
  {http://arxiv.org/abs/1707.07152} {arXiv:1707.07152 [hep-lat]} \BibitemShut
  {NoStop}%
\bibitem [{\citenamefont {Ji}\ \emph {et~al.}(2020{\natexlab{b}})\citenamefont
  {Ji}, \citenamefont {Liu}, \citenamefont {Sch\"afer}, \citenamefont {Wang},
  \citenamefont {Yang}, \citenamefont {Zhang},\ and\ \citenamefont
  {Zhao}}]{Ji:2020brr}%
  \BibitemOpen
  \bibfield  {author} {\bibinfo {author} {\bibfnamefont {X.}~\bibnamefont
  {Ji}}, \bibinfo {author} {\bibfnamefont {Y.}~\bibnamefont {Liu}}, \bibinfo
  {author} {\bibfnamefont {A.}~\bibnamefont {Sch\"afer}}, \bibinfo {author}
  {\bibfnamefont {W.}~\bibnamefont {Wang}}, \bibinfo {author} {\bibfnamefont
  {Y.-B.}\ \bibnamefont {Yang}}, \bibinfo {author} {\bibfnamefont {J.-H.}\
  \bibnamefont {Zhang}}, \ and\ \bibinfo {author} {\bibfnamefont
  {Y.}~\bibnamefont {Zhao}},\ }\href@noop {} {\  (\bibinfo {year}
  {2020}{\natexlab{b}})},\ \Eprint {http://arxiv.org/abs/2008.03886}
  {arXiv:2008.03886 [hep-ph]} \BibitemShut {NoStop}%
\bibitem [{\citenamefont {Chen}\ \emph
  {et~al.}(2018{\natexlab{b}})\citenamefont {Chen}, \citenamefont {Ishikawa},
  \citenamefont {Jin}, \citenamefont {Lin}, \citenamefont {Yang}, \citenamefont
  {Zhang},\ and\ \citenamefont {Zhao}}]{Chen:2017mzz}%
  \BibitemOpen
  \bibfield  {author} {\bibinfo {author} {\bibfnamefont {J.-W.}\ \bibnamefont
  {Chen}}, \bibinfo {author} {\bibfnamefont {T.}~\bibnamefont {Ishikawa}},
  \bibinfo {author} {\bibfnamefont {L.}~\bibnamefont {Jin}}, \bibinfo {author}
  {\bibfnamefont {H.-W.}\ \bibnamefont {Lin}}, \bibinfo {author} {\bibfnamefont
  {Y.-B.}\ \bibnamefont {Yang}}, \bibinfo {author} {\bibfnamefont {J.-H.}\
  \bibnamefont {Zhang}}, \ and\ \bibinfo {author} {\bibfnamefont
  {Y.}~\bibnamefont {Zhao}},\ }\href {\doibase 10.1103/PhysRevD.97.014505}
  {\bibfield  {journal} {\bibinfo  {journal} {Phys. Rev.}\ }\textbf {\bibinfo
  {volume} {D97}},\ \bibinfo {pages} {014505} (\bibinfo {year}
  {2018}{\natexlab{b}})},\ \Eprint {http://arxiv.org/abs/1706.01295}
  {arXiv:1706.01295 [hep-lat]} \BibitemShut {NoStop}%
\bibitem [{\citenamefont {Orginos}\ \emph {et~al.}(2017)\citenamefont
  {Orginos}, \citenamefont {Radyushkin}, \citenamefont {Karpie},\ and\
  \citenamefont {Zafeiropoulos}}]{Orginos:2017kos}%
  \BibitemOpen
  \bibfield  {author} {\bibinfo {author} {\bibfnamefont {K.}~\bibnamefont
  {Orginos}}, \bibinfo {author} {\bibfnamefont {A.}~\bibnamefont {Radyushkin}},
  \bibinfo {author} {\bibfnamefont {J.}~\bibnamefont {Karpie}}, \ and\ \bibinfo
  {author} {\bibfnamefont {S.}~\bibnamefont {Zafeiropoulos}},\ }\href {\doibase
  10.1103/PhysRevD.96.094503} {\bibfield  {journal} {\bibinfo  {journal} {Phys.
  Rev.}\ }\textbf {\bibinfo {volume} {D96}},\ \bibinfo {pages} {094503}
  (\bibinfo {year} {2017})},\ \Eprint {http://arxiv.org/abs/1706.05373}
  {arXiv:1706.05373 [hep-ph]} \BibitemShut {NoStop}%
\bibitem [{\citenamefont {Izubuchi}\ \emph {et~al.}(2018)\citenamefont
  {Izubuchi}, \citenamefont {Ji}, \citenamefont {Jin}, \citenamefont
  {Stewart},\ and\ \citenamefont {Zhao}}]{Izubuchi:2018srq}%
  \BibitemOpen
  \bibfield  {author} {\bibinfo {author} {\bibfnamefont {T.}~\bibnamefont
  {Izubuchi}}, \bibinfo {author} {\bibfnamefont {X.}~\bibnamefont {Ji}},
  \bibinfo {author} {\bibfnamefont {L.}~\bibnamefont {Jin}}, \bibinfo {author}
  {\bibfnamefont {I.~W.}\ \bibnamefont {Stewart}}, \ and\ \bibinfo {author}
  {\bibfnamefont {Y.}~\bibnamefont {Zhao}},\ }\href {\doibase
  10.1103/PhysRevD.98.056004} {\bibfield  {journal} {\bibinfo  {journal} {Phys.
  Rev.}\ }\textbf {\bibinfo {volume} {D98}},\ \bibinfo {pages} {056004}
  (\bibinfo {year} {2018})},\ \Eprint {http://arxiv.org/abs/1801.03917}
  {arXiv:1801.03917 [hep-ph]} \BibitemShut {NoStop}%
\bibitem [{\citenamefont {Chen}\ \emph {et~al.}(2017)\citenamefont {Chen},
  \citenamefont {Ji},\ and\ \citenamefont {Zhang}}]{Chen:2016fxx}%
  \BibitemOpen
  \bibfield  {author} {\bibinfo {author} {\bibfnamefont {J.-W.}\ \bibnamefont
  {Chen}}, \bibinfo {author} {\bibfnamefont {X.}~\bibnamefont {Ji}}, \ and\
  \bibinfo {author} {\bibfnamefont {J.-H.}\ \bibnamefont {Zhang}},\ }\href
  {\doibase 10.1016/j.nuclphysb.2016.12.004} {\bibfield  {journal} {\bibinfo
  {journal} {Nucl. Phys.}\ }\textbf {\bibinfo {volume} {B915}},\ \bibinfo
  {pages} {1} (\bibinfo {year} {2017})},\ \Eprint
  {http://arxiv.org/abs/1609.08102} {arXiv:1609.08102 [hep-ph]} \BibitemShut
  {NoStop}%
\bibitem [{\citenamefont {Braun}\ \emph {et~al.}(2019)\citenamefont {Braun},
  \citenamefont {Vladimirov},\ and\ \citenamefont {Zhang}}]{Braun:2018brg}%
  \BibitemOpen
  \bibfield  {author} {\bibinfo {author} {\bibfnamefont {V.~M.}\ \bibnamefont
  {Braun}}, \bibinfo {author} {\bibfnamefont {A.}~\bibnamefont {Vladimirov}}, \
  and\ \bibinfo {author} {\bibfnamefont {J.-H.}\ \bibnamefont {Zhang}},\ }\href
  {\doibase 10.1103/PhysRevD.99.014013} {\bibfield  {journal} {\bibinfo
  {journal} {Phys. Rev.}\ }\textbf {\bibinfo {volume} {D99}},\ \bibinfo {pages}
  {014013} (\bibinfo {year} {2019})},\ \Eprint
  {http://arxiv.org/abs/1810.00048} {arXiv:1810.00048 [hep-ph]} \BibitemShut
  {NoStop}%
\bibitem [{\citenamefont {Zhang}\ \emph
  {et~al.}(2020{\natexlab{c}})\citenamefont {Zhang}, \citenamefont {Li},
  \citenamefont {Huo}, \citenamefont {Sun},\ and\ \citenamefont
  {Yang}}]{Zhang:2020rsx}%
  \BibitemOpen
  \bibfield  {author} {\bibinfo {author} {\bibfnamefont {K.}~\bibnamefont
  {Zhang}}, \bibinfo {author} {\bibfnamefont {Y.-Y.}\ \bibnamefont {Li}},
  \bibinfo {author} {\bibfnamefont {Y.-K.}\ \bibnamefont {Huo}}, \bibinfo
  {author} {\bibfnamefont {P.}~\bibnamefont {Sun}}, \ and\ \bibinfo {author}
  {\bibfnamefont {Y.-B.}\ \bibnamefont {Yang}},\ }\href@noop {} {\  (\bibinfo
  {year} {2020}{\natexlab{c}})},\ \Eprint {http://arxiv.org/abs/2012.05448}
  {arXiv:2012.05448 [hep-lat]} \BibitemShut {NoStop}%
\bibitem [{\citenamefont {Bazavov}\ \emph {et~al.}(2013)\citenamefont {Bazavov}
  \emph {et~al.}}]{Bazavov:2012xda}%
  \BibitemOpen
  \bibfield  {author} {\bibinfo {author} {\bibfnamefont {A.}~\bibnamefont
  {Bazavov}} \emph {et~al.} (\bibinfo {collaboration} {MILC}),\ }\href
  {\doibase 10.1103/PhysRevD.87.054505} {\bibfield  {journal} {\bibinfo
  {journal} {Phys. Rev.}\ }\textbf {\bibinfo {volume} {D87}},\ \bibinfo {pages}
  {054505} (\bibinfo {year} {2013})},\ \Eprint {http://arxiv.org/abs/1212.4768}
  {arXiv:1212.4768 [hep-lat]} \BibitemShut {NoStop}%
\bibitem [{\citenamefont {Blum}\ \emph {et~al.}(2016)\citenamefont {Blum} \emph
  {et~al.}}]{Blum:2014tka}%
  \BibitemOpen
  \bibfield  {author} {\bibinfo {author} {\bibfnamefont {T.}~\bibnamefont
  {Blum}} \emph {et~al.} (\bibinfo {collaboration} {RBC, UKQCD}),\ }\href
  {\doibase 10.1103/PhysRevD.93.074505} {\bibfield  {journal} {\bibinfo
  {journal} {Phys. Rev.}\ }\textbf {\bibinfo {volume} {D93}},\ \bibinfo {pages}
  {074505} (\bibinfo {year} {2016})},\ \Eprint {http://arxiv.org/abs/1411.7017}
  {arXiv:1411.7017 [hep-lat]} \BibitemShut {NoStop}%
\bibitem [{\citenamefont {Collins}(1986)}]{Collins:1984xc}%
  \BibitemOpen
  \bibfield  {author} {\bibinfo {author} {\bibfnamefont {J.~C.}\ \bibnamefont
  {Collins}},\ }\href {\doibase 10.1017/CBO9780511622656} {\emph {\bibinfo
  {title} {{Renormalization}: {An Introduction to Renormalization, The
  Renormalization Group, and the Operator Product Expansion}}}},\ \bibinfo
  {series} {Cambridge Monographs on Mathematical Physics}, Vol.~\bibinfo
  {volume} {26}\ (\bibinfo  {publisher} {Cambridge University Press},\ \bibinfo
  {address} {Cambridge},\ \bibinfo {year} {1986})\BibitemShut {NoStop}%
\bibitem [{\citenamefont {Ji}(1995)}]{Ji:1995tm}%
  \BibitemOpen
  \bibfield  {author} {\bibinfo {author} {\bibfnamefont {X.-D.}\ \bibnamefont
  {Ji}},\ }\href@noop {} {\  (\bibinfo {year} {1995})},\ \Eprint
  {http://arxiv.org/abs/hep-ph/9507322} {arXiv:hep-ph/9507322} \BibitemShut
  {NoStop}%
\bibitem [{\citenamefont {Beneke}(1999)}]{Beneke:1998ui}%
  \BibitemOpen
  \bibfield  {author} {\bibinfo {author} {\bibfnamefont {M.}~\bibnamefont
  {Beneke}},\ }\href {\doibase 10.1016/S0370-1573(98)00130-6} {\bibfield
  {journal} {\bibinfo  {journal} {Phys. Rept.}\ }\textbf {\bibinfo {volume}
  {317}},\ \bibinfo {pages} {1} (\bibinfo {year} {1999})},\ \Eprint
  {http://arxiv.org/abs/hep-ph/9807443} {arXiv:hep-ph/9807443} \BibitemShut
  {NoStop}%
\bibitem [{\citenamefont {Bauer}\ \emph {et~al.}(2012)\citenamefont {Bauer},
  \citenamefont {Bali},\ and\ \citenamefont {Pineda}}]{Bauer:2011ws}%
  \BibitemOpen
  \bibfield  {author} {\bibinfo {author} {\bibfnamefont {C.}~\bibnamefont
  {Bauer}}, \bibinfo {author} {\bibfnamefont {G.~S.}\ \bibnamefont {Bali}}, \
  and\ \bibinfo {author} {\bibfnamefont {A.}~\bibnamefont {Pineda}},\ }\href
  {\doibase 10.1103/PhysRevLett.108.242002} {\bibfield  {journal} {\bibinfo
  {journal} {Phys. Rev. Lett.}\ }\textbf {\bibinfo {volume} {108}},\ \bibinfo
  {pages} {242002} (\bibinfo {year} {2012})},\ \Eprint
  {http://arxiv.org/abs/1111.3946} {arXiv:1111.3946 [hep-ph]} \BibitemShut
  {NoStop}%
\bibitem [{\citenamefont {Bali}\ \emph {et~al.}(2013)\citenamefont {Bali},
  \citenamefont {Bauer}, \citenamefont {Pineda},\ and\ \citenamefont
  {Torrero}}]{Bali:2013pla}%
  \BibitemOpen
  \bibfield  {author} {\bibinfo {author} {\bibfnamefont {G.~S.}\ \bibnamefont
  {Bali}}, \bibinfo {author} {\bibfnamefont {C.}~\bibnamefont {Bauer}},
  \bibinfo {author} {\bibfnamefont {A.}~\bibnamefont {Pineda}}, \ and\ \bibinfo
  {author} {\bibfnamefont {C.}~\bibnamefont {Torrero}},\ }\href {\doibase
  10.1103/PhysRevD.87.094517} {\bibfield  {journal} {\bibinfo  {journal} {Phys.
  Rev. D}\ }\textbf {\bibinfo {volume} {87}},\ \bibinfo {pages} {094517}
  (\bibinfo {year} {2013})},\ \Eprint {http://arxiv.org/abs/1303.3279}
  {arXiv:1303.3279 [hep-lat]} \BibitemShut {NoStop}%
\bibitem [{\citenamefont {Constantinou}\ and\ \citenamefont
  {Panagopoulos}(2017)}]{Constantinou:2017sej}%
  \BibitemOpen
  \bibfield  {author} {\bibinfo {author} {\bibfnamefont {M.}~\bibnamefont
  {Constantinou}}\ and\ \bibinfo {author} {\bibfnamefont {H.}~\bibnamefont
  {Panagopoulos}},\ }\href {\doibase 10.1103/PhysRevD.96.054506} {\bibfield
  {journal} {\bibinfo  {journal} {Phys. Rev. D}\ }\textbf {\bibinfo {volume}
  {96}},\ \bibinfo {pages} {054506} (\bibinfo {year} {2017})},\ \Eprint
  {http://arxiv.org/abs/1705.11193} {arXiv:1705.11193 [hep-lat]} \BibitemShut
  {NoStop}%
\bibitem [{\citenamefont {Ji}\ and\ \citenamefont {Musolf}(1991)}]{Ji:1991pr}%
  \BibitemOpen
  \bibfield  {author} {\bibinfo {author} {\bibfnamefont {X.-D.}\ \bibnamefont
  {Ji}}\ and\ \bibinfo {author} {\bibfnamefont {M.}~\bibnamefont {Musolf}},\
  }\href {\doibase 10.1016/0370-2693(91)91916-J} {\bibfield  {journal}
  {\bibinfo  {journal} {Phys. Lett. B}\ }\textbf {\bibinfo {volume} {257}},\
  \bibinfo {pages} {409} (\bibinfo {year} {1991})}\BibitemShut {NoStop}%
\bibitem [{\citenamefont {Martinelli}\ \emph {et~al.}(1995)\citenamefont
  {Martinelli}, \citenamefont {Pittori}, \citenamefont {Sachrajda},
  \citenamefont {Testa},\ and\ \citenamefont {Vladikas}}]{Martinelli:1994ty}%
  \BibitemOpen
  \bibfield  {author} {\bibinfo {author} {\bibfnamefont {G.}~\bibnamefont
  {Martinelli}}, \bibinfo {author} {\bibfnamefont {C.}~\bibnamefont {Pittori}},
  \bibinfo {author} {\bibfnamefont {C.~T.}\ \bibnamefont {Sachrajda}}, \bibinfo
  {author} {\bibfnamefont {M.}~\bibnamefont {Testa}}, \ and\ \bibinfo {author}
  {\bibfnamefont {A.}~\bibnamefont {Vladikas}},\ }\href {\doibase
  10.1016/0550-3213(95)00126-D} {\bibfield  {journal} {\bibinfo  {journal}
  {Nucl. Phys.}\ }\textbf {\bibinfo {volume} {B445}},\ \bibinfo {pages} {81}
  (\bibinfo {year} {1995})},\ \Eprint {http://arxiv.org/abs/hep-lat/9411010}
  {arXiv:hep-lat/9411010 [hep-lat]} \BibitemShut {NoStop}%
\bibitem [{\citenamefont {Aoki}\ \emph {et~al.}(2008)\citenamefont {Aoki} \emph
  {et~al.}}]{Aoki:2007xm}%
  \BibitemOpen
  \bibfield  {author} {\bibinfo {author} {\bibfnamefont {Y.}~\bibnamefont
  {Aoki}} \emph {et~al.},\ }\href {\doibase 10.1103/PhysRevD.78.054510}
  {\bibfield  {journal} {\bibinfo  {journal} {Phys. Rev.}\ }\textbf {\bibinfo
  {volume} {D78}},\ \bibinfo {pages} {054510} (\bibinfo {year} {2008})},\
  \Eprint {http://arxiv.org/abs/0712.1061} {arXiv:0712.1061 [hep-lat]}
  \BibitemShut {NoStop}%
\bibitem [{\citenamefont {Stewart}\ and\ \citenamefont
  {Zhao}(2018)}]{Stewart:2017tvs}%
  \BibitemOpen
  \bibfield  {author} {\bibinfo {author} {\bibfnamefont {I.~W.}\ \bibnamefont
  {Stewart}}\ and\ \bibinfo {author} {\bibfnamefont {Y.}~\bibnamefont {Zhao}},\
  }\href {\doibase 10.1103/PhysRevD.97.054512} {\bibfield  {journal} {\bibinfo
  {journal} {Phys. Rev. D}\ }\textbf {\bibinfo {volume} {97}},\ \bibinfo
  {pages} {054512} (\bibinfo {year} {2018})},\ \Eprint
  {http://arxiv.org/abs/1709.04933} {arXiv:1709.04933 [hep-ph]} \BibitemShut
  {NoStop}%
\bibitem [{\citenamefont {Chen}\ \emph
  {et~al.}(2020{\natexlab{b}})\citenamefont {Chen}, \citenamefont {Wang},\ and\
  \citenamefont {Zhu}}]{Chen:2020ody}%
  \BibitemOpen
  \bibfield  {author} {\bibinfo {author} {\bibfnamefont {L.-B.}\ \bibnamefont
  {Chen}}, \bibinfo {author} {\bibfnamefont {W.}~\bibnamefont {Wang}}, \ and\
  \bibinfo {author} {\bibfnamefont {R.}~\bibnamefont {Zhu}},\ }\href@noop {} {\
   (\bibinfo {year} {2020}{\natexlab{b}})},\ \Eprint
  {http://arxiv.org/abs/2006.14825} {arXiv:2006.14825 [hep-ph]} \BibitemShut
  {NoStop}%
\bibitem [{\citenamefont {Musch}\ \emph {et~al.}(2011)\citenamefont {Musch},
  \citenamefont {Hagler}, \citenamefont {Negele},\ and\ \citenamefont
  {Schafer}}]{Musch:2010ka}%
  \BibitemOpen
  \bibfield  {author} {\bibinfo {author} {\bibfnamefont {B.~U.}\ \bibnamefont
  {Musch}}, \bibinfo {author} {\bibfnamefont {P.}~\bibnamefont {Hagler}},
  \bibinfo {author} {\bibfnamefont {J.~W.}\ \bibnamefont {Negele}}, \ and\
  \bibinfo {author} {\bibfnamefont {A.}~\bibnamefont {Schafer}},\ }\href
  {\doibase 10.1103/PhysRevD.83.094507} {\bibfield  {journal} {\bibinfo
  {journal} {Phys. Rev. D}\ }\textbf {\bibinfo {volume} {83}},\ \bibinfo
  {pages} {094507} (\bibinfo {year} {2011})},\ \Eprint
  {http://arxiv.org/abs/1011.1213} {arXiv:1011.1213 [hep-lat]} \BibitemShut
  {NoStop}%
\bibitem [{\citenamefont {Edwards}\ and\ \citenamefont
  {Joo}(2005)}]{Edwards:2004sx}%
  \BibitemOpen
  \bibfield  {author} {\bibinfo {author} {\bibfnamefont {R.~G.}\ \bibnamefont
  {Edwards}}\ and\ \bibinfo {author} {\bibfnamefont {B.}~\bibnamefont {Joo}}
  (\bibinfo {collaboration} {SciDAC, LHPC, UKQCD}),\ }\bibfield  {booktitle}
  {\emph {\bibinfo {booktitle} {{Lattice field theory. Proceedings, 22nd
  International Symposium, Lattice 2004, Batavia, USA, June 21-26, 2004}}},\
  }\href {\doibase 10.1016/j.nuclphysbps.2004.11.254} {\bibfield  {journal}
  {\bibinfo  {journal} {Nucl. Phys. Proc. Suppl.}\ }\textbf {\bibinfo {volume}
  {140}},\ \bibinfo {pages} {832} (\bibinfo {year} {2005})},\ \bibinfo {note}
  {[,832(2004)]},\ \Eprint {http://arxiv.org/abs/hep-lat/0409003}
  {arXiv:hep-lat/0409003 [hep-lat]} \BibitemShut {NoStop}%
\bibitem [{\citenamefont {Clark}\ \emph {et~al.}(2010)\citenamefont {Clark},
  \citenamefont {Babich}, \citenamefont {Barros}, \citenamefont {Brower},\ and\
  \citenamefont {Rebbi}}]{Clark:2009wm}%
  \BibitemOpen
  \bibfield  {author} {\bibinfo {author} {\bibfnamefont {M.~A.}\ \bibnamefont
  {Clark}}, \bibinfo {author} {\bibfnamefont {R.}~\bibnamefont {Babich}},
  \bibinfo {author} {\bibfnamefont {K.}~\bibnamefont {Barros}}, \bibinfo
  {author} {\bibfnamefont {R.~C.}\ \bibnamefont {Brower}}, \ and\ \bibinfo
  {author} {\bibfnamefont {C.}~\bibnamefont {Rebbi}},\ }\href {\doibase
  10.1016/j.cpc.2010.05.002} {\bibfield  {journal} {\bibinfo  {journal}
  {Comput. Phys. Commun.}\ }\textbf {\bibinfo {volume} {181}},\ \bibinfo
  {pages} {1517} (\bibinfo {year} {2010})},\ \Eprint
  {http://arxiv.org/abs/0911.3191} {arXiv:0911.3191 [hep-lat]} \BibitemShut
  {NoStop}%
\bibitem [{\citenamefont {Babich}\ \emph {et~al.}(2011)\citenamefont {Babich},
  \citenamefont {Clark}, \citenamefont {Joo}, \citenamefont {Shi},
  \citenamefont {Brower},\ and\ \citenamefont {Gottlieb}}]{Babich:2011np}%
  \BibitemOpen
  \bibfield  {author} {\bibinfo {author} {\bibfnamefont {R.}~\bibnamefont
  {Babich}}, \bibinfo {author} {\bibfnamefont {M.~A.}\ \bibnamefont {Clark}},
  \bibinfo {author} {\bibfnamefont {B.}~\bibnamefont {Joo}}, \bibinfo {author}
  {\bibfnamefont {G.}~\bibnamefont {Shi}}, \bibinfo {author} {\bibfnamefont
  {R.~C.}\ \bibnamefont {Brower}}, \ and\ \bibinfo {author} {\bibfnamefont
  {S.}~\bibnamefont {Gottlieb}},\ }in\ \href {\doibase 10.1145/2063384.2063478}
  {\emph {\bibinfo {booktitle} {{SC11 International Conference for High
  Performance Computing, Networking, Storage and Analysis Seattle, Washington,
  November 12-18, 2011}}}}\ (\bibinfo {year} {2011})\ \Eprint
  {http://arxiv.org/abs/1109.2935} {arXiv:1109.2935 [hep-lat]} \BibitemShut
  {NoStop}%
\bibitem [{\citenamefont {Clark}\ \emph {et~al.}(2016)\citenamefont {Clark},
  \citenamefont {Jo}, \citenamefont {Strelchenko}, \citenamefont {Cheng},
  \citenamefont {Gambhir},\ and\ \citenamefont {Brower}}]{Clark:2016rdz}%
  \BibitemOpen
  \bibfield  {author} {\bibinfo {author} {\bibfnamefont {M.~A.}\ \bibnamefont
  {Clark}}, \bibinfo {author} {\bibfnamefont {B.}~\bibnamefont {Jo}}, \bibinfo
  {author} {\bibfnamefont {A.}~\bibnamefont {Strelchenko}}, \bibinfo {author}
  {\bibfnamefont {M.}~\bibnamefont {Cheng}}, \bibinfo {author} {\bibfnamefont
  {A.}~\bibnamefont {Gambhir}}, \ and\ \bibinfo {author} {\bibfnamefont
  {R.}~\bibnamefont {Brower}},\ }\href@noop {} {\  (\bibinfo {year} {2016})},\
  \Eprint {http://arxiv.org/abs/1612.07873} {arXiv:1612.07873 [hep-lat]}
  \BibitemShut {NoStop}%
\bibitem [{\citenamefont {Bi}\ \emph {et~al.}(2020)\citenamefont {Bi},
  \citenamefont {Xiao}, \citenamefont {Gong}, \citenamefont {Guo},
  \citenamefont {Sun}, \citenamefont {Xu},\ and\ \citenamefont
  {Yang}}]{Bi:2020wpt}%
  \BibitemOpen
  \bibfield  {author} {\bibinfo {author} {\bibfnamefont {Y.-J.}\ \bibnamefont
  {Bi}}, \bibinfo {author} {\bibfnamefont {Y.}~\bibnamefont {Xiao}}, \bibinfo
  {author} {\bibfnamefont {M.}~\bibnamefont {Gong}}, \bibinfo {author}
  {\bibfnamefont {W.-Y.}\ \bibnamefont {Guo}}, \bibinfo {author} {\bibfnamefont
  {P.}~\bibnamefont {Sun}}, \bibinfo {author} {\bibfnamefont {S.}~\bibnamefont
  {Xu}}, \ and\ \bibinfo {author} {\bibfnamefont {Y.-B.}\ \bibnamefont
  {Yang}},\ }\bibfield  {booktitle} {\emph {\bibinfo {booktitle} {{Proceedings,
  37th International Symposium on Lattice Field Theory (Lattice 2019): Wuhan,
  China, June 16-22 2019}}},\ }\href {\doibase 10.22323/1.363.0286} {\bibfield
  {journal} {\bibinfo  {journal} {PoS}\ }\textbf {\bibinfo {volume}
  {LATTICE2019}},\ \bibinfo {pages} {286} (\bibinfo {year} {2020})},\ \Eprint
  {http://arxiv.org/abs/2001.05706} {arXiv:2001.05706 [hep-lat]} \BibitemShut
  {NoStop}%
\bibitem [{\citenamefont {Follana}\ \emph {et~al.}(2007)\citenamefont
  {Follana}, \citenamefont {Mason}, \citenamefont {Davies}, \citenamefont
  {Hornbostel}, \citenamefont {Lepage}, \citenamefont {Shigemitsu},
  \citenamefont {Trottier},\ and\ \citenamefont {Wong}}]{Follana:2006rc}%
  \BibitemOpen
  \bibfield  {author} {\bibinfo {author} {\bibfnamefont {E.}~\bibnamefont
  {Follana}}, \bibinfo {author} {\bibfnamefont {Q.}~\bibnamefont {Mason}},
  \bibinfo {author} {\bibfnamefont {C.}~\bibnamefont {Davies}}, \bibinfo
  {author} {\bibfnamefont {K.}~\bibnamefont {Hornbostel}}, \bibinfo {author}
  {\bibfnamefont {G.~P.}\ \bibnamefont {Lepage}}, \bibinfo {author}
  {\bibfnamefont {J.}~\bibnamefont {Shigemitsu}}, \bibinfo {author}
  {\bibfnamefont {H.}~\bibnamefont {Trottier}}, \ and\ \bibinfo {author}
  {\bibfnamefont {K.}~\bibnamefont {Wong}} (\bibinfo {collaboration} {HPQCD,
  UKQCD}),\ }\href {\doibase 10.1103/PhysRevD.75.054502} {\bibfield  {journal}
  {\bibinfo  {journal} {Phys. Rev.}\ }\textbf {\bibinfo {volume} {D75}},\
  \bibinfo {pages} {054502} (\bibinfo {year} {2007})},\ \Eprint
  {http://arxiv.org/abs/hep-lat/0610092} {arXiv:hep-lat/0610092 [hep-lat]}
  \BibitemShut {NoStop}%
\bibitem [{\citenamefont {Miller}\ \emph {et~al.}(2020)\citenamefont {Miller}
  \emph {et~al.}}]{Miller:2020evg}%
  \BibitemOpen
  \bibfield  {author} {\bibinfo {author} {\bibfnamefont {N.}~\bibnamefont
  {Miller}} \emph {et~al.},\ }\href@noop {} {\  (\bibinfo {year} {2020})},\
  \Eprint {http://arxiv.org/abs/2011.12166} {arXiv:2011.12166 [hep-lat]}
  \BibitemShut {NoStop}%
\bibitem [{\citenamefont {Hasenfratz}\ and\ \citenamefont
  {Knechtli}(2001)}]{Hasenfratz:2001hp}%
  \BibitemOpen
  \bibfield  {author} {\bibinfo {author} {\bibfnamefont {A.}~\bibnamefont
  {Hasenfratz}}\ and\ \bibinfo {author} {\bibfnamefont {F.}~\bibnamefont
  {Knechtli}},\ }\href {\doibase 10.1103/PhysRevD.64.034504} {\bibfield
  {journal} {\bibinfo  {journal} {Phys. Rev.}\ }\textbf {\bibinfo {volume}
  {D64}},\ \bibinfo {pages} {034504} (\bibinfo {year} {2001})},\ \Eprint
  {http://arxiv.org/abs/hep-lat/0103029} {arXiv:hep-lat/0103029 [hep-lat]}
  \BibitemShut {NoStop}%
\bibitem [{\citenamefont {Chiu}(1999)}]{Chiu:1998eu}%
  \BibitemOpen
  \bibfield  {author} {\bibinfo {author} {\bibfnamefont {T.-W.}\ \bibnamefont
  {Chiu}},\ }\href {\doibase 10.1103/PhysRevD.60.034503} {\bibfield  {journal}
  {\bibinfo  {journal} {Phys. Rev. D}\ }\textbf {\bibinfo {volume} {60}},\
  \bibinfo {pages} {034503} (\bibinfo {year} {1999})},\ \Eprint
  {http://arxiv.org/abs/hep-lat/9810052} {arXiv:hep-lat/9810052} \BibitemShut
  {NoStop}%
\bibitem [{\citenamefont {Liu}(2005)}]{Liu:2002qu}%
  \BibitemOpen
  \bibfield  {author} {\bibinfo {author} {\bibfnamefont {K.-F.}\ \bibnamefont
  {Liu}},\ }\href {\doibase 10.1142/S0217751X05022366} {\bibfield  {journal}
  {\bibinfo  {journal} {Int. J. Mod. Phys. A}\ }\textbf {\bibinfo {volume}
  {20}},\ \bibinfo {pages} {7241} (\bibinfo {year} {2005})},\ \Eprint
  {http://arxiv.org/abs/hep-lat/0206002} {arXiv:hep-lat/0206002} \BibitemShut
  {NoStop}%
\bibitem [{\citenamefont {Lepage}\ and\ \citenamefont
  {Mackenzie}(1993)}]{Lepage:1992xa}%
  \BibitemOpen
  \bibfield  {author} {\bibinfo {author} {\bibfnamefont {G.}~\bibnamefont
  {Lepage}}\ and\ \bibinfo {author} {\bibfnamefont {P.~B.}\ \bibnamefont
  {Mackenzie}},\ }\href {\doibase 10.1103/PhysRevD.48.2250} {\bibfield
  {journal} {\bibinfo  {journal} {Phys. Rev. D}\ }\textbf {\bibinfo {volume}
  {48}},\ \bibinfo {pages} {2250} (\bibinfo {year} {1993})},\ \Eprint
  {http://arxiv.org/abs/hep-lat/9209022} {arXiv:hep-lat/9209022} \BibitemShut
  {NoStop}%
\bibitem [{\citenamefont {Bruno}\ \emph {et~al.}(2015)\citenamefont {Bruno}
  \emph {et~al.}}]{Bruno:2014jqa}%
  \BibitemOpen
  \bibfield  {author} {\bibinfo {author} {\bibfnamefont {M.}~\bibnamefont
  {Bruno}} \emph {et~al.},\ }\href {\doibase 10.1007/JHEP02(2015)043}
  {\bibfield  {journal} {\bibinfo  {journal} {JHEP}\ }\textbf {\bibinfo
  {volume} {02}},\ \bibinfo {pages} {043} (\bibinfo {year} {2015})},\ \Eprint
  {http://arxiv.org/abs/1411.3982} {arXiv:1411.3982 [hep-lat]} \BibitemShut
  {NoStop}%
\end{thebibliography}

\end{document}